\documentclass[10pt,paper,aps,prab,twocolumn,amsmath,superscriptaddress,preprintnumbers,floatfix,nofootinbib,showpacs]{revtex4-2}
\usepackage{siunitx}
\usepackage{graphicx}
\usepackage[colorlinks]{hyperref}
\hypersetup{
	colorlinks=true, %set true if you want colored links
	linktoc=all,     %set to all if you want both sections aa^\daggernd subsections
	linkcolor=blue,  %choose some color if you want links to stand out
	citecolor=magenta,
	urlcolor=blue	
	}
\def\bfg #1{{\mbox{\boldmath $#1$}}}

\usepackage[dvipsnames]{xcolor}
\usepackage{tikz}
\usepackage{booktabs}
\usepackage{dcolumn}
\usepackage{mathtools}
\usepackage{braket}
\usepackage[utf8]{inputenc}
\usepackage{placeins}
\usepackage[T1]{fontenc}
\usepackage{natbib}
\usepackage{amssymb,latexsym}
\usepackage{wasysym}
\usepackage{soul}
\usepackage{xfrac}

\usepackage{algorithmic}

\newcommand{\matr}[1]{\mathbf{#1}}

\begin{document}
\preprint{Draft for PRAB \today}	
	
		\author{B. Gou}
	\affiliation{Institute of Modern Physics, Chinese Academy of Sciences, Lanzhou 730000, China}
	\date {\today}
	
	\author{V.P.\,Ladygin}
	\affiliation{Baldin-Veksler Laboratory of High Energy Physics, Joint Institute for Nuclear Research, 141980 Dubna, Russia}

\author{N.N.\,Nikolaev}
\affiliation{L.D. Landau Institute for Theoretical Physics, 142432 Chernogolovka, Russia}
	\affiliation{Moscow Institute for Physics and Technology, 141700 Dolgoprudny, 		Russia}
	\affiliation{Bogoliubov Laboratory of Theoretical Physics, Joint Institute for Nuclear Research, 141980 Dubna, Russia}
	
\author{F.\,Rathmann}
%\email[Corresponding author: ]{frathmann@bnl.gov}
\affiliation{ Brookhaven National Laboratory, Upton, NY 11973, USA}
%\affiliation{Institut f\"ur Kernphysik, Forschungszentrum J\"ulich, 52425 		J\"ulich, Germany}
	
	\author{A.J.\,Silenko}
	\affiliation{Bogoliubov Laboratory of Theoretical Physics, Joint Institute for Nuclear Research, 141980 Dubna, Russia}
\affiliation{Research Institute for Nuclear Problems, Belarusian State University,
Minsk 220030, Belarus}	

	\author{Y.N.\,Uzikov}
\affiliation{Dzhelepov Laboratory of Nuclear Problems, Joint Institute for Nuclear Research, 141980 Dubna, Russia}
	\affiliation{Moscow Institute for Physics and Technology, 141700 Dolgoprudny, 		Russia}
\affiliation{Dubna State University, 141980 Dubna, Russia}
\affiliation{M.V. Lomonosov Moscow State University, Faculty of Physics, 119991, Moscow, Russia}
\date {\today}
	
\title{Fresh Look at Polarized Deuterons at the Nuclotron/NICA \& HIAF and Beyond}
	
%%%%%%%%%%%%%%%%%%%%%%%%%%%%%%%%%%%%%%%%%%%%%%%%%%%%%%%%%%%%%%%%%%%%%%%%%%%%%%%%%%%%%
%%%%%%%%%%%%%%%%%%%%%%%%%%%%%%%%%%%%%%%%%%%%%%%%%%%%%%%%%%%%%%%%%%%%%%%%%%%%%%%%%%%%%
\begin{abstract}
	Being a spin-1 particle, the deuteron, when polarized, offers access to a wide range of interesting tensor observables. A quest for tensor asymmetries is at the heart of the spin physics with polarized deuterons at future colliders, and calls for frequent tensor polarization flips in storage rings or in deuterium targets. 
	
	Here we focus on the JEDI developed technique of the  Fourier analysis of the horizontal polarization oscillations under continuous radiofrequency driven spin flips. The so generated time-stamped oscillating helicity of stored particles paves the way to the nucleon/nucleus spin structure studies and to the search for the parity violation. Simultaneously a whole set of tensor spin asymmetries, including the off-diagonal ones needed for searches for the T-violation beyond the Standard Model, will be produced. 
	
	A crucial ingredient is a long polarization lifetime, and we report the first analytic treatment of decoherence of the tensor polarization under continuous radiofrequency spin flips. Our  principal conclusion is that  at low- and intermediate-energy accelerators the deuteron polarization lifetime could be in the ballpark of several hours, making the oscillating polarization approach viable at Nuclotron/NICA and HIAF. 
	
	We comment on a utility of the flattop spin-tensor dichroism via oscillations in the cross section of interaction of deuterons in the internal target as a monitor of the tensor polarization. Simultaneously, the dichroism effect can be used for measuring the spin precession frequences and as a comagnetometer to synchronize spin phases of colliding bunches in a collider. 
\end{abstract}

%	the interpretation of the experimental data on the novel pilot bunch approach to control the spin-resonance condition during the operation of the radiofrequency-driven Wien filter that is used as a spin rotator in the first direct deuteron electric dipole moment measurement at COSY. We emphasize the potential importance of the hitherto unexplored phase of the envelope of the horizontal polarization as an indicator of the stability of the RF-driven spin rotations in storage rings. The work presented here serves as a satellite publication to the work published concurrently on the proof of principle experiment about the so-called pilot bunch approach that was  developed to provide co-magnetometry for the deuteron electric dipole moment experiment at COSY.
%\end{abstract}
%%%%%%%%%%%%%%%%%%%%%%%%%%%%%%%%%%%%%%%%%%%%%%%%%%%%%%%%%%%%%%%%%%%%%%%%%%%%%%%%%%%%%%%%%%%%%%
%\pacs{13.40.Em, 11.30.Er, 29.20.Dh, 29.27.Hj}
%%%%%%%%%%%%%%%%%%%%%%%%%%%%%%%%%%%%%%%%%%%%%%%%%%%%%%%%%%%%%%%%%%%%%%%%%%%%%%%%%%%%%%%%%%%%%%

\maketitle

%\tableofcontents
	
\section{Introduction} 
\label{sec:introduction}
%Controlled spin rotations, notably the spin flips (SF), are imperative for particle and nuclear physics experiments that involve polarized particles, for extensive reviews, see\,\cite{SYLee, Yokoya}). 
%
Spin physics has been one of specialties of the Joint Institute of Nuclear Research since early 1950's \cite{Nagaytsev-History,SpinDubna2023,Teryaev-NSR2026}. A unique potential of the Nuclotron/NICA facility at JINR is its capability of storage and acceleration of polarized deuterons, this 
paves the way to a study for the first time of the tensor polarization observables in the collider mode at the NICA and in the fixed-target mode at the Nuclotron
\cite{Arbuzov:NICA-SPD,fimushkinNSR2026,filatov2026polarization,zelenski2026polarimetry}.
Acceleration of
polarized deuterons is contemplated also at the HIAF facility of IMP, CAS \cite{IMP_SPIS, Lanzhou-Tensor, Gou_SPIN2025}.

 Many aspects of future polarized deuteron physics at the Nuclotron/NICA facility were given a thorough treatment in the very recent reviews \cite{BaryshevskyNSR2026,fimushkinNSR2026,Teryaev-NSR2026}. We also mention a  recent development of the potentially very powerful approach to the deuteron spin rotations in the so-called spin transparency regime in the vicinity of integer spin resonances 
 \cite{FilatovTransparent2020,Filatov2020EPJ,FilatovNavigator2022,FilatovOrbitSteer2023,FilatovImperfections2024,filatov2026polarization}. 	Possibilities of an upgrade of the Nuclotron/NICA facility towards searches for physics beyond the Standard Model, notably searches for axions and the electric dipole moments of protons and deuterons, are being actively explored \cite{Senichev:2022ide,Melnikov:2024gcc,Senichev:2025jqf,Melnikov:2025ytd}.

 	 A focus of this review is on the novel approach to tensor observables based on the radiofrequency (RF) driven continuous spin flips, and its potential in search for a spin physics beyond the Standard Model. This particular aspect derives from a legacy of more than a decade of experimentation with a deuteron oscillating polarization by the JEDI collaboration at the COSY synchrotron of Forshungszentrun Juelich, as exposed in Sec. \ref{sec:JEDI}. Crux of the matter is that in conjunction with the idle spin precession  in the guiding field of a storage ring, the RF-driven spin flips would allow a full control of a time-stamped 3D orientation of stored spins. An extension of the JEDI developed Fourier analysis technique would allow  to measure in one go the whole set of vector and tensor spin asymmetries which have distinct Fourier footprints \cite{NikolaevPrecessingD}.  Notable examples of the utility of this technique are the  oscillating longitudinal vector polarization giving an access to the Standard Model predicted time-reversal conserving (TC) parity violation (PV) \cite{Koop2020parity,Abramov_2021} and oscillating tensor polarizations, including the  off-diagonal ones, giving an access to the
 vector-tensor signal of T-violation (TV) in the {\it pd} and {\it dd}    collisions at NICA and in the fixed target
  {\it pd} interactions at Nuclotron  and HIAF \cite{NikolaevPrecessingD,Temerbayev:2015foa,Uzikov:2015aua,Uzikov:2016lsc,Uzikov_pD_NICA,Platonova_T-odd-DD,Gou_SPIN2025}.

 A current interest in tensor observables spans from searches for the long sought  millistrong CP-violation beyond the Standard Model \cite{OkunMillistrong,PrentkiMillistrong,LeeMillistrong,UFNreview} in storage ring experiments \cite{NikolaevPrecessingD,TIVOLI:STORI11,LenisaTIVOLI,Uzikov_pD_NICA,Platonova_T-odd-DD}, to the tensor structure functions at future colliders \cite{Arbuzov:NICA-SPD,TensorDeuteron-eIC,eIC_Yellow_Report,Nikolaev:TensorSF}, to the shear forces in tensor polarized deuterons  \cite{Teryaev:ShearForces}, to tensor polarized deuteron breakup in {\it ed} interactions \cite{Teryaev:DeuteronSumRules,CEBAF_TensorD}, to the spin dichroism \cite{Baryshevsky_2008,BaryshevskyNSR2026}
	studies at Nuclotron in Dubna, to tensor polarizability effects in studies of fundamental symmetries  
	\cite{Baryshevsky_2008,SilenkoSpin1Proca},
	%\cite{Вaryshevsky_2008,SilenkoSpin1},
 to plans of T-violation  beyond the Standard Model experiments at HIAF \cite{Gou_SPIN2025},  
 	to the tensor polarization effects in the inelastic deuteron forward
 		scattering, $(d, d^\prime)X$, and the excitation of the baryonic resonances \cite{DDAlpha1995,DDSphere2000},
 and to a  long standing open issues with the short-range tensor structure of the deuteron 
 		\cite{Dubna_TensorD,Dubna_TensorD1,Anomalon1994,Anomalon1995,DubnaTensorD,Uzikov:1998qk,DubnaAyy2002,DubnaAyy2004,DubnaAyy2005,Imambekov:1990ru,Uzikov:2001uc,Teryaev:DeuteronSumRules}.

A discussion of the early results on  tensor asymmetries in {\it pd}  elastic scattering at intermediate energies  \cite{Arvieux1984,Ghazikhanian1991} is found in \cite{Uzikov:1998qk,Platonova_Kukulin}. Experimental studies of interactions of accelerated high-energy tensor polarized deuterons, as well as the first observation of the deuteron spin dichroism at high energies  at JINR Dubna, were reviewed in \cite{fimushkinNSR2026} and in \cite{azhgirey2008tensor,azhgirey2010tensor}, respectively, pioneering experiments with the tensor polarized deuterons at COSY storage ring were reported in \cite{Morozov:Tensor,ChiladzeTensorD,Mchedlishvili:2013bja,Uzikov:2015vta,Mchedlishvili:2018uur}.
 
 However, a storage of polarized deuterons \cite{FilatovTransparent2020,FilatovNavigator2022,filatov2026polarization} and the tensor polarimetry  \cite{fimushkinNSR2026,zelenski2026polarimetry,Uzikov_TensorPolarimetry,NikolaevPrecessingD} at high energy colliders are in as yet formative stage. 

High precision measurements of small spin asymmetries during several hour cycles in the storage ring and collider experiments call for a very slow decoherence of the beam polarization. This decoherence stems from the phase-space spread of the idle-precession and of the spin-flip frequencies over the ensemble of beam particles.  A decohernce of the idly precessing deuteron spins by betatron oscillations \cite{KoopShatunov} has been studied by JEDI collaboration at COSY, culminating in a substabtual enhancement of the spin decoherence time to 1000 s or longer \cite{SCT1000sJEDI}, for a review see \cite{AbusaifCYR}. The synchrotron oscillation driven decoherence of vector polarization under continuous spin flips was treated in \cite{SO-Decoherence-2025}, the experimental situation is not conclusive yet \cite{PilotBunch-2025,JEDIaxion}. A decoherence of the tensor polarization in storage rings remains an as yet open issue.

A further presentation is organized as follows. In Sec. \ref{sec:JEDI} we summarize a legacy of the JEDI collaboration in the physics of oscillating polarization in storage rings. In Sec. \ref{sec:vector} we describe operation of the RF spin flipper and the corresponding generation of oscillating helicities of stored beams. A new subject, presented here for the first time in subsection \ref{sec:matching}, is a synchronization of helicities of the two colliding bunches and a technique of switching of relative helicities. Our matrix approach to RF driven oscillations of the tensor  polarization is presented  in Sec. \ref{sec:tensor}. A subject of Sec. \ref{sec:decoherence} is a synchrotron-oscillation dominated  decoherence of the deuteron vector and tensor polarizations. Here in subsection \ref{sec:Tensor decoherence} we report for the first time new results for a connection between decoherence of the vector and tensor polarizations. In Sec. \ref{sec:initial}  we present evolution of the tensor polarization for the initial conditions typical of magnetic storage rings. In Sec. \ref{sec:NICAdepolarization} we argue that short bunches at NICA entail a strong suppression of the synchrotron oscillation driven depolarization -- this is a new result. Parity and time reversal propeties of single and double polarization components of the {\it pd}  total cross section are treated in Sec. \ref{sec:pd}. A focus of Sec. \ref{sec:Tin pd} is a dynamical multiple scattering theory evaluation of the P-even and T-odd vector-tensor asymmetry in the {\it pd}  total cross section. New ideas on applications ot the oscillating tensor deformation of deuterons (dichroism) to the tensor polarimetry and the related single-species comagnetometry are discussed in Sec. \ref{sec:dichroism}. Alongside with magnetic and electric quadrupole moments, spin-1 deuterons are  endowed with four more fundamental characteristics: the electric and magnetic scalar and tensor polarizabilities, and in Sec. \ref{sec:Polarizability} we comment on a possibility  of generation of the vector polarization from the tensor one, which depends on the dynamics of the tensor and vector polarizations in the specific magnetic storage ring. In Summary and Conclusions we comment on a modest infrastructure required to implement the oscillating polarization regime.

	\section{Oscillating polarization: a legacy of JEDI at COSY}
	\label{sec:JEDI}

	Before the start of the spin program at the NICA collider, the COSY Synchrotron in J\"ulich has been the only machine in the world fully equipped for the precision polarization experiments \cite{MaierCOSY,FeldenCOSY,WilkinCOSYlegacy}. In a series of experiments at COSY, the JEDI collaboration obtained the record-breaking results in the physics of oscillating polarizations  %{\footnote{The %early ideas on the utility of oscillating polarization were published by the Dubna group in 2002 \cite{Sitnik}, but this paper was overlooked by the community.}}
	\begin{itemize}
		\item {A demonstration of measuring the deuteron beam polarization to an accuracy of $10^{-6}$ \cite{BrantjesPolarimetry}.}
		\item {A technique has been developed for measuring the deuteron spin tune with an accuracy of $10^{-10}$ \cite{JEDIspintune2}.}
		\item {A feedback technique for a continuous control of the spin idle precession phase to an accuracy of 0.15 rad has been developed \cite{JEDIphase,PhaseLock}.}  
		\item {The coherence time exceeding 1000~s for deuteron spins idly precessing in the horizontal plane has been achieved \cite{SCT1000sJEDI,SCTchromaticityJEDI}. The previous record result of 0.5~s for electrons and positrons was obtained at the Budker INP \cite{VassermanSCT}.}
		\item {A new method for spin tune mapping was developed, allowing for the first time to evaluate experimentally the integral systematic impact on the spin precession of the unwanted magnetic fields due to the imperfection of the magnetic system in the COSY plane \cite{SpinTuneMapping}.}
		\item {The beam based alignment  to ensure a passage of the orbit  through the center of the quadrupole magnets has been realized \cite{BBAJEDI}.}
		\item {The RF Wien filter as a novel spin rotator has been proposed, commissioned and  was in operation at COSY  \cite{SlimWFDesign,SlimWFCircuit,SlimComissioning}.}
		\item {Collective betatron oscillations close to the quantum unceratinty limit have been detected \cite{SlimWFoscillations,QuantumOscillations}.}
		\item {The first RF scan search for the axion driven spin rotations has been performed \cite{JEDIaxion}.}
		\item The first single-species comagnetometry for the RF driven spin flips in storage rings has been developed \cite{PilotBunch-2025,JEDI_Feedback-2025}. 
			\item {First experimental limit on the permanent electric dipole moment of the deuteron was published \cite{EDM-JEDI}.}
	\end{itemize}
This is a very incomplete list of important JEDI results that were recognized as a sufficient foundation for  preparation of the PTR project \cite{AbusaifCYR}.  Incidentally, the early ideas on the utility of oscillating polarization were published by the Dubna group in 2002 \cite{Sitnik}, but were overlooked by the community.

\section{Evolution of the vector polarization and oscillating helicities}
\label{sec:vector}
\subsection{Stroboscopic spin evolution}
\label{sec:Stroboscopic}

 We consider a storage ring with the RF solenoid as a spin-flipper  following Ref. \cite{SO-Decoherence-2025}. 
Frequences of spin idle precession $f_\text{s}$, synchrotron oscillations  $f_\text{sy}$, spin-flip $f_\text{sf}$ and inverse spin decoherence time $\tau_\text{SCT}^{-1}$ satisfy a strong hierarchy,   
\begin{equation}
	f_\text{s} \gg f_\text{sy} \gg f_\text{sf} \gg \tau_\text{SCT}^{-1}\, , \label{hierarchy}
	\end{equation}
entailing the Bogoliubov-Krylov averaging approach.

  We use a customary reference frame with ${\bf e}_\text{y}$ normal to the ring plane, radial ${\bf e}_\text{x}$ and ${\bf e}_\text{z}$ tangential to the orbit. One turn stroboscopic  evolution of a spin consists of the idle precession around the vertical  guiding field in conjunction with a kick  $\chi_\text{RF}(t)=\chi_\text{sf}\cos(2\pi f_\text{sol}t)$ in the longitudinal RF field of the solenoidal spin flipper or the radial magnetic field of the Wien filter \cite{SlimWFDesign,SlimComissioning}. Here we focus on the RF solenoid option.
  
  We start with the reference particle with zero synchrotron amplitude. The spin resonance condition is $f_\text{sol} = f_\text{s} + Kf_\text{rev}$, where $f_\text{rev}$ is the beam revolution frequency, and the side band $K=0,\pm 1,\pm 2\, \ldots .$ 
The idle precession one-turn spin transfer unitary matrix equals
\begin{equation}
	\matr{R}_\text{s}(\theta_\text{s})=
	\begin{pmatrix} \cos\theta_\text{s} & 0 &  \sin\theta_\text{s} \\
		0 &  1 & 0 \\
		-\sin\theta_\text{s}  & 0 &  \cos\theta_\text{s}
		\label{IdleMatrix}
	\end{pmatrix} \, ,
\end{equation}
where  $\theta_\text{s} = 2\pi \nu_\text{s}$ is a spin precession angle with respect to the particles momentum,  and the spin tune $\nu_\text{s} = f_\text{s}/f_\text{rev}$. In an ideal storage ring $\nu_\text{s}= G\gamma$, where $G$ is the anomalous $g$-factor and $\gamma$ the beam relativistic factor. 

 In the interaction representation, ${\bf S} (n)=\matr{R}_\text{s}(n\theta_\text{s}){\bf p}(n)$   with a  spin envelope ${\bf  p}(n)$, where n is the turn number. In view of the hierarchy (\ref{hierarchy}), a sequential  Bogoliubov-Krylov averaging over rapid spin precessions yields \cite{NikolaevPrecessingD,SO-Decoherence-2025} the adiabatic spin envelope rotation ${\bf p}(n)=\matr{E}(x) {\bf p}(0)$ with the unitary 
\begin{equation}
	%  \begin{equation}
		\matr{E}(x) = 
		\begin{pmatrix}
			\cos x  & \sin x & 0 \\
			-\sin x & \cos x & 0 \\
			0 & 0 & 1
		\end{pmatrix}\, , \label{Envelope}
		%  \end{equation}
\end{equation}
where the spin-flip phase $x=2\pi \nu_\text{sf} n$ and $\nu_\text{sf}=\chi_\text{sf}/4\pi$ is the spin-flip tune (the resonance strength). The longitudinal ${\text p}_\text{z}$ is a spectator w.r.t. the spin flip. A design of the RF flipper is found, for instance, in \cite{Koop2020parity},  typical spin flip frequencies in the JEDI experiments were of the order of 0.1 Hz \cite{PilotBunch-2025,JEDIaxion}.

Stored beams have an average polarization ${\text S}_\text{y}(0)$ along the guiding field of the storage ring, and the in plane polarizations average to zero: ${\text S}_\text{x}(0)={\text S}_\text{z}(0)=0$. Subject to this initial 
condition,
\begin{equation}
	\begin{split}
		{\text S}_\text{y}(n)& = {\text S}_\text{y}(0)\cos x\, , \\
		{\text S}_\text{x}(n) &= {\text S}_\text{y}(0)\cos \theta \sin x\, , \\
			{\text S}_\text{z}(n) &= -{\text S}_\text{y}(0)\sin \theta \sin x\, ,\\  \label{VectorEvolution}
	\end{split}
		\end{equation}
where $\theta = n \theta_s	$, the individual spin evolution has an obvious preservaton of the spin magnitude.

\subsection{Fine tuning the oscillating beam helicities}
\label{sec:helicity}

\subsubsection{Single-spin asymmetries in the fixed target mode}

Eq. (\ref{VectorEvolution}) shows how the continuously oscillating longitudinal polarization ${\text S}_\text{z}(n)$ is generated from the initial vertical polarization of the stored beam. The phases $\theta$ and $x$  for each particle  are tightly controlled by the continuous time  stamp, thus making a spin 3D orientation well under control. For instance, in the atomic cell internal target, each  stored bunch will cross the cell with the turn-by-turn changing but known helicity. In the unpolarized fixed target mode, oscillating single-spin asymmetry amounts to a signal of the parity violation \cite{milstein2020parity,NikolaevPrecessingD},ch can be extracted by the Fourier analysis. A discussion of the one-turn extraction of bunches with time-stamped helicity onto external target is found in \cite{Koop2020parity,Koop2021tests}. 

\subsubsection{Double-spin asymmetries in the collider mode}
\label{sec:matching}

In the collider mode, beam revolution frequencies are kept synchronous, which synchronizes the spin precession as well. Still, two rings may have different imperfections \cite{SpinTuneMapping}, and the further control of the spin precession is a must. To furnish the JEDI  approach to the single-species comagnetometry one has to  upgrade the CNI polarimeters by  adding a capability to detect the up-down asymmetry \cite{JEDIspintune2,PilotBunch-2025,JEDI_Feedback-2025}. A novel approach to be presented in Sec. \ref{sec:DichroismComagnetometer} is based on the spin dichroism specific of spin-1 deuterons. Under these conditions, the longitudinal polarizations	${\text S}_\text{z}(n)$ of all colliding pairs, each consisting of a colliding  clockwise (CW) bunch and anticlockwise (ACW) bunch,  will be kept either parallel or antiparallel.  In order to isolate the double-spin asymmetries switching a sign of the product of two helicities, one has to execute perodic  slips of the spin precession phase $\theta$ of all the bunches in one of the rings by $\pi$. 

Such a phase slip, without perturbing a beam orbit, can be accomplished via shifting the spin tune by periodically switching on/off a static Wien filter operated with the vertical magnetic field in the selected ring \cite{NikolaevPTR,Silenko:2021qgc}. In order to preserve the spin resonance condition, the frequency of the RF power supply of the Wien filter has to be varied in tandem. Such a scheme is custom tailored, for instance,  for measuring gluon helicity distributions in double-polarized proton-proton and deuteron-deuteron collisions.

\section{Evolution of the tensor polarization}
	\label{sec:tensor}
	
	The deuteron spin dynamics of relevance to magnetic storage rings of interest in this review, is dominated by interaction of the deuteron magnetic moment with the static guiding magnetic field and RF magnetic field of the solenoid spin flipper  \cite{SilenkoTensor2015,NikolaevPrecessingD}.   Indeed,  	
	the tensor polarization enters the spin Hamiltonian only via an interaction of the deuteron electric quadrupole moment $Q_\text{d}\approx 0.3\ \text{fm}^2$ with the electric field gradient,  and via  tensor polarizabilities, to be discussed to more details in Sec. \ref{sec:Polarizability}. In the former case the small suppression parameter is of the order of 
	$Q_\text{d}/(\mu_\text{N} L) \sim {\text fm}/L  < 10^{-13}$, where $\mu_\text{N}$ is the nuclear magneton and L is a  range of variation of the electric field on the orbit. According to Baryshevsky, the tensor polarizability driven instantaneous spin rotation frequency in the 1 T static magnetic field is slower that $10^{-5}\ \text{Hz}$ \cite{Baryshevsky_2008}.

The envelope of the traceless symmetric tensor-polarization matrix,
\begin{equation}
	\matr{T}_\text{ab} = \frac{1}{2}  ({\text S}_\text{a}{\text S}_\text{b} + {\text S}_\text{b}{\text S}_\text{a} - \frac{2}{3} \delta_\text{ab} {\bf S} ^2)\, ,\label{Tensor}
\end{equation}  
is defined as $\matr{t}(n) = \matr{R}_\text{s}^{-1}(\theta)\matr{T}(n) \matr{R}_\text{s}(\theta)$ and evolves as
\begin{equation}
		\matr{t}(n) = \matr{E}(x)\matr{t}(0)\matr{E}^\text{T}(x)\,.  \label{Tensor-from-Vector}
\end{equation}
The initial condition is $\matr{t}(0)=\matr{T}(0)$. 

The evolution of the five independent tensor polarization envelopes, 
\begin{widetext}
\begin{equation}
	\begin{split}
			&\frac{1}{2}(\text{t}_\text{xx}(n)-\text{t}_\text{yy }(n) )= 
			\frac{1}{2}(\text{t}_\text{xx}(0)-\text{t}_\text{yy}(0) )\cos 2x + \text{t}_\text{xy}(0)\sin 2x\,,\\
			&\text{t}_\text{xy }(n) =  -\frac{1}{2}(\text{t}_\text{xx}(0)-\text{t}_\text{yy}(0) )\sin 2x + \text{t}_\text{xy}(0)\cos 2x\,,\\
					    	&\text{t}_\text{xz}(n) = \text{t}_\text{xz}(0)\cos x + \text{t}_\text{yz}(0)\sin x\,,\\
			&\text{t}_\text{yz}(n) = -\text{t}_\text{xz}(0)\sin x +\text{t}_\text{yz}(0)\cos x\,,\\
			&\text{t}_\text{zz}(n) = \text{t}_\text{zz}(0)\,,   \label{TensorEnvelopeExpansion}
		\end{split}
	\end{equation}
	\end{widetext}
 nicely exhibits a block structure of the tensor envelope evolution matrix \cite{NikolaevPrecessingD}: 
\begin{itemize}
	\item {
${\text t}_\text{zz}(n)$ is a spectator component immune to the RF-driven spin-flip;}
\item  ${\text t}_\text{yz}(n)$ and  ${\text t}_\text{xz}(0)$ form a two-dimensional vector ${\bf V}   _1$ which exhibits spin-flip like rotation with frequency $f_\text{sf}=\nu_\text{sf}f_\text{rev}$ in the XY plane;
\item { Simlilar spin-flip behavior holds for  $\frac{1}{2}({\text t}_\text{xx}(n)-{\text t}_\text{yy}(n))$ and ${\text t}_\text{xy}(n)$, which form a vector ${\bf V}   _2$ rotating  with twice the spin-flip frequency $2f_\text{sf}$ in  the XY plane }. 
\end{itemize}	

	\section{Decoherence of the vector and tensor polarizations}
	\label{sec:decoherence}
	
Unitary rotations of individual spins preserve the magnitide of the polarization. Depolarization  stems from a spread of spin rotations in the ensemble of beam particles.  Betatron oscillations affect the idle precession via the spin tune spread, which can be corrected for by fine-tuning sextupole families to minmize the chromaticity  \cite{KoopShatunov,SCT1000sJEDI,SCTchromaticityJEDI,Chromaticity_Kolokolchikov:2023}.  Still another source of depolarization is the synchrotron oscillation, which leads to a spread of frequencies of the RF-driven spin-flips \cite{SO-Decoherence-2025}.

\subsection{Decoherence of the vector polarization}
\label{sec:VectorDecoherence}

Noteworthy is a very nice result  from the first observation of the polarization loss in a long sequence of  RF-driven
proton spin flips at a beam kinetic energy of \SI{49.3}{MeV} at COSY \cite{MachineDevelopmentSF2}. With the  RF  spin flipper turned off, the vertical polarization was found to have a lifetime of $\tau_\text{p} = (2.7 \pm 0.8)\cdot\SI{e5}{s}$. 
However, when  99 successive spin flips driven by an RF  solenoid were performed within \SI{300}{s}, a clear signal of depolarization was observed. Assuming an exponential attenuation of polarization, the average spin flip efficiency was found to be $\epsilon_\text{flip} = 0.9872\pm 0.0001$\,\cite{MachineDevelopmentSF2}, corresponding to a lifetime of the continuously flipping spin of $\tau_\text{flip} = \SI{240}{s}$.  Such a change of polarization lifetime by 3 orders in magnitude is a clear indication of a close link between depolarization and spin-flip dynamics.

A fundamental JEDI observation is that fine tuning of the chromaticity enabled a very strong suppression of the decoherence of the idly precessing vector polarization. As a spin off of the search  for the axion driven spin resonance, the JEDI collaboration reported a strong dependence of depolarization rate on the rf-driven spin-flip frequency under otherwise identical ring settings \cite{JEDIaxion}. This behavior fits expectations from the synchrotron oscillation (SO) driven depolarization \cite{SO-Decoherence-2025}.
.
\subsection{Depolarization by the synchrotron oscillations}
\label{sec:SO}

 The basic impact of SOs is a periodic spread of the beam momentum, which defines the longitudinal profile of the bunch density. It also gives rise to the spin decoherence via frequency modulation of the  idle precession  and  the spin-flipper phases. A convenient parameter is 
\begin{equation}
	Q_\text{sy} = \frac{1}{2}(K+G\gamma)^2 \sigma_\text{sy}^2\, ,
	\label{ApproxBessel2}
	\end{equation}
where $\sigma_\text{sy}\ll 1$ is the rms length of a bunch in the unit of the ring circumference.  For an individual particle with the relative SO amplitude $\xi$ the spin-flip frequency modulation amplitude equals $2\sqrt{Q_\text{sy}}\xi \ll 1$. Bogoliubov-Krylov average over rapid SOs yields a suppression of the spin-flip tune,
\begin{equation}
	\nu_\text{sf}(\xi)=\nu_\text{sf}J_0(2\xi\sqrt{Q_\text{sy}})\,, \label{NuXi}
\end{equation} 
entailing a substituton of the spin-flip phase, 
\begin{equation}
	x\rightarrow x(Q_\text{sy},\xi)=J_0(2\xi\sqrt{Q_\text{sy}})x \approx (1-Q_\text{sy}\xi^2)x\, , \label{Bessel}
\end{equation}
 where $J_0(z)$ is the Bessel function.

Next step is an averaging over the SO  phase space:  
\begin{equation}
	\begin{split}
	\langle \exp(ix(Q_\text{sy},\xi)\rangle =D(x)\exp i(x-\phi(x))\,.
		\label{PhaseSpace}
\end{split}
\end{equation}		
The experimentally observed bunch longitudinal profile is close to a Gaussian distribution \cite{PilotBunch-2025}, so that one can approximate the SO amplitude distribution by the Rayleigh function,  $F(\xi)=2\xi\exp(-\xi^2)$, which yields a simple analytic results 		
\begin{equation}
	\begin{split}					
&	D(x) =\frac{1}{\sqrt{1+Q_\text{sy}^2 x^2}}\\
&\phi(x)=\arctan(Q_\text{sy}x)\,.		\label{PhaseWalk}
	\end{split}
\end{equation}
With allowance for the depolarization, the envelope evolution matrix takes the form
\begin{equation}
	%  \begin{equation}
		\matr{E}_\text{D}(x) = 
		\begin{pmatrix}
			D(x)\cos (x -\phi(x)) & D(x)\sin (x-\phi(x)) & 0 \\
			- D(x)\sin (x -\phi(x)) &  D(x)\cos (x -\phi(x))& 0 \\
			0 & 0 & 1
		\end{pmatrix}\, , \label{EnvelopeD}
		%  \end{equation}
\end{equation} 
Note the non-exponential attenuation  factor $D(x)$ of the envelope, and the nonlinear walk, $\phi(x)$, of the envelope rotation phase. Depolarization does not affect the  tangential spectator ${\text p}_\text{z}(x)={\text p}_\text{z}(0)$. 

\subsection{Decoherence of the tensor polarization}
\label{sec:Tensor decoherence}

\begin{equation}
	\begin{split}
		\langle \exp(i2x(Q_\text{sy},\xi))\rangle =D(2x)\exp i(2x-\phi(2x))\,.	\label{TensorPhaseSpace}
	\end{split}
\end{equation}
This unified description of the vector and tensor depolarizations holds for any bunch profile.
Then the phase-space averaged tensor polartization envelope (\ref{Tensor-from-Vector}) will take the form 
\begin{widetext}
\begin{equation}
	\begin{split}
			&	\frac{1}{2}(\text{t}_\text{xx}(n)-\text{t}_\text{yy }(n) )=
			D(2x) [(\frac{1}{2}(\text{t}_\text{xx}(0)-\text{t}_\text{yy}(0) )\cos (2x-\phi(2x) )+ \text{t}_\text{xy}(0)\sin (2x-\phi(2x))]\, ,\\
	&\text{t}_\text{xy}(n)=D(2x)[-\frac{1}{2}(\text{t}_\text{xx}(0)-\text{t}_\text{yy}(0) )\sin (2x-\phi(2x)) + \text{t}_\text{xy}(0)\cos (2x-\phi(2x))]\, ,\\
			&\text{t}_\text{xz}(n) = D(x)[\text{t}_\text{xz}(0)\cos (x -\phi(x))+ \text{t}_\text{yz}(0)\sin (x-\phi(x))]\,,\\
		&\text{t}_\text{yz}(n) =	D(x)[-\text{t}_\text{xz}(0)\sin (x -\phi(x))+ \text{t}_\text{yz}(0)\cos (x-\phi(x))]\, ,\\
						&\text{t}_\text{zz}(n) = \text{t}_\text{zz}(0)\, .   \label{TensorAttenuation}
	\end{split}
\end{equation}
\end{widetext}
Note that the tensor envelope vector ${\bf V}   _2$ attenuates more rapidlly than ${\bf V}   _1$, and has a more rapid nonlinear phase walk, while the spectator component $\text{t}_\text{zz}(n)$ stays constant.
Evidently, this distinction derives from the spectator property 
of the p$_z$ component. These results exemplify a power of the approach of Eq.  (\ref{Tensor-from-Vector}) in the language of $3 \times 3$ matrices, which can be readily  extended to spin transfer in other magnetic elements in storage rings.

\section{Storage ring initial conditions}
\label{sec:initial}

Now we are in the position to invert Eq. (\ref{Tensor-from-Vector}) for the tensor envelope, and to report the tensor polarization $\matr{T}(n)$ in the comoving frame of interest in the storage ring experiments. In a typical experimental situation, particles are injected with vertical $\langle {\bf S} (0)\rangle = {\text S}_\text{y}(0){\bf e}_\text{y}$,  while  $\langle \text{S}_\text{x,z} (0) \rangle =0$.  Likewise, all the off-diagonal tensor polarizations average to zero, 
\begin{equation}
\langle \text{T}_  \text{yx}(0)\rangle = \langle  \text{T}_  \text{yz}(0)\rangle = \langle  \text{T}_ \text{xz}(0) \rangle =0\, ,
\label{off-diagonal}
\end{equation}
which means the tensor envelope  component  ${\bf V}   _1(0)=0$. For each individual particle its tensor polarization is traceless,  so that at the start of the spin flipper operation
\begin{equation}
\langle  \text{T}_  \text{xx}(0) \rangle  =  \langle \text{T}_  \text{zz} (0)\rangle = -\frac{1}{2}\langle \text{T}_ \text{yy}(0) \rangle\, . \label{InitialTensor} 
\end{equation}
Subject to this initial condition,  we obtain
\begin{widetext}
\begin{equation}
	\begin{split}
	\text{T}_  \text{yy}(n) &=   \frac{1}{2} \langle \text{T}_  \text{yy}(0)\rangle  \{-1+\frac{3}{2} [ 1-D(2x)\cos(2x-\phi(2x))]\} \, ,\\
    \text{T}_  \text{xx}(n) &=   \frac{1}{2} \langle \text{T}_  \text{yy}(0)\rangle \{-1 +\frac{3}{2} \cos^2(\theta)[ 1+D(2x)\cos(2x-\phi(2x))]  \}\,  , \\
	\text{T}_  \text{zz}(n) &=   \frac{1}{2} \langle \text{T}_  \text{yy}(0)\rangle  \{-1 +\frac{3}{2} \sin^2(\theta)[ 1+D(2x)\cos(2x-\phi(2x))] \}   \, ,  \\
	\text{T}_  \text{xy}(n) &=   \frac{3}{4} \langle \text{T}_  \text{yy}(0)\rangle  D(2x) \cos(\theta) \sin(2x-\phi(2x))\, ,  \\
    \text{T}_  \text{xz}(n) & = -\frac{3}{8} \langle \text{T}_  \text{yy}(0)\rangle        \sin (2\theta)  [ 1-D(2x)\cos(2x-\phi(2x))]\, ,\\
	\text{T}_  \text{yz}(n) & = -\frac{3}{4} \langle \text{T}_  \text{yy}(0)\rangle D(2x)  \sin(\theta) \sin(2x-\phi(2x))\, .\label{StorageRing}
 \end{split}
\end{equation}
\end{widetext}
Remarkably, all components of the tensor poilarization have distinct Fourier signatures.
\bibliographystyle{apsrev4-2}

\section{Deuteron spin depolarization time at NICA}
\label{sec:NICAdepolarization}

We presume that depolarization of the idle precession can be taken care of by fine tuning the chromaticity \cite{KoopShatunov,SCT1000sJEDI,SCTchromaticityJEDI,Chromaticity_Kolokolchikov:2023}, and focus on depolarization of spin flips by the synchrotron oscillations. Specifically, at the same field integral in the spin flipper, the spin-flip tune scales with $1/\gamma$. Judging from the JEDI experience at COSY \cite{PilotBunch-2025,JEDIaxion}, for deuterons with $\gamma =6.5$  at NICA $\gamma =3.2$  at HIAF, it is feasible to have a solenoid flippers with spin-flip frequencies  $f_\text{sf} \sim 0.02 $\,Hz. Taking this for a reference value, we expect 144 spin-flips of the vector polarization and the total spin-flip phase  $x =452.4$ for 1 h fill.

A crucial point is that the NICA  bunch  length of $\approx 60$ cm \cite{NICA_BEAM_MESHKOV} corresponds to  $\sigma_\text{sy} \approx 1.2\cdot 10^{-3}$, which is about two orders in magnitude shorter than in the JEDI experiments at COSY \cite{PilotBunch-2025}. 
If the RF solenoid at NICA were operating at $K=0$, we would expect  $Q_\text{sy} \approx  7\cdot 10^{-7}$, leading to $Q_\text{sy}x \sim 0.00028$ for an 1 h collision time  and $Q_\text{sy}x \sim 0.0009$ for a 3 h fill. 

Very compact hadron beam bunches of several centimeters are pursued at the EicC facility~\cite{EicC_wp}, so it is reasonable to expect that deuteron bunches at the booster ring (BRing) of HIAF could be compressed shorter than 10~cm. This leads to an optimistic estimation of $\sigma_\text{sy} \leq 1.8\cdot 10^{-4}$ and $Q_\text{sy} \leq   3.4\cdot 10^{-9}$ at the 569-meter-long BRing.
While for deuterons of momentum 0.97 GeV/c at COSY  \cite{PilotBunch-2025,JEDIaxion} the synchrotron oscillation effects were prominent, they are  expected to be basically negligible at NICA and HIAF  for both the vector and tensor polarization of deuterons. We conclude that it makes worthwhile to pursue the continuous spin-flip scheme for deuterons  to create longitudinal polarization \cite{NikolaevPrecessingD}.

Let us emphasize that we assumed an  exact RF spin resonance, which is the reason for the resulting depolarization being much slower than in the Froissart-Stora frequency scan approach \cite{FroissartScan}. To this end, an implementation of the JEDI developed continuous comagnetomtry 
of the idle precessing frequency  by the up-down asymmetry \cite{PilotBunch-2025,JEDI_Feedback-2025} in the upgraded CNI polarimeter would be most useful (see, however, a new idea of spin dichroism based comagnetometry in Sec. XB).  Of course, whether the above  optimistic theoretical estimates are viable or not, is a matter of the experimental scrutiny.

\section{Tensor and vector observables in {\it pd}  interactions}
\label{sec:pd}

Now  we focus on the fundamental symmetry tests with with oscillating vector and tensor polarizations of stored deuterons.  
A generic structure of the polarization-dependent {\it pd}  total cross section reads 
\begin{equation}
	\begin{split}
		\sigma_{tot} = \sigma_{0}
		& + \sigma_\text{tt} (({\bf P} ^d \cdot {\bf P} ^p) - ({\bf P} ^d \cdot {\bf k} )({\bf P} ^p\cdot{\bf k} )) \\
		&+\sigma_\text{zz} ({\bf P} ^d \cdot {\bf k} )({\bf P} ^p \cdot{\bf k} )\\
		& +\sigma^\text{T} \text{T}_  \text{ab}{\text k}_ \text{a} {\text k}_ \text{b}\\
		& + \sigma_\text{PV}^p({\bf P} ^p\cdot {\bf k} ) \\
		&+\sigma_\text{PV}^\text{d} ({\bf P} ^d \cdot {\bf k} )\\
		&+\sigma_\text{PV}^\text{T} ({\bf P} ^p\cdot {\bf k} )\text{T}_\text{ab}{\text k}_ \text{a} {\text k}_ \text{b} \\
		&+\sigma_\text{TVPV} ({\bf k} \cdot[{\bf P} ^p\times {\bf P} ^d])\\
		&+\sigma_\text{TVPC} {\text k}_ \text{a} \text{T}_  \text{ab}\epsilon_\text{bcd}{\text P}^p_\text{c} {\text k}_ \text{d}\, ,	\end{split}
\end{equation} 
where ${\bf P}^{h,d}$ are the vector polarizations, and ${\bf k} ={\bf e}_z$ is a unit vector along the collision axis. The superscrit  T labels  cross sections (see below Sec. 8.1 and Sec.8.2) and should not be confused with the T which stands for the time reversal in the subscripts TVPV and TVPC.We recall that the vector polarization is a T-odd axial vector, while the tensor polarization is the P- and T-even TCPC observable.

\subsection{Parity and time  reversal invariant scattering}
\label{sec:PCTC}

There are four PC- and TC  components:  $\sigma_0$ is the unpolarized cross section,  $\sigma_\text{tt}$ and $\sigma_\text{zz}$ describe the transverse-transverse (tt) and longitudinal-longitudinal (zz) double spin asymmetries, respectively; the  tensor cross section, $\sigma^\text{T}$, derives from the tensor alignement of the dumbbell  deuteron quantified by the D-wave admixture in the deuteron wave function, and the  tensor polarization  $ \text{T}_  \text{ab} {\text k}_ \text{a} {\text k}_ \text{b}=\text{T}_  \text{zz}$ \cite{FrancoGlauber1969}. We shall revisit it in Sec. XB in connection to the tensor polarimetry and spin precession comagnetometry.

\subsection{Parity and T-violation in the Standard Model} 
\label{PTinSM}

We will not dwell on the oscillating helicity approach to the Standard Model predicted \cite{milstein2020parity} parity violating (PV) and T-conserving (TC) components $\sigma_\text{PV}^\text{p}$ and $\sigma_\text{PV}^\text{d}$, they were exposed to great detail in \cite{Koop2020parity,Koop2021tests}.  We only mention that the parity-violating tensor asymmetry described by $\sigma_\text{PV}^\text{T}$, has never been searched for experimentally. From the theoretical viewpoint, it is a natural consequence of the spin alignement of the  dumbbell-shaped deuteron. 

The Standard model has all the ingredients for the PV  and  TV nuclear interactions, a nice review of extensive theoretical work is found in \cite{JordyTVPVreview}. The expected PVTV spin asymmetries are beyond the reach of high energy experiments, although strong enhancement of parity violation is possible in the resonance scattering of polarized neutrons on tensor polarized nuclei
\cite{Barabanov:1995fa,Bunakov:1998xh}.

\section{T-violation in {\it pd}  scattering}
\label{sec:Tin pd}
\subsection{Why the millistrong CP violation?}
\label{sec:millistrong}
Our principal interest is in the TVPC millistrong (MS) interaction, which by virtue of the CPT theorem violates the charge conjugation symmetry C,  and CP too \cite{OkunMillistrong,PrentkiMillistrong,LeeMillistrong,UFNreview}. The flavor-breaking C-even weak interaction has a dimenionless strength $\sim 10^{-7}$, and the observed CP violation in the $K_\text{L}$ decays entails a strength of the strangenes-changing CP-odd interaction  $\sim 10^{-10}$.  If we attribute $K_\text{L} \to \pi\pi$ decays to CP-odd MS interaction in conjunction with CP-even weak interaction, then we ask for the dimensionless strength of MS interaction $ g_\text{MS} \sim 10^{-3}$. This must be regarded as an upper bound for $g_\text{MS}$, as the MS CP-violation could well be a small complement to the Standard Model flavor-non-diagonal Kobayashi-Maskawa mechanism of  CP-violation. For instance, it may be an exclusive feature of the light-flavor sector, thus not affecting the Kobayashi-Maskawa CP-phenomenology of heavy flavor decays. Still, such a MS CP-violation is substantially stronger than that in the Standard Model and, in conjuntion with siutable baryon number nonconservation, may  contribute to a resolution of the  mystery of the unexpectedly large baryon asymmetry of the Universe  \cite{UFNreview}. So far, this remains an entirely unexplored option.

The importance of searches for the MS CP violation has been recognized right after conception of the MS interaction  \cite{OkunMillistrong,PrentkiMillistrong,LeeMillistrong}. A brief summary of the early experimental efforts is found in the recent review \cite{UFNreview}. Much preparatory work for the TIVOLI (formerly TRIC) experiment  on T-violation in {\it pd}  scattering at COSY has been done in earnest
\cite{EversheimTV,TIVOLI:STORI11,Aksentyev:2017dnk,LenisaTIVOLI}, but unfortunately the COSY facility was decomissioned much too soon. 

\subsection{Double polarized {\it pd}  scattering}
		
	The double polarized {\it pd}  interactions can be studied in either the fixed target or the asymmetric collider mode. A relevant TVPC observable in the {\it pd}  total cross section has the spin structure 
 \begin{equation}
 {\text k}_ \text{a} \text{T}_  \text{ab} \epsilon_\text{bcd} {\text P}^\text{p}_\text{c} {\text k}_ \text{d} = \text{T}_  \text{xz} {\text P}^\text{p}_\text{y} - \text{T}_  \text{yz}\text{P}^\text{p}_\text{x}\, . \label{TVPC}
 \end{equation}
 
 \subsubsection{Accelerated polarized deuterons}
 
 For instance, in the asymmetric {\it pd}  
 collider mode one can work with oscillating deuteron spin in  the CW beam and vertically polarized protons in ACW beam. Similarly, in storage rings like Nuclotron and HIAF  one can use deuteron beams interacting with a vertically polarized internal hydrogen gas target. We recite the corresponding Fourier signature of the TVPC signal,
\begin{equation}
  \text{T}_  \text{xz}(n)  = -\frac{3}{8} \langle \text{T}_  \text{yy}(0)\rangle       \sin (2\theta)  [ 1-Q(2x)\cos(2x-\phi(2x))]\, . \label{TVPCxz}
\end{equation} 
 It is unique and, within the statistical sensitivity of the experiment, free of the systematic background.

\subsubsection{Polarized deuteron targets}

Here one can work with vertically polarized proton beams, a challenge is to generate a continuous flip of the tensor polarization  $\text{T}_  \text{xz}$. A practically sound approach would be the adiabatic spin rotation by the RF-driven three-coil technique \cite{Gou_SPIN2025}. Specifically, Eq. (\ref{TVPCxz}) is fully applicable to the target deuterons. In the  target frame, the spin precessioin angle $\theta$ is the azimuthal angle, and the spin-flip angle $x$ is the polar angle of the vector polarization.  Earlier a similar technique based on three DC polarization-holding coil pairs  has been applied at IUCF and PAX to switch the proton polarization in the internal gas target \cite{Rathmann-TargetSpinFlip,PAXtarget2} . In the case of the  polarized deuteron target, a switch of the oscillating component of $\text{T}_  \text{xz}$ takes place after the increment of the spin-flip phase $x$ by $\Delta x = \pi/2$, a dependence on the azimuthal angle $\theta$ at fixed $x$ is an extra knob to eliminate the systematic background.  To this end, it is instructive to recall a search for the T-violation in the interaction of polarized neutrons with the magnetized $^{165}\text{Ho}$  target, where the flips of the tensor polarization were accomplished by a mechanical target rotation  \cite{Huffman:1996ix}.

At HIAF, two inverse-kinematic experiments with a polarized H/D atomic gas target~\cite{IMP_PIT} are being proposed~\cite{Gou_SPIN2025}. The first one uses polarized electrons in atoms as targets and searches for new bosons beyond the Standard Model in the elastic $p\vec{e}$ scattering. The second one probes the isovector force in the breakup of longitudinally polarized deuteron target induced by radioactive nuclear beams. The technique of spin-rotating polarized target could be exploited in these experiments to remove background asymmetries by the Fourier analysis.

\subsection{T-violation in {\it pd}  scattering through a prism of multiple scattering theory}
\label{sec:MST}
The cross section $\sigma_\text{TVPC}$  would vanish unless there is a manifestly time-reversal violating interaction. As such it is often referred to as a null-test observable. A standard treatment of {\it pd}  scattering at intermedite and high energies involves multiple scattering theory, and brief exposition of how the T-violation in {\it pd}  scattering derives from the T-violation in NN scattering is in order.

In the standard formalism the spin-1, deuterons are described by the 3-vector wave function $\bfg{\epsilon}$ (in analogy to spin-1 photons it is often called the deuteron polarization vector, but must not be confused for the deuteron spin!). In this representation \cite{Rekalo:1997fh}, the TVPC operator  ${\text k}_\text{m} \text{T}_ \text{mn} \epsilon_\text{nlr}{\text P}^p_\text{l} {\text k}_ \text{r}$ %, sandwiched between the initial and final state vectors ${\bfg \epsilon_\text{in} }$ and ${\bfg\epsilon_\text{f}}$, respectively, 
 leads to the TVPC amplitude \cite{Uzikov:2015aua}
\begin{eqnarray}
	\label{Foper}
	{\hat F}(0) ={\tilde g} \{ {\bfg\sigma} [({\bf k} \times  {\bfg \epsilon_\text{f}}^*]({\bf k} {\bfg\epsilon_\text{in}})+
	\bfg \sigma [({\bf k} \times  {\bfg \epsilon_\text{in}}] ({\bf k} {\bfg \epsilon_\text{f}}^*)\}\, , 
\end{eqnarray}
where the Pauli operator  ${\bfg \sigma }$ describes the proton spin. It is easy to check that under the time reversal, which implies a series of operations:
${\bf k}\to  -{\bf k} ,\,
\bfg \sigma \to -\bfg\sigma,\, {\bfg \epsilon_\text{in}}\to {-\bfg \epsilon_\text{in}}$,
${\bfg \epsilon_\text{f}}\to -{\bfg \epsilon_\text{f}}$, followed by the permutation of the initial and final deuteron wave function vectors, this P-even operator changes its sign, as has to be for the true TVPC operator.

One must be careful about the permutation of the initial and final states: the P-even and  superfluously T-odd operator, $(\bfg\sigma  {\bf  k} )({\bf k}  [{\bfg \epsilon_\text{in}} \times {\bfg  \epsilon_\text{f}}^*])$,  changes its sign under combined operations
${\bf k}\to  -{\bf k} ,
{\bfg \sigma} \to -\bfg\sigma, {\bfg \epsilon_\text{in}}\to {-\bfg \epsilon_\text{in}}$,
and ${\bfg \epsilon_\text{f}}\to -{\bfg \epsilon_\text{f}}$, but upon the mandatory
permutation  of the initial and final state vectors 
it recovers its initial form, i.e., it is P- and T-conserving. 
In Eq. (\ref{Foper}) ${\tilde g}$ is a complex function of the collision
energy, see below. Its magnitude depends on the specific dynamics of the T-violation.

One more illustration of the time-reversal property of 	${\hat F}(0)$ is useful \cite{Uzikov:2015aua}:
\begin{equation}
	\begin{split}
		\label{TVPCme}
		&\langle\mu'=\frac{1}{2},\lambda'=0|{\hat F}|\mu'=-\frac{1}{2},\lambda'=1\rangle =i\sqrt{2}\tilde g, \\
		&\langle \mu'=\frac{1}{2},\lambda'=-1|{\hat F}|\mu'=-\frac{1}{2},\lambda'=0\rangle =-i\sqrt{2}\tilde g,
	\end{split}
\end{equation}
where $\mu$ ($\mu'$ ) and $\lambda$  ($\lambda'$ ) are helicities of the initial (final)
proton and deuteron, respectively. This switch of sign shows that once a nonvanishing  amplitude $\tilde g$ pops up from the dynamical calculation, it 
will be an unequivocal signal of the T-violation.

The integrated cross section $\sigma_\text{TVPC}$ is given by the optical theorem
\cite{Temerbayev:2015foa,Uzikov:2015aua,Uzikov:2016lsc}:
\begin{eqnarray}
	\sigma_\text{TVPC} = -\frac{8}{3}\sqrt{\pi} {\text{Im}}{\tilde g}({\bf q}=0),
	\label{opt}
\end{eqnarray}
where ${\bf q}$ is the momentum transfer in elastic scattering. Needless to say, dressing by  
the initial- and final-state interactions (ISI and FSI), which are dictated by unitarity,  is important for the complex valued $\tilde{g}(q)$. Still,  taken per se,  FSI and ISI can not generate non-vanishing $\sigma_\text{TVPC}$ and affect the null-effect unless a true 
T-violating interaction is present. In this respect, the integrated TVPC  cross section 
is different from the decay processes, in which FSI gives rise to spurious T-violation effects, and where the true T-violation exhibits itself as a deviation of the  phase of partial amplitudes from the unitarity prescribed FSI phase \cite{okun1963slaboe,OversethFSI}.

Within the Glauber multiple scattering theory, a dynamical derivation of $\tilde{g}(0)$ in terms of the T-odd {\it NN}-scattering amplitudes
proceeds as follows. The  TVPC  {\it t}-matrix of elastic {\it pN}   scattering at a momentum transfer ${\bf q}$ can be decomposed as   
\cite{SimoniusRhoNN,Beyer93}
\begin{widetext}
\begin{equation}
	\begin{split}
	t_{pN}
	&=h_N[({\bfg \sigma}_p\cdot {\bf \hat k})(\bfg \sigma_N\cdot {\bf \hat q})+
	({\bfg \sigma}_p\cdot {\bf \hat q})({\bfg \sigma}_N\cdot {\bf \hat k}) 
	-\frac{2}{3}({\bfg \sigma}_N\cdot {\bfg \sigma}_p)({\bf q} \cdot {\bf \hat  k)}]\\
	&+g_N[{\bfg \sigma}_p\times{\bfg\sigma}_N]\cdot [{\bf\hat  q}\times {\bf \hat k}] 
	+i g_N'({\bfg \sigma}_p-{\bfg \sigma}_N)\cdot [{\bf \hat q}\times{\bf\hat  k}]
	[{\bfg \tau}_p\times{\bfg \tau}_N]_\text{z}\, ,
	\label{tpN}
	\end{split}
\end{equation}
\end{widetext}
where $h_N$, $g_{N}$,  $g_{N}'$ are still unknown q-dependent TVPC  {\it pN}  amplitudes,
${\bfg \tau}_p$ (${\bfg \tau}_{N}$) are the Pauli isospin matrices,
${\bfg \sigma}_p$ (${\bfg \sigma}_{N})$ are the Pauli  spin matrices, the unit vectors  ${\bf \hat  k}, {\bf \hat  q}$ and   ${\bf \hat  n}$ are defined  via the initial   ${\bf p}$ and  final   ${\bf p'}$  momenta  of the scatterred  proton:
% \begin{eqnarray}
	$\bf \hat k=  ({\bf p+\bf p'})/{|\bf p+\bf p'|}$, $\bf \hat q=  (\bf p'-\bf p)/{|\bf p'-\bf p|}$, and the normal to the scattering plane
	$\bf \hat n=[ \bf \hat k\times  \bf \hat q]$.
	%  \label{orty}
	%\end{eqnarray}
	%
	
	The corresponding P- and  T-conserving  TCPC operator of elastic 
	{\it  pN}-scattering  has a structure \cite{Platonova:2010wjt}
	\begin{widetext}
	\begin{equation}
		\begin{split}
			M_{N}&=A_{N}+C_{N}{\bfg \sigma}_{p}\cdot {\bf \hat n}
		+C'_{N}{\bfg \sigma}_{N} \cdot {\bf \hat n}
			+ B_{N}({\bfg \sigma}_{p}\cdot {\bf \hat  k})({\bfg \sigma}_{N}\cdot {\bf \hat  k}) \\
			&+(G_{N}+H_{N})(\bfg \sigma_{p}\cdot {\bf \hat q})(\bfg \sigma_{N}\cdot {\bf \hat q})
			+(G_{N}-H_{N})(\bfg \sigma_{p}\cdot {\bf \hat  n})(\bfg \sigma_{N}\cdot {\bf \hat n})\, ,
			\label{MpN}
		\end{split}
	\end{equation}
	\end{widetext}
	with q-dependent coefficient functions, where  \cite{Sorensen:1978vk}.
	\begin{equation}
		C_{N}'=C_{N}+i A_{N} \frac{q}{2m_{N}}\, .
		\end{equation}

	Note that for the spin-$\frac{1}{2}$ nucleons the TVPC amplitude (\ref{tpN}) vanishes kinematically at zero scattering angle, entailing vanishing impulse approximation contribution to $\tilde{g}(0)$. In terms of the pN amplitudes, the double-scattering contribution  to the pd scattering has the well known convolution-like structure, and its TVPC component $\tilde{g}(q)$ derives from the sequential TVPC and TCPC spin-flip transitions (and vice versa). In the case of ${\bf q}=0$  of our interest, it takes a simpler product form with an averaging over the intrinsic momentum Q in the deuteron, 
	\begin{widetext}
	\begin{equation}
		\begin{split}
								{\tilde g}(0)= \frac{i}{4\pi m_{p}}\int_0^\infty dQ\,
								 Q^2
								[ F_\text{SS}(Q)
					+F_\text{SD}(Q ) + F_\text{DD}(Q )] \{-C'_{n}(Q )h_{p}(Q )+C'_{p}(Q )	[g_{n}(Q )-h_{n}(Q )]\}\,,
					\label{gtilde}
				\end{split}
			\end{equation}
				\end{widetext}
	where subscripts SS, DD and SD stand for the deuteron S-wave, D-wave and the SD interference contributions respectvly.  More detailed description of  their structure in terms of the S- and D-wave functions of the deuteron is found in 	\cite{Uzikov:2016lsc}.

	Salient features of separate contributions to the  TVPC signal are as follows. In the effective field theories,  TVPC interaction first emerges as a high dimesnion d=7 four-fermion  operator \cite{Kurylov:2000}, which is  basically a short distance version of the Simonius $\rho$-exchange interaction  \cite{SimoniusRhoNN}. It is important that by the power counting rules such a four-fermion interaction has a potential of a fast rise with energy up to a large TVPC scale. Factorizing the effective four-fermion effective interactions in terms of meson exchanges is not imperative, but it is still convenient as a mnemonic rule for the classification purposes:
	\begin{enumerate}
		\item{The largest distance contrinution to the TVPC $g'$-term in the NN amplitude comes from the $\rho$-meson exchange  \cite{SimoniusRhoNN}.
			However,
			as was shown in
				\cite{Uzikov:2015aua,Uzikov:2016lsc,Uzikov:2016bja}, within the Glauber theory,
				a contribution from  the $g'$-term
				to the TVPC signal in {\it pd}  scattering 
			vanishes  due  the
			symmetry property of the charge-exchange transitions:
			\begin{equation}
				\label{rho}
			 \langle pn |[{\bfg \tau}_{p}\times{\bfg \tau}_{N}]_\text{z}|np\rangle
				=
				-\langle np |[{\bfg \tau}_{p}\times{\bfg \tau}_{N}]_\text{}|pn\rangle\, .
			\end{equation}
			Correspondingly, there is no contribution from $g'$ to the expansion (\ref{gtilde}) for $\tilde{g}(0)$.
			
			To this end, we recall that a strong suppression of the $\rho$-exchange contribution vs. the axial  $h_1$-exchange contribution was found numerically in the
			Faddeev calculations \cite{Song:2011jh} of the null-test signal for the {\it nd}
			scattering at 100 keV. However,  no explanation of this result was
			offered in the published paper. A plausible cause for this suppression may be the
			same spin-isospin structure of the $\rho$-exchange  which
			leads to the vanishing $\rho$-meson contribution in the Glauber approach.}
		\item{ The contributions of the $g_{N}$ and $h_{N}$-terms are not suppressed in the
			$pd$ scattering 
			%%XXX \cite{Uzikov2015NullTest,UzikovPDTV}.
			\cite{Uzikov:2015aua,Uzikov:2016lsc}.}
		\item {However, in  the $^3He-d$ scattering  the contribution of the $g_{N}$-type interaction
			is suppresed \cite{UzikovHe3D}.}
		\item{Furthermore, in the  double polarized {\it dd} scattering,
			a contribution from the $g_{N}$-term  was found \cite{Platonova_T-odd-DD,Uzikov_T-odd-DD} to  be
			 suppressed as well as  from the $g'$ term, making
			the {\it dd}  channel a favorite one for a search for the  $h_{N}$ type
			TVPC interaction.}
	\end{enumerate}

	Still another issue is an impact of the Coulomb
	interaction in the null-test signal  $\sigma_\text{TVPC}$. Being  T-even and  P-even,
	the pure Coulomb ${\it pp}$ interaction taken per se can not generate the TVPC signal in the  impulse approximation amplitude in the {\it pd}  collision. However, interaction of the beam proton's anomalous magnetic moment with the Coulomb field of the target proton gives a very large contribution to the spin flip parameter $C_{p}'(q)$. This observation by Kopeliovich and Lapidus \cite{KopLap1974,CNI-Theory} is at the heart of the CNI polarimetry at RHIC \cite{CNI-at-RHIC} and will stay such at all high energy facilities. To this end, the Coulomb interaction has, via Eq. (\ref{gtilde}), a sizeable indirect impact on $\tilde{g}(0)$ via Coulomb contribution to $C_{p}'$ (see also \cite{Uzikov:2015aua}).  However, in practice the hadronic and Coulomb spin-flips are inseparable, and one should not bother about Coulomb effects once the direct experimental values of $C_{p}'$ and $C_{n}'$ are used in the evaluation of  $\tilde{g}(0)$.

		\section{Two virtues of nuclear dichroism}
		\label{sec:dichroism}
		\subsection{The tensor polarimetry}
		\label{sec:tensorpolarimetry}

	The conventional storage of polarized particles entails the initial conditions, by which magnitudes of all the components of the tensor polarization are uniquely set by $\langle \text{T}_  \text{yy}(0)\rangle$, see Eq. (\ref{StorageRing}). In principle,  all the T  components  can be determined in storage rings by the  JEDI developed Fourier signature technique (\cite{JEDIspintune2}, see also \cite{Sitnik}). As far as monitoring the tensor polarization at flattop  under RF driven spin flips is concerned,  of special  interest is $\text{T}_  \text{zz}$. Deuteron is a   dumbbell, and the alignement of the dumbbell and the related Glauber eclipse effect in the total cross section depend on the beam tensor polarization \cite{FrancoGlauber1969,Franco1977dC}. Precisely this eclipse effect is behind the spin dichroism \cite{Baryshevsky_2008,BaryshevskyNSR2026} in the deuteron-nucleon and deuteron-nucleus interactions. We emphasize that dichroism in deuteron-carbon interactions was well established experimentally at low energy in \cite{TensorCarbon} and at high energy in the Dubna experiments \cite{azhgirey2008tensor,azhgirey2010tensor}. The tensor polarimetry at intermediate  energies is based on the two-body nuclear reactions \cite{ChiladzeTensorD}. At higher energies, a basically energy independent tensor polarimetry is possible in elastic {\it pd}  scattering \cite{Platonova_Kukulin} and/or the deuteron stripping (breakup) reaction. In the case of collisions at NICA and HIAF, one can isolate the impulse approximation dominanted region of the breakup ${\it dd}
	 \rightarrow {\it pn} d$ and relate it to the ${\it pd}   \rightarrow {\it pd}  $ scattering \cite{Uzikov_TensorPolarimetry}.
	
	Still another basically energy-independent, and presumably technically simpler option, is offered by the nuclear dichroism. The internal  target is indispensable for the continuous Coulomb-nuclear interference (CNI) polarimetry of the beam vector polarization. The CNI setup can be complemented by a relatively simple forward counters for measuring the inelastic interaction rate. One can equally 
	anticipate a separate detector working on the beam halo. The count rate could be even higher than in the CNI recoil detectors.  
	
	We recall the oscillation feature of the diagonal tensor polarization
	\begin{widetext}
		\begin{equation}
			\begin{split}
		\text{T}_  \text{zz}(n) =   \frac{1}{2} \langle \text{T}_  \text{yy}(0)\rangle \{-1 +\frac{3}{2} \sin^2(\theta)[ 1+Q(2x)\cos(2x-\phi(2x))] \} \,.
		\end{split}
		\end{equation}
		\end{widetext}
		 The time stamp would allow the bunch-by-bunch extraction of the oscillating nuclear dichroism component of the deuteron  cross section,  
	\begin{equation}
\sigma_\text{tot} =\sigma_0 (1+ A_\text{zz} \text{T}_  \text{zz})\, , \label{Dichroism}
\end{equation}
 where $A_\text{zz}$ is the tensor analyzing power, which can be evaluated in the Glauber theory. Technically, the Glauber theory is readily applicable to the total and elastic cross sections,  while for a partial inelastic cross section measured using forward detectors with limited acceptance, we will have only semiquantitative guess for the analyzing power. More theoretical scrutiny is in order, still this way a continuous relative tensor polarimetry could  be accomplished and a theoretical prediction of a link between the tensor and vector depolarizations can be tested experimentally.
	
	\subsection{Dichroism as a comagnetometer of spin precession?}
	\label{sec:DichroismComagnetometer}
	
	Synchronized revolution of two colliding bunches is imperative for maintaining of stable luminosity. As we emphasized before, once  the double-spin asymmetries are concerned, independent control of the spin precession phases $\theta_\text{CW}$ and  $\theta_\text{ACW}$ of the clockwise and anticlockwise beams is a must. The single-species comagnetormetry developed by the JEDI collaboration is based on a continuous measurement of the oscillating up-down asymmetry caused by  the precessing radial polarization of the beam \cite{PilotBunch-2025,JEDI_Feedback-2025}.  Here we comment that the oscillating tensor asymmetry (dichroism) offers an equally good handle on the spin precession phase and frequency, and emerges as a novel comagnetometry tool. The matching of precession tunes preserving  spin-flip tunes can be accomplished by the static Wien filter technique presented in Sec. (\ref{sec:matching}). 	
	Detailed simulations of this novel option are called for.
	
	\section{Deuteron tensor polarizability}
	\label{sec:Polarizability}
	External electric and magnetic fields distort wave functions of atoms and molecules. The resulting energy shifts are described in terms of the electric and magnetic polarizabilities.  The same is true of nuclei and particles \cite{ramsey1953polarizability}, the early development of this subject was reviewed in \cite{Petrunkin:1981me}.  Besides the magnetic moment and electric quadrupole moment, spin-1 deuterons are  endowed with four more fundamental attributes: the electric and magnetic scalar and tensor polarizabilities. For the sake of completness, here we expand on the early comments in Sec. \ref{sec:tensor} on possible manifestations of the tensor polarizability on spin dynamics in storage rings. 
	
	The deuteron tensor polarizability is  proportional to its tensor polarization, which is quadratic in the spin operator. For spin-1 particle at rest in the laboratory fields ${\bf E}$ and ${\bf B}$, the corresponding tensor polarizability  Hamiltonian has the form (\cite{Baryshevsky_2008,Silenko2007Dpol,BaryshevskyNSR2026} and references therein)
		\begin{equation}
			{H}=-\alpha_\text{T}(\bf{S}\cdot\bf{E})^2-\beta_\text{T}(\bf S\cdot\bf B)^2. \label{refcomove}
		\end{equation}

	Nonvanishing tensor polarizabilities do exist even for a pointlike spin-1 particle \cite{SilenkoPointlike}
			\begin{equation}
				\begin{split}
				\alpha_T&=\frac{e^2\hbar^2(g-1)^2}{2m^3}, \\
				\beta_T&=\frac{e^2\hbar^2}{8m^3}\left[(g-1)^2+3\right]\,
			 \label{Pointlike}
			 \end{split}
		\end{equation} 
		where $m$ and $g$ are the particle's mass and the Lande factor. A rigorous derivation of the  corresponding Foldy-Wouthuysen transformed relativistic Hamiltonian for a pointlike Proca particle is found in  Ref. \cite{SilenkoSpin1Proca}).
		
			Tensor   polarizabilities, $\alpha_\text{T}$ and $\beta_\text{T}$, have never been measured experimentally. If the deuteron were a pointlike Proca particle with the Lande factor $g$, then we would have expected  the tensor electric polarizability 
		$\alpha_T \approx 4\times10^{-6}\ \text{fm}^{3}$, and the tensor magnetic polarizability $\beta_T$ of the same order of magnitude. The major point is that for the deuteron which is a neutron-proton  bound state, the quantum-mechanical considerations yield values about 4 orders of magnitude larger:  $\alpha_T=-6.2\times10^{-2}\ \text{fm}^{3}$ \cite{CGS}, $\alpha_T=-6.8\times10^{-2}\ \text{fm}^{3}$ \cite{JL}, $\alpha_T=3.2\times10^{-2}\ \text{fm}^{3}$ \cite{FP}, and $\beta_T=1.95\times10^{-1}\ \text{fm}^{3}$ \cite{CGS,JL}. The implication is that the deuteron tensor polarizabilities are ``anomalous''
		ones , with a negligibly small Proca correction, and for a deuteron circulating in a storage ring one can use  the relativistic polarizability Hamiltonian  \cite{Silenko:2006polarzability},
		\begin{equation}
			{\cal H}=-\frac{\alpha_T}{\gamma}\left({\bf S}\cdot{\bf E}^{(0)}\right)^2-\frac{\beta_T}{\gamma}\left(\bf S\cdot\bf B^{(0)}\right)^2\,, 
			\label{Hamspip}
			\end{equation}
			where ${\bf E}^{(0)}$ and ${\bf B}^{(0)}$ are the fields acting on particles in motion.
			
		A generic field transformations read   
		\begin{equation}
			\begin{split} 
				\bf E^{(0)}&=\gamma\left[\bf E-\frac{\gamma}{\gamma+1}\boldsymbol\beta(\boldsymbol\beta\cdot\bf E)
				+\boldsymbol\beta\times\bf B\right],\\
				\bf B^{(0)}&=\gamma\left[\bf B-\frac{\gamma}{\gamma+1}\boldsymbol\beta(\boldsymbol\beta\cdot\bf B)
				-\boldsymbol\beta\times\bf E\right]\, ,
			\end{split} \label{meffinal} \end{equation}
		but for particles circulating on a closed orbit in a storage ring one has to be more specific. There are several major cases of practical interest:
		\begin{itemize}
			\item{Up to now our focus was on a conventional all-magnetic storage ring with vertical guiding field ${\bf B}$ and ${\bf E}\equiv 0$;}
								\item {A hybrid ring  with the vertical ${\bf B}$ and the radial ${\bf E}$ guiding fields, matched to eliminate spin rotations associated with the magnetic moment of the deuteron, so that spin rotations are governed by  the beyond the Standard Model electric dipole moment \cite{AbusaifCYR}, has the frozen spin property, introduced initially for protons in the all electric ring  \cite{srEDM}; }
				\item {In the closely related quasi-frozen spin approach \cite{Senichev:2022ide,Melnikov:2024gcc,Senichev:2025jqf,Melnikov:2025ytd}, deuteron vector spin rotations by the magnetic moment in the arcs are eliminated by electric field elements like the Wien filters in the straight sections. }
		\end{itemize}

		A signature of the deuteron tensor polarizabilities in storage ring experiments is their impact on the  spin dynamics. 		
		 Specifically, the equation  of the spin motion of  circulating deuterons  by the Hamiltonian (\ref{Hamspip}) is  
		\begin{widetext}
		\begin{equation}\begin{split}
				\frac{d\bf S}{dt}=&-\frac{\alpha_T}{\gamma}\left\{\left({\bf S}\cdot{\bf E}^{(0)}\right),\left({\bf E}^{(0)}\times\bf S\right) \right\}
				-\frac{\beta_T}{\gamma}\left\{\left({\bf S}\cdot{\bf B}^{(0)}\right),\left({\bf B}^{(0)}\times\bf S\right) \right\}\,, 
			\end{split}\label{Lamspip}\end{equation}\end{widetext}
		where the curly brackets $\{\dots,\dots\}$ denote anticommutators.  The right hand side of Eq. (\ref{Lamspip}) has a nonvanishing contributions from the in-plane components, $({\rm \text{T}_  \text{xx}, \text{T}_  \text{zz}, \text{T}_  \text{xz}})$, of the tensor polarization, and a deuteron beam with an initial horizontal polarization could acquire a vertical vector polarization driven by ${\rm \text{T}_  {xz}}$ and/or $({\rm \text{T}_  {xx}-\text{T}_  {zz}})$. This manifestation of birefringence was first predicted by Baryshevsky \cite{BaryshevskyBirefringence,Baryshevsky2005}. As we have seen in Sec.\ref{sec:initial}, the initial condition and the subsequent evolution of the vector and tensor polarizations depend on how polarized particles are injected into the storage ring. A further scrutiny and revisiting the early considerations in \cite{Silenko:2006polarzability,Baryshevsky_2008,Silenko2007Dpol,Silenko:2009zq,Baryshevsky2005} are in order to ascertain that conditions for a  resonance growth of the vertical vector polarization can be met, and the tensor electric and magnetic polarizabilities can be measured in magnetic storage rings like Nuclotron/NICA and HIAF or in the hybrid quasi-frozen spin rings.

	\section{Summary and outlook}
	\label{sec:summary}
	We explored a potential of the spin physics with the oscillating deuteron polarization. This paradigm is based on a solid foundation of experimental studies at the Laboratory of High Energy Physics, Dubna and  a prolific  JEDI collaboration work at the COSY storage ring of  Forschungszentrum J\"ulich. As far as tests of fundamental symmetries at Nuclotron/NICA and HIAF are concerned, of particular interest is a relatively easy generation of the longitudinal polarization from a pure vector polarization initially in the verical direction. We argued that with short bunches at NICA and HIAF, the deuteron beam polarization lifetime could be in the 1 hour ballpark and plausibly longer. The principal challenge is an experimental confirmation of this optimistic theoretical expectation of a slow depolarization of the continuously flipping vector polarization at NICA and HIAF.
	
	A perfect identity of the CW and ACW rings is beyond the reach. Even identical beam circulation frequences do not ensure the identical spin precession in the CW and ACW rings. To eliminate the unwanted relative walk of the spin phases one has to implement certain knobs.
	
	A modest infrastructure for  implermentation of the oscillating polarization mode at the Nuclotron/NICA includes per ring:
	\begin{itemize}
	\item {The RF solenoid  flipper operating with  spin-flip frequencies in the 0.02 Hz ballpark will provide  $\sim 150$ vector polarization  flips per 1 h fill, a sufficiently large number for supression of systematic effects in the observed spin asymmetries};
	\item {A measurement of the spin-flip frequency is a simple task for the mandatory CNI polarimeter;}
	\item {Spin precession polarimeter is a must for estabishing the time-stamped polarization orientation. Here one can rely on either the JEDI developed oscillating up-down asymmetry at the CNI polarimeter,  or a novel approach of an oscillating tensor polarization dichroism in inelastic cross section.}
	\item {Static Wien filter for switching the relative helicity of colliding bunches.}
	\item {The same static Wien filters can be used as a comagnetometer knob to fine tune the spin precessions in pairs of colliding bunches}
\end{itemize} 
		
	A fixed target option at Nuclotron and HIAF is less demanding regarding the depolarization rate. In the fixed target search for the T-violation there are two options. One is the stored vertically polarized proton beam interacting with the internal polarized deuterium gas target. In this case one needs an atomic beam source \cite{Erhard_Steffens_2003}. The deuteron tensor polarization can be generated by proper rotations of the target polarization. The second option is the above described oscillating polarization of stored deuterons interacting with the polarized proton target. Of particular interest in this mode is an access  to parity violation from the time-tagged oscillating longitudinal polarization of the deuteron beam. One turn extraction of the deuteron beam of the known helicity onto an  external unpolarized dense target has a merit of a reach to very small parity violating signal by measuring the beam charge upstream and downstream the target by the beam current transformers, see Refs. \cite{Koop2020parity,Koop2021tests}. 
	
	To summarize, the oscillating deuteron polarization approach has a unique potential worth of pursuing at Nuclotron/NICA and HIAF.
	
	{\bf{Conflict of interest:}} The authors declare that they have no conflicts of interest.
	
	{\bfseries{Acknowledgements:}}  N.\,Nikolaev and Yu.\,Uzikov acknowledge a support by a Russian Science Foundation grant 25-72-30005  (https://rscf.ru/project/25-72-30005).   B.~Gou acknowledges support by the National Key Research and Development Program of China under Contract No.~2023YFA1606800 and No.~2024YFA1611003, the Natural Science Foundation of Beijing, China under Grant No.~JQ22002.

	\bibliographystyle{apsrev4-2}

	\bibliography{PilotBunch_03.07.2026}

\begin{thebibliography}{141}%
\makeatletter
\providecommand \@ifxundefined [1]{%
 \@ifx{#1\undefined}
}%
\providecommand \@ifnum [1]{%
 \ifnum #1\expandafter \@firstoftwo
 \else \expandafter \@secondoftwo
 \fi
}%
\providecommand \@ifx [1]{%
 \ifx #1\expandafter \@firstoftwo
 \else \expandafter \@secondoftwo
 \fi
}%
\providecommand \natexlab [1]{#1}%
\providecommand \enquote  [1]{``#1''}%
\providecommand \bibnamefont  [1]{#1}%
\providecommand \bibfnamefont [1]{#1}%
\providecommand \citenamefont [1]{#1}%
\providecommand \href@noop [0]{\@secondoftwo}%
\providecommand \href [0]{\begingroup \@sanitize@url \@href}%
\providecommand \@href[1]{\@@startlink{#1}\@@href}%
\providecommand \@@href[1]{\endgroup#1\@@endlink}%
\providecommand \@sanitize@url [0]{\catcode `\\12\catcode `\$12\catcode
  `\&12\catcode `\#12\catcode `\^12\catcode `\_12\catcode `\%12\relax}%
\providecommand \@@startlink[1]{}%
\providecommand \@@endlink[0]{}%
\providecommand \url  [0]{\begingroup\@sanitize@url \@url }%
\providecommand \@url [1]{\endgroup\@href {#1}{\urlprefix }}%
\providecommand \urlprefix  [0]{URL }%
\providecommand \Eprint [0]{\href }%
\providecommand \doibase [0]{https://doi.org/}%
\providecommand \selectlanguage [0]{\@gobble}%
\providecommand \bibinfo  [0]{\@secondoftwo}%
\providecommand \bibfield  [0]{\@secondoftwo}%
\providecommand \translation [1]{[#1]}%
\providecommand \BibitemOpen [0]{}%
\providecommand \bibitemStop [0]{}%
\providecommand \bibitemNoStop [0]{.\EOS\space}%
\providecommand \EOS [0]{\spacefactor3000\relax}%
\providecommand \BibitemShut  [1]{\csname bibitem#1\endcsname}%
\let\auto@bib@innerbib\@empty
%</preamble>
\bibitem [{\citenamefont {Nagaytsev}(2013)}]{Nagaytsev-History}%
  \BibitemOpen
  \bibfield  {author} {\bibinfo {author} {\bibfnamefont {A.~P.}\ \bibnamefont
  {Nagaytsev}},\ }\href {https://doi.org/10.1134/S1063779613060166} {\bibfield
  {journal} {\bibinfo  {journal} {Phys. Part. Nucl.}\ }\textbf {\bibinfo
  {volume} {44}},\ \bibinfo {pages} {937} (\bibinfo {year} {2013})}\BibitemShut
  {NoStop}%
\bibitem [{\citenamefont {Anishchenko}(2023)}]{SpinDubna2023}%
  \BibitemOpen
  \bibfield  {author} {\bibinfo {author} {\bibfnamefont {N.~G.}\ \bibnamefont
  {Anishchenko}},\ }\href@noop {} {\bibfield  {journal} {\bibinfo  {journal}
  {Advances in Historical Studies}\ }\textbf {\bibinfo {volume} {12}},\
  \bibinfo {pages} {115} (\bibinfo {year} {2023})}\BibitemShut {NoStop}%
\bibitem [{\citenamefont {Teryaev}(2026)}]{Teryaev-NSR2026}%
  \BibitemOpen
  \bibfield  {author} {\bibinfo {author} {\bibfnamefont {O.~V.}\ \bibnamefont
  {Teryaev}},\ }\href {https://doi.org/10.54546/NaturalSciRev.200606}
  {\bibfield  {journal} {\bibinfo  {journal} {Natural Sci. Rev.}\ }\textbf
  {\bibinfo {volume} {3}},\ \bibinfo {pages} {200606} (\bibinfo {year}
  {2026})}\BibitemShut {NoStop}%
\bibitem [{\citenamefont {Arbuzov}\ \emph {et~al.}(2021)\citenamefont {Arbuzov}
  \emph {et~al.}}]{Arbuzov:NICA-SPD}%
  \BibitemOpen
  \bibfield  {author} {\bibinfo {author} {\bibfnamefont {A.}~\bibnamefont
  {Arbuzov}} \emph {et~al.},\ }\href
  {https://doi.org/10.1016/j.ppnp.2021.103858} {\bibfield  {journal} {\bibinfo
  {journal} {Prog. Part. Nucl. Phys.}\ }\textbf {\bibinfo {volume} {119}},\
  \bibinfo {pages} {103858} (\bibinfo {year} {2021})},\ \Eprint
  {https://arxiv.org/abs/2011.15005} {arXiv:2011.15005 [hep-ex]} \BibitemShut
  {NoStop}%
\bibitem [{\citenamefont {Fimushkin}\ \emph {et~al.}(2026)\citenamefont
  {Fimushkin}, \citenamefont {Piskunov}, \citenamefont {Strokovsky},
  \citenamefont {Ladygin}, \citenamefont {Filatov},\ and\ \citenamefont
  {Syresin}}]{fimushkinNSR2026}%
  \BibitemOpen
  \bibfield  {author} {\bibinfo {author} {\bibfnamefont {V.}~\bibnamefont
  {Fimushkin}}, \bibinfo {author} {\bibfnamefont {N.}~\bibnamefont {Piskunov}},
  \bibinfo {author} {\bibfnamefont {E.}~\bibnamefont {Strokovsky}}, \bibinfo
  {author} {\bibfnamefont {V.}~\bibnamefont {Ladygin}}, \bibinfo {author}
  {\bibfnamefont {Y.}~\bibnamefont {Filatov}},\ and\ \bibinfo {author}
  {\bibfnamefont {E.}~\bibnamefont {Syresin}},\ }\href@noop {} {\bibfield
  {journal} {\bibinfo  {journal} {Natural Science Review}\ }\textbf {\bibinfo
  {volume} {3}},\ \bibinfo {pages} {200608} (\bibinfo {year}
  {2026})}\BibitemShut {NoStop}%
\bibitem [{\citenamefont {Filatov}\ \emph {et~al.}(2026)\citenamefont
  {Filatov}, \citenamefont {Kondratenko}, \citenamefont {Kondratenko},
  \citenamefont {Tsyplakov}, \citenamefont {Butenko}, \citenamefont {Ladygin},
  \citenamefont {Lebedev}, \citenamefont {Syresin}, \citenamefont {Butenko},\
  and\ \citenamefont {Chernyshov}}]{filatov2026polarization}%
  \BibitemOpen
  \bibfield  {author} {\bibinfo {author} {\bibfnamefont {Y.~N.}\ \bibnamefont
  {Filatov}}, \bibinfo {author} {\bibfnamefont {A.}~\bibnamefont
  {Kondratenko}}, \bibinfo {author} {\bibfnamefont {M.}~\bibnamefont
  {Kondratenko}}, \bibinfo {author} {\bibfnamefont {E.}~\bibnamefont
  {Tsyplakov}}, \bibinfo {author} {\bibfnamefont {A.}~\bibnamefont {Butenko}},
  \bibinfo {author} {\bibfnamefont {V.}~\bibnamefont {Ladygin}}, \bibinfo
  {author} {\bibfnamefont {V.}~\bibnamefont {Lebedev}}, \bibinfo {author}
  {\bibfnamefont {E.}~\bibnamefont {Syresin}}, \bibinfo {author} {\bibfnamefont
  {E.}~\bibnamefont {Butenko}},\ and\ \bibinfo {author} {\bibfnamefont
  {A.}~\bibnamefont {Chernyshov}},\ }\href
  {https://doi.org/10.1134/S1547477125702085} {\bibfield  {journal} {\bibinfo
  {journal} {Physics of Particles and Nuclei Letters}\ }\textbf {\bibinfo
  {volume} {23}},\ \bibinfo {pages} {185} (\bibinfo {year} {2026})}\BibitemShut
  {NoStop}%
\bibitem [{\citenamefont {Zelenski}\ \emph {et~al.}(2026)\citenamefont
  {Zelenski}, \citenamefont {Ladygin}, \citenamefont {Fimushkin}, \citenamefont
  {Dunin}, \citenamefont {Ivshin}, \citenamefont {Solovev}, \citenamefont
  {Shindin}, \citenamefont {Zhabin}, \citenamefont {Filatov}, \citenamefont
  {Chernikova},\ and\ \citenamefont {Belov}}]{zelenski2026polarimetry}%
  \BibitemOpen
  \bibfield  {author} {\bibinfo {author} {\bibfnamefont {A.}~\bibnamefont
  {Zelenski}}, \bibinfo {author} {\bibfnamefont {V.}~\bibnamefont {Ladygin}},
  \bibinfo {author} {\bibfnamefont {V.}~\bibnamefont {Fimushkin}}, \bibinfo
  {author} {\bibfnamefont {V.}~\bibnamefont {Dunin}}, \bibinfo {author}
  {\bibfnamefont {K.}~\bibnamefont {Ivshin}}, \bibinfo {author} {\bibfnamefont
  {A.}~\bibnamefont {Solovev}}, \bibinfo {author} {\bibfnamefont
  {R.}~\bibnamefont {Shindin}}, \bibinfo {author} {\bibfnamefont
  {S.}~\bibnamefont {Zhabin}}, \bibinfo {author} {\bibfnamefont {Y.~N.}\
  \bibnamefont {Filatov}}, \bibinfo {author} {\bibfnamefont {A.}~\bibnamefont
  {Chernikova}},\ and\ \bibinfo {author} {\bibfnamefont {A.}~\bibnamefont
  {Belov}},\ }\href {https://doi.org/10.1134/S1547477125702103} {\bibfield
  {journal} {\bibinfo  {journal} {Physics of Particles and Nuclei Letters}\
  }\textbf {\bibinfo {volume} {23}},\ \bibinfo {pages} {196} (\bibinfo {year}
  {2026})}\BibitemShut {NoStop}%
\bibitem [{\citenamefont {Zhai}\ \emph {et~al.}(2025)\citenamefont {Zhai} \emph
  {et~al.}}]{IMP_SPIS}%
  \BibitemOpen
  \bibfield  {author} {\bibinfo {author} {\bibfnamefont {Y.}~\bibnamefont
  {Zhai}} \emph {et~al.},\ }in\ \href@noop {} {\emph {\bibinfo {booktitle}
  {Proceedings of 26th International Symposium on Spin Physics --
  PoS(SPIN2025)}}}\ (\bibinfo {address} {Tsingtao, China},\ \bibinfo {year}
  {2025})\BibitemShut {NoStop}%
\bibitem [{\citenamefont {Li}\ \emph {et~al.}(2025)\citenamefont {Li},
  \citenamefont {Zhao}, \citenamefont {Mao}, \citenamefont {Yao},\ and\
  \citenamefont {Yang}}]{Lanzhou-Tensor}%
  \BibitemOpen
  \bibfield  {author} {\bibinfo {author} {\bibfnamefont {M.}~\bibnamefont
  {Li}}, \bibinfo {author} {\bibfnamefont {H.}~\bibnamefont {Zhao}}, \bibinfo
  {author} {\bibfnamefont {L.}~\bibnamefont {Mao}}, \bibinfo {author}
  {\bibfnamefont {Z.}~\bibnamefont {Yao}},\ and\ \bibinfo {author}
  {\bibfnamefont {J.}~\bibnamefont {Yang}},\ }\href
  {https://doi.org/10.1103/99jh-vtd1} {\bibfield  {journal} {\bibinfo
  {journal} {Phys. Rev. Accel. Beams}\ }\textbf {\bibinfo {volume} {28}},\
  \bibinfo {pages} {094002} (\bibinfo {year} {2025})}\BibitemShut {NoStop}%
\bibitem [{\citenamefont {Gou}\ \emph {et~al.}(2025)\citenamefont {Gou} \emph
  {et~al.}}]{Gou_SPIN2025}%
  \BibitemOpen
  \bibfield  {author} {\bibinfo {author} {\bibfnamefont {B.}~\bibnamefont
  {Gou}} \emph {et~al.},\ }in\ \href@noop {} {\emph {\bibinfo {booktitle}
  {Proceedings of 26th International Symposium on Spin Physics --
  PoS(SPIN2025)}}}\ (\bibinfo {address} {Tsingtao, China},\ \bibinfo {year}
  {2025})\BibitemShut {NoStop}%
\bibitem [{\citenamefont {Baryshevsky}(2026)}]{BaryshevskyNSR2026}%
  \BibitemOpen
  \bibfield  {author} {\bibinfo {author} {\bibfnamefont {V.~G.}\ \bibnamefont
  {Baryshevsky}},\ }\href {https://doi.org/10.54546/NaturalSciRev.200607}
  {\bibfield  {journal} {\bibinfo  {journal} {Natural Science Review}\ }\textbf
  {\bibinfo {volume} {3}},\ \bibinfo {pages} {200607} (\bibinfo {year}
  {2026})}\BibitemShut {NoStop}%
\bibitem [{\citenamefont {Filatov}\ \emph
  {et~al.}(2020{\natexlab{a}})\citenamefont {Filatov}, \citenamefont
  {Kondratenko}, \citenamefont {Kondratenko}, \citenamefont {Derbenev},\ and\
  \citenamefont {Morozov}}]{FilatovTransparent2020}%
  \BibitemOpen
  \bibfield  {author} {\bibinfo {author} {\bibfnamefont {Y.~N.}\ \bibnamefont
  {Filatov}}, \bibinfo {author} {\bibfnamefont {A.~M.}\ \bibnamefont
  {Kondratenko}}, \bibinfo {author} {\bibfnamefont {M.~A.}\ \bibnamefont
  {Kondratenko}}, \bibinfo {author} {\bibfnamefont {Y.~S.}\ \bibnamefont
  {Derbenev}},\ and\ \bibinfo {author} {\bibfnamefont {V.~S.}\ \bibnamefont
  {Morozov}},\ }\href {https://doi.org/10.1103/PhysRevLett.124.194801}
  {\bibfield  {journal} {\bibinfo  {journal} {Phys. Rev. Lett.}\ }\textbf
  {\bibinfo {volume} {124}},\ \bibinfo {pages} {194801} (\bibinfo {year}
  {2020}{\natexlab{a}})}\BibitemShut {NoStop}%
\bibitem [{\citenamefont {Filatov}\ \emph
  {et~al.}(2020{\natexlab{b}})\citenamefont {Filatov}, \citenamefont
  {Kondratenko}, \citenamefont {Kondratenko}, \citenamefont {Derbenev},
  \citenamefont {Morozov},\ and\ \citenamefont {Kovalenko}}]{Filatov2020EPJ}%
  \BibitemOpen
  \bibfield  {author} {\bibinfo {author} {\bibfnamefont {Y.~N.}\ \bibnamefont
  {Filatov}}, \bibinfo {author} {\bibfnamefont {A.~M.}\ \bibnamefont
  {Kondratenko}}, \bibinfo {author} {\bibfnamefont {M.~A.}\ \bibnamefont
  {Kondratenko}}, \bibinfo {author} {\bibfnamefont {Y.~S.}\ \bibnamefont
  {Derbenev}}, \bibinfo {author} {\bibfnamefont {V.~S.}\ \bibnamefont
  {Morozov}},\ and\ \bibinfo {author} {\bibfnamefont {A.~D.}\ \bibnamefont
  {Kovalenko}},\ }\href {https://doi.org/10.1140/epjc/s10052-020-8344-5}
  {\bibfield  {journal} {\bibinfo  {journal} {Eur. Phys. J.}\ }\textbf
  {\bibinfo {volume} {C80}},\ \bibinfo {pages} {778} (\bibinfo {year}
  {2020}{\natexlab{b}})}\BibitemShut {NoStop}%
%%CITATION = EPHJA,C80,778;%%
\bibitem [{\citenamefont {Filatov}\ \emph {et~al.}(2022)\citenamefont
  {Filatov}, \citenamefont {Kondratenko}, \citenamefont {Kondratenko},
  \citenamefont {Tsyplakov}, \citenamefont {Butenko}, \citenamefont
  {Kostromin}, \citenamefont {Ladygin}, \citenamefont {Syresin}, \citenamefont
  {Guryleva}, \citenamefont {Melnikov} \emph {et~al.}}]{FilatovNavigator2022}%
  \BibitemOpen
  \bibfield  {author} {\bibinfo {author} {\bibfnamefont {Y.~N.}\ \bibnamefont
  {Filatov}}, \bibinfo {author} {\bibfnamefont {A.~M.}\ \bibnamefont
  {Kondratenko}}, \bibinfo {author} {\bibfnamefont {M.~A.}\ \bibnamefont
  {Kondratenko}}, \bibinfo {author} {\bibfnamefont {E.~D.}\ \bibnamefont
  {Tsyplakov}}, \bibinfo {author} {\bibfnamefont {A.~V.}\ \bibnamefont
  {Butenko}}, \bibinfo {author} {\bibfnamefont {S.~A.}\ \bibnamefont
  {Kostromin}}, \bibinfo {author} {\bibfnamefont {V.~P.}\ \bibnamefont
  {Ladygin}}, \bibinfo {author} {\bibfnamefont {E.}~\bibnamefont {Syresin}},
  \bibinfo {author} {\bibfnamefont {I.~L.}\ \bibnamefont {Guryleva}}, \bibinfo
  {author} {\bibfnamefont {A.~A.}\ \bibnamefont {Melnikov}}, \emph {et~al.},\
  }\href {https://doi.org/10.1134/S0021364022601762} {\bibfield  {journal}
  {\bibinfo  {journal} {JETP Letters}\ }\textbf {\bibinfo {volume} {116}},\
  \bibinfo {pages} {413} (\bibinfo {year} {2022})}\BibitemShut {NoStop}%
\bibitem [{\citenamefont {Filatov}\ \emph {et~al.}(2023)\citenamefont
  {Filatov}, \citenamefont {Kondratenko}, \citenamefont {Nikolaev},
  \citenamefont {Senichev}, \citenamefont {Kondratenko}, \citenamefont
  {Vinogradov}, \citenamefont {Tsyplakov}, \citenamefont {Butenko},
  \citenamefont {Kostromin}, \citenamefont {Ladygin}, \citenamefont {Syresin},
  \citenamefont {Guryleva}, \citenamefont {Melnikov},\ and\ \citenamefont
  {Aksentev}}]{FilatovOrbitSteer2023}%
  \BibitemOpen
  \bibfield  {author} {\bibinfo {author} {\bibfnamefont {Y.~N.}\ \bibnamefont
  {Filatov}}, \bibinfo {author} {\bibfnamefont {A.~M.}\ \bibnamefont
  {Kondratenko}}, \bibinfo {author} {\bibfnamefont {N.~N.}\ \bibnamefont
  {Nikolaev}}, \bibinfo {author} {\bibfnamefont {Y.~V.}\ \bibnamefont
  {Senichev}}, \bibinfo {author} {\bibfnamefont {M.~A.}\ \bibnamefont
  {Kondratenko}}, \bibinfo {author} {\bibfnamefont {S.~V.}\ \bibnamefont
  {Vinogradov}}, \bibinfo {author} {\bibfnamefont {E.~D.}\ \bibnamefont
  {Tsyplakov}}, \bibinfo {author} {\bibfnamefont {A.~V.}\ \bibnamefont
  {Butenko}}, \bibinfo {author} {\bibfnamefont {S.~A.}\ \bibnamefont
  {Kostromin}}, \bibinfo {author} {\bibfnamefont {V.~P.}\ \bibnamefont
  {Ladygin}}, \bibinfo {author} {\bibfnamefont {E.~M.}\ \bibnamefont
  {Syresin}}, \bibinfo {author} {\bibfnamefont {I.~L.}\ \bibnamefont
  {Guryleva}}, \bibinfo {author} {\bibfnamefont {A.~A.}\ \bibnamefont
  {Melnikov}},\ and\ \bibinfo {author} {\bibfnamefont {A.~E.}\ \bibnamefont
  {Aksentev}},\ }\href {https://doi.org/10.1134/S0021364023602695} {\bibfield
  {journal} {\bibinfo  {journal} {JETP Lett.}\ }\textbf {\bibinfo {volume}
  {118}},\ \bibinfo {pages} {387} (\bibinfo {year} {2023})}\BibitemShut
  {NoStop}%
\bibitem [{\citenamefont {Filatov}\ \emph {et~al.}(2024)\citenamefont
  {Filatov}, \citenamefont {Kondratenko}, \citenamefont {Nikolaev},
  \citenamefont {Senichev}, \citenamefont {Kondratenko}, \citenamefont
  {Vinogradov}, \citenamefont {Tsyplakov}, \citenamefont {Chernyshov},
  \citenamefont {Butenko}, \citenamefont {Kostromin}, \citenamefont {Ladygin},
  \citenamefont {Syresin}, \citenamefont {Butenko}, \citenamefont {Guryleva},
  \citenamefont {Melnikov},\ and\ \citenamefont
  {Aksentev}}]{FilatovImperfections2024}%
  \BibitemOpen
  \bibfield  {author} {\bibinfo {author} {\bibfnamefont {Y.~N.}\ \bibnamefont
  {Filatov}}, \bibinfo {author} {\bibfnamefont {A.~M.}\ \bibnamefont
  {Kondratenko}}, \bibinfo {author} {\bibfnamefont {N.~N.}\ \bibnamefont
  {Nikolaev}}, \bibinfo {author} {\bibfnamefont {Y.~V.}\ \bibnamefont
  {Senichev}}, \bibinfo {author} {\bibfnamefont {M.~A.}\ \bibnamefont
  {Kondratenko}}, \bibinfo {author} {\bibfnamefont {S.~V.}\ \bibnamefont
  {Vinogradov}}, \bibinfo {author} {\bibfnamefont {E.~D.}\ \bibnamefont
  {Tsyplakov}}, \bibinfo {author} {\bibfnamefont {A.~I.}\ \bibnamefont
  {Chernyshov}}, \bibinfo {author} {\bibfnamefont {A.~V.}\ \bibnamefont
  {Butenko}}, \bibinfo {author} {\bibfnamefont {S.~A.}\ \bibnamefont
  {Kostromin}}, \bibinfo {author} {\bibfnamefont {V.~P.}\ \bibnamefont
  {Ladygin}}, \bibinfo {author} {\bibfnamefont {E.~M.}\ \bibnamefont
  {Syresin}}, \bibinfo {author} {\bibfnamefont {E.~A.}\ \bibnamefont
  {Butenko}}, \bibinfo {author} {\bibfnamefont {I.~L.}\ \bibnamefont
  {Guryleva}}, \bibinfo {author} {\bibfnamefont {A.~A.}\ \bibnamefont
  {Melnikov}},\ and\ \bibinfo {author} {\bibfnamefont {A.~T.}\ \bibnamefont
  {Aksentev}},\ }\href {https://doi.org/10.1134/S0021364024603695} {\bibfield
  {journal} {\bibinfo  {journal} {JETP Lett.}\ }\textbf {\bibinfo {volume}
  {120}},\ \bibinfo {pages} {779} (\bibinfo {year} {2024})}\BibitemShut
  {NoStop}%
\bibitem [{\citenamefont {Senichev}\ \emph {et~al.}(2022)\citenamefont
  {Senichev}, \citenamefont {Aksentyev}, \citenamefont {Kolokolchikov},
  \citenamefont {Melnikov}, \citenamefont {Ladygin}, \citenamefont {Syresin},\
  and\ \citenamefont {Nikolaev}}]{Senichev:2022ide}%
  \BibitemOpen
  \bibfield  {author} {\bibinfo {author} {\bibfnamefont {Y.}~\bibnamefont
  {Senichev}}, \bibinfo {author} {\bibfnamefont {A.}~\bibnamefont {Aksentyev}},
  \bibinfo {author} {\bibfnamefont {S.}~\bibnamefont {Kolokolchikov}}, \bibinfo
  {author} {\bibfnamefont {A.}~\bibnamefont {Melnikov}}, \bibinfo {author}
  {\bibfnamefont {V.}~\bibnamefont {Ladygin}}, \bibinfo {author} {\bibfnamefont
  {E.}~\bibnamefont {Syresin}},\ and\ \bibinfo {author} {\bibfnamefont
  {N.}~\bibnamefont {Nikolaev}},\ }\href
  {https://doi.org/10.1088/1742-6596/2420/1/012052} {\bibfield  {journal}
  {\bibinfo  {journal} {JACoW}\ }\textbf {\bibinfo {volume} {IPAC2022}},\
  \bibinfo {pages} {MOPOTK024} (\bibinfo {year} {2022})}\BibitemShut {NoStop}%
\bibitem [{\citenamefont {Melnikov}\ \emph {et~al.}(2024)\citenamefont
  {Melnikov}, \citenamefont {Aksentyev}, \citenamefont {Senichev},\ and\
  \citenamefont {Kolokolchikov}}]{Melnikov:2024gcc}%
  \BibitemOpen
  \bibfield  {author} {\bibinfo {author} {\bibfnamefont {A.}~\bibnamefont
  {Melnikov}}, \bibinfo {author} {\bibfnamefont {A.}~\bibnamefont {Aksentyev}},
  \bibinfo {author} {\bibfnamefont {Y.}~\bibnamefont {Senichev}},\ and\
  \bibinfo {author} {\bibfnamefont {S.}~\bibnamefont {Kolokolchikov}},\ }\href
  {https://doi.org/10.1134/S1063778824700285} {\bibfield  {journal} {\bibinfo
  {journal} {Phys. Atom. Nucl.}\ }\textbf {\bibinfo {volume} {87}},\ \bibinfo
  {pages} {447} (\bibinfo {year} {2024})}\BibitemShut {NoStop}%
\bibitem [{\citenamefont {Senichev}\ \emph {et~al.}(2025)\citenamefont
  {Senichev}, \citenamefont {Melnikov}, \citenamefont {Aksentyev},
  \citenamefont {Syresin}, \citenamefont {Palamarchuka}, \citenamefont
  {Kolokolchikov},\ and\ \citenamefont {Ladygin}}]{Senichev:2025jqf}%
  \BibitemOpen
  \bibfield  {author} {\bibinfo {author} {\bibfnamefont {Y.}~\bibnamefont
  {Senichev}}, \bibinfo {author} {\bibfnamefont {A.}~\bibnamefont {Melnikov}},
  \bibinfo {author} {\bibfnamefont {A.}~\bibnamefont {Aksentyev}}, \bibinfo
  {author} {\bibfnamefont {E.}~\bibnamefont {Syresin}}, \bibinfo {author}
  {\bibfnamefont {P.}~\bibnamefont {Palamarchuka}}, \bibinfo {author}
  {\bibfnamefont {S.}~\bibnamefont {Kolokolchikov}},\ and\ \bibinfo {author}
  {\bibfnamefont {V.}~\bibnamefont {Ladygin}},\ }\href
  {https://doi.org/10.18429/JACoW-IPAC2025-TUPS032} {\bibfield  {journal}
  {\bibinfo  {journal} {JACoW}\ }\textbf {\bibinfo {volume} {IPAC2025}},\
  \bibinfo {pages} {TUPS032} (\bibinfo {year} {2025})}\BibitemShut {NoStop}%
\bibitem [{\citenamefont {Melnikov}\ \emph {et~al.}(2025)\citenamefont
  {Melnikov}, \citenamefont {Aksentyev}, \citenamefont {Nikolaev},
  \citenamefont {Kolokolchikov},\ and\ \citenamefont
  {Senichev}}]{Melnikov:2025ytd}%
  \BibitemOpen
  \bibfield  {author} {\bibinfo {author} {\bibfnamefont {A.}~\bibnamefont
  {Melnikov}}, \bibinfo {author} {\bibfnamefont {A.}~\bibnamefont {Aksentyev}},
  \bibinfo {author} {\bibfnamefont {N.}~\bibnamefont {Nikolaev}}, \bibinfo
  {author} {\bibfnamefont {S.}~\bibnamefont {Kolokolchikov}},\ and\ \bibinfo
  {author} {\bibfnamefont {Y.}~\bibnamefont {Senichev}},\ }\href
  {https://doi.org/10.18429/JACoW-IPAC2025-TUPS033} {\bibfield  {journal}
  {\bibinfo  {journal} {JACoW}\ }\textbf {\bibinfo {volume} {IPAC2025}},\
  \bibinfo {pages} {TUPS033} (\bibinfo {year} {2025})}\BibitemShut {NoStop}%
\bibitem [{\citenamefont {Nikolaev}\ \emph {et~al.}(2020)\citenamefont
  {Nikolaev}, \citenamefont {Rathmann}, \citenamefont {Silenko},\ and\
  \citenamefont {Uzikov}}]{NikolaevPrecessingD}%
  \BibitemOpen
  \bibfield  {author} {\bibinfo {author} {\bibfnamefont {N.}~\bibnamefont
  {Nikolaev}}, \bibinfo {author} {\bibfnamefont {F.}~\bibnamefont {Rathmann}},
  \bibinfo {author} {\bibfnamefont {A.}~\bibnamefont {Silenko}},\ and\ \bibinfo
  {author} {\bibfnamefont {Y.}~\bibnamefont {Uzikov}},\ }\href
  {https://doi.org/10.1016/j.physletb.2020.135983} {\bibfield  {journal}
  {\bibinfo  {journal} {Phys. Lett. B}\ }\textbf {\bibinfo {volume} {811}},\
  \bibinfo {pages} {135983} (\bibinfo {year} {2020})}\BibitemShut {NoStop}%
\bibitem [{\citenamefont {Koop}\ \emph {et~al.}(2020)\citenamefont {Koop},
  \citenamefont {Milstein}, \citenamefont {Nikolaev}, \citenamefont {Popov},
  \citenamefont {Salnikov}, \citenamefont {Shatunov},\ and\ \citenamefont
  {Shatunov}}]{Koop2020parity}%
  \BibitemOpen
  \bibfield  {author} {\bibinfo {author} {\bibfnamefont {I.~A.}\ \bibnamefont
  {Koop}}, \bibinfo {author} {\bibfnamefont {A.~I.}\ \bibnamefont {Milstein}},
  \bibinfo {author} {\bibfnamefont {N.~N.}\ \bibnamefont {Nikolaev}}, \bibinfo
  {author} {\bibfnamefont {A.~S.}\ \bibnamefont {Popov}}, \bibinfo {author}
  {\bibfnamefont {S.~G.}\ \bibnamefont {Salnikov}}, \bibinfo {author}
  {\bibfnamefont {P.~{\relax Yu}.}\ \bibnamefont {Shatunov}},\ and\ \bibinfo
  {author} {\bibfnamefont {{\relax Yu}.~M.}\ \bibnamefont {Shatunov}},\ }\href
  {https://doi.org/10.1134/S1547477120020107} {\bibfield  {journal} {\bibinfo
  {journal} {Phys. Part. Nucl. Lett.}\ }\textbf {\bibinfo {volume} {17}},\
  \bibinfo {pages} {154} (\bibinfo {year} {2020})}\BibitemShut {NoStop}%
\bibitem [{\citenamefont {Abramov}\ \emph {et~al.}(2021)\citenamefont
  {Abramov}, \citenamefont {Aleshko}, \citenamefont {Baskov}, \citenamefont
  {Boos}, \citenamefont {Bunichev}, \citenamefont {Dalkarov}, \citenamefont
  {El-Kholy}, \citenamefont {Galoyan}, \citenamefont {Guskov}, \citenamefont
  {Kim}, \citenamefont {Kokoulina}, \citenamefont {Koop}, \citenamefont
  {Kostenko}, \citenamefont {Kovalenko}, \citenamefont {Ladygin}, \citenamefont
  {Larionov}, \citenamefont {L’vov}, \citenamefont {Milstein}, \citenamefont
  {Nikitin}, \citenamefont {Nikolaev}, \citenamefont {Popov}, \citenamefont
  {Polyanskiy}, \citenamefont {Richard}, \citenamefont {Salnikov},
  \citenamefont {Shavrin}, \citenamefont {Shatunov}, \citenamefont {Shatunov},
  \citenamefont {Selyugin}, \citenamefont {Strikman}, \citenamefont
  {Tomasi-Gustafsson}, \citenamefont {Uzhinsky}, \citenamefont {Uzikov},
  \citenamefont {Wang}, \citenamefont {Zhao},\ and\ \citenamefont
  {Zelenov}}]{Abramov_2021}%
  \BibitemOpen
  \bibfield  {author} {\bibinfo {author} {\bibfnamefont {V.~V.}\ \bibnamefont
  {Abramov}}, \bibinfo {author} {\bibfnamefont {A.}~\bibnamefont {Aleshko}},
  \bibinfo {author} {\bibfnamefont {V.~A.}\ \bibnamefont {Baskov}}, \bibinfo
  {author} {\bibfnamefont {E.}~\bibnamefont {Boos}}, \bibinfo {author}
  {\bibfnamefont {V.}~\bibnamefont {Bunichev}}, \bibinfo {author}
  {\bibfnamefont {O.~D.}\ \bibnamefont {Dalkarov}}, \bibinfo {author}
  {\bibfnamefont {R.}~\bibnamefont {El-Kholy}}, \bibinfo {author}
  {\bibfnamefont {A.}~\bibnamefont {Galoyan}}, \bibinfo {author} {\bibfnamefont
  {A.~V.}\ \bibnamefont {Guskov}}, \bibinfo {author} {\bibfnamefont {V.~T.}\
  \bibnamefont {Kim}}, \bibinfo {author} {\bibfnamefont {E.}~\bibnamefont
  {Kokoulina}}, \bibinfo {author} {\bibfnamefont {I.~A.}\ \bibnamefont {Koop}},
  \bibinfo {author} {\bibfnamefont {B.~F.}\ \bibnamefont {Kostenko}}, \bibinfo
  {author} {\bibfnamefont {A.~D.}\ \bibnamefont {Kovalenko}}, \bibinfo {author}
  {\bibfnamefont {V.~P.}\ \bibnamefont {Ladygin}}, \bibinfo {author}
  {\bibfnamefont {A.~B.}\ \bibnamefont {Larionov}}, \bibinfo {author}
  {\bibfnamefont {A.~I.}\ \bibnamefont {L’vov}}, \bibinfo {author}
  {\bibfnamefont {A.~I.}\ \bibnamefont {Milstein}}, \bibinfo {author}
  {\bibfnamefont {V.~A.}\ \bibnamefont {Nikitin}}, \bibinfo {author}
  {\bibfnamefont {N.~N.}\ \bibnamefont {Nikolaev}}, \bibinfo {author}
  {\bibfnamefont {A.~S.}\ \bibnamefont {Popov}}, \bibinfo {author}
  {\bibfnamefont {V.~V.}\ \bibnamefont {Polyanskiy}}, \bibinfo {author}
  {\bibfnamefont {J.-M.}\ \bibnamefont {Richard}}, \bibinfo {author}
  {\bibfnamefont {S.~G.}\ \bibnamefont {Salnikov}}, \bibinfo {author}
  {\bibfnamefont {A.~A.}\ \bibnamefont {Shavrin}}, \bibinfo {author}
  {\bibfnamefont {P.~Y.}\ \bibnamefont {Shatunov}}, \bibinfo {author}
  {\bibfnamefont {Y.~M.}\ \bibnamefont {Shatunov}}, \bibinfo {author}
  {\bibfnamefont {O.~V.}\ \bibnamefont {Selyugin}}, \bibinfo {author}
  {\bibfnamefont {M.}~\bibnamefont {Strikman}}, \bibinfo {author}
  {\bibfnamefont {E.}~\bibnamefont {Tomasi-Gustafsson}}, \bibinfo {author}
  {\bibfnamefont {V.~V.}\ \bibnamefont {Uzhinsky}}, \bibinfo {author}
  {\bibfnamefont {Y.~N.}\ \bibnamefont {Uzikov}}, \bibinfo {author}
  {\bibfnamefont {Q.}~\bibnamefont {Wang}}, \bibinfo {author} {\bibfnamefont
  {Q.}~\bibnamefont {Zhao}},\ and\ \bibinfo {author} {\bibfnamefont {A.~V.}\
  \bibnamefont {Zelenov}},\ }\href {https://doi.org/10.1134/s1063779621060022}
  {\bibfield  {journal} {\bibinfo  {journal} {Physics of Particles and Nuclei}\
  }\textbf {\bibinfo {volume} {52}},\ \bibinfo {pages} {1044–1119} (\bibinfo
  {year} {2021})}\BibitemShut {NoStop}%
\bibitem [{\citenamefont {Temerbayev}\ and\ \citenamefont
  {Uzikov}(2015)}]{Temerbayev:2015foa}%
  \BibitemOpen
  \bibfield  {author} {\bibinfo {author} {\bibfnamefont {A.}~\bibnamefont
  {Temerbayev}}\ and\ \bibinfo {author} {\bibfnamefont {Y.}~\bibnamefont
  {Uzikov}},\ }\href {https://doi.org/10.1134/S1063778815010184} {\bibfield
  {journal} {\bibinfo  {journal} {Phys. Atom. Nucl.}\ }\textbf {\bibinfo
  {volume} {78}},\ \bibinfo {pages} {35} (\bibinfo {year} {2015})}\BibitemShut
  {NoStop}%
\bibitem [{\citenamefont {Uzikov}\ and\ \citenamefont
  {Temerbayev}(2015)}]{Uzikov:2015aua}%
  \BibitemOpen
  \bibfield  {author} {\bibinfo {author} {\bibfnamefont {Y.~N.}\ \bibnamefont
  {Uzikov}}\ and\ \bibinfo {author} {\bibfnamefont {A.}~\bibnamefont
  {Temerbayev}},\ }\href {https://doi.org/10.1103/PhysRevC.92.014002}
  {\bibfield  {journal} {\bibinfo  {journal} {Phys. Rev. C}\ }\textbf {\bibinfo
  {volume} {92}},\ \bibinfo {pages} {014002} (\bibinfo {year} {2015})},\
  \Eprint {https://arxiv.org/abs/1506.08303} {arXiv:1506.08303 [nucl-th]}
  \BibitemShut {NoStop}%
\bibitem [{\citenamefont {Uzikov}\ and\ \citenamefont
  {Haidenbauer}(2016)}]{Uzikov:2016lsc}%
  \BibitemOpen
  \bibfield  {author} {\bibinfo {author} {\bibfnamefont {Y.~N.}\ \bibnamefont
  {Uzikov}}\ and\ \bibinfo {author} {\bibfnamefont {J.}~\bibnamefont
  {Haidenbauer}},\ }\href {https://doi.org/10.1103/PhysRevC.94.035501}
  {\bibfield  {journal} {\bibinfo  {journal} {Phys. Rev. C}\ }\textbf {\bibinfo
  {volume} {94}},\ \bibinfo {pages} {035501} (\bibinfo {year} {2016})},\
  \Eprint {https://arxiv.org/abs/1607.04409} {arXiv:1607.04409 [nucl-th]}
  \BibitemShut {NoStop}%
\bibitem [{\citenamefont {Uzikov}\ and\ \citenamefont
  {Platonova}(2025)}]{Uzikov_pD_NICA}%
  \BibitemOpen
  \bibfield  {author} {\bibinfo {author} {\bibfnamefont {Y.~N.}\ \bibnamefont
  {Uzikov}}\ and\ \bibinfo {author} {\bibfnamefont {M.~N.}\ \bibnamefont
  {Platonova}},\ }\href {https://doi.org/10.1134/S1063779624701910} {\bibfield
  {journal} {\bibinfo  {journal} {Phys. Part. Nucl.}\ }\textbf {\bibinfo
  {volume} {56}},\ \bibinfo {pages} {533} (\bibinfo {year} {2025})},\ \Eprint
  {https://arxiv.org/abs/2410.15337} {arXiv:2410.15337 [nucl-th]} \BibitemShut
  {NoStop}%
\bibitem [{\citenamefont {Platonova}\ and\ \citenamefont
  {Uzikov}(2025)}]{Platonova_T-odd-DD}%
  \BibitemOpen
  \bibfield  {author} {\bibinfo {author} {\bibfnamefont {M.~N.}\ \bibnamefont
  {Platonova}}\ and\ \bibinfo {author} {\bibfnamefont {Y.~N.}\ \bibnamefont
  {Uzikov}},\ }\href {https://doi.org/10.1088/1674-1137/ad9b9f} {\bibfield
  {journal} {\bibinfo  {journal} {Chin. Phys. C}\ }\textbf {\bibinfo {volume}
  {49}},\ \bibinfo {pages} {034108} (\bibinfo {year} {2025})},\ \Eprint
  {https://arxiv.org/abs/2410.03262} {arXiv:2410.03262 [nucl-th]} \BibitemShut
  {NoStop}%
\bibitem [{\citenamefont {Okun}(1965)}]{OkunMillistrong}%
  \BibitemOpen
  \bibfield  {author} {\bibinfo {author} {\bibfnamefont {L.}~\bibnamefont
  {Okun}},\ }\href@noop {} {\bibfield  {journal} {\bibinfo  {journal} {Sov. J.
  Nucl. Phys.}\ }\textbf {\bibinfo {volume} {1}},\ \bibinfo {pages} {670}
  (\bibinfo {year} {1965})}\BibitemShut {NoStop}%
\bibitem [{\citenamefont {Prentki}\ and\ \citenamefont
  {Veltman}(1965)}]{PrentkiMillistrong}%
  \BibitemOpen
  \bibfield  {author} {\bibinfo {author} {\bibfnamefont {J.}~\bibnamefont
  {Prentki}}\ and\ \bibinfo {author} {\bibfnamefont {M.~J.~G.}\ \bibnamefont
  {Veltman}},\ }\href {https://doi.org/10.1016/0031-9163(65)91141-8} {\bibfield
   {journal} {\bibinfo  {journal} {Phys. Lett.}\ }\textbf {\bibinfo {volume}
  {15}},\ \bibinfo {pages} {88} (\bibinfo {year} {1965})}\BibitemShut {NoStop}%
%%CITATION = PHLTA,15,88;%%
\bibitem [{\citenamefont {Lee}\ and\ \citenamefont
  {Wolfenstein}(1965)}]{LeeMillistrong}%
  \BibitemOpen
  \bibfield  {author} {\bibinfo {author} {\bibfnamefont {T.~D.}\ \bibnamefont
  {Lee}}\ and\ \bibinfo {author} {\bibfnamefont {L.}~\bibnamefont
  {Wolfenstein}},\ }\href {https://doi.org/10.1103/PhysRev.138.B1490}
  {\bibfield  {journal} {\bibinfo  {journal} {Phys. Rev.}\ }\textbf {\bibinfo
  {volume} {138}},\ \bibinfo {pages} {B1490} (\bibinfo {year}
  {1965})}\BibitemShut {NoStop}%
%%CITATION = PHRVA,138,B1490;%%
\bibitem [{\citenamefont {Vergeles}\ \emph {et~al.}(2023)\citenamefont
  {Vergeles}, \citenamefont {Nikolaev}, \citenamefont {Obukhov}, \citenamefont
  {Silenko},\ and\ \citenamefont {Teryaev}}]{UFNreview}%
  \BibitemOpen
  \bibfield  {author} {\bibinfo {author} {\bibfnamefont {S.~N.}\ \bibnamefont
  {Vergeles}}, \bibinfo {author} {\bibfnamefont {N.~N.}\ \bibnamefont
  {Nikolaev}}, \bibinfo {author} {\bibfnamefont {Y.~N.}\ \bibnamefont
  {Obukhov}}, \bibinfo {author} {\bibfnamefont {A.~Y.}\ \bibnamefont
  {Silenko}},\ and\ \bibinfo {author} {\bibfnamefont {O.~V.}\ \bibnamefont
  {Teryaev}},\ }\href {https://doi.org/10.3367/UFNe.2021.09.039074} {\bibfield
  {journal} {\bibinfo  {journal} {Phys. Usp.}\ }\textbf {\bibinfo {volume}
  {66}},\ \bibinfo {pages} {109} (\bibinfo {year} {2023})}\BibitemShut
  {NoStop}%
\bibitem [{\citenamefont {Valdau}(2012)}]{TIVOLI:STORI11}%
  \BibitemOpen
  \bibfield  {author} {\bibinfo {author} {\bibfnamefont {Y.}~\bibnamefont
  {Valdau}},\ }in\ \href {https://doi.org/10.22323/1.150.0013} {\emph {\bibinfo
  {booktitle} {Proceedings of 8th International Conference on Nuclear Physics
  at Storage Rings {\textemdash} PoS(STORI11)}}},\ Vol.\ \bibinfo {volume}
  {150}\ (\bibinfo {year} {2012})\ p.\ \bibinfo {pages} {013}\BibitemShut
  {NoStop}%
\bibitem [{\citenamefont {Lenisa}\ \emph {et~al.}(2019)\citenamefont {Lenisa},
  \citenamefont {Rathmann}, \citenamefont {Barion}, \citenamefont {Barsov},
  \citenamefont {Bertelli}, \citenamefont {Carassiti}, \citenamefont {Ciullo},
  \citenamefont {Contalbrigo}, \citenamefont {Ramusino}, \citenamefont {Dymov},
  \citenamefont {Engels}, \citenamefont {Eversheim}, \citenamefont {Gebel},
  \citenamefont {Grigoryev}, \citenamefont {Haidenbauer}, \citenamefont
  {Hejny}, \citenamefont {Jagdfeld}, \citenamefont {Kacharava}, \citenamefont
  {Keshelashvili}, \citenamefont {Kononov}, \citenamefont {Krings},
  \citenamefont {Kulikov}, \citenamefont {Lehrach}, \citenamefont {Lorentz},
  \citenamefont {Lomidze}, \citenamefont {Macharashvili}, \citenamefont
  {Malaguti}, \citenamefont {Martin}, \citenamefont {Merzliakov}, \citenamefont
  {Mikirtychiants}, \citenamefont {Nass}, \citenamefont {Nikolaev},
  \citenamefont {Pesce}, \citenamefont {Prasuhn}, \citenamefont {Semke},
  \citenamefont {Squerzanti}, \citenamefont {Soltner}, \citenamefont {Statera},
  \citenamefont {Steffens}, \citenamefont {Str{\"o}her}, \citenamefont
  {Tabidze}, \citenamefont {Tagliente}, \citenamefont {Th{\"o}rngren-Engblom},
  \citenamefont {Trusov}, \citenamefont {Uzikov}, \citenamefont {Valdau},
  \citenamefont {Weidemann}, \citenamefont {W{\"u}stner},\ and\ \citenamefont
  {Zupranski}}]{LenisaTIVOLI}%
  \BibitemOpen
  \bibfield  {author} {\bibinfo {author} {\bibfnamefont {P.}~\bibnamefont
  {Lenisa}}, \bibinfo {author} {\bibfnamefont {F.}~\bibnamefont {Rathmann}},
  \bibinfo {author} {\bibfnamefont {L.}~\bibnamefont {Barion}}, \bibinfo
  {author} {\bibfnamefont {S.}~\bibnamefont {Barsov}}, \bibinfo {author}
  {\bibfnamefont {S.}~\bibnamefont {Bertelli}}, \bibinfo {author}
  {\bibfnamefont {V.}~\bibnamefont {Carassiti}}, \bibinfo {author}
  {\bibfnamefont {G.}~\bibnamefont {Ciullo}}, \bibinfo {author} {\bibfnamefont
  {M.}~\bibnamefont {Contalbrigo}}, \bibinfo {author} {\bibfnamefont {A.~C.}\
  \bibnamefont {Ramusino}}, \bibinfo {author} {\bibfnamefont {S.}~\bibnamefont
  {Dymov}}, \bibinfo {author} {\bibfnamefont {R.}~\bibnamefont {Engels}},
  \bibinfo {author} {\bibfnamefont {D.}~\bibnamefont {Eversheim}}, \bibinfo
  {author} {\bibfnamefont {R.}~\bibnamefont {Gebel}}, \bibinfo {author}
  {\bibfnamefont {K.}~\bibnamefont {Grigoryev}}, \bibinfo {author}
  {\bibfnamefont {J.}~\bibnamefont {Haidenbauer}}, \bibinfo {author}
  {\bibfnamefont {V.}~\bibnamefont {Hejny}}, \bibinfo {author} {\bibfnamefont
  {H.}~\bibnamefont {Jagdfeld}}, \bibinfo {author} {\bibfnamefont
  {A.}~\bibnamefont {Kacharava}}, \bibinfo {author} {\bibfnamefont
  {I.}~\bibnamefont {Keshelashvili}}, \bibinfo {author} {\bibfnamefont
  {A.}~\bibnamefont {Kononov}}, \bibinfo {author} {\bibfnamefont
  {T.}~\bibnamefont {Krings}}, \bibinfo {author} {\bibfnamefont
  {A.}~\bibnamefont {Kulikov}}, \bibinfo {author} {\bibfnamefont
  {A.}~\bibnamefont {Lehrach}}, \bibinfo {author} {\bibfnamefont
  {B.}~\bibnamefont {Lorentz}}, \bibinfo {author} {\bibfnamefont
  {N.}~\bibnamefont {Lomidze}}, \bibinfo {author} {\bibfnamefont
  {G.}~\bibnamefont {Macharashvili}}, \bibinfo {author} {\bibfnamefont
  {R.}~\bibnamefont {Malaguti}}, \bibinfo {author} {\bibfnamefont
  {S.}~\bibnamefont {Martin}}, \bibinfo {author} {\bibfnamefont
  {S.}~\bibnamefont {Merzliakov}}, \bibinfo {author} {\bibfnamefont
  {S.}~\bibnamefont {Mikirtychiants}}, \bibinfo {author} {\bibfnamefont
  {A.}~\bibnamefont {Nass}}, \bibinfo {author} {\bibfnamefont {N.}~\bibnamefont
  {Nikolaev}}, \bibinfo {author} {\bibfnamefont {A.}~\bibnamefont {Pesce}},
  \bibinfo {author} {\bibfnamefont {D.}~\bibnamefont {Prasuhn}}, \bibinfo
  {author} {\bibfnamefont {L.}~\bibnamefont {Semke}}, \bibinfo {author}
  {\bibfnamefont {S.}~\bibnamefont {Squerzanti}}, \bibinfo {author}
  {\bibfnamefont {H.}~\bibnamefont {Soltner}}, \bibinfo {author} {\bibfnamefont
  {M.}~\bibnamefont {Statera}}, \bibinfo {author} {\bibfnamefont
  {E.}~\bibnamefont {Steffens}}, \bibinfo {author} {\bibfnamefont
  {H.}~\bibnamefont {Str{\"o}her}}, \bibinfo {author} {\bibfnamefont
  {M.}~\bibnamefont {Tabidze}}, \bibinfo {author} {\bibfnamefont
  {G.}~\bibnamefont {Tagliente}}, \bibinfo {author} {\bibfnamefont
  {P.}~\bibnamefont {Th{\"o}rngren-Engblom}}, \bibinfo {author} {\bibfnamefont
  {S.}~\bibnamefont {Trusov}}, \bibinfo {author} {\bibfnamefont
  {Y.}~\bibnamefont {Uzikov}}, \bibinfo {author} {\bibfnamefont
  {Y.}~\bibnamefont {Valdau}}, \bibinfo {author} {\bibfnamefont
  {C.}~\bibnamefont {Weidemann}}, \bibinfo {author} {\bibfnamefont
  {P.}~\bibnamefont {W{\"u}stner}},\ and\ \bibinfo {author} {\bibfnamefont
  {P.}~\bibnamefont {Zupranski}},\ }\href
  {https://doi.org/10.1140/epjti/s40485-019-0051-y} {\bibfield  {journal}
  {\bibinfo  {journal} {EPJ Techniques and Instrumentation}\ }\textbf {\bibinfo
  {volume} {6}},\ \bibinfo {pages} {2} (\bibinfo {year} {2019})}\BibitemShut
  {NoStop}%
\bibitem [{\citenamefont {Kalantarians}(2014)}]{TensorDeuteron-eIC}%
  \BibitemOpen
  \bibfield  {author} {\bibinfo {author} {\bibfnamefont {N.}~\bibnamefont
  {Kalantarians}},\ }\href {https://doi.org/10.1088/1742-6596/543/1/012008}
  {\bibfield  {journal} {\bibinfo  {journal} {Journal of Physics: Conference
  Series}\ }\textbf {\bibinfo {volume} {543}},\ \bibinfo {pages} {012008}
  (\bibinfo {year} {2014})}\BibitemShut {NoStop}%
\bibitem [{\citenamefont {{Abdul Khalek}}\ \emph {et~al.}(2022)\citenamefont
  {{Abdul Khalek}}, \citenamefont {Accardi}, \citenamefont {Adam},
  \citenamefont {Adamiak}, \citenamefont {Akers}, \citenamefont {Albaladejo},
  \citenamefont {Al-bataineh}, \citenamefont {Alexeev}, \citenamefont {Ameli},
  \citenamefont {Antonioli}, \citenamefont {Armesto}, \citenamefont
  {Armstrong}, \citenamefont {Arratia}, \citenamefont {Arrington},
  \citenamefont {Asaturyan}, \citenamefont {Asai}, \citenamefont {Aschenauer},
  \citenamefont {Aune}, \citenamefont {Avagyan}, \citenamefont {{Ayerbe
  Gayoso}}, \citenamefont {Azmoun}, \citenamefont {Bacchetta}, \citenamefont
  {Baker}, \citenamefont {Barbosa}, \citenamefont {Barion}, \citenamefont
  {Barish}, \citenamefont {Barry}, \citenamefont {Battaglieri}, \citenamefont
  {Bazilevsky}, \citenamefont {Behera}, \citenamefont {Benmokhtar},
  \citenamefont {Berdnikov}, \citenamefont {Bernauer}, \citenamefont {Bertone},
  \citenamefont {Bhattacharya}, \citenamefont {Bissolotti}, \citenamefont
  {Boer}, \citenamefont {Boglione}, \citenamefont {Bondì}, \citenamefont
  {Boora}, \citenamefont {Borsa}, \citenamefont {Bossù}, \citenamefont
  {Bozzi}, \citenamefont {Brandenburg}, \citenamefont {Brei}, \citenamefont
  {Bressan}, \citenamefont {Brooks}, \citenamefont {Bufalino}, \citenamefont
  {Bukhari}, \citenamefont {Burkert}, \citenamefont {Buttimore}, \citenamefont
  {Camsonne}, \citenamefont {Celentano}, \citenamefont {Celiberto},
  \citenamefont {Chang}, \citenamefont {Chatterjee}, \citenamefont {Chen},
  \citenamefont {Chetry}, \citenamefont {Chiarusi}, \citenamefont {Chien},
  \citenamefont {Chiosso}, \citenamefont {Chu}, \citenamefont {Chudakov},
  \citenamefont {Cicala}, \citenamefont {Cisbani}, \citenamefont {Cloet},
  \citenamefont {Cocuzza}, \citenamefont {Cole}, \citenamefont {Colella},
  \citenamefont {Collins}, \citenamefont {Constantinou}, \citenamefont
  {Contalbrigo}, \citenamefont {Contin}, \citenamefont {Corliss}, \citenamefont
  {Cosyn}, \citenamefont {Courtoy}, \citenamefont {Crafts}, \citenamefont
  {Cruz-Torres}, \citenamefont {Cuevas}, \citenamefont {D'Alesio},
  \citenamefont {{Dalla Torre}}, \citenamefont {Das}, \citenamefont {Dasgupta},
  \citenamefont {{Da Silva}}, \citenamefont {Deconinck}, \citenamefont
  {Defurne}, \citenamefont {DeGraw}, \citenamefont {Dehmelt}, \citenamefont
  {{Del Dotto}}, \citenamefont {Delcarro}, \citenamefont {Deshpande},
  \citenamefont {Detmold}, \citenamefont {{De Vita}}, \citenamefont
  {Diefenthaler}, \citenamefont {Dilks}, \citenamefont {Dixit}, \citenamefont
  {Dulat}, \citenamefont {Dumitru}, \citenamefont {Dupré}, \citenamefont
  {Durham}, \citenamefont {Echevarria}, \citenamefont {{El Fassi}},
  \citenamefont {Elia}, \citenamefont {Ent}, \citenamefont {Esha},
  \citenamefont {Ethier}, \citenamefont {Evdokimov}, \citenamefont {Eyser},
  \citenamefont {Fanelli}, \citenamefont {Fatemi}, \citenamefont {Fazio},
  \citenamefont {Fernandez-Ramirez}, \citenamefont {Finger}, \citenamefont
  {Finger}, \citenamefont {Fitzgerald}, \citenamefont {Flore}, \citenamefont
  {Frederico}, \citenamefont {Friščić}, \citenamefont {Fucini},
  \citenamefont {Furletov}, \citenamefont {Furletova}, \citenamefont {Gal},
  \citenamefont {Gamberg}, \citenamefont {Gao}, \citenamefont {Garg},
  \citenamefont {Gaskell}, \citenamefont {Gates}, \citenamefont {{Gay Ducati}},
  \citenamefont {Gericke}, \citenamefont {{Gil Da Silveira}}, \citenamefont
  {Girod}, \citenamefont {Glazier}, \citenamefont {Gnanvo}, \citenamefont
  {Goncalves}, \citenamefont {Gonella}, \citenamefont {{Gonzalez Hernandez}},
  \citenamefont {Goto}, \citenamefont {Grancagnolo}, \citenamefont {Greiner},
  \citenamefont {Guryn}, \citenamefont {Guzey}, \citenamefont {Hatta},
  \citenamefont {Hattawy}, \citenamefont {Hauenstein}, \citenamefont {He},
  \citenamefont {Hemmick}, \citenamefont {Hen}, \citenamefont {Heyes},
  \citenamefont {Higinbotham}, \citenamefont {{Hiller Blin}}, \citenamefont
  {Hobbs}, \citenamefont {Hohlmann}, \citenamefont {Horn}, \citenamefont {Hou},
  \citenamefont {Huang}, \citenamefont {Huang}, \citenamefont {Huber},
  \citenamefont {Hyde}, \citenamefont {Iakovidis}, \citenamefont {Ilieva},
  \citenamefont {Jacak}, \citenamefont {Jacobs}, \citenamefont {Jadhav},
  \citenamefont {Janoska}, \citenamefont {Jentsch}, \citenamefont {Jezo},
  \citenamefont {Jing}, \citenamefont {Jones}, \citenamefont {Joo},
  \citenamefont {Joosten}, \citenamefont {Kafka}, \citenamefont {Kalantarians},
  \citenamefont {Kalicy}, \citenamefont {Kang}, \citenamefont {Kang},
  \citenamefont {Kauder}, \citenamefont {Kay}, \citenamefont {Keppel},
  \citenamefont {Kim}, \citenamefont {Kiselev}, \citenamefont {Klasen},
  \citenamefont {Klein}, \citenamefont {Klest}, \citenamefont {Korchak},
  \citenamefont {Kostina}, \citenamefont {Kotko}, \citenamefont {Kovchegov},
  \citenamefont {Krelina}, \citenamefont {Kuleshov}, \citenamefont {Kumano},
  \citenamefont {Kumar}, \citenamefont {Kumar}, \citenamefont {Kumar},
  \citenamefont {Kumerički}, \citenamefont {Kusina}, \citenamefont {Kutak},
  \citenamefont {Lai}, \citenamefont {Lalwani}, \citenamefont {Lappi},
  \citenamefont {Lauret}, \citenamefont {Lavinsky}, \citenamefont {Lawrence},
  \citenamefont {Lednicky}, \citenamefont {Lee}, \citenamefont {Lee},
  \citenamefont {Lee}, \citenamefont {Levorato}, \citenamefont {Li},
  \citenamefont {Li}, \citenamefont {Li}, \citenamefont {Li}, \citenamefont
  {Li}, \citenamefont {Li}, \citenamefont {Ligonzo}, \citenamefont {Liu},
  \citenamefont {Liu}, \citenamefont {Liu}, \citenamefont {Liuti},
  \citenamefont {Liyanage}, \citenamefont {Lorcé}, \citenamefont {Lu},
  \citenamefont {Lucero}, \citenamefont {Lukow}, \citenamefont {Lunghi},
  \citenamefont {Majka}, \citenamefont {Makris}, \citenamefont {Mandjavidze},
  \citenamefont {Mantry}, \citenamefont {Mäntysaari}, \citenamefont
  {Marhauser}, \citenamefont {Markowitz}, \citenamefont {Marsicano},
  \citenamefont {Mastroserio}, \citenamefont {Mathieu}, \citenamefont
  {Mehtar-Tani}, \citenamefont {Melnitchouk}, \citenamefont {Mendez},
  \citenamefont {Metz}, \citenamefont {Meziani}, \citenamefont {Mezrag},
  \citenamefont {Mihovilovič}, \citenamefont {Milner}, \citenamefont
  {Mirazita}, \citenamefont {Mkrtchyan}, \citenamefont {Mkrtchyan},
  \citenamefont {Mochalov}, \citenamefont {Moiseev}, \citenamefont {Mondal},
  \citenamefont {Morreale}, \citenamefont {Morrison}, \citenamefont {Motyka},
  \citenamefont {Moutarde}, \citenamefont {{Muñoz Camacho}}, \citenamefont
  {Murgia}, \citenamefont {Murray}, \citenamefont {Musico}, \citenamefont
  {Nadel-Turonski}, \citenamefont {Nadolsky}, \citenamefont {Nam},
  \citenamefont {Newman}, \citenamefont {Neyret}, \citenamefont {Nguyen},
  \citenamefont {Nocera}, \citenamefont {Noferini}, \citenamefont {Noto},
  \citenamefont {Nunes}, \citenamefont {Okorokov}, \citenamefont {Olness},
  \citenamefont {Osborn}, \citenamefont {Page}, \citenamefont {Park},
  \citenamefont {Parker}, \citenamefont {Paschke}, \citenamefont {Pasquini},
  \citenamefont {Paukkunen}, \citenamefont {Paul}, \citenamefont {Pecar},
  \citenamefont {Pegg}, \citenamefont {Pellegrino}, \citenamefont {Peng},
  \citenamefont {Pentchev}, \citenamefont {Perrino}, \citenamefont {Petriello},
  \citenamefont {Petti}, \citenamefont {Pilloni}, \citenamefont {Pinkenburg},
  \citenamefont {Pire}, \citenamefont {Pisano}, \citenamefont {Pitonyak},
  \citenamefont {Poblaguev}, \citenamefont {Polakovic}, \citenamefont {Posik},
  \citenamefont {Potekhin}, \citenamefont {Preghenella}, \citenamefont
  {Preins}, \citenamefont {Prokudin}, \citenamefont {Pujahari}, \citenamefont
  {Purschke}, \citenamefont {Pybus}, \citenamefont {Radici}, \citenamefont
  {Rajput-Ghoshal}, \citenamefont {Reimer}, \citenamefont {Rinaldi},
  \citenamefont {Ringer}, \citenamefont {Roberts}, \citenamefont {Rodini},
  \citenamefont {Rojo}, \citenamefont {Romanov}, \citenamefont {Rossi},
  \citenamefont {Santopinto}, \citenamefont {Sarsour}, \citenamefont {Sassot},
  \citenamefont {Sato}, \citenamefont {Schenke}, \citenamefont {Schmidke},
  \citenamefont {Schmidt}, \citenamefont {Schmidt}, \citenamefont {Schmookler},
  \citenamefont {Schnell}, \citenamefont {Schweitzer}, \citenamefont
  {Schwiening}, \citenamefont {Scimemi}, \citenamefont {Scopetta},
  \citenamefont {Segovia}, \citenamefont {Seidl}, \citenamefont {Sekula},
  \citenamefont {Semenov-Tian-Shanskiy}, \citenamefont {Shao}, \citenamefont
  {Sherrill}, \citenamefont {Sichtermann}, \citenamefont {Siddikov},
  \citenamefont {Signori}, \citenamefont {Singh}, \citenamefont {Širca},
  \citenamefont {Slifer}, \citenamefont {Slominski}, \citenamefont {Sokhan},
  \citenamefont {Sondheim}, \citenamefont {Song}, \citenamefont {Soto},
  \citenamefont {Spiesberger}, \citenamefont {Stasto}, \citenamefont
  {Stepanov}, \citenamefont {Sterman}, \citenamefont {Stevens}, \citenamefont
  {Stewart}, \citenamefont {Strakovsky}, \citenamefont {Strikman},
  \citenamefont {Sturm}, \citenamefont {Stutzman}, \citenamefont {Sullivan},
  \citenamefont {Surrow}, \citenamefont {Svihra}, \citenamefont {Syritsyn},
  \citenamefont {Szczepaniak}, \citenamefont {Sznajder}, \citenamefont
  {Szumila-Vance}, \citenamefont {Szymanowski}, \citenamefont {Tadepalli},
  \citenamefont {{Tapia Takaki}}, \citenamefont {Tassielli}, \citenamefont
  {Terry}, \citenamefont {Tessarotto}, \citenamefont {Tezgin}, \citenamefont
  {Tomasek}, \citenamefont {{Torales Acosta}}, \citenamefont {Tribedy},
  \citenamefont {Tricoli}, \citenamefont {Triloki}, \citenamefont {Tripathi},
  \citenamefont {Trotta}, \citenamefont {Tsai}, \citenamefont {Tu},
  \citenamefont {Tuvè}, \citenamefont {Ullrich}, \citenamefont {Ungaro},
  \citenamefont {Urciuoli}, \citenamefont {Valentini}, \citenamefont {Vancura},
  \citenamefont {Vandenbroucke}, \citenamefont {{Van Hulse}}, \citenamefont
  {Varner}, \citenamefont {Venugopalan}, \citenamefont {Vitev}, \citenamefont
  {Vladimirov}, \citenamefont {Volpe}, \citenamefont {Vossen}, \citenamefont
  {Voutier}, \citenamefont {Wagner}, \citenamefont {Wallon}, \citenamefont
  {Wang}, \citenamefont {Wang}, \citenamefont {Wang}, \citenamefont {Wei},
  \citenamefont {Weiss}, \citenamefont {Wenaus}, \citenamefont {Wennlöf},
  \citenamefont {Wickramaarachchi}, \citenamefont {Wikramanayake},
  \citenamefont {Winney}, \citenamefont {Wong}, \citenamefont {Woody},
  \citenamefont {Xia}, \citenamefont {Xiao}, \citenamefont {Xie}, \citenamefont
  {Xing}, \citenamefont {Xu}, \citenamefont {Zhang}, \citenamefont {Zhang},
  \citenamefont {Zhang}, \citenamefont {Zhao}, \citenamefont {Zhao},
  \citenamefont {Zheng}, \citenamefont {Zhou},\ and\ \citenamefont
  {Zurita}}]{eIC_Yellow_Report}%
  \BibitemOpen
  \bibfield  {author} {\bibinfo {author} {\bibfnamefont {R.}~\bibnamefont
  {{Abdul Khalek}}}, \bibinfo {author} {\bibfnamefont {A.}~\bibnamefont
  {Accardi}}, \bibinfo {author} {\bibfnamefont {J.}~\bibnamefont {Adam}},
  \bibinfo {author} {\bibfnamefont {D.}~\bibnamefont {Adamiak}}, \bibinfo
  {author} {\bibfnamefont {W.}~\bibnamefont {Akers}}, \bibinfo {author}
  {\bibfnamefont {M.}~\bibnamefont {Albaladejo}}, \bibinfo {author}
  {\bibfnamefont {A.}~\bibnamefont {Al-bataineh}}, \bibinfo {author}
  {\bibfnamefont {M.}~\bibnamefont {Alexeev}}, \bibinfo {author} {\bibfnamefont
  {F.}~\bibnamefont {Ameli}}, \bibinfo {author} {\bibfnamefont
  {P.}~\bibnamefont {Antonioli}}, \bibinfo {author} {\bibfnamefont
  {N.}~\bibnamefont {Armesto}}, \bibinfo {author} {\bibfnamefont
  {W.}~\bibnamefont {Armstrong}}, \bibinfo {author} {\bibfnamefont
  {M.}~\bibnamefont {Arratia}}, \bibinfo {author} {\bibfnamefont
  {J.}~\bibnamefont {Arrington}}, \bibinfo {author} {\bibfnamefont
  {A.}~\bibnamefont {Asaturyan}}, \bibinfo {author} {\bibfnamefont
  {M.}~\bibnamefont {Asai}}, \bibinfo {author} {\bibfnamefont {E.}~\bibnamefont
  {Aschenauer}}, \bibinfo {author} {\bibfnamefont {S.}~\bibnamefont {Aune}},
  \bibinfo {author} {\bibfnamefont {H.}~\bibnamefont {Avagyan}}, \bibinfo
  {author} {\bibfnamefont {C.}~\bibnamefont {{Ayerbe Gayoso}}}, \bibinfo
  {author} {\bibfnamefont {B.}~\bibnamefont {Azmoun}}, \bibinfo {author}
  {\bibfnamefont {A.}~\bibnamefont {Bacchetta}}, \bibinfo {author}
  {\bibfnamefont {M.}~\bibnamefont {Baker}}, \bibinfo {author} {\bibfnamefont
  {F.}~\bibnamefont {Barbosa}}, \bibinfo {author} {\bibfnamefont
  {L.}~\bibnamefont {Barion}}, \bibinfo {author} {\bibfnamefont
  {K.}~\bibnamefont {Barish}}, \bibinfo {author} {\bibfnamefont
  {P.}~\bibnamefont {Barry}}, \bibinfo {author} {\bibfnamefont
  {M.}~\bibnamefont {Battaglieri}}, \bibinfo {author} {\bibfnamefont
  {A.}~\bibnamefont {Bazilevsky}}, \bibinfo {author} {\bibfnamefont
  {N.}~\bibnamefont {Behera}}, \bibinfo {author} {\bibfnamefont
  {F.}~\bibnamefont {Benmokhtar}}, \bibinfo {author} {\bibfnamefont
  {V.}~\bibnamefont {Berdnikov}}, \bibinfo {author} {\bibfnamefont
  {J.}~\bibnamefont {Bernauer}}, \bibinfo {author} {\bibfnamefont
  {V.}~\bibnamefont {Bertone}}, \bibinfo {author} {\bibfnamefont
  {S.}~\bibnamefont {Bhattacharya}}, \bibinfo {author} {\bibfnamefont
  {C.}~\bibnamefont {Bissolotti}}, \bibinfo {author} {\bibfnamefont
  {D.}~\bibnamefont {Boer}}, \bibinfo {author} {\bibfnamefont {M.}~\bibnamefont
  {Boglione}}, \bibinfo {author} {\bibfnamefont {M.}~\bibnamefont {Bondì}},
  \bibinfo {author} {\bibfnamefont {P.}~\bibnamefont {Boora}}, \bibinfo
  {author} {\bibfnamefont {I.}~\bibnamefont {Borsa}}, \bibinfo {author}
  {\bibfnamefont {F.}~\bibnamefont {Bossù}}, \bibinfo {author} {\bibfnamefont
  {G.}~\bibnamefont {Bozzi}}, \bibinfo {author} {\bibfnamefont
  {J.}~\bibnamefont {Brandenburg}}, \bibinfo {author} {\bibfnamefont
  {N.}~\bibnamefont {Brei}}, \bibinfo {author} {\bibfnamefont {A.}~\bibnamefont
  {Bressan}}, \bibinfo {author} {\bibfnamefont {W.}~\bibnamefont {Brooks}},
  \bibinfo {author} {\bibfnamefont {S.}~\bibnamefont {Bufalino}}, \bibinfo
  {author} {\bibfnamefont {M.}~\bibnamefont {Bukhari}}, \bibinfo {author}
  {\bibfnamefont {V.}~\bibnamefont {Burkert}}, \bibinfo {author} {\bibfnamefont
  {N.}~\bibnamefont {Buttimore}}, \bibinfo {author} {\bibfnamefont
  {A.}~\bibnamefont {Camsonne}}, \bibinfo {author} {\bibfnamefont
  {A.}~\bibnamefont {Celentano}}, \bibinfo {author} {\bibfnamefont
  {F.}~\bibnamefont {Celiberto}}, \bibinfo {author} {\bibfnamefont
  {W.}~\bibnamefont {Chang}}, \bibinfo {author} {\bibfnamefont
  {C.}~\bibnamefont {Chatterjee}}, \bibinfo {author} {\bibfnamefont
  {K.}~\bibnamefont {Chen}}, \bibinfo {author} {\bibfnamefont {T.}~\bibnamefont
  {Chetry}}, \bibinfo {author} {\bibfnamefont {T.}~\bibnamefont {Chiarusi}},
  \bibinfo {author} {\bibfnamefont {Y.-T.}\ \bibnamefont {Chien}}, \bibinfo
  {author} {\bibfnamefont {M.}~\bibnamefont {Chiosso}}, \bibinfo {author}
  {\bibfnamefont {X.}~\bibnamefont {Chu}}, \bibinfo {author} {\bibfnamefont
  {E.}~\bibnamefont {Chudakov}}, \bibinfo {author} {\bibfnamefont
  {G.}~\bibnamefont {Cicala}}, \bibinfo {author} {\bibfnamefont
  {E.}~\bibnamefont {Cisbani}}, \bibinfo {author} {\bibfnamefont
  {I.}~\bibnamefont {Cloet}}, \bibinfo {author} {\bibfnamefont
  {C.}~\bibnamefont {Cocuzza}}, \bibinfo {author} {\bibfnamefont
  {P.}~\bibnamefont {Cole}}, \bibinfo {author} {\bibfnamefont {D.}~\bibnamefont
  {Colella}}, \bibinfo {author} {\bibfnamefont {J.}~\bibnamefont {Collins}},
  \bibinfo {author} {\bibfnamefont {M.}~\bibnamefont {Constantinou}}, \bibinfo
  {author} {\bibfnamefont {M.}~\bibnamefont {Contalbrigo}}, \bibinfo {author}
  {\bibfnamefont {G.}~\bibnamefont {Contin}}, \bibinfo {author} {\bibfnamefont
  {R.}~\bibnamefont {Corliss}}, \bibinfo {author} {\bibfnamefont
  {W.}~\bibnamefont {Cosyn}}, \bibinfo {author} {\bibfnamefont
  {A.}~\bibnamefont {Courtoy}}, \bibinfo {author} {\bibfnamefont
  {J.}~\bibnamefont {Crafts}}, \bibinfo {author} {\bibfnamefont
  {R.}~\bibnamefont {Cruz-Torres}}, \bibinfo {author} {\bibfnamefont
  {R.}~\bibnamefont {Cuevas}}, \bibinfo {author} {\bibfnamefont
  {U.}~\bibnamefont {D'Alesio}}, \bibinfo {author} {\bibfnamefont
  {S.}~\bibnamefont {{Dalla Torre}}}, \bibinfo {author} {\bibfnamefont
  {D.}~\bibnamefont {Das}}, \bibinfo {author} {\bibfnamefont {S.}~\bibnamefont
  {Dasgupta}}, \bibinfo {author} {\bibfnamefont {C.}~\bibnamefont {{Da
  Silva}}}, \bibinfo {author} {\bibfnamefont {W.}~\bibnamefont {Deconinck}},
  \bibinfo {author} {\bibfnamefont {M.}~\bibnamefont {Defurne}}, \bibinfo
  {author} {\bibfnamefont {W.}~\bibnamefont {DeGraw}}, \bibinfo {author}
  {\bibfnamefont {K.}~\bibnamefont {Dehmelt}}, \bibinfo {author} {\bibfnamefont
  {A.}~\bibnamefont {{Del Dotto}}}, \bibinfo {author} {\bibfnamefont
  {F.}~\bibnamefont {Delcarro}}, \bibinfo {author} {\bibfnamefont
  {A.}~\bibnamefont {Deshpande}}, \bibinfo {author} {\bibfnamefont
  {W.}~\bibnamefont {Detmold}}, \bibinfo {author} {\bibfnamefont
  {R.}~\bibnamefont {{De Vita}}}, \bibinfo {author} {\bibfnamefont
  {M.}~\bibnamefont {Diefenthaler}}, \bibinfo {author} {\bibfnamefont
  {C.}~\bibnamefont {Dilks}}, \bibinfo {author} {\bibfnamefont
  {D.}~\bibnamefont {Dixit}}, \bibinfo {author} {\bibfnamefont
  {S.}~\bibnamefont {Dulat}}, \bibinfo {author} {\bibfnamefont
  {A.}~\bibnamefont {Dumitru}}, \bibinfo {author} {\bibfnamefont
  {R.}~\bibnamefont {Dupré}}, \bibinfo {author} {\bibfnamefont
  {J.}~\bibnamefont {Durham}}, \bibinfo {author} {\bibfnamefont
  {M.}~\bibnamefont {Echevarria}}, \bibinfo {author} {\bibfnamefont
  {L.}~\bibnamefont {{El Fassi}}}, \bibinfo {author} {\bibfnamefont
  {D.}~\bibnamefont {Elia}}, \bibinfo {author} {\bibfnamefont {R.}~\bibnamefont
  {Ent}}, \bibinfo {author} {\bibfnamefont {R.}~\bibnamefont {Esha}}, \bibinfo
  {author} {\bibfnamefont {J.}~\bibnamefont {Ethier}}, \bibinfo {author}
  {\bibfnamefont {O.}~\bibnamefont {Evdokimov}}, \bibinfo {author}
  {\bibfnamefont {K.}~\bibnamefont {Eyser}}, \bibinfo {author} {\bibfnamefont
  {C.}~\bibnamefont {Fanelli}}, \bibinfo {author} {\bibfnamefont
  {R.}~\bibnamefont {Fatemi}}, \bibinfo {author} {\bibfnamefont
  {S.}~\bibnamefont {Fazio}}, \bibinfo {author} {\bibfnamefont
  {C.}~\bibnamefont {Fernandez-Ramirez}}, \bibinfo {author} {\bibfnamefont
  {M.}~\bibnamefont {Finger}}, \bibinfo {author} {\bibfnamefont
  {M.}~\bibnamefont {Finger}}, \bibinfo {author} {\bibfnamefont
  {D.}~\bibnamefont {Fitzgerald}}, \bibinfo {author} {\bibfnamefont
  {C.}~\bibnamefont {Flore}}, \bibinfo {author} {\bibfnamefont
  {T.}~\bibnamefont {Frederico}}, \bibinfo {author} {\bibfnamefont
  {I.}~\bibnamefont {Friščić}}, \bibinfo {author} {\bibfnamefont
  {S.}~\bibnamefont {Fucini}}, \bibinfo {author} {\bibfnamefont
  {S.}~\bibnamefont {Furletov}}, \bibinfo {author} {\bibfnamefont
  {Y.}~\bibnamefont {Furletova}}, \bibinfo {author} {\bibfnamefont
  {C.}~\bibnamefont {Gal}}, \bibinfo {author} {\bibfnamefont {L.}~\bibnamefont
  {Gamberg}}, \bibinfo {author} {\bibfnamefont {H.}~\bibnamefont {Gao}},
  \bibinfo {author} {\bibfnamefont {P.}~\bibnamefont {Garg}}, \bibinfo {author}
  {\bibfnamefont {D.}~\bibnamefont {Gaskell}}, \bibinfo {author} {\bibfnamefont
  {K.}~\bibnamefont {Gates}}, \bibinfo {author} {\bibfnamefont
  {M.}~\bibnamefont {{Gay Ducati}}}, \bibinfo {author} {\bibfnamefont
  {M.}~\bibnamefont {Gericke}}, \bibinfo {author} {\bibfnamefont
  {G.}~\bibnamefont {{Gil Da Silveira}}}, \bibinfo {author} {\bibfnamefont
  {F.-X.}\ \bibnamefont {Girod}}, \bibinfo {author} {\bibfnamefont
  {D.}~\bibnamefont {Glazier}}, \bibinfo {author} {\bibfnamefont
  {K.}~\bibnamefont {Gnanvo}}, \bibinfo {author} {\bibfnamefont
  {V.}~\bibnamefont {Goncalves}}, \bibinfo {author} {\bibfnamefont
  {L.}~\bibnamefont {Gonella}}, \bibinfo {author} {\bibfnamefont
  {J.}~\bibnamefont {{Gonzalez Hernandez}}}, \bibinfo {author} {\bibfnamefont
  {Y.}~\bibnamefont {Goto}}, \bibinfo {author} {\bibfnamefont {F.}~\bibnamefont
  {Grancagnolo}}, \bibinfo {author} {\bibfnamefont {L.}~\bibnamefont
  {Greiner}}, \bibinfo {author} {\bibfnamefont {W.}~\bibnamefont {Guryn}},
  \bibinfo {author} {\bibfnamefont {V.}~\bibnamefont {Guzey}}, \bibinfo
  {author} {\bibfnamefont {Y.}~\bibnamefont {Hatta}}, \bibinfo {author}
  {\bibfnamefont {M.}~\bibnamefont {Hattawy}}, \bibinfo {author} {\bibfnamefont
  {F.}~\bibnamefont {Hauenstein}}, \bibinfo {author} {\bibfnamefont
  {X.}~\bibnamefont {He}}, \bibinfo {author} {\bibfnamefont {T.}~\bibnamefont
  {Hemmick}}, \bibinfo {author} {\bibfnamefont {O.}~\bibnamefont {Hen}},
  \bibinfo {author} {\bibfnamefont {G.}~\bibnamefont {Heyes}}, \bibinfo
  {author} {\bibfnamefont {D.}~\bibnamefont {Higinbotham}}, \bibinfo {author}
  {\bibfnamefont {A.}~\bibnamefont {{Hiller Blin}}}, \bibinfo {author}
  {\bibfnamefont {T.}~\bibnamefont {Hobbs}}, \bibinfo {author} {\bibfnamefont
  {M.}~\bibnamefont {Hohlmann}}, \bibinfo {author} {\bibfnamefont
  {T.}~\bibnamefont {Horn}}, \bibinfo {author} {\bibfnamefont {T.-J.}\
  \bibnamefont {Hou}}, \bibinfo {author} {\bibfnamefont {J.}~\bibnamefont
  {Huang}}, \bibinfo {author} {\bibfnamefont {Q.}~\bibnamefont {Huang}},
  \bibinfo {author} {\bibfnamefont {G.}~\bibnamefont {Huber}}, \bibinfo
  {author} {\bibfnamefont {C.}~\bibnamefont {Hyde}}, \bibinfo {author}
  {\bibfnamefont {G.}~\bibnamefont {Iakovidis}}, \bibinfo {author}
  {\bibfnamefont {Y.}~\bibnamefont {Ilieva}}, \bibinfo {author} {\bibfnamefont
  {B.}~\bibnamefont {Jacak}}, \bibinfo {author} {\bibfnamefont
  {P.}~\bibnamefont {Jacobs}}, \bibinfo {author} {\bibfnamefont
  {M.}~\bibnamefont {Jadhav}}, \bibinfo {author} {\bibfnamefont
  {Z.}~\bibnamefont {Janoska}}, \bibinfo {author} {\bibfnamefont
  {A.}~\bibnamefont {Jentsch}}, \bibinfo {author} {\bibfnamefont
  {T.}~\bibnamefont {Jezo}}, \bibinfo {author} {\bibfnamefont {X.}~\bibnamefont
  {Jing}}, \bibinfo {author} {\bibfnamefont {P.}~\bibnamefont {Jones}},
  \bibinfo {author} {\bibfnamefont {K.}~\bibnamefont {Joo}}, \bibinfo {author}
  {\bibfnamefont {S.}~\bibnamefont {Joosten}}, \bibinfo {author} {\bibfnamefont
  {V.}~\bibnamefont {Kafka}}, \bibinfo {author} {\bibfnamefont
  {N.}~\bibnamefont {Kalantarians}}, \bibinfo {author} {\bibfnamefont
  {G.}~\bibnamefont {Kalicy}}, \bibinfo {author} {\bibfnamefont
  {D.}~\bibnamefont {Kang}}, \bibinfo {author} {\bibfnamefont {Z.}~\bibnamefont
  {Kang}}, \bibinfo {author} {\bibfnamefont {K.}~\bibnamefont {Kauder}},
  \bibinfo {author} {\bibfnamefont {S.}~\bibnamefont {Kay}}, \bibinfo {author}
  {\bibfnamefont {C.}~\bibnamefont {Keppel}}, \bibinfo {author} {\bibfnamefont
  {J.}~\bibnamefont {Kim}}, \bibinfo {author} {\bibfnamefont {A.}~\bibnamefont
  {Kiselev}}, \bibinfo {author} {\bibfnamefont {M.}~\bibnamefont {Klasen}},
  \bibinfo {author} {\bibfnamefont {S.}~\bibnamefont {Klein}}, \bibinfo
  {author} {\bibfnamefont {H.}~\bibnamefont {Klest}}, \bibinfo {author}
  {\bibfnamefont {O.}~\bibnamefont {Korchak}}, \bibinfo {author} {\bibfnamefont
  {A.}~\bibnamefont {Kostina}}, \bibinfo {author} {\bibfnamefont
  {P.}~\bibnamefont {Kotko}}, \bibinfo {author} {\bibfnamefont
  {Y.}~\bibnamefont {Kovchegov}}, \bibinfo {author} {\bibfnamefont
  {M.}~\bibnamefont {Krelina}}, \bibinfo {author} {\bibfnamefont
  {S.}~\bibnamefont {Kuleshov}}, \bibinfo {author} {\bibfnamefont
  {S.}~\bibnamefont {Kumano}}, \bibinfo {author} {\bibfnamefont
  {K.}~\bibnamefont {Kumar}}, \bibinfo {author} {\bibfnamefont
  {R.}~\bibnamefont {Kumar}}, \bibinfo {author} {\bibfnamefont
  {L.}~\bibnamefont {Kumar}}, \bibinfo {author} {\bibfnamefont
  {K.}~\bibnamefont {Kumerički}}, \bibinfo {author} {\bibfnamefont
  {A.}~\bibnamefont {Kusina}}, \bibinfo {author} {\bibfnamefont
  {K.}~\bibnamefont {Kutak}}, \bibinfo {author} {\bibfnamefont
  {Y.}~\bibnamefont {Lai}}, \bibinfo {author} {\bibfnamefont {K.}~\bibnamefont
  {Lalwani}}, \bibinfo {author} {\bibfnamefont {T.}~\bibnamefont {Lappi}},
  \bibinfo {author} {\bibfnamefont {J.}~\bibnamefont {Lauret}}, \bibinfo
  {author} {\bibfnamefont {M.}~\bibnamefont {Lavinsky}}, \bibinfo {author}
  {\bibfnamefont {D.}~\bibnamefont {Lawrence}}, \bibinfo {author}
  {\bibfnamefont {D.}~\bibnamefont {Lednicky}}, \bibinfo {author}
  {\bibfnamefont {C.}~\bibnamefont {Lee}}, \bibinfo {author} {\bibfnamefont
  {K.}~\bibnamefont {Lee}}, \bibinfo {author} {\bibfnamefont {S.}~\bibnamefont
  {Lee}}, \bibinfo {author} {\bibfnamefont {S.}~\bibnamefont {Levorato}},
  \bibinfo {author} {\bibfnamefont {H.}~\bibnamefont {Li}}, \bibinfo {author}
  {\bibfnamefont {S.}~\bibnamefont {Li}}, \bibinfo {author} {\bibfnamefont
  {W.}~\bibnamefont {Li}}, \bibinfo {author} {\bibfnamefont {X.}~\bibnamefont
  {Li}}, \bibinfo {author} {\bibfnamefont {X.}~\bibnamefont {Li}}, \bibinfo
  {author} {\bibfnamefont {W.}~\bibnamefont {Li}}, \bibinfo {author}
  {\bibfnamefont {T.}~\bibnamefont {Ligonzo}}, \bibinfo {author} {\bibfnamefont
  {H.}~\bibnamefont {Liu}}, \bibinfo {author} {\bibfnamefont {M.}~\bibnamefont
  {Liu}}, \bibinfo {author} {\bibfnamefont {X.}~\bibnamefont {Liu}}, \bibinfo
  {author} {\bibfnamefont {S.}~\bibnamefont {Liuti}}, \bibinfo {author}
  {\bibfnamefont {N.}~\bibnamefont {Liyanage}}, \bibinfo {author}
  {\bibfnamefont {C.}~\bibnamefont {Lorcé}}, \bibinfo {author} {\bibfnamefont
  {Z.}~\bibnamefont {Lu}}, \bibinfo {author} {\bibfnamefont {G.}~\bibnamefont
  {Lucero}}, \bibinfo {author} {\bibfnamefont {N.}~\bibnamefont {Lukow}},
  \bibinfo {author} {\bibfnamefont {E.}~\bibnamefont {Lunghi}}, \bibinfo
  {author} {\bibfnamefont {R.}~\bibnamefont {Majka}}, \bibinfo {author}
  {\bibfnamefont {Y.}~\bibnamefont {Makris}}, \bibinfo {author} {\bibfnamefont
  {I.}~\bibnamefont {Mandjavidze}}, \bibinfo {author} {\bibfnamefont
  {S.}~\bibnamefont {Mantry}}, \bibinfo {author} {\bibfnamefont
  {H.}~\bibnamefont {Mäntysaari}}, \bibinfo {author} {\bibfnamefont
  {F.}~\bibnamefont {Marhauser}}, \bibinfo {author} {\bibfnamefont
  {P.}~\bibnamefont {Markowitz}}, \bibinfo {author} {\bibfnamefont
  {L.}~\bibnamefont {Marsicano}}, \bibinfo {author} {\bibfnamefont
  {A.}~\bibnamefont {Mastroserio}}, \bibinfo {author} {\bibfnamefont
  {V.}~\bibnamefont {Mathieu}}, \bibinfo {author} {\bibfnamefont
  {Y.}~\bibnamefont {Mehtar-Tani}}, \bibinfo {author} {\bibfnamefont
  {W.}~\bibnamefont {Melnitchouk}}, \bibinfo {author} {\bibfnamefont
  {L.}~\bibnamefont {Mendez}}, \bibinfo {author} {\bibfnamefont
  {A.}~\bibnamefont {Metz}}, \bibinfo {author} {\bibfnamefont {Z.-E.}\
  \bibnamefont {Meziani}}, \bibinfo {author} {\bibfnamefont {C.}~\bibnamefont
  {Mezrag}}, \bibinfo {author} {\bibfnamefont {M.}~\bibnamefont
  {Mihovilovič}}, \bibinfo {author} {\bibfnamefont {R.}~\bibnamefont
  {Milner}}, \bibinfo {author} {\bibfnamefont {M.}~\bibnamefont {Mirazita}},
  \bibinfo {author} {\bibfnamefont {H.}~\bibnamefont {Mkrtchyan}}, \bibinfo
  {author} {\bibfnamefont {A.}~\bibnamefont {Mkrtchyan}}, \bibinfo {author}
  {\bibfnamefont {V.}~\bibnamefont {Mochalov}}, \bibinfo {author}
  {\bibfnamefont {V.}~\bibnamefont {Moiseev}}, \bibinfo {author} {\bibfnamefont
  {M.}~\bibnamefont {Mondal}}, \bibinfo {author} {\bibfnamefont
  {A.}~\bibnamefont {Morreale}}, \bibinfo {author} {\bibfnamefont
  {D.}~\bibnamefont {Morrison}}, \bibinfo {author} {\bibfnamefont
  {L.}~\bibnamefont {Motyka}}, \bibinfo {author} {\bibfnamefont
  {H.}~\bibnamefont {Moutarde}}, \bibinfo {author} {\bibfnamefont
  {C.}~\bibnamefont {{Muñoz Camacho}}}, \bibinfo {author} {\bibfnamefont
  {F.}~\bibnamefont {Murgia}}, \bibinfo {author} {\bibfnamefont
  {M.}~\bibnamefont {Murray}}, \bibinfo {author} {\bibfnamefont
  {P.}~\bibnamefont {Musico}}, \bibinfo {author} {\bibfnamefont
  {P.}~\bibnamefont {Nadel-Turonski}}, \bibinfo {author} {\bibfnamefont
  {P.}~\bibnamefont {Nadolsky}}, \bibinfo {author} {\bibfnamefont
  {J.}~\bibnamefont {Nam}}, \bibinfo {author} {\bibfnamefont {P.}~\bibnamefont
  {Newman}}, \bibinfo {author} {\bibfnamefont {D.}~\bibnamefont {Neyret}},
  \bibinfo {author} {\bibfnamefont {D.}~\bibnamefont {Nguyen}}, \bibinfo
  {author} {\bibfnamefont {E.}~\bibnamefont {Nocera}}, \bibinfo {author}
  {\bibfnamefont {F.}~\bibnamefont {Noferini}}, \bibinfo {author}
  {\bibfnamefont {F.}~\bibnamefont {Noto}}, \bibinfo {author} {\bibfnamefont
  {A.}~\bibnamefont {Nunes}}, \bibinfo {author} {\bibfnamefont
  {V.}~\bibnamefont {Okorokov}}, \bibinfo {author} {\bibfnamefont
  {F.}~\bibnamefont {Olness}}, \bibinfo {author} {\bibfnamefont
  {J.}~\bibnamefont {Osborn}}, \bibinfo {author} {\bibfnamefont
  {B.}~\bibnamefont {Page}}, \bibinfo {author} {\bibfnamefont {S.}~\bibnamefont
  {Park}}, \bibinfo {author} {\bibfnamefont {A.}~\bibnamefont {Parker}},
  \bibinfo {author} {\bibfnamefont {K.}~\bibnamefont {Paschke}}, \bibinfo
  {author} {\bibfnamefont {B.}~\bibnamefont {Pasquini}}, \bibinfo {author}
  {\bibfnamefont {H.}~\bibnamefont {Paukkunen}}, \bibinfo {author}
  {\bibfnamefont {S.}~\bibnamefont {Paul}}, \bibinfo {author} {\bibfnamefont
  {C.}~\bibnamefont {Pecar}}, \bibinfo {author} {\bibfnamefont
  {I.}~\bibnamefont {Pegg}}, \bibinfo {author} {\bibfnamefont {C.}~\bibnamefont
  {Pellegrino}}, \bibinfo {author} {\bibfnamefont {C.}~\bibnamefont {Peng}},
  \bibinfo {author} {\bibfnamefont {L.}~\bibnamefont {Pentchev}}, \bibinfo
  {author} {\bibfnamefont {R.}~\bibnamefont {Perrino}}, \bibinfo {author}
  {\bibfnamefont {F.}~\bibnamefont {Petriello}}, \bibinfo {author}
  {\bibfnamefont {R.}~\bibnamefont {Petti}}, \bibinfo {author} {\bibfnamefont
  {A.}~\bibnamefont {Pilloni}}, \bibinfo {author} {\bibfnamefont
  {C.}~\bibnamefont {Pinkenburg}}, \bibinfo {author} {\bibfnamefont
  {B.}~\bibnamefont {Pire}}, \bibinfo {author} {\bibfnamefont {C.}~\bibnamefont
  {Pisano}}, \bibinfo {author} {\bibfnamefont {D.}~\bibnamefont {Pitonyak}},
  \bibinfo {author} {\bibfnamefont {A.}~\bibnamefont {Poblaguev}}, \bibinfo
  {author} {\bibfnamefont {T.}~\bibnamefont {Polakovic}}, \bibinfo {author}
  {\bibfnamefont {M.}~\bibnamefont {Posik}}, \bibinfo {author} {\bibfnamefont
  {M.}~\bibnamefont {Potekhin}}, \bibinfo {author} {\bibfnamefont
  {R.}~\bibnamefont {Preghenella}}, \bibinfo {author} {\bibfnamefont
  {S.}~\bibnamefont {Preins}}, \bibinfo {author} {\bibfnamefont
  {A.}~\bibnamefont {Prokudin}}, \bibinfo {author} {\bibfnamefont
  {P.}~\bibnamefont {Pujahari}}, \bibinfo {author} {\bibfnamefont
  {M.}~\bibnamefont {Purschke}}, \bibinfo {author} {\bibfnamefont
  {J.}~\bibnamefont {Pybus}}, \bibinfo {author} {\bibfnamefont
  {M.}~\bibnamefont {Radici}}, \bibinfo {author} {\bibfnamefont
  {R.}~\bibnamefont {Rajput-Ghoshal}}, \bibinfo {author} {\bibfnamefont
  {P.}~\bibnamefont {Reimer}}, \bibinfo {author} {\bibfnamefont
  {M.}~\bibnamefont {Rinaldi}}, \bibinfo {author} {\bibfnamefont
  {F.}~\bibnamefont {Ringer}}, \bibinfo {author} {\bibfnamefont
  {C.}~\bibnamefont {Roberts}}, \bibinfo {author} {\bibfnamefont
  {S.}~\bibnamefont {Rodini}}, \bibinfo {author} {\bibfnamefont
  {J.}~\bibnamefont {Rojo}}, \bibinfo {author} {\bibfnamefont {D.}~\bibnamefont
  {Romanov}}, \bibinfo {author} {\bibfnamefont {P.}~\bibnamefont {Rossi}},
  \bibinfo {author} {\bibfnamefont {E.}~\bibnamefont {Santopinto}}, \bibinfo
  {author} {\bibfnamefont {M.}~\bibnamefont {Sarsour}}, \bibinfo {author}
  {\bibfnamefont {R.}~\bibnamefont {Sassot}}, \bibinfo {author} {\bibfnamefont
  {N.}~\bibnamefont {Sato}}, \bibinfo {author} {\bibfnamefont {B.}~\bibnamefont
  {Schenke}}, \bibinfo {author} {\bibfnamefont {W.}~\bibnamefont {Schmidke}},
  \bibinfo {author} {\bibfnamefont {I.}~\bibnamefont {Schmidt}}, \bibinfo
  {author} {\bibfnamefont {A.}~\bibnamefont {Schmidt}}, \bibinfo {author}
  {\bibfnamefont {B.}~\bibnamefont {Schmookler}}, \bibinfo {author}
  {\bibfnamefont {G.}~\bibnamefont {Schnell}}, \bibinfo {author} {\bibfnamefont
  {P.}~\bibnamefont {Schweitzer}}, \bibinfo {author} {\bibfnamefont
  {J.}~\bibnamefont {Schwiening}}, \bibinfo {author} {\bibfnamefont
  {I.}~\bibnamefont {Scimemi}}, \bibinfo {author} {\bibfnamefont
  {S.}~\bibnamefont {Scopetta}}, \bibinfo {author} {\bibfnamefont
  {J.}~\bibnamefont {Segovia}}, \bibinfo {author} {\bibfnamefont
  {R.}~\bibnamefont {Seidl}}, \bibinfo {author} {\bibfnamefont
  {S.}~\bibnamefont {Sekula}}, \bibinfo {author} {\bibfnamefont
  {K.}~\bibnamefont {Semenov-Tian-Shanskiy}}, \bibinfo {author} {\bibfnamefont
  {D.}~\bibnamefont {Shao}}, \bibinfo {author} {\bibfnamefont {N.}~\bibnamefont
  {Sherrill}}, \bibinfo {author} {\bibfnamefont {E.}~\bibnamefont
  {Sichtermann}}, \bibinfo {author} {\bibfnamefont {M.}~\bibnamefont
  {Siddikov}}, \bibinfo {author} {\bibfnamefont {A.}~\bibnamefont {Signori}},
  \bibinfo {author} {\bibfnamefont {B.}~\bibnamefont {Singh}}, \bibinfo
  {author} {\bibfnamefont {S.}~\bibnamefont {Širca}}, \bibinfo {author}
  {\bibfnamefont {K.}~\bibnamefont {Slifer}}, \bibinfo {author} {\bibfnamefont
  {W.}~\bibnamefont {Slominski}}, \bibinfo {author} {\bibfnamefont
  {D.}~\bibnamefont {Sokhan}}, \bibinfo {author} {\bibfnamefont
  {W.}~\bibnamefont {Sondheim}}, \bibinfo {author} {\bibfnamefont
  {Y.}~\bibnamefont {Song}}, \bibinfo {author} {\bibfnamefont {O.}~\bibnamefont
  {Soto}}, \bibinfo {author} {\bibfnamefont {H.}~\bibnamefont {Spiesberger}},
  \bibinfo {author} {\bibfnamefont {A.}~\bibnamefont {Stasto}}, \bibinfo
  {author} {\bibfnamefont {P.}~\bibnamefont {Stepanov}}, \bibinfo {author}
  {\bibfnamefont {G.}~\bibnamefont {Sterman}}, \bibinfo {author} {\bibfnamefont
  {J.}~\bibnamefont {Stevens}}, \bibinfo {author} {\bibfnamefont
  {I.}~\bibnamefont {Stewart}}, \bibinfo {author} {\bibfnamefont
  {I.}~\bibnamefont {Strakovsky}}, \bibinfo {author} {\bibfnamefont
  {M.}~\bibnamefont {Strikman}}, \bibinfo {author} {\bibfnamefont
  {M.}~\bibnamefont {Sturm}}, \bibinfo {author} {\bibfnamefont
  {M.}~\bibnamefont {Stutzman}}, \bibinfo {author} {\bibfnamefont
  {M.}~\bibnamefont {Sullivan}}, \bibinfo {author} {\bibfnamefont
  {B.}~\bibnamefont {Surrow}}, \bibinfo {author} {\bibfnamefont
  {P.}~\bibnamefont {Svihra}}, \bibinfo {author} {\bibfnamefont
  {S.}~\bibnamefont {Syritsyn}}, \bibinfo {author} {\bibfnamefont
  {A.}~\bibnamefont {Szczepaniak}}, \bibinfo {author} {\bibfnamefont
  {P.}~\bibnamefont {Sznajder}}, \bibinfo {author} {\bibfnamefont
  {H.}~\bibnamefont {Szumila-Vance}}, \bibinfo {author} {\bibfnamefont
  {L.}~\bibnamefont {Szymanowski}}, \bibinfo {author} {\bibfnamefont
  {A.}~\bibnamefont {Tadepalli}}, \bibinfo {author} {\bibfnamefont
  {J.}~\bibnamefont {{Tapia Takaki}}}, \bibinfo {author} {\bibfnamefont
  {G.}~\bibnamefont {Tassielli}}, \bibinfo {author} {\bibfnamefont
  {J.}~\bibnamefont {Terry}}, \bibinfo {author} {\bibfnamefont
  {F.}~\bibnamefont {Tessarotto}}, \bibinfo {author} {\bibfnamefont
  {K.}~\bibnamefont {Tezgin}}, \bibinfo {author} {\bibfnamefont
  {L.}~\bibnamefont {Tomasek}}, \bibinfo {author} {\bibfnamefont
  {F.}~\bibnamefont {{Torales Acosta}}}, \bibinfo {author} {\bibfnamefont
  {P.}~\bibnamefont {Tribedy}}, \bibinfo {author} {\bibfnamefont
  {A.}~\bibnamefont {Tricoli}}, \bibinfo {author} {\bibnamefont {Triloki}},
  \bibinfo {author} {\bibfnamefont {S.}~\bibnamefont {Tripathi}}, \bibinfo
  {author} {\bibfnamefont {R.}~\bibnamefont {Trotta}}, \bibinfo {author}
  {\bibfnamefont {O.}~\bibnamefont {Tsai}}, \bibinfo {author} {\bibfnamefont
  {Z.}~\bibnamefont {Tu}}, \bibinfo {author} {\bibfnamefont {C.}~\bibnamefont
  {Tuvè}}, \bibinfo {author} {\bibfnamefont {T.}~\bibnamefont {Ullrich}},
  \bibinfo {author} {\bibfnamefont {M.}~\bibnamefont {Ungaro}}, \bibinfo
  {author} {\bibfnamefont {G.}~\bibnamefont {Urciuoli}}, \bibinfo {author}
  {\bibfnamefont {A.}~\bibnamefont {Valentini}}, \bibinfo {author}
  {\bibfnamefont {P.}~\bibnamefont {Vancura}}, \bibinfo {author} {\bibfnamefont
  {M.}~\bibnamefont {Vandenbroucke}}, \bibinfo {author} {\bibfnamefont
  {C.}~\bibnamefont {{Van Hulse}}}, \bibinfo {author} {\bibfnamefont
  {G.}~\bibnamefont {Varner}}, \bibinfo {author} {\bibfnamefont
  {R.}~\bibnamefont {Venugopalan}}, \bibinfo {author} {\bibfnamefont
  {I.}~\bibnamefont {Vitev}}, \bibinfo {author} {\bibfnamefont
  {A.}~\bibnamefont {Vladimirov}}, \bibinfo {author} {\bibfnamefont
  {G.}~\bibnamefont {Volpe}}, \bibinfo {author} {\bibfnamefont
  {A.}~\bibnamefont {Vossen}}, \bibinfo {author} {\bibfnamefont
  {E.}~\bibnamefont {Voutier}}, \bibinfo {author} {\bibfnamefont
  {J.}~\bibnamefont {Wagner}}, \bibinfo {author} {\bibfnamefont
  {S.}~\bibnamefont {Wallon}}, \bibinfo {author} {\bibfnamefont
  {H.}~\bibnamefont {Wang}}, \bibinfo {author} {\bibfnamefont {Q.}~\bibnamefont
  {Wang}}, \bibinfo {author} {\bibfnamefont {X.}~\bibnamefont {Wang}}, \bibinfo
  {author} {\bibfnamefont {S.}~\bibnamefont {Wei}}, \bibinfo {author}
  {\bibfnamefont {C.}~\bibnamefont {Weiss}}, \bibinfo {author} {\bibfnamefont
  {T.}~\bibnamefont {Wenaus}}, \bibinfo {author} {\bibfnamefont
  {H.}~\bibnamefont {Wennlöf}}, \bibinfo {author} {\bibfnamefont
  {N.}~\bibnamefont {Wickramaarachchi}}, \bibinfo {author} {\bibfnamefont
  {A.}~\bibnamefont {Wikramanayake}}, \bibinfo {author} {\bibfnamefont
  {D.}~\bibnamefont {Winney}}, \bibinfo {author} {\bibfnamefont
  {C.}~\bibnamefont {Wong}}, \bibinfo {author} {\bibfnamefont {C.}~\bibnamefont
  {Woody}}, \bibinfo {author} {\bibfnamefont {L.}~\bibnamefont {Xia}}, \bibinfo
  {author} {\bibfnamefont {B.}~\bibnamefont {Xiao}}, \bibinfo {author}
  {\bibfnamefont {J.}~\bibnamefont {Xie}}, \bibinfo {author} {\bibfnamefont
  {H.}~\bibnamefont {Xing}}, \bibinfo {author} {\bibfnamefont {Q.}~\bibnamefont
  {Xu}}, \bibinfo {author} {\bibfnamefont {J.}~\bibnamefont {Zhang}}, \bibinfo
  {author} {\bibfnamefont {S.}~\bibnamefont {Zhang}}, \bibinfo {author}
  {\bibfnamefont {Z.}~\bibnamefont {Zhang}}, \bibinfo {author} {\bibfnamefont
  {Z.}~\bibnamefont {Zhao}}, \bibinfo {author} {\bibfnamefont {Y.}~\bibnamefont
  {Zhao}}, \bibinfo {author} {\bibfnamefont {L.}~\bibnamefont {Zheng}},
  \bibinfo {author} {\bibfnamefont {Y.}~\bibnamefont {Zhou}},\ and\ \bibinfo
  {author} {\bibfnamefont {P.}~\bibnamefont {Zurita}},\ }\href
  {https://doi.org/https://doi.org/10.1016/j.nuclphysa.2022.122447} {\bibfield
  {journal} {\bibinfo  {journal} {Nuclear Physics A}\ }\textbf {\bibinfo
  {volume} {1026}},\ \bibinfo {pages} {122447} (\bibinfo {year}
  {2022})}\BibitemShut {NoStop}%
\bibitem [{\citenamefont {Nikolaev}\ and\ \citenamefont
  {Schafer}(1997)}]{Nikolaev:TensorSF}%
  \BibitemOpen
  \bibfield  {author} {\bibinfo {author} {\bibfnamefont {N.~N.}\ \bibnamefont
  {Nikolaev}}\ and\ \bibinfo {author} {\bibfnamefont {W.}~\bibnamefont
  {Schafer}},\ }\href {https://doi.org/10.1016/S0370-2693(97)00250-5}
  {\bibfield  {journal} {\bibinfo  {journal} {Phys. Lett. B}\ }\textbf
  {\bibinfo {volume} {398}},\ \bibinfo {pages} {245} (\bibinfo {year}
  {1997})},\ \bibinfo {note} {[Erratum: Phys.Lett.B 407, 453 (1997)]},\ \Eprint
  {https://arxiv.org/abs/hep-ph/9611460} {arXiv:hep-ph/9611460} \BibitemShut
  {NoStop}%
\bibitem [{\citenamefont {Teryaev}(2019)}]{Teryaev:ShearForces}%
  \BibitemOpen
  \bibfield  {author} {\bibinfo {author} {\bibfnamefont {O.}~\bibnamefont
  {Teryaev}},\ }\href {https://doi.org/10.22323/1.352.0240} {\bibfield
  {journal} {\bibinfo  {journal} {PoS}\ }\textbf {\bibinfo {volume}
  {DIS2019}},\ \bibinfo {pages} {240} (\bibinfo {year} {2019})}\BibitemShut
  {NoStop}%
\bibitem [{\citenamefont {Teryaev}(2003)}]{Teryaev:DeuteronSumRules}%
  \BibitemOpen
  \bibfield  {author} {\bibinfo {author} {\bibfnamefont {O.~V.}\ \bibnamefont
  {Teryaev}},\ }in\ \href@noop {} {\emph {\bibinfo {booktitle} {{16th
  International Baldin Seminar on High Energy Physics Problems}: {Relativistic
  Nuclear Physics and Quantum Chromodynamics}}}}\ (\bibinfo {year} {2003})\
  \Eprint {https://arxiv.org/abs/hep-ph/0303003} {arXiv:hep-ph/0303003}
  \BibitemShut {NoStop}%
\bibitem [{\citenamefont {Dalton}\ \emph {et~al.}(2025)\citenamefont {Dalton},
  \citenamefont {Deur}, \citenamefont {Keith},\ and\ \citenamefont
  {Keith}}]{CEBAF_TensorD}%
  \BibitemOpen
  \bibfield  {author} {\bibinfo {author} {\bibfnamefont {M.~M.}\ \bibnamefont
  {Dalton}}, \bibinfo {author} {\bibfnamefont {A.}~\bibnamefont {Deur}},
  \bibinfo {author} {\bibfnamefont {C.~D.}\ \bibnamefont {Keith}},\ and\
  \bibinfo {author} {\bibfnamefont {C.}~\bibnamefont {Keith}},\ }\href
  {https://doi.org/10.1140/epja/s10050-025-01580-y} {\bibfield  {journal}
  {\bibinfo  {journal} {Eur. Phys. J. A}\ }\textbf {\bibinfo {volume} {61}},\
  \bibinfo {pages} {111} (\bibinfo {year} {2025})},\ \Eprint
  {https://arxiv.org/abs/2504.21177} {arXiv:2504.21177 [nucl-ex]} \BibitemShut
  {NoStop}%
\bibitem [{\citenamefont {Baryshevsky}(2008)}]{Baryshevsky_2008}%
  \BibitemOpen
  \bibfield  {author} {\bibinfo {author} {\bibfnamefont {V.~G.}\ \bibnamefont
  {Baryshevsky}},\ }\href {https://doi.org/10.1088/0954-3899/35/3/035102}
  {\bibfield  {journal} {\bibinfo  {journal} {Journal of Physics G: Nuclear and
  Particle Physics}\ }\textbf {\bibinfo {volume} {35}},\ \bibinfo {pages}
  {035102} (\bibinfo {year} {2008})}\BibitemShut {NoStop}%
\bibitem [{\citenamefont {Silenko}(2014)}]{SilenkoSpin1Proca}%
  \BibitemOpen
  \bibfield  {author} {\bibinfo {author} {\bibfnamefont {A.~J.}\ \bibnamefont
  {Silenko}},\ }\href {https://doi.org/10.1103/PhysRevD.89.121701} {\bibfield
  {journal} {\bibinfo  {journal} {Phys. Rev. D}\ }\textbf {\bibinfo {volume}
  {89}},\ \bibinfo {pages} {121701} (\bibinfo {year} {2014})},\ \Eprint
  {https://arxiv.org/abs/1404.4953} {arXiv:1404.4953 [quant-ph]} \BibitemShut
  {NoStop}%
\bibitem [{\citenamefont {Azhgirey}\ \emph {et~al.}(1995)\citenamefont
  {Azhgirey}, \citenamefont {Chernykh}, \citenamefont {Kobushkin},
  \citenamefont {Korovin}, \citenamefont {Kuehn}, \citenamefont {Ladygin},
  \citenamefont {Nedev}, \citenamefont {Perdrisat}, \citenamefont {Piskunov},
  \citenamefont {Punjabi}, \citenamefont {Sitnik}, \citenamefont {Stoletov},
  \citenamefont {Strokovsky}, \citenamefont {Syamtomov},\ and\ \citenamefont
  {Zaporozhets}}]{DDAlpha1995}%
  \BibitemOpen
  \bibfield  {author} {\bibinfo {author} {\bibfnamefont {L.}~\bibnamefont
  {Azhgirey}}, \bibinfo {author} {\bibfnamefont {E.}~\bibnamefont {Chernykh}},
  \bibinfo {author} {\bibfnamefont {A.}~\bibnamefont {Kobushkin}}, \bibinfo
  {author} {\bibfnamefont {P.}~\bibnamefont {Korovin}}, \bibinfo {author}
  {\bibfnamefont {B.}~\bibnamefont {Kuehn}}, \bibinfo {author} {\bibfnamefont
  {V.}~\bibnamefont {Ladygin}}, \bibinfo {author} {\bibfnamefont
  {S.}~\bibnamefont {Nedev}}, \bibinfo {author} {\bibfnamefont
  {C.}~\bibnamefont {Perdrisat}}, \bibinfo {author} {\bibfnamefont
  {N.}~\bibnamefont {Piskunov}}, \bibinfo {author} {\bibfnamefont
  {V.}~\bibnamefont {Punjabi}}, \bibinfo {author} {\bibfnamefont
  {I.}~\bibnamefont {Sitnik}}, \bibinfo {author} {\bibfnamefont
  {G.}~\bibnamefont {Stoletov}}, \bibinfo {author} {\bibfnamefont
  {E.}~\bibnamefont {Strokovsky}}, \bibinfo {author} {\bibfnamefont
  {A.}~\bibnamefont {Syamtomov}},\ and\ \bibinfo {author} {\bibfnamefont
  {S.}~\bibnamefont {Zaporozhets}},\ }\href
  {https://doi.org/https://doi.org/10.1016/0370-2693(95)01160-R} {\bibfield
  {journal} {\bibinfo  {journal} {Physics Letters B}\ }\textbf {\bibinfo
  {volume} {361}},\ \bibinfo {pages} {21} (\bibinfo {year} {1995})}\BibitemShut
  {NoStop}%
\bibitem [{\citenamefont {Ladygin}\ \emph {et~al.}(2000)\citenamefont
  {Ladygin}, \citenamefont {Azhgirey}, \citenamefont {Afanasiev}, \citenamefont
  {Arkhipov}, \citenamefont {Bondarev}, \citenamefont {Filipov}, \citenamefont
  {Isupov}, \citenamefont {Ivanov}, \citenamefont {Kartamyshev}, \citenamefont
  {Kashirin} \emph {et~al.}}]{DDSphere2000}%
  \BibitemOpen
  \bibfield  {author} {\bibinfo {author} {\bibfnamefont {V.}~\bibnamefont
  {Ladygin}}, \bibinfo {author} {\bibfnamefont {L.}~\bibnamefont {Azhgirey}},
  \bibinfo {author} {\bibfnamefont {S.}~\bibnamefont {Afanasiev}}, \bibinfo
  {author} {\bibfnamefont {V.}~\bibnamefont {Arkhipov}}, \bibinfo {author}
  {\bibfnamefont {V.}~\bibnamefont {Bondarev}}, \bibinfo {author}
  {\bibfnamefont {G.}~\bibnamefont {Filipov}}, \bibinfo {author} {\bibfnamefont
  {A.~Y.}\ \bibnamefont {Isupov}}, \bibinfo {author} {\bibfnamefont
  {V.}~\bibnamefont {Ivanov}}, \bibinfo {author} {\bibfnamefont
  {A.}~\bibnamefont {Kartamyshev}}, \bibinfo {author} {\bibfnamefont
  {V.}~\bibnamefont {Kashirin}}, \emph {et~al.},\ }\href@noop {} {\bibfield
  {journal} {\bibinfo  {journal} {The European Physical Journal A}\ }\textbf
  {\bibinfo {volume} {8}},\ \bibinfo {pages} {409} (\bibinfo {year}
  {2000})}\BibitemShut {NoStop}%
\bibitem [{\citenamefont {Azhgirey}\ \emph {et~al.}(1996)\citenamefont
  {Azhgirey}, \citenamefont {Afanasyev}, \citenamefont {Chernykh},
  \citenamefont {Kobushkin}, \citenamefont {Ladygin}, \citenamefont {Nedev},
  \citenamefont {Penchev}, \citenamefont {Perdrisat}, \citenamefont {Piskunov},
  \citenamefont {Punjabi}, \citenamefont {Sitnik}, \citenamefont {Stoletov},
  \citenamefont {Strokovsky}, \citenamefont {Syamtomov},\ and\ \citenamefont
  {Zaporozhets}}]{Dubna_TensorD}%
  \BibitemOpen
  \bibfield  {author} {\bibinfo {author} {\bibfnamefont {L.}~\bibnamefont
  {Azhgirey}}, \bibinfo {author} {\bibfnamefont {S.}~\bibnamefont {Afanasyev}},
  \bibinfo {author} {\bibfnamefont {E.}~\bibnamefont {Chernykh}}, \bibinfo
  {author} {\bibfnamefont {A.}~\bibnamefont {Kobushkin}}, \bibinfo {author}
  {\bibfnamefont {V.}~\bibnamefont {Ladygin}}, \bibinfo {author} {\bibfnamefont
  {S.}~\bibnamefont {Nedev}}, \bibinfo {author} {\bibfnamefont
  {L.}~\bibnamefont {Penchev}}, \bibinfo {author} {\bibfnamefont
  {C.}~\bibnamefont {Perdrisat}}, \bibinfo {author} {\bibfnamefont
  {N.}~\bibnamefont {Piskunov}}, \bibinfo {author} {\bibfnamefont
  {V.}~\bibnamefont {Punjabi}}, \bibinfo {author} {\bibfnamefont
  {I.}~\bibnamefont {Sitnik}}, \bibinfo {author} {\bibfnamefont
  {G.}~\bibnamefont {Stoletov}}, \bibinfo {author} {\bibfnamefont
  {E.}~\bibnamefont {Strokovsky}}, \bibinfo {author} {\bibfnamefont
  {A.}~\bibnamefont {Syamtomov}},\ and\ \bibinfo {author} {\bibfnamefont
  {S.}~\bibnamefont {Zaporozhets}},\ }\href
  {https://doi.org/https://doi.org/10.1016/0370-2693(96)01007-6} {\bibfield
  {journal} {\bibinfo  {journal} {Physics Letters B}\ }\textbf {\bibinfo
  {volume} {387}},\ \bibinfo {pages} {37} (\bibinfo {year} {1996})}\BibitemShut
  {NoStop}%
\bibitem [{\citenamefont {Azhgirey}\ \emph {et~al.}(1997)\citenamefont
  {Azhgirey}, \citenamefont {Chernykh}, \citenamefont {Kobushkin},
  \citenamefont {Ladygin}, \citenamefont {Nedev}, \citenamefont {Penchev},
  \citenamefont {Perdrisat}, \citenamefont {Piskunov}, \citenamefont {Punjabi},
  \citenamefont {Sitnik}, \citenamefont {Stoletov}, \citenamefont {Strokovsky},
  \citenamefont {Syamtomov}, \citenamefont {Vikhrov}, \citenamefont
  {Vizireva},\ and\ \citenamefont {Zaporozhets}}]{Dubna_TensorD1}%
  \BibitemOpen
  \bibfield  {author} {\bibinfo {author} {\bibfnamefont {L.}~\bibnamefont
  {Azhgirey}}, \bibinfo {author} {\bibfnamefont {E.}~\bibnamefont {Chernykh}},
  \bibinfo {author} {\bibfnamefont {A.}~\bibnamefont {Kobushkin}}, \bibinfo
  {author} {\bibfnamefont {V.}~\bibnamefont {Ladygin}}, \bibinfo {author}
  {\bibfnamefont {S.}~\bibnamefont {Nedev}}, \bibinfo {author} {\bibfnamefont
  {L.}~\bibnamefont {Penchev}}, \bibinfo {author} {\bibfnamefont
  {C.}~\bibnamefont {Perdrisat}}, \bibinfo {author} {\bibfnamefont
  {N.}~\bibnamefont {Piskunov}}, \bibinfo {author} {\bibfnamefont
  {V.}~\bibnamefont {Punjabi}}, \bibinfo {author} {\bibfnamefont
  {I.}~\bibnamefont {Sitnik}}, \bibinfo {author} {\bibfnamefont
  {G.}~\bibnamefont {Stoletov}}, \bibinfo {author} {\bibfnamefont
  {E.}~\bibnamefont {Strokovsky}}, \bibinfo {author} {\bibfnamefont
  {A.}~\bibnamefont {Syamtomov}}, \bibinfo {author} {\bibfnamefont
  {V.}~\bibnamefont {Vikhrov}}, \bibinfo {author} {\bibfnamefont
  {L.}~\bibnamefont {Vizireva}},\ and\ \bibinfo {author} {\bibfnamefont
  {S.}~\bibnamefont {Zaporozhets}},\ }\href
  {https://doi.org/https://doi.org/10.1016/S0370-2693(96)01455-4} {\bibfield
  {journal} {\bibinfo  {journal} {Physics Letters B}\ }\textbf {\bibinfo
  {volume} {391}},\ \bibinfo {pages} {22} (\bibinfo {year} {1997})}\BibitemShut
  {NoStop}%
\bibitem [{\citenamefont {Nomofilov}\ \emph {et~al.}(1994)\citenamefont
  {Nomofilov}, \citenamefont {Perelygin}, \citenamefont {Peresedov},
  \citenamefont {Senner}, \citenamefont {Sharov}, \citenamefont {Sotnikov},
  \citenamefont {Strunov}, \citenamefont {Zarubin}, \citenamefont {Zolin},
  \citenamefont {Belostotsky}, \citenamefont {Izotov}, \citenamefont {Nelubin},
  \citenamefont {Sulimov}, \citenamefont {Vikhrov}, \citenamefont {Dzikowski},\
  and\ \citenamefont {Korejwo}}]{Anomalon1994}%
  \BibitemOpen
  \bibfield  {author} {\bibinfo {author} {\bibfnamefont {A.}~\bibnamefont
  {Nomofilov}}, \bibinfo {author} {\bibfnamefont {V.}~\bibnamefont
  {Perelygin}}, \bibinfo {author} {\bibfnamefont {V.}~\bibnamefont
  {Peresedov}}, \bibinfo {author} {\bibfnamefont {A.}~\bibnamefont {Senner}},
  \bibinfo {author} {\bibfnamefont {V.}~\bibnamefont {Sharov}}, \bibinfo
  {author} {\bibfnamefont {V.}~\bibnamefont {Sotnikov}}, \bibinfo {author}
  {\bibfnamefont {L.}~\bibnamefont {Strunov}}, \bibinfo {author} {\bibfnamefont
  {A.}~\bibnamefont {Zarubin}}, \bibinfo {author} {\bibfnamefont
  {L.}~\bibnamefont {Zolin}}, \bibinfo {author} {\bibfnamefont
  {S.}~\bibnamefont {Belostotsky}}, \bibinfo {author} {\bibfnamefont
  {A.}~\bibnamefont {Izotov}}, \bibinfo {author} {\bibfnamefont
  {V.}~\bibnamefont {Nelubin}}, \bibinfo {author} {\bibfnamefont
  {V.}~\bibnamefont {Sulimov}}, \bibinfo {author} {\bibfnamefont
  {V.}~\bibnamefont {Vikhrov}}, \bibinfo {author} {\bibfnamefont
  {T.}~\bibnamefont {Dzikowski}},\ and\ \bibinfo {author} {\bibfnamefont
  {A.}~\bibnamefont {Korejwo}},\ }\href
  {https://doi.org/https://doi.org/10.1016/0370-2693(94)90020-5} {\bibfield
  {journal} {\bibinfo  {journal} {Physics Letters B}\ }\textbf {\bibinfo
  {volume} {325}},\ \bibinfo {pages} {327} (\bibinfo {year}
  {1994})}\BibitemShut {NoStop}%
\bibitem [{\citenamefont {Aono}\ \emph {et~al.}(1995)\citenamefont {Aono},
  \citenamefont {Chernykh}, \citenamefont {Dzikowski}, \citenamefont
  {Hasegawa}, \citenamefont {Horikawa}, \citenamefont {Iwata}, \citenamefont
  {Izotov}, \citenamefont {Nomofilov}, \citenamefont {Ogawa}, \citenamefont
  {Perelygin}, \citenamefont {Sasaki}, \citenamefont {Sharov}, \citenamefont
  {Smolin}, \citenamefont {Sotnikov}, \citenamefont {Strunov}, \citenamefont
  {Toyoda}, \citenamefont {Yamada}, \citenamefont {Zaporozhets}, \citenamefont
  {Zarubin}, \citenamefont {Zhiltsov},\ and\ \citenamefont
  {Zolin}}]{Anomalon1995}%
  \BibitemOpen
  \bibfield  {author} {\bibinfo {author} {\bibfnamefont {T.}~\bibnamefont
  {Aono}}, \bibinfo {author} {\bibfnamefont {E.~V.}\ \bibnamefont {Chernykh}},
  \bibinfo {author} {\bibfnamefont {T.}~\bibnamefont {Dzikowski}}, \bibinfo
  {author} {\bibfnamefont {T.}~\bibnamefont {Hasegawa}}, \bibinfo {author}
  {\bibfnamefont {N.}~\bibnamefont {Horikawa}}, \bibinfo {author}
  {\bibfnamefont {T.}~\bibnamefont {Iwata}}, \bibinfo {author} {\bibfnamefont
  {A.~A.}\ \bibnamefont {Izotov}}, \bibinfo {author} {\bibfnamefont {A.~A.}\
  \bibnamefont {Nomofilov}}, \bibinfo {author} {\bibfnamefont {A.}~\bibnamefont
  {Ogawa}}, \bibinfo {author} {\bibfnamefont {V.~V.}\ \bibnamefont
  {Perelygin}}, \bibinfo {author} {\bibfnamefont {T.}~\bibnamefont {Sasaki}},
  \bibinfo {author} {\bibfnamefont {V.~I.}\ \bibnamefont {Sharov}}, \bibinfo
  {author} {\bibfnamefont {D.~A.}\ \bibnamefont {Smolin}}, \bibinfo {author}
  {\bibfnamefont {V.~N.}\ \bibnamefont {Sotnikov}}, \bibinfo {author}
  {\bibfnamefont {L.~N.}\ \bibnamefont {Strunov}}, \bibinfo {author}
  {\bibfnamefont {S.}~\bibnamefont {Toyoda}}, \bibinfo {author} {\bibfnamefont
  {T.}~\bibnamefont {Yamada}}, \bibinfo {author} {\bibfnamefont {S.~A.}\
  \bibnamefont {Zaporozhets}}, \bibinfo {author} {\bibfnamefont {A.~V.}\
  \bibnamefont {Zarubin}}, \bibinfo {author} {\bibfnamefont {V.~E.}\
  \bibnamefont {Zhiltsov}},\ and\ \bibinfo {author} {\bibfnamefont {L.~S.}\
  \bibnamefont {Zolin}},\ }\href {https://doi.org/10.1103/PhysRevLett.74.4997}
  {\bibfield  {journal} {\bibinfo  {journal} {Phys. Rev. Lett.}\ }\textbf
  {\bibinfo {volume} {74}},\ \bibinfo {pages} {4997} (\bibinfo {year}
  {1995})}\BibitemShut {NoStop}%
\bibitem [{\citenamefont {Afanasiev}\ \emph {et~al.}(1998)\citenamefont
  {Afanasiev}, \citenamefont {Arkhipov}, \citenamefont {Azhgirey},
  \citenamefont {Bondarev}, \citenamefont {Chernykh}, \citenamefont {Ehara},
  \citenamefont {Ershov}, \citenamefont {Filipov}, \citenamefont {Fimushkin},
  \citenamefont {Fukui} \emph {et~al.}}]{DubnaTensorD}%
  \BibitemOpen
  \bibfield  {author} {\bibinfo {author} {\bibfnamefont {S.}~\bibnamefont
  {Afanasiev}}, \bibinfo {author} {\bibfnamefont {V.}~\bibnamefont {Arkhipov}},
  \bibinfo {author} {\bibfnamefont {L.}~\bibnamefont {Azhgirey}}, \bibinfo
  {author} {\bibfnamefont {V.}~\bibnamefont {Bondarev}}, \bibinfo {author}
  {\bibfnamefont {E.}~\bibnamefont {Chernykh}}, \bibinfo {author}
  {\bibfnamefont {M.}~\bibnamefont {Ehara}}, \bibinfo {author} {\bibfnamefont
  {V.}~\bibnamefont {Ershov}}, \bibinfo {author} {\bibfnamefont
  {G.}~\bibnamefont {Filipov}}, \bibinfo {author} {\bibfnamefont
  {V.}~\bibnamefont {Fimushkin}}, \bibinfo {author} {\bibfnamefont
  {S.}~\bibnamefont {Fukui}}, \emph {et~al.},\ }\href
  {https://doi.org/https://doi.org/10.1016/S0370-2693(98)00733-3} {\bibfield
  {journal} {\bibinfo  {journal} {Physics Letters B}\ }\textbf {\bibinfo
  {volume} {434}},\ \bibinfo {pages} {21} (\bibinfo {year} {1998})}\BibitemShut
  {NoStop}%
\bibitem [{\citenamefont {Uzikov}(1998)}]{Uzikov:1998qk}%
  \BibitemOpen
  \bibfield  {author} {\bibinfo {author} {\bibfnamefont {Y.~N.}\ \bibnamefont
  {Uzikov}},\ }\href {https://doi.org/10.1134/1.953092} {\bibfield  {journal}
  {\bibinfo  {journal} {Phys. Part. Nucl.}\ }\textbf {\bibinfo {volume} {29}},\
  \bibinfo {pages} {583} (\bibinfo {year} {1998})}\BibitemShut {NoStop}%
\bibitem [{\citenamefont {Ladygin}\ \emph {et~al.}(2002)\citenamefont
  {Ladygin}, \citenamefont {Azhgirey}, \citenamefont {Afanasiev}, \citenamefont
  {Arkhipov}, , \citenamefont {Bondarev}, \citenamefont {Filipov},
  \citenamefont {Isupov}, \citenamefont {VI}, \citenamefont {Kartamyshev},
  \citenamefont {Kashirin} \emph {et~al.}}]{DubnaAyy2002}%
  \BibitemOpen
  \bibfield  {author} {\bibinfo {author} {\bibfnamefont {V.}~\bibnamefont
  {Ladygin}}, \bibinfo {author} {\bibfnamefont {L.}~\bibnamefont {Azhgirey}},
  \bibinfo {author} {\bibfnamefont {S.}~\bibnamefont {Afanasiev}}, \bibinfo
  {author} {\bibfnamefont {V.}~\bibnamefont {Arkhipov}}, , \bibinfo {author}
  {\bibfnamefont {V.}~\bibnamefont {Bondarev}}, \bibinfo {author}
  {\bibfnamefont {G.}~\bibnamefont {Filipov}}, \bibinfo {author} {\bibfnamefont
  {A.}~\bibnamefont {Isupov}}, \bibinfo {author} {\bibfnamefont
  {I.}~\bibnamefont {VI}}, \bibinfo {author} {\bibfnamefont {A.}~\bibnamefont
  {Kartamyshev}}, \bibinfo {author} {\bibfnamefont {V.}~\bibnamefont
  {Kashirin}}, \emph {et~al.},\ }\href
  {https://doi.org/https://doi.org/10.1007/s00601-002-0115-3} {\bibfield
  {journal} {\bibinfo  {journal} {Few-Body Systems}\ }\textbf {\bibinfo
  {volume} {32}},\ \bibinfo {pages} {127} (\bibinfo {year} {2002})}\BibitemShut
  {NoStop}%
\bibitem [{\citenamefont {Azhgirey}\ \emph {et~al.}(2004)\citenamefont
  {Azhgirey}, \citenamefont {Afanasiev}, \citenamefont {Isupov}, \citenamefont
  {Ivanov}, \citenamefont {Khrenov}, \citenamefont {Ladygin}, \citenamefont
  {Litvinenko}, \citenamefont {Peresedov}, \citenamefont {Yudin}, \citenamefont
  {Zhmyrov},\ and\ \citenamefont {Zolin}}]{DubnaAyy2004}%
  \BibitemOpen
  \bibfield  {author} {\bibinfo {author} {\bibfnamefont {L.}~\bibnamefont
  {Azhgirey}}, \bibinfo {author} {\bibfnamefont {S.}~\bibnamefont {Afanasiev}},
  \bibinfo {author} {\bibfnamefont {A.}~\bibnamefont {Isupov}}, \bibinfo
  {author} {\bibfnamefont {V.}~\bibnamefont {Ivanov}}, \bibinfo {author}
  {\bibfnamefont {A.}~\bibnamefont {Khrenov}}, \bibinfo {author} {\bibfnamefont
  {V.}~\bibnamefont {Ladygin}}, \bibinfo {author} {\bibfnamefont
  {A.}~\bibnamefont {Litvinenko}}, \bibinfo {author} {\bibfnamefont
  {V.}~\bibnamefont {Peresedov}}, \bibinfo {author} {\bibfnamefont
  {N.}~\bibnamefont {Yudin}}, \bibinfo {author} {\bibfnamefont
  {V.}~\bibnamefont {Zhmyrov}},\ and\ \bibinfo {author} {\bibfnamefont
  {L.}~\bibnamefont {Zolin}},\ }\href
  {https://doi.org/https://doi.org/10.1016/j.physletb.2004.05.057} {\bibfield
  {journal} {\bibinfo  {journal} {Physics Letters B}\ }\textbf {\bibinfo
  {volume} {595}},\ \bibinfo {pages} {151} (\bibinfo {year}
  {2004})}\BibitemShut {NoStop}%
\bibitem [{\citenamefont {Ladygin}\ \emph {et~al.}(2005)\citenamefont
  {Ladygin}, \citenamefont {Azhgirey}, \citenamefont {Afanasiev}, \citenamefont
  {Arkhipov}, \citenamefont {Bondarev}, \citenamefont {Borzounov},
  \citenamefont {Filipov}, \citenamefont {Golovanov}, \citenamefont {Isupov},
  \citenamefont {Ivanov}, \citenamefont {Kartamyshev}, \citenamefont
  {Kashirin}, \citenamefont {Khrenov}, \citenamefont {Kolesnikov},
  \citenamefont {Kuznetsov}, \citenamefont {Litvinenko}, \citenamefont
  {Reznikov}, \citenamefont {Rukoyatkin}, \citenamefont {Semenov},
  \citenamefont {Semenova}, \citenamefont {Stoletov}, \citenamefont {Tzvinev},
  \citenamefont {Yudin}, \citenamefont {Zhmyrov},\ and\ \citenamefont
  {Zolin}}]{DubnaAyy2005}%
  \BibitemOpen
  \bibfield  {author} {\bibinfo {author} {\bibfnamefont {V.}~\bibnamefont
  {Ladygin}}, \bibinfo {author} {\bibfnamefont {L.}~\bibnamefont {Azhgirey}},
  \bibinfo {author} {\bibfnamefont {S.}~\bibnamefont {Afanasiev}}, \bibinfo
  {author} {\bibfnamefont {V.}~\bibnamefont {Arkhipov}}, \bibinfo {author}
  {\bibfnamefont {V.}~\bibnamefont {Bondarev}}, \bibinfo {author}
  {\bibfnamefont {Y.}~\bibnamefont {Borzounov}}, \bibinfo {author}
  {\bibfnamefont {G.}~\bibnamefont {Filipov}}, \bibinfo {author} {\bibfnamefont
  {L.}~\bibnamefont {Golovanov}}, \bibinfo {author} {\bibfnamefont
  {A.}~\bibnamefont {Isupov}}, \bibinfo {author} {\bibfnamefont
  {V.}~\bibnamefont {Ivanov}}, \bibinfo {author} {\bibfnamefont
  {A.}~\bibnamefont {Kartamyshev}}, \bibinfo {author} {\bibfnamefont
  {V.}~\bibnamefont {Kashirin}}, \bibinfo {author} {\bibfnamefont
  {A.}~\bibnamefont {Khrenov}}, \bibinfo {author} {\bibfnamefont
  {V.}~\bibnamefont {Kolesnikov}}, \bibinfo {author} {\bibfnamefont
  {V.}~\bibnamefont {Kuznetsov}}, \bibinfo {author} {\bibfnamefont
  {A.}~\bibnamefont {Litvinenko}}, \bibinfo {author} {\bibfnamefont
  {S.}~\bibnamefont {Reznikov}}, \bibinfo {author} {\bibfnamefont
  {P.}~\bibnamefont {Rukoyatkin}}, \bibinfo {author} {\bibfnamefont
  {A.}~\bibnamefont {Semenov}}, \bibinfo {author} {\bibfnamefont
  {I.}~\bibnamefont {Semenova}}, \bibinfo {author} {\bibfnamefont
  {G.}~\bibnamefont {Stoletov}}, \bibinfo {author} {\bibfnamefont
  {A.}~\bibnamefont {Tzvinev}}, \bibinfo {author} {\bibfnamefont
  {N.}~\bibnamefont {Yudin}}, \bibinfo {author} {\bibfnamefont
  {V.}~\bibnamefont {Zhmyrov}},\ and\ \bibinfo {author} {\bibfnamefont
  {L.}~\bibnamefont {Zolin}},\ }\href
  {https://doi.org/https://doi.org/10.1016/j.physletb.2005.09.072} {\bibfield
  {journal} {\bibinfo  {journal} {Physics Letters B}\ }\textbf {\bibinfo
  {volume} {629}},\ \bibinfo {pages} {60} (\bibinfo {year} {2005})}\BibitemShut
  {NoStop}%
\bibitem [{\citenamefont {Imambekov}\ and\ \citenamefont
  {Uzikov}(1990)}]{Imambekov:1990ru}%
  \BibitemOpen
  \bibfield  {author} {\bibinfo {author} {\bibfnamefont {O.}~\bibnamefont
  {Imambekov}}\ and\ \bibinfo {author} {\bibfnamefont {Y.~N.}\ \bibnamefont
  {Uzikov}},\ }\href@noop {} {\bibfield  {journal} {\bibinfo  {journal} {Sov.
  J. Nucl. Phys.}\ }\textbf {\bibinfo {volume} {52}},\ \bibinfo {pages} {862}
  (\bibinfo {year} {1990})}\BibitemShut {NoStop}%
\bibitem [{\citenamefont {Uzikov}(2002)}]{Uzikov:2001uc}%
  \BibitemOpen
  \bibfield  {author} {\bibinfo {author} {\bibfnamefont {Y.~N.}\ \bibnamefont
  {Uzikov}},\ }\href {https://doi.org/10.1088/0954-3899/28/4/401} {\bibfield
  {journal} {\bibinfo  {journal} {J. Phys. G}\ }\textbf {\bibinfo {volume}
  {28}},\ \bibinfo {pages} {B13} (\bibinfo {year} {2002})},\ \Eprint
  {https://arxiv.org/abs/nucl-th/0111081} {arXiv:nucl-th/0111081} \BibitemShut
  {NoStop}%
\bibitem [{\citenamefont {Arvieux}\ \emph {et~al.}(1984)\citenamefont
  {Arvieux}, \citenamefont {Baker}, \citenamefont {Beurtey}, \citenamefont
  {Boivin}, \citenamefont {Cameron}, \citenamefont {Hasegawa}, \citenamefont
  {Hutcheon}, \citenamefont {Banaigs}, \citenamefont {Berger}, \citenamefont
  {Codino}, \citenamefont {Duflo}, \citenamefont {Goldzahl}, \citenamefont
  {Plouin}, \citenamefont {Boudard}, \citenamefont {Gaillard}, \citenamefont
  {{Van Sen}},\ and\ \citenamefont {Perdrisat}}]{Arvieux1984}%
  \BibitemOpen
  \bibfield  {author} {\bibinfo {author} {\bibfnamefont {J.}~\bibnamefont
  {Arvieux}}, \bibinfo {author} {\bibfnamefont {S.}~\bibnamefont {Baker}},
  \bibinfo {author} {\bibfnamefont {R.}~\bibnamefont {Beurtey}}, \bibinfo
  {author} {\bibfnamefont {M.}~\bibnamefont {Boivin}}, \bibinfo {author}
  {\bibfnamefont {J.}~\bibnamefont {Cameron}}, \bibinfo {author} {\bibfnamefont
  {T.}~\bibnamefont {Hasegawa}}, \bibinfo {author} {\bibfnamefont
  {D.}~\bibnamefont {Hutcheon}}, \bibinfo {author} {\bibfnamefont
  {J.}~\bibnamefont {Banaigs}}, \bibinfo {author} {\bibfnamefont
  {J.}~\bibnamefont {Berger}}, \bibinfo {author} {\bibfnamefont
  {A.}~\bibnamefont {Codino}}, \bibinfo {author} {\bibfnamefont
  {J.}~\bibnamefont {Duflo}}, \bibinfo {author} {\bibfnamefont
  {L.}~\bibnamefont {Goldzahl}}, \bibinfo {author} {\bibfnamefont
  {F.}~\bibnamefont {Plouin}}, \bibinfo {author} {\bibfnamefont
  {A.}~\bibnamefont {Boudard}}, \bibinfo {author} {\bibfnamefont
  {G.}~\bibnamefont {Gaillard}}, \bibinfo {author} {\bibfnamefont
  {N.}~\bibnamefont {{Van Sen}}},\ and\ \bibinfo {author} {\bibfnamefont
  {C.}~\bibnamefont {Perdrisat}},\ }\href
  {https://doi.org/https://doi.org/10.1016/0375-9474(84)90272-0} {\bibfield
  {journal} {\bibinfo  {journal} {Nuclear Physics A}\ }\textbf {\bibinfo
  {volume} {431}},\ \bibinfo {pages} {613} (\bibinfo {year}
  {1984})}\BibitemShut {NoStop}%
\bibitem [{\citenamefont {Ghazikhanian}\ \emph {et~al.}(1991)\citenamefont
  {Ghazikhanian}, \citenamefont {Aas}, \citenamefont {Adams}, \citenamefont
  {Bleszynski}, \citenamefont {Bleszynski}, \citenamefont {Bystricky},
  \citenamefont {Igo}, \citenamefont {Jaroszewicz}, \citenamefont {Sperisen},
  \citenamefont {Whitten}, \citenamefont {Chaumette}, \citenamefont {Deregel},
  \citenamefont {Fabre}, \citenamefont {Lehar}, \citenamefont {de~Lesquen},
  \citenamefont {van Rossum}, \citenamefont {Arvieux}, \citenamefont {Ball},
  \citenamefont {Boudard},\ and\ \citenamefont {Perrot}}]{Ghazikhanian1991}%
  \BibitemOpen
  \bibfield  {author} {\bibinfo {author} {\bibfnamefont {V.}~\bibnamefont
  {Ghazikhanian}}, \bibinfo {author} {\bibfnamefont {B.}~\bibnamefont {Aas}},
  \bibinfo {author} {\bibfnamefont {D.}~\bibnamefont {Adams}}, \bibinfo
  {author} {\bibfnamefont {E.}~\bibnamefont {Bleszynski}}, \bibinfo {author}
  {\bibfnamefont {M.}~\bibnamefont {Bleszynski}}, \bibinfo {author}
  {\bibfnamefont {J.}~\bibnamefont {Bystricky}}, \bibinfo {author}
  {\bibfnamefont {G.~J.}\ \bibnamefont {Igo}}, \bibinfo {author} {\bibfnamefont
  {T.}~\bibnamefont {Jaroszewicz}}, \bibinfo {author} {\bibfnamefont
  {F.}~\bibnamefont {Sperisen}}, \bibinfo {author} {\bibfnamefont {C.~A.}\
  \bibnamefont {Whitten}}, \bibinfo {author} {\bibfnamefont {P.}~\bibnamefont
  {Chaumette}}, \bibinfo {author} {\bibfnamefont {J.}~\bibnamefont {Deregel}},
  \bibinfo {author} {\bibfnamefont {J.}~\bibnamefont {Fabre}}, \bibinfo
  {author} {\bibfnamefont {F.}~\bibnamefont {Lehar}}, \bibinfo {author}
  {\bibfnamefont {A.}~\bibnamefont {de~Lesquen}}, \bibinfo {author}
  {\bibfnamefont {L.}~\bibnamefont {van Rossum}}, \bibinfo {author}
  {\bibfnamefont {J.}~\bibnamefont {Arvieux}}, \bibinfo {author} {\bibfnamefont
  {J.}~\bibnamefont {Ball}}, \bibinfo {author} {\bibfnamefont {A.}~\bibnamefont
  {Boudard}},\ and\ \bibinfo {author} {\bibfnamefont {F.}~\bibnamefont
  {Perrot}},\ }\href {https://doi.org/10.1103/PhysRevC.43.1532} {\bibfield
  {journal} {\bibinfo  {journal} {Phys. Rev. C}\ }\textbf {\bibinfo {volume}
  {43}},\ \bibinfo {pages} {1532} (\bibinfo {year} {1991})}\BibitemShut
  {NoStop}%
\bibitem [{\citenamefont {Platonova}\ and\ \citenamefont
  {Kukulin}(2020)}]{Platonova_Kukulin}%
  \BibitemOpen
  \bibfield  {author} {\bibinfo {author} {\bibfnamefont {M.~N.}\ \bibnamefont
  {Platonova}}\ and\ \bibinfo {author} {\bibfnamefont {V.~I.}\ \bibnamefont
  {Kukulin}},\ }\href {https://doi.org/10.1140/epja/s10050-020-00126-8}
  {\bibfield  {journal} {\bibinfo  {journal} {The European Physical Journal A}\
  }\textbf {\bibinfo {volume} {56}},\ \bibinfo {pages} {132} (\bibinfo {year}
  {2020})}\BibitemShut {NoStop}%
\bibitem [{\citenamefont {Azhgirey}\ \emph {et~al.}(2008)\citenamefont
  {Azhgirey}, \citenamefont {Gurchin}, \citenamefont {Isupov}, \citenamefont
  {Khrenov}, \citenamefont {Kiselev}, \citenamefont {Kurilkin}, \citenamefont
  {Kurilkin}, \citenamefont {Ladygin}, \citenamefont {Litvinenko},
  \citenamefont {Peresedov} \emph {et~al.}}]{azhgirey2008tensor}%
  \BibitemOpen
  \bibfield  {author} {\bibinfo {author} {\bibfnamefont {L.}~\bibnamefont
  {Azhgirey}}, \bibinfo {author} {\bibfnamefont {Y.~V.}\ \bibnamefont
  {Gurchin}}, \bibinfo {author} {\bibfnamefont {A.~Y.}\ \bibnamefont {Isupov}},
  \bibinfo {author} {\bibfnamefont {A.}~\bibnamefont {Khrenov}}, \bibinfo
  {author} {\bibfnamefont {A.}~\bibnamefont {Kiselev}}, \bibinfo {author}
  {\bibfnamefont {A.}~\bibnamefont {Kurilkin}}, \bibinfo {author}
  {\bibfnamefont {P.}~\bibnamefont {Kurilkin}}, \bibinfo {author}
  {\bibfnamefont {V.}~\bibnamefont {Ladygin}}, \bibinfo {author} {\bibfnamefont
  {A.}~\bibnamefont {Litvinenko}}, \bibinfo {author} {\bibfnamefont
  {V.}~\bibnamefont {Peresedov}}, \emph {et~al.},\ }\href
  {https://doi.org/10.1134/S1547477108050051} {\bibfield  {journal} {\bibinfo
  {journal} {Physics of Particles and Nuclei Letters}\ }\textbf {\bibinfo
  {volume} {5}},\ \bibinfo {pages} {432} (\bibinfo {year} {2008})}\BibitemShut
  {NoStop}%
\bibitem [{\citenamefont {Azhgirei}\ \emph {et~al.}(2010)\citenamefont
  {Azhgirei}, \citenamefont {Vasiliev}, \citenamefont {Gurchin}, \citenamefont
  {Zhmyrov}, \citenamefont {Zolin}, \citenamefont {Isupov}, \citenamefont
  {Kurilkin}, \citenamefont {Kurilkin}, \citenamefont {Ladygin}, \citenamefont
  {Litvinenko} \emph {et~al.}}]{azhgirey2010tensor}%
  \BibitemOpen
  \bibfield  {author} {\bibinfo {author} {\bibfnamefont {L.}~\bibnamefont
  {Azhgirei}}, \bibinfo {author} {\bibfnamefont {T.}~\bibnamefont {Vasiliev}},
  \bibinfo {author} {\bibfnamefont {Y.~V.}\ \bibnamefont {Gurchin}}, \bibinfo
  {author} {\bibfnamefont {V.}~\bibnamefont {Zhmyrov}}, \bibinfo {author}
  {\bibfnamefont {L.}~\bibnamefont {Zolin}}, \bibinfo {author} {\bibfnamefont
  {A.~Y.}\ \bibnamefont {Isupov}}, \bibinfo {author} {\bibfnamefont
  {A.}~\bibnamefont {Kurilkin}}, \bibinfo {author} {\bibfnamefont
  {P.}~\bibnamefont {Kurilkin}}, \bibinfo {author} {\bibfnamefont
  {V.}~\bibnamefont {Ladygin}}, \bibinfo {author} {\bibfnamefont
  {A.}~\bibnamefont {Litvinenko}}, \emph {et~al.},\ }\href
  {https://doi.org/10.1134/S1547477110010073} {\bibfield  {journal} {\bibinfo
  {journal} {Physics of Particles and Nuclei Letters}\ }\textbf {\bibinfo
  {volume} {7}},\ \bibinfo {pages} {27} (\bibinfo {year} {2010})}\BibitemShut
  {NoStop}%
\bibitem [{\citenamefont {{V. S. Morozov, A. D. Krisch, M. A. Leonova, R. S.
  Raymond, D. W. Sivers, V. K. Wong, R. Gebel, A. Lehrach, B. Lorentz, R.
  Maier, D. Prasuhn, A. Schnase, H. Stockhorst, D. Eversheim, F. Hinterberger,
  H. Rohdjeß, K. Ulbrich, and K. Yonehara}}(2005)}]{Morozov:Tensor}%
  \BibitemOpen
  \bibfield  {author} {\bibinfo {author} {\bibnamefont {{V. S. Morozov, A. D.
  Krisch, M. A. Leonova, R. S. Raymond, D. W. Sivers, V. K. Wong, R. Gebel, A.
  Lehrach, B. Lorentz, R. Maier, D. Prasuhn, A. Schnase, H. Stockhorst, D.
  Eversheim, F. Hinterberger, H. Rohdjeß, K. Ulbrich, and K. Yonehara}}},\
  }\href {https://doi.org/10.1103/PhysRevSTAB.8.061001} {\bibfield  {journal}
  {\bibinfo  {journal} {Phys. Rev. ST Accel. Beams}\ }\textbf {\bibinfo
  {volume} {8}},\ \bibinfo {pages} {061001} (\bibinfo {year}
  {2005})}\BibitemShut {NoStop}%
%%CITATION = PRSTA,8,061001;%%
\bibitem [{\citenamefont {Chiladze}\ \emph {et~al.}(2006)\citenamefont
  {Chiladze}, \citenamefont {Kacharava}, \citenamefont {Rathmann},
  \citenamefont {Wilkin}, \citenamefont {Barsov}, \citenamefont {Carbonell},
  \citenamefont {Dymov}, \citenamefont {Engels}, \citenamefont {Eversheim},
  \citenamefont {Felden}, \citenamefont {Gebel}, \citenamefont {Glagolev},
  \citenamefont {Grigoriev}, \citenamefont {Gusev}, \citenamefont {Hartmann},
  \citenamefont {Hinterberger}, \citenamefont {Hejny}, \citenamefont {Khoukaz},
  \citenamefont {Keshelashvili}, \citenamefont {Koch}, \citenamefont {Komarov},
  \citenamefont {Kulessa}, \citenamefont {Kulikov}, \citenamefont {Lehrach},
  \citenamefont {Lorentz}, \citenamefont {Macharashvili}, \citenamefont
  {Maier}, \citenamefont {Maeda}, \citenamefont {Menke}, \citenamefont
  {Mersmann}, \citenamefont {Merzliakov}, \citenamefont {Mikirtytchiants},
  \citenamefont {Mikirtytchiants}, \citenamefont {Mussgiller}, \citenamefont
  {Nioradze}, \citenamefont {Ohm}, \citenamefont {Prasuhn}, \citenamefont
  {Rohdje\ss{}}, \citenamefont {Schleichert}, \citenamefont {Seyfarth},
  \citenamefont {Steffens}, \citenamefont {Stein}, \citenamefont {Str\"oher},
  \citenamefont {Trusov}, \citenamefont {Ulbrich}, \citenamefont {Uzikov},
  \citenamefont {Wro\ifmmode~\acute{n}\else \'{n}\fi{}ska},\ and\ \citenamefont
  {Yaschenko}}]{ChiladzeTensorD}%
  \BibitemOpen
  \bibfield  {author} {\bibinfo {author} {\bibfnamefont {D.}~\bibnamefont
  {Chiladze}}, \bibinfo {author} {\bibfnamefont {A.}~\bibnamefont {Kacharava}},
  \bibinfo {author} {\bibfnamefont {F.}~\bibnamefont {Rathmann}}, \bibinfo
  {author} {\bibfnamefont {C.}~\bibnamefont {Wilkin}}, \bibinfo {author}
  {\bibfnamefont {S.}~\bibnamefont {Barsov}}, \bibinfo {author} {\bibfnamefont
  {J.}~\bibnamefont {Carbonell}}, \bibinfo {author} {\bibfnamefont
  {S.}~\bibnamefont {Dymov}}, \bibinfo {author} {\bibfnamefont
  {R.}~\bibnamefont {Engels}}, \bibinfo {author} {\bibfnamefont {P.~D.}\
  \bibnamefont {Eversheim}}, \bibinfo {author} {\bibfnamefont {O.}~\bibnamefont
  {Felden}}, \bibinfo {author} {\bibfnamefont {R.}~\bibnamefont {Gebel}},
  \bibinfo {author} {\bibfnamefont {V.}~\bibnamefont {Glagolev}}, \bibinfo
  {author} {\bibfnamefont {K.}~\bibnamefont {Grigoriev}}, \bibinfo {author}
  {\bibfnamefont {D.}~\bibnamefont {Gusev}}, \bibinfo {author} {\bibfnamefont
  {M.}~\bibnamefont {Hartmann}}, \bibinfo {author} {\bibfnamefont
  {F.}~\bibnamefont {Hinterberger}}, \bibinfo {author} {\bibfnamefont
  {V.}~\bibnamefont {Hejny}}, \bibinfo {author} {\bibfnamefont
  {A.}~\bibnamefont {Khoukaz}}, \bibinfo {author} {\bibfnamefont
  {I.}~\bibnamefont {Keshelashvili}}, \bibinfo {author} {\bibfnamefont {H.~R.}\
  \bibnamefont {Koch}}, \bibinfo {author} {\bibfnamefont {V.}~\bibnamefont
  {Komarov}}, \bibinfo {author} {\bibfnamefont {P.}~\bibnamefont {Kulessa}},
  \bibinfo {author} {\bibfnamefont {A.}~\bibnamefont {Kulikov}}, \bibinfo
  {author} {\bibfnamefont {A.}~\bibnamefont {Lehrach}}, \bibinfo {author}
  {\bibfnamefont {B.}~\bibnamefont {Lorentz}}, \bibinfo {author} {\bibfnamefont
  {G.}~\bibnamefont {Macharashvili}}, \bibinfo {author} {\bibfnamefont
  {R.}~\bibnamefont {Maier}}, \bibinfo {author} {\bibfnamefont
  {Y.}~\bibnamefont {Maeda}}, \bibinfo {author} {\bibfnamefont
  {R.}~\bibnamefont {Menke}}, \bibinfo {author} {\bibfnamefont
  {T.}~\bibnamefont {Mersmann}}, \bibinfo {author} {\bibfnamefont
  {S.}~\bibnamefont {Merzliakov}}, \bibinfo {author} {\bibfnamefont
  {M.}~\bibnamefont {Mikirtytchiants}}, \bibinfo {author} {\bibfnamefont
  {S.}~\bibnamefont {Mikirtytchiants}}, \bibinfo {author} {\bibfnamefont
  {A.}~\bibnamefont {Mussgiller}}, \bibinfo {author} {\bibfnamefont
  {M.}~\bibnamefont {Nioradze}}, \bibinfo {author} {\bibfnamefont
  {H.}~\bibnamefont {Ohm}}, \bibinfo {author} {\bibfnamefont {D.}~\bibnamefont
  {Prasuhn}}, \bibinfo {author} {\bibfnamefont {H.}~\bibnamefont
  {Rohdje\ss{}}}, \bibinfo {author} {\bibfnamefont {R.}~\bibnamefont
  {Schleichert}}, \bibinfo {author} {\bibfnamefont {H.}~\bibnamefont
  {Seyfarth}}, \bibinfo {author} {\bibfnamefont {E.}~\bibnamefont {Steffens}},
  \bibinfo {author} {\bibfnamefont {H.~J.}\ \bibnamefont {Stein}}, \bibinfo
  {author} {\bibfnamefont {H.}~\bibnamefont {Str\"oher}}, \bibinfo {author}
  {\bibfnamefont {S.}~\bibnamefont {Trusov}}, \bibinfo {author} {\bibfnamefont
  {K.}~\bibnamefont {Ulbrich}}, \bibinfo {author} {\bibfnamefont
  {Y.}~\bibnamefont {Uzikov}}, \bibinfo {author} {\bibfnamefont
  {A.}~\bibnamefont {Wro\ifmmode~\acute{n}\else \'{n}\fi{}ska}},\ and\ \bibinfo
  {author} {\bibfnamefont {S.}~\bibnamefont {Yaschenko}},\ }\href
  {https://doi.org/10.1103/PhysRevSTAB.9.050101} {\bibfield  {journal}
  {\bibinfo  {journal} {Phys. Rev. ST Accel. Beams}\ }\textbf {\bibinfo
  {volume} {9}},\ \bibinfo {pages} {050101} (\bibinfo {year}
  {2006})}\BibitemShut {NoStop}%
\bibitem [{\citenamefont {Mchedlishvili}\ \emph {et~al.}(2013)\citenamefont
  {Mchedlishvili}, \citenamefont {Barsov}, \citenamefont {Carbonell},
  \citenamefont {Chiladze}, \citenamefont {Dymov}, \citenamefont {Dzyuba},
  \citenamefont {Engels}, \citenamefont {Gebel}, \citenamefont {Glagolev},
  \citenamefont {Grigoryev}, \citenamefont {Goslawski}, \citenamefont
  {Hartmann}, \citenamefont {Imambekov}, \citenamefont {Kacharava},
  \citenamefont {Kamerdzhiev}, \citenamefont {Keshelashvili}, \citenamefont
  {Khoukaz}, \citenamefont {Komarov}, \citenamefont {Kulessa}, \citenamefont
  {Kulikov}, \citenamefont {Lehrach}, \citenamefont {Lomidze}, \citenamefont
  {Lorentz}, \citenamefont {Macharashvili}, \citenamefont {Maier},
  \citenamefont {Merzliakov}, \citenamefont {Mielke}, \citenamefont
  {Mikirtychyants}, \citenamefont {Mikirtychyants}, \citenamefont {Nioradze},
  \citenamefont {Ohm}, \citenamefont {Papenbrock}, \citenamefont {Prasuhn},
  \citenamefont {Rathmann}, \citenamefont {Serdyuk}, \citenamefont {Seyfarth},
  \citenamefont {Stein}, \citenamefont {Steffens}, \citenamefont {Stockhorst},
  \citenamefont {Ströher}, \citenamefont {Tabidze}, \citenamefont {Trusov},
  \citenamefont {Uzikov}, \citenamefont {Valdau},\ and\ \citenamefont
  {Wilkin}}]{Mchedlishvili:2013bja}%
  \BibitemOpen
  \bibfield  {author} {\bibinfo {author} {\bibfnamefont {D.}~\bibnamefont
  {Mchedlishvili}}, \bibinfo {author} {\bibfnamefont {S.}~\bibnamefont
  {Barsov}}, \bibinfo {author} {\bibfnamefont {J.}~\bibnamefont {Carbonell}},
  \bibinfo {author} {\bibfnamefont {D.}~\bibnamefont {Chiladze}}, \bibinfo
  {author} {\bibfnamefont {S.}~\bibnamefont {Dymov}}, \bibinfo {author}
  {\bibfnamefont {A.}~\bibnamefont {Dzyuba}}, \bibinfo {author} {\bibfnamefont
  {R.}~\bibnamefont {Engels}}, \bibinfo {author} {\bibfnamefont
  {R.}~\bibnamefont {Gebel}}, \bibinfo {author} {\bibfnamefont
  {V.}~\bibnamefont {Glagolev}}, \bibinfo {author} {\bibfnamefont
  {K.}~\bibnamefont {Grigoryev}}, \bibinfo {author} {\bibfnamefont
  {P.}~\bibnamefont {Goslawski}}, \bibinfo {author} {\bibfnamefont
  {M.}~\bibnamefont {Hartmann}}, \bibinfo {author} {\bibfnamefont
  {O.}~\bibnamefont {Imambekov}}, \bibinfo {author} {\bibfnamefont
  {A.}~\bibnamefont {Kacharava}}, \bibinfo {author} {\bibfnamefont
  {V.}~\bibnamefont {Kamerdzhiev}}, \bibinfo {author} {\bibfnamefont
  {I.}~\bibnamefont {Keshelashvili}}, \bibinfo {author} {\bibfnamefont
  {A.}~\bibnamefont {Khoukaz}}, \bibinfo {author} {\bibfnamefont
  {V.}~\bibnamefont {Komarov}}, \bibinfo {author} {\bibfnamefont
  {P.}~\bibnamefont {Kulessa}}, \bibinfo {author} {\bibfnamefont
  {A.}~\bibnamefont {Kulikov}}, \bibinfo {author} {\bibfnamefont
  {A.}~\bibnamefont {Lehrach}}, \bibinfo {author} {\bibfnamefont
  {N.}~\bibnamefont {Lomidze}}, \bibinfo {author} {\bibfnamefont
  {B.}~\bibnamefont {Lorentz}}, \bibinfo {author} {\bibfnamefont
  {G.}~\bibnamefont {Macharashvili}}, \bibinfo {author} {\bibfnamefont
  {R.}~\bibnamefont {Maier}}, \bibinfo {author} {\bibfnamefont
  {S.}~\bibnamefont {Merzliakov}}, \bibinfo {author} {\bibfnamefont
  {M.}~\bibnamefont {Mielke}}, \bibinfo {author} {\bibfnamefont
  {M.}~\bibnamefont {Mikirtychyants}}, \bibinfo {author} {\bibfnamefont
  {S.}~\bibnamefont {Mikirtychyants}}, \bibinfo {author} {\bibfnamefont
  {M.}~\bibnamefont {Nioradze}}, \bibinfo {author} {\bibfnamefont
  {H.}~\bibnamefont {Ohm}}, \bibinfo {author} {\bibfnamefont {M.}~\bibnamefont
  {Papenbrock}}, \bibinfo {author} {\bibfnamefont {D.}~\bibnamefont {Prasuhn}},
  \bibinfo {author} {\bibfnamefont {F.}~\bibnamefont {Rathmann}}, \bibinfo
  {author} {\bibfnamefont {V.}~\bibnamefont {Serdyuk}}, \bibinfo {author}
  {\bibfnamefont {H.}~\bibnamefont {Seyfarth}}, \bibinfo {author}
  {\bibfnamefont {H.}~\bibnamefont {Stein}}, \bibinfo {author} {\bibfnamefont
  {E.}~\bibnamefont {Steffens}}, \bibinfo {author} {\bibfnamefont
  {H.}~\bibnamefont {Stockhorst}}, \bibinfo {author} {\bibfnamefont
  {H.}~\bibnamefont {Ströher}}, \bibinfo {author} {\bibfnamefont
  {M.}~\bibnamefont {Tabidze}}, \bibinfo {author} {\bibfnamefont
  {S.}~\bibnamefont {Trusov}}, \bibinfo {author} {\bibfnamefont
  {Y.}~\bibnamefont {Uzikov}}, \bibinfo {author} {\bibfnamefont
  {Y.}~\bibnamefont {Valdau}},\ and\ \bibinfo {author} {\bibfnamefont
  {C.}~\bibnamefont {Wilkin}},\ }\href
  {https://doi.org/https://doi.org/10.1016/j.physletb.2013.08.018} {\bibfield
  {journal} {\bibinfo  {journal} {Physics Letters B}\ }\textbf {\bibinfo
  {volume} {726}},\ \bibinfo {pages} {145} (\bibinfo {year}
  {2013})}\BibitemShut {NoStop}%
\bibitem [{\citenamefont {Uzikov}\ \emph {et~al.}(2015)\citenamefont {Uzikov},
  \citenamefont {Haidenbauer},\ and\ \citenamefont {Wilkin}}]{Uzikov:2015vta}%
  \BibitemOpen
  \bibfield  {author} {\bibinfo {author} {\bibfnamefont {Y.~N.}\ \bibnamefont
  {Uzikov}}, \bibinfo {author} {\bibfnamefont {J.}~\bibnamefont
  {Haidenbauer}},\ and\ \bibinfo {author} {\bibfnamefont {C.}~\bibnamefont
  {Wilkin}},\ }\href {https://doi.org/10.22323/1.225.0093} {\bibfield
  {journal} {\bibinfo  {journal} {PoS}\ }\textbf {\bibinfo {volume}
  {BaldinISHEPPXXII}},\ \bibinfo {pages} {093} (\bibinfo {year} {2015})},\
  \Eprint {https://arxiv.org/abs/1502.04675} {arXiv:1502.04675 [nucl-th]}
  \BibitemShut {NoStop}%
\bibitem [{\citenamefont {Mchedlishvili}\ \emph {et~al.}(2018)\citenamefont
  {Mchedlishvili}, \citenamefont {Bagdasarian}, \citenamefont {Barsov},
  \citenamefont {Dymov}, \citenamefont {Engels}, \citenamefont {Gebel},
  \citenamefont {Grigoryev}, \citenamefont {Haidenbauer}, \citenamefont
  {Hartmann}, \citenamefont {Kacharava}, \citenamefont {Keshelashvili},
  \citenamefont {Khoukaz}, \citenamefont {Komarov}, \citenamefont {Kulessa},
  \citenamefont {Kulikov}, \citenamefont {Lehrach}, \citenamefont {Lomidze},
  \citenamefont {Lorentz}, \citenamefont {Macharashvili}, \citenamefont
  {Merzliakov}, \citenamefont {Mikirtychyants}, \citenamefont {Nioradze},
  \citenamefont {Ohm}, \citenamefont {Papenbrock}, \citenamefont {Prasuhn},
  \citenamefont {Rathmann}, \citenamefont {Serdyuk}, \citenamefont {Shmakova},
  \citenamefont {Ströher}, \citenamefont {Tabidze}, \citenamefont {Tsirkov},
  \citenamefont {Uzikov}, \citenamefont {Valdau},\ and\ \citenamefont
  {Wilkin}}]{Mchedlishvili:2018uur}%
  \BibitemOpen
  \bibfield  {author} {\bibinfo {author} {\bibfnamefont {D.}~\bibnamefont
  {Mchedlishvili}}, \bibinfo {author} {\bibfnamefont {Z.}~\bibnamefont
  {Bagdasarian}}, \bibinfo {author} {\bibfnamefont {S.}~\bibnamefont {Barsov}},
  \bibinfo {author} {\bibfnamefont {S.}~\bibnamefont {Dymov}}, \bibinfo
  {author} {\bibfnamefont {R.}~\bibnamefont {Engels}}, \bibinfo {author}
  {\bibfnamefont {R.}~\bibnamefont {Gebel}}, \bibinfo {author} {\bibfnamefont
  {K.}~\bibnamefont {Grigoryev}}, \bibinfo {author} {\bibfnamefont
  {J.}~\bibnamefont {Haidenbauer}}, \bibinfo {author} {\bibfnamefont
  {M.}~\bibnamefont {Hartmann}}, \bibinfo {author} {\bibfnamefont
  {A.}~\bibnamefont {Kacharava}}, \bibinfo {author} {\bibfnamefont
  {I.}~\bibnamefont {Keshelashvili}}, \bibinfo {author} {\bibfnamefont
  {A.}~\bibnamefont {Khoukaz}}, \bibinfo {author} {\bibfnamefont
  {V.}~\bibnamefont {Komarov}}, \bibinfo {author} {\bibfnamefont
  {P.}~\bibnamefont {Kulessa}}, \bibinfo {author} {\bibfnamefont
  {A.}~\bibnamefont {Kulikov}}, \bibinfo {author} {\bibfnamefont
  {A.}~\bibnamefont {Lehrach}}, \bibinfo {author} {\bibfnamefont
  {N.}~\bibnamefont {Lomidze}}, \bibinfo {author} {\bibfnamefont
  {B.}~\bibnamefont {Lorentz}}, \bibinfo {author} {\bibfnamefont
  {G.}~\bibnamefont {Macharashvili}}, \bibinfo {author} {\bibfnamefont
  {S.}~\bibnamefont {Merzliakov}}, \bibinfo {author} {\bibfnamefont
  {S.}~\bibnamefont {Mikirtychyants}}, \bibinfo {author} {\bibfnamefont
  {M.}~\bibnamefont {Nioradze}}, \bibinfo {author} {\bibfnamefont
  {H.}~\bibnamefont {Ohm}}, \bibinfo {author} {\bibfnamefont {M.}~\bibnamefont
  {Papenbrock}}, \bibinfo {author} {\bibfnamefont {D.}~\bibnamefont {Prasuhn}},
  \bibinfo {author} {\bibfnamefont {F.}~\bibnamefont {Rathmann}}, \bibinfo
  {author} {\bibfnamefont {V.}~\bibnamefont {Serdyuk}}, \bibinfo {author}
  {\bibfnamefont {V.}~\bibnamefont {Shmakova}}, \bibinfo {author}
  {\bibfnamefont {H.}~\bibnamefont {Ströher}}, \bibinfo {author}
  {\bibfnamefont {M.}~\bibnamefont {Tabidze}}, \bibinfo {author} {\bibfnamefont
  {D.}~\bibnamefont {Tsirkov}}, \bibinfo {author} {\bibfnamefont
  {Y.}~\bibnamefont {Uzikov}}, \bibinfo {author} {\bibfnamefont
  {Y.}~\bibnamefont {Valdau}},\ and\ \bibinfo {author} {\bibfnamefont
  {C.}~\bibnamefont {Wilkin}},\ }\href
  {https://doi.org/https://doi.org/10.1016/j.nuclphysa.2018.05.006} {\bibfield
  {journal} {\bibinfo  {journal} {Nuclear Physics A}\ }\textbf {\bibinfo
  {volume} {977}},\ \bibinfo {pages} {14} (\bibinfo {year} {2018})}\BibitemShut
  {NoStop}%
\bibitem [{\citenamefont {Uzikov}(2025)}]{Uzikov_TensorPolarimetry}%
  \BibitemOpen
  \bibfield  {author} {\bibinfo {author} {\bibfnamefont {Y.~N.}\ \bibnamefont
  {Uzikov}},\ }\href {https://doi.org/10.1134/S154747712570178X} {\bibfield
  {journal} {\bibinfo  {journal} {Phys. Part. Nucl. Lett.}\ }\textbf {\bibinfo
  {volume} {22}},\ \bibinfo {pages} {1392} (\bibinfo {year} {2025})},\ \Eprint
  {https://arxiv.org/abs/2506.17799} {arXiv:2506.17799 [nucl-th]} \BibitemShut
  {NoStop}%
\bibitem [{\citenamefont {Koop}\ and\ \citenamefont
  {Shatunov}(1988)}]{KoopShatunov}%
  \BibitemOpen
  \bibfield  {author} {\bibinfo {author} {\bibfnamefont {I.}~\bibnamefont
  {Koop}}\ and\ \bibinfo {author} {\bibfnamefont {Y.}~\bibnamefont
  {Shatunov}},\ }in\ \href {https://doi.org/10.1109/PAC.2007.4440278} {\emph
  {\bibinfo {booktitle} {{ Particle accelerator. Proceedings, 1st EPAC
  Conference, Rome, Italy, June 7-11, 1988. Vol. 1, 2}}}},\ \bibinfo {editor}
  {edited by\ \bibinfo {editor} {\bibfnamefont {S.}~\bibnamefont {Tazzari}}}\
  (\bibinfo  {publisher} {Singapore: World Scientific},\ \bibinfo {year}
  {1988})\ pp.\ \bibinfo {pages} {738--739}\BibitemShut {NoStop}%
\bibitem [{\citenamefont {Guidoboni}\ \emph {et~al.}(2016)\citenamefont
  {Guidoboni}, \citenamefont {Stephenson}, \citenamefont {Andrianov},
  \citenamefont {Augustyniak}, \citenamefont {Bagdasarian}, \citenamefont
  {Bai}, \citenamefont {Baylac}, \citenamefont {Bernreuther}, \citenamefont
  {Bertelli}, \citenamefont {Berz}, \citenamefont {B\"oker}, \citenamefont
  {B\"ohme}, \citenamefont {Bsaisou}, \citenamefont {Chekmenev}, \citenamefont
  {Chiladze}, \citenamefont {Ciullo}, \citenamefont {Contalbrigo},
  \citenamefont {de~Conto}, \citenamefont {Dymov}, \citenamefont {Engels},
  \citenamefont {Esser}, \citenamefont {Eversmann}, \citenamefont {Felden},
  \citenamefont {Gaisser}, \citenamefont {Gebel}, \citenamefont {Gl\"uckler},
  \citenamefont {Goldenbaum}, \citenamefont {Grigoryev}, \citenamefont
  {Grzonka}, \citenamefont {Hahnraths}, \citenamefont {Heberling},
  \citenamefont {Hejny}, \citenamefont {Hempelmann}, \citenamefont {Hetzel},
  \citenamefont {Hinder}, \citenamefont {Hipple}, \citenamefont {H\"olscher},
  \citenamefont {Ivanov}, \citenamefont {Kacharava}, \citenamefont
  {Kamerdzhiev}, \citenamefont {Kamys}, \citenamefont {Keshelashvili},
  \citenamefont {Khoukaz}, \citenamefont {Koop}, \citenamefont {Krause},
  \citenamefont {Krewald}, \citenamefont {Kulikov}, \citenamefont {Lehrach},
  \citenamefont {Lenisa}, \citenamefont {Lomidze}, \citenamefont {Lorentz},
  \citenamefont {Maanen}, \citenamefont {Macharashvili}, \citenamefont
  {Magiera}, \citenamefont {Maier}, \citenamefont {Makino}, \citenamefont
  {Maria\ifmmode~\acute{n}\else \'{n}\fi{}ski}, \citenamefont {Mchedlishvili},
  \citenamefont {Mei\ss{}ner}, \citenamefont {Mey}, \citenamefont {Morse},
  \citenamefont {M\"uller}, \citenamefont {Nass}, \citenamefont {Natour},
  \citenamefont {Nikolaev}, \citenamefont {Nioradze}, \citenamefont
  {Nowakowski}, \citenamefont {Orlov}, \citenamefont {Pesce}, \citenamefont
  {Prasuhn}, \citenamefont {Pretz}, \citenamefont {Rathmann}, \citenamefont
  {Ritman}, \citenamefont {Rosenthal}, \citenamefont {Rudy}, \citenamefont
  {Saleev}, \citenamefont {Sefzick}, \citenamefont {Semertzidis}, \citenamefont
  {Senichev}, \citenamefont {Shmakova}, \citenamefont {Silenko}, \citenamefont
  {Simon}, \citenamefont {Slim}, \citenamefont {Soltner}, \citenamefont
  {Stahl}, \citenamefont {Stassen}, \citenamefont {Statera}, \citenamefont
  {Stockhorst}, \citenamefont {Straatmann}, \citenamefont {Str\"oher},
  \citenamefont {Tabidze}, \citenamefont {Talman}, \citenamefont
  {Th\"orngren~Engblom}, \citenamefont {Trinkel}, \citenamefont
  {Trzci\ifmmode~\acute{n}\else \'{n}\fi{}ski}, \citenamefont {Uzikov},
  \citenamefont {Valdau}, \citenamefont {Valetov}, \citenamefont {Vassiliev},
  \citenamefont {Weidemann}, \citenamefont {Wilkin}, \citenamefont
  {Wro\ifmmode~\acute{n}\else \'{n}\fi{}ska}, \citenamefont {W\"ustner},
  \citenamefont {Zakrzewska}, \citenamefont {Zupra\ifmmode~\acute{n}\else
  \'{n}\fi{}ski},\ and\ \citenamefont {Zyuzin}}]{SCT1000sJEDI}%
  \BibitemOpen
  \bibfield  {author} {\bibinfo {author} {\bibfnamefont {G.}~\bibnamefont
  {Guidoboni}}, \bibinfo {author} {\bibfnamefont {E.}~\bibnamefont
  {Stephenson}}, \bibinfo {author} {\bibfnamefont {S.}~\bibnamefont
  {Andrianov}}, \bibinfo {author} {\bibfnamefont {W.}~\bibnamefont
  {Augustyniak}}, \bibinfo {author} {\bibfnamefont {Z.}~\bibnamefont
  {Bagdasarian}}, \bibinfo {author} {\bibfnamefont {M.}~\bibnamefont {Bai}},
  \bibinfo {author} {\bibfnamefont {M.}~\bibnamefont {Baylac}}, \bibinfo
  {author} {\bibfnamefont {W.}~\bibnamefont {Bernreuther}}, \bibinfo {author}
  {\bibfnamefont {S.}~\bibnamefont {Bertelli}}, \bibinfo {author}
  {\bibfnamefont {M.}~\bibnamefont {Berz}}, \bibinfo {author} {\bibfnamefont
  {J.}~\bibnamefont {B\"oker}}, \bibinfo {author} {\bibfnamefont
  {C.}~\bibnamefont {B\"ohme}}, \bibinfo {author} {\bibfnamefont
  {J.}~\bibnamefont {Bsaisou}}, \bibinfo {author} {\bibfnamefont
  {S.}~\bibnamefont {Chekmenev}}, \bibinfo {author} {\bibfnamefont
  {D.}~\bibnamefont {Chiladze}}, \bibinfo {author} {\bibfnamefont
  {G.}~\bibnamefont {Ciullo}}, \bibinfo {author} {\bibfnamefont
  {M.}~\bibnamefont {Contalbrigo}}, \bibinfo {author} {\bibfnamefont {J.-M.}\
  \bibnamefont {de~Conto}}, \bibinfo {author} {\bibfnamefont {S.}~\bibnamefont
  {Dymov}}, \bibinfo {author} {\bibfnamefont {R.}~\bibnamefont {Engels}},
  \bibinfo {author} {\bibfnamefont {F.~M.}\ \bibnamefont {Esser}}, \bibinfo
  {author} {\bibfnamefont {D.}~\bibnamefont {Eversmann}}, \bibinfo {author}
  {\bibfnamefont {O.}~\bibnamefont {Felden}}, \bibinfo {author} {\bibfnamefont
  {M.}~\bibnamefont {Gaisser}}, \bibinfo {author} {\bibfnamefont
  {R.}~\bibnamefont {Gebel}}, \bibinfo {author} {\bibfnamefont
  {H.}~\bibnamefont {Gl\"uckler}}, \bibinfo {author} {\bibfnamefont
  {F.}~\bibnamefont {Goldenbaum}}, \bibinfo {author} {\bibfnamefont
  {K.}~\bibnamefont {Grigoryev}}, \bibinfo {author} {\bibfnamefont
  {D.}~\bibnamefont {Grzonka}}, \bibinfo {author} {\bibfnamefont
  {T.}~\bibnamefont {Hahnraths}}, \bibinfo {author} {\bibfnamefont
  {D.}~\bibnamefont {Heberling}}, \bibinfo {author} {\bibfnamefont
  {V.}~\bibnamefont {Hejny}}, \bibinfo {author} {\bibfnamefont
  {N.}~\bibnamefont {Hempelmann}}, \bibinfo {author} {\bibfnamefont
  {J.}~\bibnamefont {Hetzel}}, \bibinfo {author} {\bibfnamefont
  {F.}~\bibnamefont {Hinder}}, \bibinfo {author} {\bibfnamefont
  {R.}~\bibnamefont {Hipple}}, \bibinfo {author} {\bibfnamefont
  {D.}~\bibnamefont {H\"olscher}}, \bibinfo {author} {\bibfnamefont
  {A.}~\bibnamefont {Ivanov}}, \bibinfo {author} {\bibfnamefont
  {A.}~\bibnamefont {Kacharava}}, \bibinfo {author} {\bibfnamefont
  {V.}~\bibnamefont {Kamerdzhiev}}, \bibinfo {author} {\bibfnamefont
  {B.}~\bibnamefont {Kamys}}, \bibinfo {author} {\bibfnamefont
  {I.}~\bibnamefont {Keshelashvili}}, \bibinfo {author} {\bibfnamefont
  {A.}~\bibnamefont {Khoukaz}}, \bibinfo {author} {\bibfnamefont
  {I.}~\bibnamefont {Koop}}, \bibinfo {author} {\bibfnamefont {H.-J.}\
  \bibnamefont {Krause}}, \bibinfo {author} {\bibfnamefont {S.}~\bibnamefont
  {Krewald}}, \bibinfo {author} {\bibfnamefont {A.}~\bibnamefont {Kulikov}},
  \bibinfo {author} {\bibfnamefont {A.}~\bibnamefont {Lehrach}}, \bibinfo
  {author} {\bibfnamefont {P.}~\bibnamefont {Lenisa}}, \bibinfo {author}
  {\bibfnamefont {N.}~\bibnamefont {Lomidze}}, \bibinfo {author} {\bibfnamefont
  {B.}~\bibnamefont {Lorentz}}, \bibinfo {author} {\bibfnamefont
  {P.}~\bibnamefont {Maanen}}, \bibinfo {author} {\bibfnamefont
  {G.}~\bibnamefont {Macharashvili}}, \bibinfo {author} {\bibfnamefont
  {A.}~\bibnamefont {Magiera}}, \bibinfo {author} {\bibfnamefont
  {R.}~\bibnamefont {Maier}}, \bibinfo {author} {\bibfnamefont
  {K.}~\bibnamefont {Makino}}, \bibinfo {author} {\bibfnamefont
  {B.}~\bibnamefont {Maria\ifmmode~\acute{n}\else \'{n}\fi{}ski}}, \bibinfo
  {author} {\bibfnamefont {D.}~\bibnamefont {Mchedlishvili}}, \bibinfo {author}
  {\bibfnamefont {U.-G.}\ \bibnamefont {Mei\ss{}ner}}, \bibinfo {author}
  {\bibfnamefont {S.}~\bibnamefont {Mey}}, \bibinfo {author} {\bibfnamefont
  {W.}~\bibnamefont {Morse}}, \bibinfo {author} {\bibfnamefont
  {F.}~\bibnamefont {M\"uller}}, \bibinfo {author} {\bibfnamefont
  {A.}~\bibnamefont {Nass}}, \bibinfo {author} {\bibfnamefont {G.}~\bibnamefont
  {Natour}}, \bibinfo {author} {\bibfnamefont {N.}~\bibnamefont {Nikolaev}},
  \bibinfo {author} {\bibfnamefont {M.}~\bibnamefont {Nioradze}}, \bibinfo
  {author} {\bibfnamefont {K.}~\bibnamefont {Nowakowski}}, \bibinfo {author}
  {\bibfnamefont {Y.}~\bibnamefont {Orlov}}, \bibinfo {author} {\bibfnamefont
  {A.}~\bibnamefont {Pesce}}, \bibinfo {author} {\bibfnamefont
  {D.}~\bibnamefont {Prasuhn}}, \bibinfo {author} {\bibfnamefont
  {J.}~\bibnamefont {Pretz}}, \bibinfo {author} {\bibfnamefont
  {F.}~\bibnamefont {Rathmann}}, \bibinfo {author} {\bibfnamefont
  {J.}~\bibnamefont {Ritman}}, \bibinfo {author} {\bibfnamefont
  {M.}~\bibnamefont {Rosenthal}}, \bibinfo {author} {\bibfnamefont
  {Z.}~\bibnamefont {Rudy}}, \bibinfo {author} {\bibfnamefont {A.}~\bibnamefont
  {Saleev}}, \bibinfo {author} {\bibfnamefont {T.}~\bibnamefont {Sefzick}},
  \bibinfo {author} {\bibfnamefont {Y.}~\bibnamefont {Semertzidis}}, \bibinfo
  {author} {\bibfnamefont {Y.}~\bibnamefont {Senichev}}, \bibinfo {author}
  {\bibfnamefont {V.}~\bibnamefont {Shmakova}}, \bibinfo {author}
  {\bibfnamefont {A.}~\bibnamefont {Silenko}}, \bibinfo {author} {\bibfnamefont
  {M.}~\bibnamefont {Simon}}, \bibinfo {author} {\bibfnamefont
  {J.}~\bibnamefont {Slim}}, \bibinfo {author} {\bibfnamefont {H.}~\bibnamefont
  {Soltner}}, \bibinfo {author} {\bibfnamefont {A.}~\bibnamefont {Stahl}},
  \bibinfo {author} {\bibfnamefont {R.}~\bibnamefont {Stassen}}, \bibinfo
  {author} {\bibfnamefont {M.}~\bibnamefont {Statera}}, \bibinfo {author}
  {\bibfnamefont {H.}~\bibnamefont {Stockhorst}}, \bibinfo {author}
  {\bibfnamefont {H.}~\bibnamefont {Straatmann}}, \bibinfo {author}
  {\bibfnamefont {H.}~\bibnamefont {Str\"oher}}, \bibinfo {author}
  {\bibfnamefont {M.}~\bibnamefont {Tabidze}}, \bibinfo {author} {\bibfnamefont
  {R.}~\bibnamefont {Talman}}, \bibinfo {author} {\bibfnamefont
  {P.}~\bibnamefont {Th\"orngren~Engblom}}, \bibinfo {author} {\bibfnamefont
  {F.}~\bibnamefont {Trinkel}}, \bibinfo {author} {\bibfnamefont
  {A.}~\bibnamefont {Trzci\ifmmode~\acute{n}\else \'{n}\fi{}ski}}, \bibinfo
  {author} {\bibfnamefont {Y.}~\bibnamefont {Uzikov}}, \bibinfo {author}
  {\bibfnamefont {Y.}~\bibnamefont {Valdau}}, \bibinfo {author} {\bibfnamefont
  {E.}~\bibnamefont {Valetov}}, \bibinfo {author} {\bibfnamefont
  {A.}~\bibnamefont {Vassiliev}}, \bibinfo {author} {\bibfnamefont
  {C.}~\bibnamefont {Weidemann}}, \bibinfo {author} {\bibfnamefont
  {C.}~\bibnamefont {Wilkin}}, \bibinfo {author} {\bibfnamefont
  {A.}~\bibnamefont {Wro\ifmmode~\acute{n}\else \'{n}\fi{}ska}}, \bibinfo
  {author} {\bibfnamefont {P.}~\bibnamefont {W\"ustner}}, \bibinfo {author}
  {\bibfnamefont {M.}~\bibnamefont {Zakrzewska}}, \bibinfo {author}
  {\bibfnamefont {P.}~\bibnamefont {Zupra\ifmmode~\acute{n}\else
  \'{n}\fi{}ski}},\ and\ \bibinfo {author} {\bibfnamefont {D.}~\bibnamefont
  {Zyuzin}} (\bibinfo {collaboration} {JEDI Collaboration}),\ }\href
  {https://doi.org/10.1103/PhysRevLett.117.054801} {\bibfield  {journal}
  {\bibinfo  {journal} {Phys. Rev. Lett.}\ }\textbf {\bibinfo {volume} {117}},\
  \bibinfo {pages} {054801} (\bibinfo {year} {2016})}\BibitemShut {NoStop}%
\bibitem [{\citenamefont {Abusaif}\ \emph {et~al.}(2021)\citenamefont
  {Abusaif}, \citenamefont {Aggarwal}, \citenamefont {Aksentev}, \citenamefont
  {Alberdi-Esuain}, \citenamefont {Atanasov}, \citenamefont {Barion},
  \citenamefont {Basile}, \citenamefont {Berz}, \citenamefont {Beyß},
  \citenamefont {Böhme}, \citenamefont {Böker}, \citenamefont {Borburgh},
  \citenamefont {Carli}, \citenamefont {Ciepał}, \citenamefont {Ciullo},
  \citenamefont {Contalbrigo}, \citenamefont {De~Conto}, \citenamefont {Dymov},
  \citenamefont {Felden}, \citenamefont {Gagoshidze}, \citenamefont {Gaisser},
  \citenamefont {Gebel}, \citenamefont {Giese}, \citenamefont {Grigoryev},
  \citenamefont {Grzonka}, \citenamefont {Haj~Tahar}, \citenamefont
  {Hahnraths}, \citenamefont {Heberling}, \citenamefont {Hejny}, \citenamefont
  {Hetzel}, \citenamefont {Hölscher}, \citenamefont {Javakhishvili},
  \citenamefont {Jorat}, \citenamefont {Kacharava}, \citenamefont
  {Kamerdzhiev}, \citenamefont {Karanth}, \citenamefont {Käseberg},
  \citenamefont {Keshelashvili}, \citenamefont {Koop}, \citenamefont {Kulikov},
  \citenamefont {Laihem}, \citenamefont {Lamont}, \citenamefont {Lehrach},
  \citenamefont {Lenisa}, \citenamefont {Lomidze}, \citenamefont {Lorentz},
  \citenamefont {Macharashvili}, \citenamefont {Magiera}, \citenamefont
  {Makino}, \citenamefont {Martin}, \citenamefont {Mchedlishvili},
  \citenamefont {Meißner}, \citenamefont {Metreveli}, \citenamefont {Michaud},
  \citenamefont {Müller}, \citenamefont {Nass}, \citenamefont {Natour},
  \citenamefont {Nikolaev}, \citenamefont {Nogga}, \citenamefont {Pesce},
  \citenamefont {Poncza}, \citenamefont {Prasuhn}, \citenamefont {Pretz},
  \citenamefont {Rathmann}, \citenamefont {Ritman}, \citenamefont {Rosenthal},
  \citenamefont {Saleev}, \citenamefont {Schott}, \citenamefont {Sefzick},
  \citenamefont {Senichev}, \citenamefont {Shergelashvili}, \citenamefont
  {Shmakova}, \citenamefont {Siddique}, \citenamefont {Silenko}, \citenamefont
  {Simon}, \citenamefont {Slim}, \citenamefont {Soltner}, \citenamefont
  {Stahl}, \citenamefont {Stassen}, \citenamefont {Stephenson}, \citenamefont
  {Straatmann}, \citenamefont {Ströher}, \citenamefont {Tabidze},
  \citenamefont {Tagliente}, \citenamefont {Talman}, \citenamefont {Uzikov},
  \citenamefont {Valdau}, \citenamefont {Valetov}, \citenamefont {Wagner},
  \citenamefont {Weidemann}, \citenamefont {Wirzba}, \citenamefont {Wrońska},
  \citenamefont {Wüstner}, \citenamefont {Zupranski},\ and\ \citenamefont
  {Zurek.}}]{AbusaifCYR}%
  \BibitemOpen
  \bibfield  {author} {\bibinfo {author} {\bibfnamefont {F.}~\bibnamefont
  {Abusaif}}, \bibinfo {author} {\bibfnamefont {A.}~\bibnamefont {Aggarwal}},
  \bibinfo {author} {\bibfnamefont {A.}~\bibnamefont {Aksentev}}, \bibinfo
  {author} {\bibfnamefont {B.}~\bibnamefont {Alberdi-Esuain}}, \bibinfo
  {author} {\bibfnamefont {A.}~\bibnamefont {Atanasov}}, \bibinfo {author}
  {\bibfnamefont {L.}~\bibnamefont {Barion}}, \bibinfo {author} {\bibfnamefont
  {S.}~\bibnamefont {Basile}}, \bibinfo {author} {\bibfnamefont
  {M.}~\bibnamefont {Berz}}, \bibinfo {author} {\bibfnamefont {M.}~\bibnamefont
  {Beyß}}, \bibinfo {author} {\bibfnamefont {C.}~\bibnamefont {Böhme}},
  \bibinfo {author} {\bibfnamefont {J.}~\bibnamefont {Böker}}, \bibinfo
  {author} {\bibfnamefont {J.}~\bibnamefont {Borburgh}}, \bibinfo {author}
  {\bibfnamefont {C.}~\bibnamefont {Carli}}, \bibinfo {author} {\bibfnamefont
  {I.}~\bibnamefont {Ciepał}}, \bibinfo {author} {\bibfnamefont
  {G.}~\bibnamefont {Ciullo}}, \bibinfo {author} {\bibfnamefont
  {M.}~\bibnamefont {Contalbrigo}}, \bibinfo {author} {\bibfnamefont {J.-M.}\
  \bibnamefont {De~Conto}}, \bibinfo {author} {\bibfnamefont {S.}~\bibnamefont
  {Dymov}}, \bibinfo {author} {\bibfnamefont {O.}~\bibnamefont {Felden}},
  \bibinfo {author} {\bibfnamefont {M.}~\bibnamefont {Gagoshidze}}, \bibinfo
  {author} {\bibfnamefont {M.}~\bibnamefont {Gaisser}}, \bibinfo {author}
  {\bibfnamefont {R.}~\bibnamefont {Gebel}}, \bibinfo {author} {\bibfnamefont
  {N.}~\bibnamefont {Giese}}, \bibinfo {author} {\bibfnamefont
  {K.}~\bibnamefont {Grigoryev}}, \bibinfo {author} {\bibfnamefont
  {D.}~\bibnamefont {Grzonka}}, \bibinfo {author} {\bibfnamefont
  {M.}~\bibnamefont {Haj~Tahar}}, \bibinfo {author} {\bibfnamefont
  {T.}~\bibnamefont {Hahnraths}}, \bibinfo {author} {\bibfnamefont
  {D.}~\bibnamefont {Heberling}}, \bibinfo {author} {\bibfnamefont
  {V.}~\bibnamefont {Hejny}}, \bibinfo {author} {\bibfnamefont
  {J.}~\bibnamefont {Hetzel}}, \bibinfo {author} {\bibfnamefont
  {D.}~\bibnamefont {Hölscher}}, \bibinfo {author} {\bibfnamefont
  {O.}~\bibnamefont {Javakhishvili}}, \bibinfo {author} {\bibfnamefont
  {L.}~\bibnamefont {Jorat}}, \bibinfo {author} {\bibfnamefont
  {A.}~\bibnamefont {Kacharava}}, \bibinfo {author} {\bibfnamefont
  {V.}~\bibnamefont {Kamerdzhiev}}, \bibinfo {author} {\bibfnamefont
  {S.}~\bibnamefont {Karanth}}, \bibinfo {author} {\bibfnamefont
  {C.}~\bibnamefont {Käseberg}}, \bibinfo {author} {\bibfnamefont
  {I.}~\bibnamefont {Keshelashvili}}, \bibinfo {author} {\bibfnamefont
  {I.}~\bibnamefont {Koop}}, \bibinfo {author} {\bibfnamefont {A.}~\bibnamefont
  {Kulikov}}, \bibinfo {author} {\bibfnamefont {K.}~\bibnamefont {Laihem}},
  \bibinfo {author} {\bibfnamefont {M.}~\bibnamefont {Lamont}}, \bibinfo
  {author} {\bibfnamefont {A.}~\bibnamefont {Lehrach}}, \bibinfo {author}
  {\bibfnamefont {P.}~\bibnamefont {Lenisa}}, \bibinfo {author} {\bibfnamefont
  {N.}~\bibnamefont {Lomidze}}, \bibinfo {author} {\bibfnamefont
  {B.}~\bibnamefont {Lorentz}}, \bibinfo {author} {\bibfnamefont
  {G.}~\bibnamefont {Macharashvili}}, \bibinfo {author} {\bibfnamefont
  {A.}~\bibnamefont {Magiera}}, \bibinfo {author} {\bibfnamefont
  {K.}~\bibnamefont {Makino}}, \bibinfo {author} {\bibfnamefont
  {S.}~\bibnamefont {Martin}}, \bibinfo {author} {\bibfnamefont
  {D.}~\bibnamefont {Mchedlishvili}}, \bibinfo {author} {\bibfnamefont {U.-G.}\
  \bibnamefont {Meißner}}, \bibinfo {author} {\bibfnamefont {Z.}~\bibnamefont
  {Metreveli}}, \bibinfo {author} {\bibfnamefont {J.}~\bibnamefont {Michaud}},
  \bibinfo {author} {\bibfnamefont {F.}~\bibnamefont {Müller}}, \bibinfo
  {author} {\bibfnamefont {A.}~\bibnamefont {Nass}}, \bibinfo {author}
  {\bibfnamefont {G.}~\bibnamefont {Natour}}, \bibinfo {author} {\bibfnamefont
  {N.}~\bibnamefont {Nikolaev}}, \bibinfo {author} {\bibfnamefont
  {A.}~\bibnamefont {Nogga}}, \bibinfo {author} {\bibfnamefont
  {A.}~\bibnamefont {Pesce}}, \bibinfo {author} {\bibfnamefont
  {V.}~\bibnamefont {Poncza}}, \bibinfo {author} {\bibfnamefont
  {D.}~\bibnamefont {Prasuhn}}, \bibinfo {author} {\bibfnamefont
  {J.}~\bibnamefont {Pretz}}, \bibinfo {author} {\bibfnamefont
  {F.}~\bibnamefont {Rathmann}}, \bibinfo {author} {\bibfnamefont
  {J.}~\bibnamefont {Ritman}}, \bibinfo {author} {\bibfnamefont
  {M.}~\bibnamefont {Rosenthal}}, \bibinfo {author} {\bibfnamefont
  {A.}~\bibnamefont {Saleev}}, \bibinfo {author} {\bibfnamefont
  {M.}~\bibnamefont {Schott}}, \bibinfo {author} {\bibfnamefont
  {T.}~\bibnamefont {Sefzick}}, \bibinfo {author} {\bibfnamefont
  {Y.}~\bibnamefont {Senichev}}, \bibinfo {author} {\bibfnamefont
  {D.}~\bibnamefont {Shergelashvili}}, \bibinfo {author} {\bibfnamefont
  {V.}~\bibnamefont {Shmakova}}, \bibinfo {author} {\bibfnamefont
  {S.}~\bibnamefont {Siddique}}, \bibinfo {author} {\bibfnamefont
  {A.}~\bibnamefont {Silenko}}, \bibinfo {author} {\bibfnamefont
  {M.}~\bibnamefont {Simon}}, \bibinfo {author} {\bibfnamefont
  {J.}~\bibnamefont {Slim}}, \bibinfo {author} {\bibfnamefont {H.}~\bibnamefont
  {Soltner}}, \bibinfo {author} {\bibfnamefont {A.}~\bibnamefont {Stahl}},
  \bibinfo {author} {\bibfnamefont {R.}~\bibnamefont {Stassen}}, \bibinfo
  {author} {\bibfnamefont {E.}~\bibnamefont {Stephenson}}, \bibinfo {author}
  {\bibfnamefont {H.}~\bibnamefont {Straatmann}}, \bibinfo {author}
  {\bibfnamefont {H.}~\bibnamefont {Ströher}}, \bibinfo {author}
  {\bibfnamefont {M.}~\bibnamefont {Tabidze}}, \bibinfo {author} {\bibfnamefont
  {G.}~\bibnamefont {Tagliente}}, \bibinfo {author} {\bibfnamefont
  {R.}~\bibnamefont {Talman}}, \bibinfo {author} {\bibfnamefont
  {Y.}~\bibnamefont {Uzikov}}, \bibinfo {author} {\bibfnamefont
  {Y.}~\bibnamefont {Valdau}}, \bibinfo {author} {\bibfnamefont
  {E.}~\bibnamefont {Valetov}}, \bibinfo {author} {\bibfnamefont
  {T.}~\bibnamefont {Wagner}}, \bibinfo {author} {\bibfnamefont
  {C.}~\bibnamefont {Weidemann}}, \bibinfo {author} {\bibfnamefont
  {A.}~\bibnamefont {Wirzba}}, \bibinfo {author} {\bibfnamefont
  {A.}~\bibnamefont {Wrońska}}, \bibinfo {author} {\bibfnamefont
  {P.}~\bibnamefont {Wüstner}}, \bibinfo {author} {\bibfnamefont
  {P.}~\bibnamefont {Zupranski}},\ and\ \bibinfo {author} {\bibfnamefont
  {M.}~\bibnamefont {Zurek.}} (\bibinfo {collaboration} {CPEDM
  Collaboration}),\ }\href {https://cds.cern.ch/record/2654645} {\emph
  {\bibinfo {title} {{Storage ring to search for electric dipole moments of
  charged particles: Feasibility study}}}},\ CERN Yellow Reports: Monographs,
  2021-003\ (\bibinfo  {publisher} {CERN},\ \bibinfo {address} {Geneva},\
  \bibinfo {year} {2021})\ \bibinfo {note}
  {[doi:10.23731/CYRM-2021-003]}\BibitemShut {NoStop}%
\bibitem [{\citenamefont {Nikolaev}\ \emph {et~al.}(2024)\citenamefont
  {Nikolaev}, \citenamefont {Rathmann}, \citenamefont {Slim}, \citenamefont
  {Andres}, \citenamefont {Hejny}, \citenamefont {Nass}, \citenamefont
  {Kacharava}, \citenamefont {Lenisa}, \citenamefont {Pretz}, \citenamefont
  {Saleev}, \citenamefont {Shmakova}, \citenamefont {Soltner}, \citenamefont
  {Abusaif}, \citenamefont {Aggarwal}, \citenamefont {Aksentev}, \citenamefont
  {Alberdi}, \citenamefont {Barion}, \citenamefont {Bekman}, \citenamefont
  {Bey\ss{}}, \citenamefont {B\"ohme}, \citenamefont {Breitkreutz},
  \citenamefont {Canale}, \citenamefont {Ciullo}, \citenamefont {Dymov},
  \citenamefont {Fr\"ohlich}, \citenamefont {Gebel}, \citenamefont {Gaisser},
  \citenamefont {Grigoryev}, \citenamefont {Grzonka}, \citenamefont {Hetzel},
  \citenamefont {Javakhishvili}, \citenamefont {Kamerdzhiev}, \citenamefont
  {Karanth}, \citenamefont {Keshelashvili}, \citenamefont {Kononov},
  \citenamefont {Laihem}, \citenamefont {Lehrach}, \citenamefont {Lomidze},
  \citenamefont {Lorentz}, \citenamefont {Macharashvili}, \citenamefont
  {Magiera}, \citenamefont {Mchedlishvili}, \citenamefont {Melnikov},
  \citenamefont {M\"uller}, \citenamefont {Pesce}, \citenamefont {Poncza},
  \citenamefont {Prasuhn}, \citenamefont {Shergelashvili}, \citenamefont
  {Shurkhno}, \citenamefont {Siddique}, \citenamefont {Silenko}, \citenamefont
  {Stassen}, \citenamefont {Stephenson}, \citenamefont {Str\"oher},
  \citenamefont {Tabidze}, \citenamefont {Tagliente}, \citenamefont {Valdau},
  \citenamefont {Vitz}, \citenamefont {Wagner}, \citenamefont {Wirzba},
  \citenamefont {Wro\ifmmode~\acute{n}\else \'{n}\fi{}ska}, \citenamefont
  {W\"ustner},\ and\ \citenamefont {\ifmmode~\dot{Z}\else
  \.{Z}\fi{}urek}}]{SO-Decoherence-2025}%
  \BibitemOpen
  \bibfield  {author} {\bibinfo {author} {\bibfnamefont {N.~N.}\ \bibnamefont
  {Nikolaev}}, \bibinfo {author} {\bibfnamefont {F.}~\bibnamefont {Rathmann}},
  \bibinfo {author} {\bibfnamefont {J.}~\bibnamefont {Slim}}, \bibinfo {author}
  {\bibfnamefont {A.}~\bibnamefont {Andres}}, \bibinfo {author} {\bibfnamefont
  {V.}~\bibnamefont {Hejny}}, \bibinfo {author} {\bibfnamefont
  {A.}~\bibnamefont {Nass}}, \bibinfo {author} {\bibfnamefont {A.}~\bibnamefont
  {Kacharava}}, \bibinfo {author} {\bibfnamefont {P.}~\bibnamefont {Lenisa}},
  \bibinfo {author} {\bibfnamefont {J.}~\bibnamefont {Pretz}}, \bibinfo
  {author} {\bibfnamefont {A.}~\bibnamefont {Saleev}}, \bibinfo {author}
  {\bibfnamefont {V.}~\bibnamefont {Shmakova}}, \bibinfo {author}
  {\bibfnamefont {H.}~\bibnamefont {Soltner}}, \bibinfo {author} {\bibfnamefont
  {F.}~\bibnamefont {Abusaif}}, \bibinfo {author} {\bibfnamefont
  {A.}~\bibnamefont {Aggarwal}}, \bibinfo {author} {\bibfnamefont
  {A.}~\bibnamefont {Aksentev}}, \bibinfo {author} {\bibfnamefont
  {B.}~\bibnamefont {Alberdi}}, \bibinfo {author} {\bibfnamefont
  {L.}~\bibnamefont {Barion}}, \bibinfo {author} {\bibfnamefont
  {I.}~\bibnamefont {Bekman}}, \bibinfo {author} {\bibfnamefont
  {M.}~\bibnamefont {Bey\ss{}}}, \bibinfo {author} {\bibfnamefont
  {C.}~\bibnamefont {B\"ohme}}, \bibinfo {author} {\bibfnamefont
  {B.}~\bibnamefont {Breitkreutz}}, \bibinfo {author} {\bibfnamefont
  {N.}~\bibnamefont {Canale}}, \bibinfo {author} {\bibfnamefont
  {G.}~\bibnamefont {Ciullo}}, \bibinfo {author} {\bibfnamefont
  {S.}~\bibnamefont {Dymov}}, \bibinfo {author} {\bibfnamefont {N.-O.}\
  \bibnamefont {Fr\"ohlich}}, \bibinfo {author} {\bibfnamefont
  {R.}~\bibnamefont {Gebel}}, \bibinfo {author} {\bibfnamefont
  {M.}~\bibnamefont {Gaisser}}, \bibinfo {author} {\bibfnamefont
  {K.}~\bibnamefont {Grigoryev}}, \bibinfo {author} {\bibfnamefont
  {D.}~\bibnamefont {Grzonka}}, \bibinfo {author} {\bibfnamefont
  {J.}~\bibnamefont {Hetzel}}, \bibinfo {author} {\bibfnamefont
  {O.}~\bibnamefont {Javakhishvili}}, \bibinfo {author} {\bibfnamefont
  {V.}~\bibnamefont {Kamerdzhiev}}, \bibinfo {author} {\bibfnamefont
  {S.}~\bibnamefont {Karanth}}, \bibinfo {author} {\bibfnamefont
  {I.}~\bibnamefont {Keshelashvili}}, \bibinfo {author} {\bibfnamefont
  {A.}~\bibnamefont {Kononov}}, \bibinfo {author} {\bibfnamefont
  {K.}~\bibnamefont {Laihem}}, \bibinfo {author} {\bibfnamefont
  {A.}~\bibnamefont {Lehrach}}, \bibinfo {author} {\bibfnamefont
  {N.}~\bibnamefont {Lomidze}}, \bibinfo {author} {\bibfnamefont
  {B.}~\bibnamefont {Lorentz}}, \bibinfo {author} {\bibfnamefont
  {G.}~\bibnamefont {Macharashvili}}, \bibinfo {author} {\bibfnamefont
  {A.}~\bibnamefont {Magiera}}, \bibinfo {author} {\bibfnamefont
  {D.}~\bibnamefont {Mchedlishvili}}, \bibinfo {author} {\bibfnamefont
  {A.}~\bibnamefont {Melnikov}}, \bibinfo {author} {\bibfnamefont
  {F.}~\bibnamefont {M\"uller}}, \bibinfo {author} {\bibfnamefont
  {A.}~\bibnamefont {Pesce}}, \bibinfo {author} {\bibfnamefont
  {V.}~\bibnamefont {Poncza}}, \bibinfo {author} {\bibfnamefont
  {D.}~\bibnamefont {Prasuhn}}, \bibinfo {author} {\bibfnamefont
  {D.}~\bibnamefont {Shergelashvili}}, \bibinfo {author} {\bibfnamefont
  {N.}~\bibnamefont {Shurkhno}}, \bibinfo {author} {\bibfnamefont
  {S.}~\bibnamefont {Siddique}}, \bibinfo {author} {\bibfnamefont
  {A.}~\bibnamefont {Silenko}}, \bibinfo {author} {\bibfnamefont
  {S.}~\bibnamefont {Stassen}}, \bibinfo {author} {\bibfnamefont {E.~J.}\
  \bibnamefont {Stephenson}}, \bibinfo {author} {\bibfnamefont
  {H.}~\bibnamefont {Str\"oher}}, \bibinfo {author} {\bibfnamefont
  {M.}~\bibnamefont {Tabidze}}, \bibinfo {author} {\bibfnamefont
  {G.}~\bibnamefont {Tagliente}}, \bibinfo {author} {\bibfnamefont
  {Y.}~\bibnamefont {Valdau}}, \bibinfo {author} {\bibfnamefont
  {M.}~\bibnamefont {Vitz}}, \bibinfo {author} {\bibfnamefont {T.}~\bibnamefont
  {Wagner}}, \bibinfo {author} {\bibfnamefont {A.}~\bibnamefont {Wirzba}},
  \bibinfo {author} {\bibfnamefont {A.}~\bibnamefont
  {Wro\ifmmode~\acute{n}\else \'{n}\fi{}ska}}, \bibinfo {author} {\bibfnamefont
  {P.}~\bibnamefont {W\"ustner}},\ and\ \bibinfo {author} {\bibfnamefont
  {M.}~\bibnamefont {\ifmmode~\dot{Z}\else \.{Z}\fi{}urek}} (\bibinfo
  {collaboration} {JEDI Collaboration}),\ }\href
  {https://doi.org/10.1103/PhysRevAccelBeams.27.111002} {\bibfield  {journal}
  {\bibinfo  {journal} {Phys. Rev. Accel. Beams}\ }\textbf {\bibinfo {volume}
  {27}},\ \bibinfo {pages} {111002} (\bibinfo {year} {2024})}\BibitemShut
  {NoStop}%
\bibitem [{\citenamefont {Slim}\ \emph {et~al.}(2025)\citenamefont {Slim},
  \citenamefont {Rathmann}, \citenamefont {Andres}, \citenamefont {Hejny},
  \citenamefont {Nass}, \citenamefont {Kacharava}, \citenamefont {Lenisa},
  \citenamefont {Nikolaev}, \citenamefont {Pretz}, \citenamefont {Saleev},
  \citenamefont {Shmakova}, \citenamefont {Soltner}, \citenamefont {Abusaif},
  \citenamefont {Aggarwal}, \citenamefont {Aksentev}, \citenamefont {Alberdi},
  \citenamefont {Barion}, \citenamefont {Bekman}, \citenamefont {Bey\ss{}},
  \citenamefont {B\"ohme}, \citenamefont {Breitkreutz}, \citenamefont {Canale},
  \citenamefont {Ciullo}, \citenamefont {Dymov}, \citenamefont {Fr\"ohlich},
  \citenamefont {Gebel}, \citenamefont {Gaisser}, \citenamefont {Grigoryev},
  \citenamefont {Grzonka}, \citenamefont {Hetzel}, \citenamefont
  {Javakhishvili}, \citenamefont {Kamerdzhiev}, \citenamefont {Karanth},
  \citenamefont {Keshelashvili}, \citenamefont {Kononov}, \citenamefont
  {Laihem}, \citenamefont {Lehrach}, \citenamefont {Lomidze}, \citenamefont
  {Lorentz}, \citenamefont {Macharashvili}, \citenamefont {Magiera},
  \citenamefont {Mchedlishvili}, \citenamefont {Melnikov}, \citenamefont
  {M\"uller}, \citenamefont {Pesce}, \citenamefont {Poncza}, \citenamefont
  {Prasuhn}, \citenamefont {Shergelashvili}, \citenamefont {Shurkhno},
  \citenamefont {Siddique}, \citenamefont {Silenko}, \citenamefont {Stassen},
  \citenamefont {Stephenson}, \citenamefont {Str\"oher}, \citenamefont
  {Tabidze}, \citenamefont {Tagliente}, \citenamefont {Valdau}, \citenamefont
  {Vitz}, \citenamefont {Wagner}, \citenamefont {Wirzba}, \citenamefont
  {Wro\ifmmode~\acute{n}\else \'{n}\fi{}ska}, \citenamefont {W\"ustner},\ and\
  \citenamefont {\ifmmode~\dot{Z}\else \.{Z}\fi{}urek}}]{PilotBunch-2025}%
  \BibitemOpen
  \bibfield  {author} {\bibinfo {author} {\bibfnamefont {J.}~\bibnamefont
  {Slim}}, \bibinfo {author} {\bibfnamefont {F.}~\bibnamefont {Rathmann}},
  \bibinfo {author} {\bibfnamefont {A.}~\bibnamefont {Andres}}, \bibinfo
  {author} {\bibfnamefont {V.}~\bibnamefont {Hejny}}, \bibinfo {author}
  {\bibfnamefont {A.}~\bibnamefont {Nass}}, \bibinfo {author} {\bibfnamefont
  {A.}~\bibnamefont {Kacharava}}, \bibinfo {author} {\bibfnamefont
  {P.}~\bibnamefont {Lenisa}}, \bibinfo {author} {\bibfnamefont {N.~N.}\
  \bibnamefont {Nikolaev}}, \bibinfo {author} {\bibfnamefont {J.}~\bibnamefont
  {Pretz}}, \bibinfo {author} {\bibfnamefont {A.}~\bibnamefont {Saleev}},
  \bibinfo {author} {\bibfnamefont {V.}~\bibnamefont {Shmakova}}, \bibinfo
  {author} {\bibfnamefont {H.}~\bibnamefont {Soltner}}, \bibinfo {author}
  {\bibfnamefont {F.}~\bibnamefont {Abusaif}}, \bibinfo {author} {\bibfnamefont
  {A.}~\bibnamefont {Aggarwal}}, \bibinfo {author} {\bibfnamefont
  {A.}~\bibnamefont {Aksentev}}, \bibinfo {author} {\bibfnamefont
  {B.}~\bibnamefont {Alberdi}}, \bibinfo {author} {\bibfnamefont
  {L.}~\bibnamefont {Barion}}, \bibinfo {author} {\bibfnamefont
  {I.}~\bibnamefont {Bekman}}, \bibinfo {author} {\bibfnamefont
  {M.}~\bibnamefont {Bey\ss{}}}, \bibinfo {author} {\bibfnamefont
  {C.}~\bibnamefont {B\"ohme}}, \bibinfo {author} {\bibfnamefont
  {B.}~\bibnamefont {Breitkreutz}}, \bibinfo {author} {\bibfnamefont
  {N.}~\bibnamefont {Canale}}, \bibinfo {author} {\bibfnamefont
  {G.}~\bibnamefont {Ciullo}}, \bibinfo {author} {\bibfnamefont
  {S.}~\bibnamefont {Dymov}}, \bibinfo {author} {\bibfnamefont {N.-O.}\
  \bibnamefont {Fr\"ohlich}}, \bibinfo {author} {\bibfnamefont
  {R.}~\bibnamefont {Gebel}}, \bibinfo {author} {\bibfnamefont
  {M.}~\bibnamefont {Gaisser}}, \bibinfo {author} {\bibfnamefont
  {K.}~\bibnamefont {Grigoryev}}, \bibinfo {author} {\bibfnamefont
  {D.}~\bibnamefont {Grzonka}}, \bibinfo {author} {\bibfnamefont
  {J.}~\bibnamefont {Hetzel}}, \bibinfo {author} {\bibfnamefont
  {O.}~\bibnamefont {Javakhishvili}}, \bibinfo {author} {\bibfnamefont
  {V.}~\bibnamefont {Kamerdzhiev}}, \bibinfo {author} {\bibfnamefont
  {S.}~\bibnamefont {Karanth}}, \bibinfo {author} {\bibfnamefont
  {I.}~\bibnamefont {Keshelashvili}}, \bibinfo {author} {\bibfnamefont
  {A.}~\bibnamefont {Kononov}}, \bibinfo {author} {\bibfnamefont
  {K.}~\bibnamefont {Laihem}}, \bibinfo {author} {\bibfnamefont
  {A.}~\bibnamefont {Lehrach}}, \bibinfo {author} {\bibfnamefont
  {N.}~\bibnamefont {Lomidze}}, \bibinfo {author} {\bibfnamefont
  {B.}~\bibnamefont {Lorentz}}, \bibinfo {author} {\bibfnamefont
  {G.}~\bibnamefont {Macharashvili}}, \bibinfo {author} {\bibfnamefont
  {A.}~\bibnamefont {Magiera}}, \bibinfo {author} {\bibfnamefont
  {D.}~\bibnamefont {Mchedlishvili}}, \bibinfo {author} {\bibfnamefont
  {A.}~\bibnamefont {Melnikov}}, \bibinfo {author} {\bibfnamefont
  {F.}~\bibnamefont {M\"uller}}, \bibinfo {author} {\bibfnamefont
  {A.}~\bibnamefont {Pesce}}, \bibinfo {author} {\bibfnamefont
  {V.}~\bibnamefont {Poncza}}, \bibinfo {author} {\bibfnamefont
  {D.}~\bibnamefont {Prasuhn}}, \bibinfo {author} {\bibfnamefont
  {D.}~\bibnamefont {Shergelashvili}}, \bibinfo {author} {\bibfnamefont
  {N.}~\bibnamefont {Shurkhno}}, \bibinfo {author} {\bibfnamefont
  {S.}~\bibnamefont {Siddique}}, \bibinfo {author} {\bibfnamefont
  {A.}~\bibnamefont {Silenko}}, \bibinfo {author} {\bibfnamefont
  {R.}~\bibnamefont {Stassen}}, \bibinfo {author} {\bibfnamefont {E.~J.}\
  \bibnamefont {Stephenson}}, \bibinfo {author} {\bibfnamefont
  {H.}~\bibnamefont {Str\"oher}}, \bibinfo {author} {\bibfnamefont
  {M.}~\bibnamefont {Tabidze}}, \bibinfo {author} {\bibfnamefont
  {G.}~\bibnamefont {Tagliente}}, \bibinfo {author} {\bibfnamefont
  {Y.}~\bibnamefont {Valdau}}, \bibinfo {author} {\bibfnamefont
  {M.}~\bibnamefont {Vitz}}, \bibinfo {author} {\bibfnamefont {T.}~\bibnamefont
  {Wagner}}, \bibinfo {author} {\bibfnamefont {A.}~\bibnamefont {Wirzba}},
  \bibinfo {author} {\bibfnamefont {A.}~\bibnamefont
  {Wro\ifmmode~\acute{n}\else \'{n}\fi{}ska}}, \bibinfo {author} {\bibfnamefont
  {P.}~\bibnamefont {W\"ustner}},\ and\ \bibinfo {author} {\bibfnamefont
  {M.}~\bibnamefont {\ifmmode~\dot{Z}\else \.{Z}\fi{}urek}} (\bibinfo
  {collaboration} {JEDI Collaboration}),\ }\href
  {https://doi.org/10.1103/PhysRevResearch.7.023257} {\bibfield  {journal}
  {\bibinfo  {journal} {Phys. Rev. Res.}\ }\textbf {\bibinfo {volume} {7}},\
  \bibinfo {pages} {023257} (\bibinfo {year} {2025})}\BibitemShut {NoStop}%
\bibitem [{\citenamefont {Karanth}\ \emph {et~al.}(2023)\citenamefont
  {Karanth}, \citenamefont {Stephenson}, \citenamefont {Chang}, \citenamefont
  {Hejny}, \citenamefont {Park}, \citenamefont {Pretz}, \citenamefont
  {Semertzidis}, \citenamefont {Wirzba}, \citenamefont
  {Wro\ifmmode~\acute{n}\else \'{n}\fi{}ska}, \citenamefont {Abusaif},
  \citenamefont {Aggarwal}, \citenamefont {Aksentev}, \citenamefont {Alberdi},
  \citenamefont {Andres}, \citenamefont {Barion}, \citenamefont {Bekman},
  \citenamefont {Bey\ss{}}, \citenamefont {B\"ohme}, \citenamefont
  {Breitkreutz}, \citenamefont {von Byern}, \citenamefont {Canale},
  \citenamefont {Ciullo}, \citenamefont {Dymov}, \citenamefont {Fr\"ohlich},
  \citenamefont {Gebel}, \citenamefont {Grigoryev}, \citenamefont {Grzonka},
  \citenamefont {Hetzel}, \citenamefont {Javakhishvili}, \citenamefont {Jeong},
  \citenamefont {Kacharava}, \citenamefont {Kamerdzhiev}, \citenamefont
  {Keshelashvili}, \citenamefont {Kononov}, \citenamefont {Laihem},
  \citenamefont {Lehrach}, \citenamefont {Lenisa}, \citenamefont {Lomidze},
  \citenamefont {Lorentz}, \citenamefont {Magiera}, \citenamefont
  {Mchedlishvili}, \citenamefont {M\"uller}, \citenamefont {Nass},
  \citenamefont {Nikolaev}, \citenamefont {Pesce}, \citenamefont {Poncza},
  \citenamefont {Prasuhn}, \citenamefont {Rathmann}, \citenamefont {Saleev},
  \citenamefont {Shergelashvili}, \citenamefont {Shmakova}, \citenamefont
  {Shurkhno}, \citenamefont {Siddique}, \citenamefont {Slim}, \citenamefont
  {Soltner}, \citenamefont {Stassen}, \citenamefont {Str\"oher}, \citenamefont
  {Tabidze}, \citenamefont {Tagliente}, \citenamefont {Valdau}, \citenamefont
  {Vitz}, \citenamefont {Wagner},\ and\ \citenamefont {W\"ustner}}]{JEDIaxion}%
  \BibitemOpen
  \bibfield  {author} {\bibinfo {author} {\bibfnamefont {S.}~\bibnamefont
  {Karanth}}, \bibinfo {author} {\bibfnamefont {E.~J.}\ \bibnamefont
  {Stephenson}}, \bibinfo {author} {\bibfnamefont {S.~P.}\ \bibnamefont
  {Chang}}, \bibinfo {author} {\bibfnamefont {V.}~\bibnamefont {Hejny}},
  \bibinfo {author} {\bibfnamefont {S.}~\bibnamefont {Park}}, \bibinfo {author}
  {\bibfnamefont {J.}~\bibnamefont {Pretz}}, \bibinfo {author} {\bibfnamefont
  {Y.~K.}\ \bibnamefont {Semertzidis}}, \bibinfo {author} {\bibfnamefont
  {A.}~\bibnamefont {Wirzba}}, \bibinfo {author} {\bibfnamefont
  {A.}~\bibnamefont {Wro\ifmmode~\acute{n}\else \'{n}\fi{}ska}}, \bibinfo
  {author} {\bibfnamefont {F.}~\bibnamefont {Abusaif}}, \bibinfo {author}
  {\bibfnamefont {A.}~\bibnamefont {Aggarwal}}, \bibinfo {author}
  {\bibfnamefont {A.}~\bibnamefont {Aksentev}}, \bibinfo {author}
  {\bibfnamefont {B.}~\bibnamefont {Alberdi}}, \bibinfo {author} {\bibfnamefont
  {A.}~\bibnamefont {Andres}}, \bibinfo {author} {\bibfnamefont
  {L.}~\bibnamefont {Barion}}, \bibinfo {author} {\bibfnamefont
  {I.}~\bibnamefont {Bekman}}, \bibinfo {author} {\bibfnamefont
  {M.}~\bibnamefont {Bey\ss{}}}, \bibinfo {author} {\bibfnamefont
  {C.}~\bibnamefont {B\"ohme}}, \bibinfo {author} {\bibfnamefont
  {B.}~\bibnamefont {Breitkreutz}}, \bibinfo {author} {\bibfnamefont
  {C.}~\bibnamefont {von Byern}}, \bibinfo {author} {\bibfnamefont
  {N.}~\bibnamefont {Canale}}, \bibinfo {author} {\bibfnamefont
  {G.}~\bibnamefont {Ciullo}}, \bibinfo {author} {\bibfnamefont
  {S.}~\bibnamefont {Dymov}}, \bibinfo {author} {\bibfnamefont {N.-O.}\
  \bibnamefont {Fr\"ohlich}}, \bibinfo {author} {\bibfnamefont
  {R.}~\bibnamefont {Gebel}}, \bibinfo {author} {\bibfnamefont
  {K.}~\bibnamefont {Grigoryev}}, \bibinfo {author} {\bibfnamefont
  {D.}~\bibnamefont {Grzonka}}, \bibinfo {author} {\bibfnamefont
  {J.}~\bibnamefont {Hetzel}}, \bibinfo {author} {\bibfnamefont
  {O.}~\bibnamefont {Javakhishvili}}, \bibinfo {author} {\bibfnamefont
  {H.}~\bibnamefont {Jeong}}, \bibinfo {author} {\bibfnamefont
  {A.}~\bibnamefont {Kacharava}}, \bibinfo {author} {\bibfnamefont
  {V.}~\bibnamefont {Kamerdzhiev}}, \bibinfo {author} {\bibfnamefont
  {I.}~\bibnamefont {Keshelashvili}}, \bibinfo {author} {\bibfnamefont
  {A.}~\bibnamefont {Kononov}}, \bibinfo {author} {\bibfnamefont
  {K.}~\bibnamefont {Laihem}}, \bibinfo {author} {\bibfnamefont
  {A.}~\bibnamefont {Lehrach}}, \bibinfo {author} {\bibfnamefont
  {P.}~\bibnamefont {Lenisa}}, \bibinfo {author} {\bibfnamefont
  {N.}~\bibnamefont {Lomidze}}, \bibinfo {author} {\bibfnamefont
  {B.}~\bibnamefont {Lorentz}}, \bibinfo {author} {\bibfnamefont
  {A.}~\bibnamefont {Magiera}}, \bibinfo {author} {\bibfnamefont
  {D.}~\bibnamefont {Mchedlishvili}}, \bibinfo {author} {\bibfnamefont
  {F.}~\bibnamefont {M\"uller}}, \bibinfo {author} {\bibfnamefont
  {A.}~\bibnamefont {Nass}}, \bibinfo {author} {\bibfnamefont {N.~N.}\
  \bibnamefont {Nikolaev}}, \bibinfo {author} {\bibfnamefont {A.}~\bibnamefont
  {Pesce}}, \bibinfo {author} {\bibfnamefont {V.}~\bibnamefont {Poncza}},
  \bibinfo {author} {\bibfnamefont {D.}~\bibnamefont {Prasuhn}}, \bibinfo
  {author} {\bibfnamefont {F.}~\bibnamefont {Rathmann}}, \bibinfo {author}
  {\bibfnamefont {A.}~\bibnamefont {Saleev}}, \bibinfo {author} {\bibfnamefont
  {D.}~\bibnamefont {Shergelashvili}}, \bibinfo {author} {\bibfnamefont
  {V.}~\bibnamefont {Shmakova}}, \bibinfo {author} {\bibfnamefont
  {N.}~\bibnamefont {Shurkhno}}, \bibinfo {author} {\bibfnamefont
  {S.}~\bibnamefont {Siddique}}, \bibinfo {author} {\bibfnamefont
  {J.}~\bibnamefont {Slim}}, \bibinfo {author} {\bibfnamefont {H.}~\bibnamefont
  {Soltner}}, \bibinfo {author} {\bibfnamefont {R.}~\bibnamefont {Stassen}},
  \bibinfo {author} {\bibfnamefont {H.}~\bibnamefont {Str\"oher}}, \bibinfo
  {author} {\bibfnamefont {M.}~\bibnamefont {Tabidze}}, \bibinfo {author}
  {\bibfnamefont {G.}~\bibnamefont {Tagliente}}, \bibinfo {author}
  {\bibfnamefont {Y.}~\bibnamefont {Valdau}}, \bibinfo {author} {\bibfnamefont
  {M.}~\bibnamefont {Vitz}}, \bibinfo {author} {\bibfnamefont {T.}~\bibnamefont
  {Wagner}},\ and\ \bibinfo {author} {\bibfnamefont {P.}~\bibnamefont
  {W\"ustner}} (\bibinfo {collaboration} {JEDI Collaboration}),\ }\href
  {https://doi.org/10.1103/PhysRevX.13.031004} {\bibfield  {journal} {\bibinfo
  {journal} {Phys. Rev. X}\ }\textbf {\bibinfo {volume} {13}},\ \bibinfo
  {pages} {031004} (\bibinfo {year} {2023})}\BibitemShut {NoStop}%
\bibitem [{\citenamefont {Maier}(1997)}]{MaierCOSY}%
  \BibitemOpen
  \bibfield  {author} {\bibinfo {author} {\bibfnamefont {R.}~\bibnamefont
  {Maier}},\ }\href
  {https://doi.org/https://doi.org/10.1016/S0168-9002(97)00324-0} {\bibfield
  {journal} {\bibinfo  {journal} {Nuclear Instruments and Methods in Physics
  Research Section A: Accelerators, Spectrometers, Detectors and Associated
  Equipment}\ }\textbf {\bibinfo {volume} {390}},\ \bibinfo {pages} {1}
  (\bibinfo {year} {1997})}\BibitemShut {NoStop}%
\bibitem [{\citenamefont {Felden}\ \emph {et~al.}(2014)\citenamefont {Felden},
  \citenamefont {Gebel}, \citenamefont {Maier},\ and\ \citenamefont
  {Mey}}]{FeldenCOSY}%
  \BibitemOpen
  \bibfield  {author} {\bibinfo {author} {\bibfnamefont {O.}~\bibnamefont
  {Felden}}, \bibinfo {author} {\bibfnamefont {R.}~\bibnamefont {Gebel}},
  \bibinfo {author} {\bibfnamefont {R.}~\bibnamefont {Maier}},\ and\ \bibinfo
  {author} {\bibfnamefont {S.}~\bibnamefont {Mey}},\ }\href
  {https://doi.org/10.22323/1.182.0068} {\bibfield  {journal} {\bibinfo
  {journal} {PoS}\ }\textbf {\bibinfo {volume} {PSTP2013}},\ \bibinfo {pages}
  {068} (\bibinfo {year} {2014})}\BibitemShut {NoStop}%
\bibitem [{\citenamefont {Wilkin}(2017)}]{WilkinCOSYlegacy}%
  \BibitemOpen
  \bibfield  {author} {\bibinfo {author} {\bibfnamefont {C.}~\bibnamefont
  {Wilkin}},\ }\href {https://doi.org/10.1140/epja/i2017-12295-4} {\bibfield
  {journal} {\bibinfo  {journal} {Eur. Phys. J. A}\ }\textbf {\bibinfo {volume}
  {53}},\ \bibinfo {pages} {114} (\bibinfo {year} {2017})}\BibitemShut
  {NoStop}%
\bibitem [{\citenamefont {Brantjes}\ \emph {et~al.}(2012)\citenamefont
  {Brantjes}, \citenamefont {Dzordzhadze}, \citenamefont {Gebel}, \citenamefont
  {Gonnella}, \citenamefont {Gray}, \citenamefont {{van der Hoek}},
  \citenamefont {Imig}, \citenamefont {Kruithof}, \citenamefont {Lazarus},
  \citenamefont {Lehrach}, \citenamefont {Lorentz}, \citenamefont {Messi},
  \citenamefont {Moricciani}, \citenamefont {Morse}, \citenamefont {Noid},
  \citenamefont {Onderwater}, \citenamefont {Özben}, \citenamefont {Prasuhn},
  \citenamefont {{Levi Sandri}}, \citenamefont {Semertzidis}, \citenamefont
  {{da Silva e Silva}}, \citenamefont {Stephenson}, \citenamefont {Stockhorst},
  \citenamefont {Venanzoni},\ and\ \citenamefont
  {Versolato}}]{BrantjesPolarimetry}%
  \BibitemOpen
  \bibfield  {author} {\bibinfo {author} {\bibfnamefont {N.}~\bibnamefont
  {Brantjes}}, \bibinfo {author} {\bibfnamefont {V.}~\bibnamefont
  {Dzordzhadze}}, \bibinfo {author} {\bibfnamefont {R.}~\bibnamefont {Gebel}},
  \bibinfo {author} {\bibfnamefont {F.}~\bibnamefont {Gonnella}}, \bibinfo
  {author} {\bibfnamefont {F.}~\bibnamefont {Gray}}, \bibinfo {author}
  {\bibfnamefont {D.}~\bibnamefont {{van der Hoek}}}, \bibinfo {author}
  {\bibfnamefont {A.}~\bibnamefont {Imig}}, \bibinfo {author} {\bibfnamefont
  {W.}~\bibnamefont {Kruithof}}, \bibinfo {author} {\bibfnamefont
  {D.}~\bibnamefont {Lazarus}}, \bibinfo {author} {\bibfnamefont
  {A.}~\bibnamefont {Lehrach}}, \bibinfo {author} {\bibfnamefont
  {B.}~\bibnamefont {Lorentz}}, \bibinfo {author} {\bibfnamefont
  {R.}~\bibnamefont {Messi}}, \bibinfo {author} {\bibfnamefont
  {D.}~\bibnamefont {Moricciani}}, \bibinfo {author} {\bibfnamefont
  {W.}~\bibnamefont {Morse}}, \bibinfo {author} {\bibfnamefont
  {G.}~\bibnamefont {Noid}}, \bibinfo {author} {\bibfnamefont {C.}~\bibnamefont
  {Onderwater}}, \bibinfo {author} {\bibfnamefont {C.}~\bibnamefont {Özben}},
  \bibinfo {author} {\bibfnamefont {D.}~\bibnamefont {Prasuhn}}, \bibinfo
  {author} {\bibfnamefont {P.}~\bibnamefont {{Levi Sandri}}}, \bibinfo {author}
  {\bibfnamefont {Y.}~\bibnamefont {Semertzidis}}, \bibinfo {author}
  {\bibfnamefont {M.}~\bibnamefont {{da Silva e Silva}}}, \bibinfo {author}
  {\bibfnamefont {E.}~\bibnamefont {Stephenson}}, \bibinfo {author}
  {\bibfnamefont {H.}~\bibnamefont {Stockhorst}}, \bibinfo {author}
  {\bibfnamefont {G.}~\bibnamefont {Venanzoni}},\ and\ \bibinfo {author}
  {\bibfnamefont {O.}~\bibnamefont {Versolato}},\ }\href
  {https://doi.org/https://doi.org/10.1016/j.nima.2011.09.055} {\bibfield
  {journal} {\bibinfo  {journal} {Nuclear Instruments and Methods in Physics
  Research Section A: Accelerators, Spectrometers, Detectors and Associated
  Equipment}\ }\textbf {\bibinfo {volume} {664}},\ \bibinfo {pages} {49}
  (\bibinfo {year} {2012})}\BibitemShut {NoStop}%
\bibitem [{\citenamefont {Bagdasarian}\ \emph {et~al.}(2014)\citenamefont
  {Bagdasarian}, \citenamefont {Bertelli}, \citenamefont {Chiladze},
  \citenamefont {Ciullo}, \citenamefont {Dietrich}, \citenamefont {Dymov},
  \citenamefont {Eversmann}, \citenamefont {Fanourakis}, \citenamefont
  {Gaisser}, \citenamefont {Gebel}, \citenamefont {Gou}, \citenamefont
  {Guidoboni}, \citenamefont {Hejny}, \citenamefont {Kacharava}, \citenamefont
  {Kamerdzhiev}, \citenamefont {Lehrach}, \citenamefont {Lenisa}, \citenamefont
  {Lorentz}, \citenamefont {Magallanes}, \citenamefont {Maier}, \citenamefont
  {Mchedlishvili}, \citenamefont {Morse}, \citenamefont {Nass}, \citenamefont
  {Oellers}, \citenamefont {Pesce}, \citenamefont {Prasuhn}, \citenamefont
  {Pretz}, \citenamefont {Rathmann}, \citenamefont {Shmakova}, \citenamefont
  {Semertzidis}, \citenamefont {Stephenson}, \citenamefont {Stockhorst},
  \citenamefont {Str\"oher}, \citenamefont {Talman}, \citenamefont
  {Th\"orngren~Engblom}, \citenamefont {Valdau}, \citenamefont {Weidemann},\
  and\ \citenamefont {W\"ustner}}]{JEDIspintune2}%
  \BibitemOpen
  \bibfield  {author} {\bibinfo {author} {\bibfnamefont {Z.}~\bibnamefont
  {Bagdasarian}}, \bibinfo {author} {\bibfnamefont {S.}~\bibnamefont
  {Bertelli}}, \bibinfo {author} {\bibfnamefont {D.}~\bibnamefont {Chiladze}},
  \bibinfo {author} {\bibfnamefont {G.}~\bibnamefont {Ciullo}}, \bibinfo
  {author} {\bibfnamefont {J.}~\bibnamefont {Dietrich}}, \bibinfo {author}
  {\bibfnamefont {S.}~\bibnamefont {Dymov}}, \bibinfo {author} {\bibfnamefont
  {D.}~\bibnamefont {Eversmann}}, \bibinfo {author} {\bibfnamefont
  {G.}~\bibnamefont {Fanourakis}}, \bibinfo {author} {\bibfnamefont
  {M.}~\bibnamefont {Gaisser}}, \bibinfo {author} {\bibfnamefont
  {R.}~\bibnamefont {Gebel}}, \bibinfo {author} {\bibfnamefont
  {B.}~\bibnamefont {Gou}}, \bibinfo {author} {\bibfnamefont {G.}~\bibnamefont
  {Guidoboni}}, \bibinfo {author} {\bibfnamefont {V.}~\bibnamefont {Hejny}},
  \bibinfo {author} {\bibfnamefont {A.}~\bibnamefont {Kacharava}}, \bibinfo
  {author} {\bibfnamefont {V.}~\bibnamefont {Kamerdzhiev}}, \bibinfo {author}
  {\bibfnamefont {A.}~\bibnamefont {Lehrach}}, \bibinfo {author} {\bibfnamefont
  {P.}~\bibnamefont {Lenisa}}, \bibinfo {author} {\bibfnamefont
  {B.}~\bibnamefont {Lorentz}}, \bibinfo {author} {\bibfnamefont
  {L.}~\bibnamefont {Magallanes}}, \bibinfo {author} {\bibfnamefont
  {R.}~\bibnamefont {Maier}}, \bibinfo {author} {\bibfnamefont
  {D.}~\bibnamefont {Mchedlishvili}}, \bibinfo {author} {\bibfnamefont {W.~M.}\
  \bibnamefont {Morse}}, \bibinfo {author} {\bibfnamefont {A.}~\bibnamefont
  {Nass}}, \bibinfo {author} {\bibfnamefont {D.}~\bibnamefont {Oellers}},
  \bibinfo {author} {\bibfnamefont {A.}~\bibnamefont {Pesce}}, \bibinfo
  {author} {\bibfnamefont {D.}~\bibnamefont {Prasuhn}}, \bibinfo {author}
  {\bibfnamefont {J.}~\bibnamefont {Pretz}}, \bibinfo {author} {\bibfnamefont
  {F.}~\bibnamefont {Rathmann}}, \bibinfo {author} {\bibfnamefont
  {V.}~\bibnamefont {Shmakova}}, \bibinfo {author} {\bibfnamefont {Y.~K.}\
  \bibnamefont {Semertzidis}}, \bibinfo {author} {\bibfnamefont {E.~J.}\
  \bibnamefont {Stephenson}}, \bibinfo {author} {\bibfnamefont
  {H.}~\bibnamefont {Stockhorst}}, \bibinfo {author} {\bibfnamefont
  {H.}~\bibnamefont {Str\"oher}}, \bibinfo {author} {\bibfnamefont
  {R.}~\bibnamefont {Talman}}, \bibinfo {author} {\bibfnamefont
  {P.}~\bibnamefont {Th\"orngren~Engblom}}, \bibinfo {author} {\bibfnamefont
  {Y.}~\bibnamefont {Valdau}}, \bibinfo {author} {\bibfnamefont
  {C.}~\bibnamefont {Weidemann}},\ and\ \bibinfo {author} {\bibfnamefont
  {P.}~\bibnamefont {W\"ustner}},\ }\href
  {https://doi.org/10.1103/PhysRevSTAB.17.052803} {\bibfield  {journal}
  {\bibinfo  {journal} {Phys. Rev. ST Accel. Beams}\ }\textbf {\bibinfo
  {volume} {17}},\ \bibinfo {pages} {052803} (\bibinfo {year}
  {2014})}\BibitemShut {NoStop}%
\bibitem [{\citenamefont {Hempelmann}\ \emph {et~al.}(2018)\citenamefont
  {Hempelmann}, \citenamefont {Hejny}, \citenamefont {Pretz}, \citenamefont
  {Soltner}, \citenamefont {Augustyniak}, \citenamefont {Bagdasarian},
  \citenamefont {Bai}, \citenamefont {Barion}, \citenamefont {Berz},
  \citenamefont {Chekmenev}, \citenamefont {Ciullo}, \citenamefont {Dymov},
  \citenamefont {Eversmann}, \citenamefont {Gaisser}, \citenamefont {Gebel},
  \citenamefont {Grigoryev}, \citenamefont {Grzonka}, \citenamefont
  {Guidoboni}, \citenamefont {Heberling}, \citenamefont {Hetzel}, \citenamefont
  {Hinder}, \citenamefont {Kacharava}, \citenamefont {Kamerdzhiev},
  \citenamefont {Keshelashvili}, \citenamefont {Koop}, \citenamefont {Kulikov},
  \citenamefont {Lehrach}, \citenamefont {Lenisa}, \citenamefont {Lomidze},
  \citenamefont {Lorentz}, \citenamefont {Maanen}, \citenamefont
  {Macharashvili}, \citenamefont {Magiera}, \citenamefont {Mchedlishvili},
  \citenamefont {Mey}, \citenamefont {M\"uller}, \citenamefont {Nass},
  \citenamefont {Nikolaev}, \citenamefont {Nioradze}, \citenamefont {Pesce},
  \citenamefont {Prasuhn}, \citenamefont {Rathmann}, \citenamefont {Rosenthal},
  \citenamefont {Saleev}, \citenamefont {Schmidt}, \citenamefont {Semertzidis},
  \citenamefont {Senichev}, \citenamefont {Shmakova}, \citenamefont {Silenko},
  \citenamefont {Slim}, \citenamefont {Stahl}, \citenamefont {Stassen},
  \citenamefont {Stephenson}, \citenamefont {Stockhorst}, \citenamefont
  {Str\"oher}, \citenamefont {Tabidze}, \citenamefont {Tagliente},
  \citenamefont {Talman}, \citenamefont {Th\"orngren~Engblom}, \citenamefont
  {Trinkel}, \citenamefont {Uzikov}, \citenamefont {Valdau}, \citenamefont
  {Valetov}, \citenamefont {Vassiliev}, \citenamefont {Weidemann},
  \citenamefont {Wro\ifmmode~\acute{n}\else \'{n}\fi{}ska}, \citenamefont
  {W\"ustner}, \citenamefont {Zupra\ifmmode~\acute{n}\else \'{n}\fi{}ski},\
  and\ \citenamefont {\ifmmode~\dot{Z}\else \.{Z}\fi{}urek}}]{JEDIphase}%
  \BibitemOpen
  \bibfield  {author} {\bibinfo {author} {\bibfnamefont {N.}~\bibnamefont
  {Hempelmann}}, \bibinfo {author} {\bibfnamefont {V.}~\bibnamefont {Hejny}},
  \bibinfo {author} {\bibfnamefont {J.}~\bibnamefont {Pretz}}, \bibinfo
  {author} {\bibfnamefont {H.}~\bibnamefont {Soltner}}, \bibinfo {author}
  {\bibfnamefont {W.}~\bibnamefont {Augustyniak}}, \bibinfo {author}
  {\bibfnamefont {Z.}~\bibnamefont {Bagdasarian}}, \bibinfo {author}
  {\bibfnamefont {M.}~\bibnamefont {Bai}}, \bibinfo {author} {\bibfnamefont
  {L.}~\bibnamefont {Barion}}, \bibinfo {author} {\bibfnamefont
  {M.}~\bibnamefont {Berz}}, \bibinfo {author} {\bibfnamefont {S.}~\bibnamefont
  {Chekmenev}}, \bibinfo {author} {\bibfnamefont {G.}~\bibnamefont {Ciullo}},
  \bibinfo {author} {\bibfnamefont {S.}~\bibnamefont {Dymov}}, \bibinfo
  {author} {\bibfnamefont {D.}~\bibnamefont {Eversmann}}, \bibinfo {author}
  {\bibfnamefont {M.}~\bibnamefont {Gaisser}}, \bibinfo {author} {\bibfnamefont
  {R.}~\bibnamefont {Gebel}}, \bibinfo {author} {\bibfnamefont
  {K.}~\bibnamefont {Grigoryev}}, \bibinfo {author} {\bibfnamefont
  {D.}~\bibnamefont {Grzonka}}, \bibinfo {author} {\bibfnamefont
  {G.}~\bibnamefont {Guidoboni}}, \bibinfo {author} {\bibfnamefont
  {D.}~\bibnamefont {Heberling}}, \bibinfo {author} {\bibfnamefont
  {J.}~\bibnamefont {Hetzel}}, \bibinfo {author} {\bibfnamefont
  {F.}~\bibnamefont {Hinder}}, \bibinfo {author} {\bibfnamefont
  {A.}~\bibnamefont {Kacharava}}, \bibinfo {author} {\bibfnamefont
  {V.}~\bibnamefont {Kamerdzhiev}}, \bibinfo {author} {\bibfnamefont
  {I.}~\bibnamefont {Keshelashvili}}, \bibinfo {author} {\bibfnamefont
  {I.}~\bibnamefont {Koop}}, \bibinfo {author} {\bibfnamefont {A.}~\bibnamefont
  {Kulikov}}, \bibinfo {author} {\bibfnamefont {A.}~\bibnamefont {Lehrach}},
  \bibinfo {author} {\bibfnamefont {P.}~\bibnamefont {Lenisa}}, \bibinfo
  {author} {\bibfnamefont {N.}~\bibnamefont {Lomidze}}, \bibinfo {author}
  {\bibfnamefont {B.}~\bibnamefont {Lorentz}}, \bibinfo {author} {\bibfnamefont
  {P.}~\bibnamefont {Maanen}}, \bibinfo {author} {\bibfnamefont
  {G.}~\bibnamefont {Macharashvili}}, \bibinfo {author} {\bibfnamefont
  {A.}~\bibnamefont {Magiera}}, \bibinfo {author} {\bibfnamefont
  {D.}~\bibnamefont {Mchedlishvili}}, \bibinfo {author} {\bibfnamefont
  {S.}~\bibnamefont {Mey}}, \bibinfo {author} {\bibfnamefont {F.}~\bibnamefont
  {M\"uller}}, \bibinfo {author} {\bibfnamefont {A.}~\bibnamefont {Nass}},
  \bibinfo {author} {\bibfnamefont {N.~N.}\ \bibnamefont {Nikolaev}}, \bibinfo
  {author} {\bibfnamefont {M.}~\bibnamefont {Nioradze}}, \bibinfo {author}
  {\bibfnamefont {A.}~\bibnamefont {Pesce}}, \bibinfo {author} {\bibfnamefont
  {D.}~\bibnamefont {Prasuhn}}, \bibinfo {author} {\bibfnamefont
  {F.}~\bibnamefont {Rathmann}}, \bibinfo {author} {\bibfnamefont
  {M.}~\bibnamefont {Rosenthal}}, \bibinfo {author} {\bibfnamefont
  {A.}~\bibnamefont {Saleev}}, \bibinfo {author} {\bibfnamefont
  {V.}~\bibnamefont {Schmidt}}, \bibinfo {author} {\bibfnamefont
  {Y.}~\bibnamefont {Semertzidis}}, \bibinfo {author} {\bibfnamefont
  {Y.}~\bibnamefont {Senichev}}, \bibinfo {author} {\bibfnamefont
  {V.}~\bibnamefont {Shmakova}}, \bibinfo {author} {\bibfnamefont
  {A.}~\bibnamefont {Silenko}}, \bibinfo {author} {\bibfnamefont
  {J.}~\bibnamefont {Slim}}, \bibinfo {author} {\bibfnamefont {A.}~\bibnamefont
  {Stahl}}, \bibinfo {author} {\bibfnamefont {R.}~\bibnamefont {Stassen}},
  \bibinfo {author} {\bibfnamefont {E.}~\bibnamefont {Stephenson}}, \bibinfo
  {author} {\bibfnamefont {H.}~\bibnamefont {Stockhorst}}, \bibinfo {author}
  {\bibfnamefont {H.}~\bibnamefont {Str\"oher}}, \bibinfo {author}
  {\bibfnamefont {M.}~\bibnamefont {Tabidze}}, \bibinfo {author} {\bibfnamefont
  {G.}~\bibnamefont {Tagliente}}, \bibinfo {author} {\bibfnamefont
  {R.}~\bibnamefont {Talman}}, \bibinfo {author} {\bibfnamefont
  {P.}~\bibnamefont {Th\"orngren~Engblom}}, \bibinfo {author} {\bibfnamefont
  {F.}~\bibnamefont {Trinkel}}, \bibinfo {author} {\bibfnamefont
  {Y.}~\bibnamefont {Uzikov}}, \bibinfo {author} {\bibfnamefont
  {Y.}~\bibnamefont {Valdau}}, \bibinfo {author} {\bibfnamefont
  {E.}~\bibnamefont {Valetov}}, \bibinfo {author} {\bibfnamefont
  {A.}~\bibnamefont {Vassiliev}}, \bibinfo {author} {\bibfnamefont
  {C.}~\bibnamefont {Weidemann}}, \bibinfo {author} {\bibfnamefont
  {A.}~\bibnamefont {Wro\ifmmode~\acute{n}\else \'{n}\fi{}ska}}, \bibinfo
  {author} {\bibfnamefont {P.}~\bibnamefont {W\"ustner}}, \bibinfo {author}
  {\bibfnamefont {P.}~\bibnamefont {Zupra\ifmmode~\acute{n}\else
  \'{n}\fi{}ski}},\ and\ \bibinfo {author} {\bibfnamefont {M.}~\bibnamefont
  {\ifmmode~\dot{Z}\else \.{Z}\fi{}urek}} (\bibinfo {collaboration} {JEDI}),\
  }\href {https://doi.org/10.1103/PhysRevAccelBeams.21.042002} {\bibfield
  {journal} {\bibinfo  {journal} {Phys. Rev. Accel. Beams}\ }\textbf {\bibinfo
  {volume} {21}},\ \bibinfo {pages} {042002} (\bibinfo {year}
  {2018})}\BibitemShut {NoStop}%
\bibitem [{\citenamefont {Hempelmann}\ \emph {et~al.}(2017)\citenamefont
  {Hempelmann}, \citenamefont {Hejny}, \citenamefont {Pretz}, \citenamefont
  {Stephenson}, \citenamefont {Augustyniak}, \citenamefont {Bagdasarian},
  \citenamefont {Bai}, \citenamefont {Barion}, \citenamefont {Berz},
  \citenamefont {Chekmenev}, \citenamefont {Ciullo}, \citenamefont {Dymov},
  \citenamefont {Etzkorn}, \citenamefont {Eversmann}, \citenamefont {Gaisser},
  \citenamefont {Gebel}, \citenamefont {Grigoryev}, \citenamefont {Grzonka},
  \citenamefont {Guidoboni}, \citenamefont {Hanraths}, \citenamefont
  {Heberling}, \citenamefont {Hetzel}, \citenamefont {Hinder}, \citenamefont
  {Kacharava}, \citenamefont {Kamerdzhiev}, \citenamefont {Keshelashvili},
  \citenamefont {Koop}, \citenamefont {Kulikov}, \citenamefont {Lehrach},
  \citenamefont {Lenisa}, \citenamefont {Lomidze}, \citenamefont {Lorentz},
  \citenamefont {Maanen}, \citenamefont {Macharashvili}, \citenamefont
  {Magiera}, \citenamefont {Mchedlishvili}, \citenamefont {Mey}, \citenamefont
  {M\"uller}, \citenamefont {Nass}, \citenamefont {Nikolaev}, \citenamefont
  {Pesce}, \citenamefont {Prasuhn}, \citenamefont {Rathmann}, \citenamefont
  {Rosenthal}, \citenamefont {Saleev}, \citenamefont {Schmidt}, \citenamefont
  {Semertzidis}, \citenamefont {Shmakova}, \citenamefont {Silenko},
  \citenamefont {Slim}, \citenamefont {Soltner}, \citenamefont {Stahl},
  \citenamefont {Stassen}, \citenamefont {Stockhorst}, \citenamefont
  {Str\"oher}, \citenamefont {Tabidze}, \citenamefont {Tagliente},
  \citenamefont {Talman}, \citenamefont {Th\"orngren~Engblom}, \citenamefont
  {Trinkel}, \citenamefont {Uzikov}, \citenamefont {Valdau}, \citenamefont
  {Valetov}, \citenamefont {Vassiliev}, \citenamefont {Weidemann},
  \citenamefont {Wro\ifmmode~\acute{n}\else \'{n}\fi{}ska}, \citenamefont
  {W\"ustner}, \citenamefont {Zupra\ifmmode~\acute{n}\else \'{n}\fi{}ski},\
  and\ \citenamefont {\ifmmode~\dot{Z}\else \.{Z}\fi{}urek}}]{PhaseLock}%
  \BibitemOpen
  \bibfield  {author} {\bibinfo {author} {\bibfnamefont {N.}~\bibnamefont
  {Hempelmann}}, \bibinfo {author} {\bibfnamefont {V.}~\bibnamefont {Hejny}},
  \bibinfo {author} {\bibfnamefont {J.}~\bibnamefont {Pretz}}, \bibinfo
  {author} {\bibfnamefont {E.}~\bibnamefont {Stephenson}}, \bibinfo {author}
  {\bibfnamefont {W.}~\bibnamefont {Augustyniak}}, \bibinfo {author}
  {\bibfnamefont {Z.}~\bibnamefont {Bagdasarian}}, \bibinfo {author}
  {\bibfnamefont {M.}~\bibnamefont {Bai}}, \bibinfo {author} {\bibfnamefont
  {L.}~\bibnamefont {Barion}}, \bibinfo {author} {\bibfnamefont
  {M.}~\bibnamefont {Berz}}, \bibinfo {author} {\bibfnamefont {S.}~\bibnamefont
  {Chekmenev}}, \bibinfo {author} {\bibfnamefont {G.}~\bibnamefont {Ciullo}},
  \bibinfo {author} {\bibfnamefont {S.}~\bibnamefont {Dymov}}, \bibinfo
  {author} {\bibfnamefont {F.-J.}\ \bibnamefont {Etzkorn}}, \bibinfo {author}
  {\bibfnamefont {D.}~\bibnamefont {Eversmann}}, \bibinfo {author}
  {\bibfnamefont {M.}~\bibnamefont {Gaisser}}, \bibinfo {author} {\bibfnamefont
  {R.}~\bibnamefont {Gebel}}, \bibinfo {author} {\bibfnamefont
  {K.}~\bibnamefont {Grigoryev}}, \bibinfo {author} {\bibfnamefont
  {D.}~\bibnamefont {Grzonka}}, \bibinfo {author} {\bibfnamefont
  {G.}~\bibnamefont {Guidoboni}}, \bibinfo {author} {\bibfnamefont
  {T.}~\bibnamefont {Hanraths}}, \bibinfo {author} {\bibfnamefont
  {D.}~\bibnamefont {Heberling}}, \bibinfo {author} {\bibfnamefont
  {J.}~\bibnamefont {Hetzel}}, \bibinfo {author} {\bibfnamefont
  {F.}~\bibnamefont {Hinder}}, \bibinfo {author} {\bibfnamefont
  {A.}~\bibnamefont {Kacharava}}, \bibinfo {author} {\bibfnamefont
  {V.}~\bibnamefont {Kamerdzhiev}}, \bibinfo {author} {\bibfnamefont
  {I.}~\bibnamefont {Keshelashvili}}, \bibinfo {author} {\bibfnamefont
  {I.}~\bibnamefont {Koop}}, \bibinfo {author} {\bibfnamefont {A.}~\bibnamefont
  {Kulikov}}, \bibinfo {author} {\bibfnamefont {A.}~\bibnamefont {Lehrach}},
  \bibinfo {author} {\bibfnamefont {P.}~\bibnamefont {Lenisa}}, \bibinfo
  {author} {\bibfnamefont {N.}~\bibnamefont {Lomidze}}, \bibinfo {author}
  {\bibfnamefont {B.}~\bibnamefont {Lorentz}}, \bibinfo {author} {\bibfnamefont
  {P.}~\bibnamefont {Maanen}}, \bibinfo {author} {\bibfnamefont
  {G.}~\bibnamefont {Macharashvili}}, \bibinfo {author} {\bibfnamefont
  {A.}~\bibnamefont {Magiera}}, \bibinfo {author} {\bibfnamefont
  {D.}~\bibnamefont {Mchedlishvili}}, \bibinfo {author} {\bibfnamefont
  {S.}~\bibnamefont {Mey}}, \bibinfo {author} {\bibfnamefont {F.}~\bibnamefont
  {M\"uller}}, \bibinfo {author} {\bibfnamefont {A.}~\bibnamefont {Nass}},
  \bibinfo {author} {\bibfnamefont {N.~N.}\ \bibnamefont {Nikolaev}}, \bibinfo
  {author} {\bibfnamefont {A.}~\bibnamefont {Pesce}}, \bibinfo {author}
  {\bibfnamefont {D.}~\bibnamefont {Prasuhn}}, \bibinfo {author} {\bibfnamefont
  {F.}~\bibnamefont {Rathmann}}, \bibinfo {author} {\bibfnamefont
  {M.}~\bibnamefont {Rosenthal}}, \bibinfo {author} {\bibfnamefont
  {A.}~\bibnamefont {Saleev}}, \bibinfo {author} {\bibfnamefont
  {V.}~\bibnamefont {Schmidt}}, \bibinfo {author} {\bibfnamefont
  {Y.}~\bibnamefont {Semertzidis}}, \bibinfo {author} {\bibfnamefont
  {V.}~\bibnamefont {Shmakova}}, \bibinfo {author} {\bibfnamefont
  {A.}~\bibnamefont {Silenko}}, \bibinfo {author} {\bibfnamefont
  {J.}~\bibnamefont {Slim}}, \bibinfo {author} {\bibfnamefont {H.}~\bibnamefont
  {Soltner}}, \bibinfo {author} {\bibfnamefont {A.}~\bibnamefont {Stahl}},
  \bibinfo {author} {\bibfnamefont {R.}~\bibnamefont {Stassen}}, \bibinfo
  {author} {\bibfnamefont {H.}~\bibnamefont {Stockhorst}}, \bibinfo {author}
  {\bibfnamefont {H.}~\bibnamefont {Str\"oher}}, \bibinfo {author}
  {\bibfnamefont {M.}~\bibnamefont {Tabidze}}, \bibinfo {author} {\bibfnamefont
  {G.}~\bibnamefont {Tagliente}}, \bibinfo {author} {\bibfnamefont
  {R.}~\bibnamefont {Talman}}, \bibinfo {author} {\bibfnamefont
  {P.}~\bibnamefont {Th\"orngren~Engblom}}, \bibinfo {author} {\bibfnamefont
  {F.}~\bibnamefont {Trinkel}}, \bibinfo {author} {\bibfnamefont
  {Y.}~\bibnamefont {Uzikov}}, \bibinfo {author} {\bibfnamefont
  {Y.}~\bibnamefont {Valdau}}, \bibinfo {author} {\bibfnamefont
  {E.}~\bibnamefont {Valetov}}, \bibinfo {author} {\bibfnamefont
  {A.}~\bibnamefont {Vassiliev}}, \bibinfo {author} {\bibfnamefont
  {C.}~\bibnamefont {Weidemann}}, \bibinfo {author} {\bibfnamefont
  {A.}~\bibnamefont {Wro\ifmmode~\acute{n}\else \'{n}\fi{}ska}}, \bibinfo
  {author} {\bibfnamefont {P.}~\bibnamefont {W\"ustner}}, \bibinfo {author}
  {\bibfnamefont {P.}~\bibnamefont {Zupra\ifmmode~\acute{n}\else
  \'{n}\fi{}ski}},\ and\ \bibinfo {author} {\bibfnamefont {M.}~\bibnamefont
  {\ifmmode~\dot{Z}\else \.{Z}\fi{}urek}} (\bibinfo {collaboration} {JEDI
  Collaboration}),\ }\href {https://doi.org/10.1103/PhysRevLett.119.014801}
  {\bibfield  {journal} {\bibinfo  {journal} {Phys. Rev. Lett.}\ }\textbf
  {\bibinfo {volume} {119}},\ \bibinfo {pages} {014801} (\bibinfo {year}
  {2017})}\BibitemShut {NoStop}%
\bibitem [{\citenamefont {Guidoboni}\ \emph {et~al.}(2018)\citenamefont
  {Guidoboni}, \citenamefont {Stephenson}, \citenamefont
  {Wro\ifmmode~\acute{n}\else \'{n}\fi{}ska}, \citenamefont {Bagdasarian},
  \citenamefont {Bsaisou}, \citenamefont {Chekmenev}, \citenamefont {Ciullo},
  \citenamefont {Dymov}, \citenamefont {Eversmann}, \citenamefont {Gaisser},
  \citenamefont {Gebel}, \citenamefont {Hejny}, \citenamefont {Hempelmann},
  \citenamefont {Hinder}, \citenamefont {Kacharava}, \citenamefont
  {Keshelashvili}, \citenamefont {Kulessa}, \citenamefont {Lenisa},
  \citenamefont {Lehrach}, \citenamefont {Lorentz}, \citenamefont {Maanen},
  \citenamefont {Maier}, \citenamefont {Mchedlishvili}, \citenamefont {Mey},
  \citenamefont {Nass}, \citenamefont {Pesce}, \citenamefont {Orlov},
  \citenamefont {Pretz}, \citenamefont {Prasuhn}, \citenamefont {Rathmann},
  \citenamefont {Rosenthal}, \citenamefont {Saleev}, \citenamefont
  {Semertzidis}, \citenamefont {Senichev}, \citenamefont {Shmakova},
  \citenamefont {Stockhorst}, \citenamefont {Str\"oher}, \citenamefont
  {Talman}, \citenamefont {Th\"orngren~Engblom}, \citenamefont {Trinkel},
  \citenamefont {Valdau}, \citenamefont {Weidemann}, \citenamefont {W\"ustner},
  \citenamefont {\ifmmode~\dot{Z}\else \.{Z}\fi{}urek},\ and\ \citenamefont
  {Zyuzin}}]{SCTchromaticityJEDI}%
  \BibitemOpen
  \bibfield  {author} {\bibinfo {author} {\bibfnamefont {G.}~\bibnamefont
  {Guidoboni}}, \bibinfo {author} {\bibfnamefont {E.~J.}\ \bibnamefont
  {Stephenson}}, \bibinfo {author} {\bibfnamefont {A.}~\bibnamefont
  {Wro\ifmmode~\acute{n}\else \'{n}\fi{}ska}}, \bibinfo {author} {\bibfnamefont
  {Z.}~\bibnamefont {Bagdasarian}}, \bibinfo {author} {\bibfnamefont
  {J.}~\bibnamefont {Bsaisou}}, \bibinfo {author} {\bibfnamefont
  {S.}~\bibnamefont {Chekmenev}}, \bibinfo {author} {\bibfnamefont
  {G.}~\bibnamefont {Ciullo}}, \bibinfo {author} {\bibfnamefont
  {S.}~\bibnamefont {Dymov}}, \bibinfo {author} {\bibfnamefont
  {D.}~\bibnamefont {Eversmann}}, \bibinfo {author} {\bibfnamefont
  {M.}~\bibnamefont {Gaisser}}, \bibinfo {author} {\bibfnamefont
  {R.}~\bibnamefont {Gebel}}, \bibinfo {author} {\bibfnamefont
  {V.}~\bibnamefont {Hejny}}, \bibinfo {author} {\bibfnamefont
  {N.}~\bibnamefont {Hempelmann}}, \bibinfo {author} {\bibfnamefont
  {F.}~\bibnamefont {Hinder}}, \bibinfo {author} {\bibfnamefont
  {A.}~\bibnamefont {Kacharava}}, \bibinfo {author} {\bibfnamefont
  {I.}~\bibnamefont {Keshelashvili}}, \bibinfo {author} {\bibfnamefont
  {P.}~\bibnamefont {Kulessa}}, \bibinfo {author} {\bibfnamefont
  {P.}~\bibnamefont {Lenisa}}, \bibinfo {author} {\bibfnamefont
  {A.}~\bibnamefont {Lehrach}}, \bibinfo {author} {\bibfnamefont
  {B.}~\bibnamefont {Lorentz}}, \bibinfo {author} {\bibfnamefont
  {P.}~\bibnamefont {Maanen}}, \bibinfo {author} {\bibfnamefont
  {R.}~\bibnamefont {Maier}}, \bibinfo {author} {\bibfnamefont
  {D.}~\bibnamefont {Mchedlishvili}}, \bibinfo {author} {\bibfnamefont
  {S.}~\bibnamefont {Mey}}, \bibinfo {author} {\bibfnamefont {A.}~\bibnamefont
  {Nass}}, \bibinfo {author} {\bibfnamefont {A.}~\bibnamefont {Pesce}},
  \bibinfo {author} {\bibfnamefont {Y.}~\bibnamefont {Orlov}}, \bibinfo
  {author} {\bibfnamefont {J.}~\bibnamefont {Pretz}}, \bibinfo {author}
  {\bibfnamefont {D.}~\bibnamefont {Prasuhn}}, \bibinfo {author} {\bibfnamefont
  {F.}~\bibnamefont {Rathmann}}, \bibinfo {author} {\bibfnamefont
  {M.}~\bibnamefont {Rosenthal}}, \bibinfo {author} {\bibfnamefont
  {A.}~\bibnamefont {Saleev}}, \bibinfo {author} {\bibfnamefont {Y.~K.}\
  \bibnamefont {Semertzidis}}, \bibinfo {author} {\bibfnamefont
  {Y.}~\bibnamefont {Senichev}}, \bibinfo {author} {\bibfnamefont
  {V.}~\bibnamefont {Shmakova}}, \bibinfo {author} {\bibfnamefont
  {H.}~\bibnamefont {Stockhorst}}, \bibinfo {author} {\bibfnamefont
  {H.}~\bibnamefont {Str\"oher}}, \bibinfo {author} {\bibfnamefont
  {R.}~\bibnamefont {Talman}}, \bibinfo {author} {\bibfnamefont
  {P.}~\bibnamefont {Th\"orngren~Engblom}}, \bibinfo {author} {\bibfnamefont
  {F.}~\bibnamefont {Trinkel}}, \bibinfo {author} {\bibfnamefont
  {Y.}~\bibnamefont {Valdau}}, \bibinfo {author} {\bibfnamefont
  {C.}~\bibnamefont {Weidemann}}, \bibinfo {author} {\bibfnamefont
  {P.}~\bibnamefont {W\"ustner}}, \bibinfo {author} {\bibfnamefont
  {M.}~\bibnamefont {\ifmmode~\dot{Z}\else \.{Z}\fi{}urek}},\ and\ \bibinfo
  {author} {\bibfnamefont {D.}~\bibnamefont {Zyuzin}} (\bibinfo {collaboration}
  {JEDI Collaboration}),\ }\href
  {https://doi.org/10.1103/PhysRevAccelBeams.21.024201} {\bibfield  {journal}
  {\bibinfo  {journal} {Phys. Rev. Accel. Beams}\ }\textbf {\bibinfo {volume}
  {21}},\ \bibinfo {pages} {024201} (\bibinfo {year} {2018})}\BibitemShut
  {NoStop}%
\bibitem [{\citenamefont {Vasserman}\ \emph {et~al.}(1987)\citenamefont
  {Vasserman} \emph {et~al.}}]{VassermanSCT}%
  \BibitemOpen
  \bibfield  {author} {\bibinfo {author} {\bibfnamefont {I.~B.}\ \bibnamefont
  {Vasserman}} \emph {et~al.},\ }\href
  {https://doi.org/http://dx.doi.org/10.1016/0370-2693(87)90094-3} {\bibfield
  {journal} {\bibinfo  {journal} {Phys. Lett. B}\ }\textbf {\bibinfo {volume}
  {187}},\ \bibinfo {pages} {172 } (\bibinfo {year} {1987})}\BibitemShut
  {NoStop}%
\bibitem [{\citenamefont {Saleev}\ \emph {et~al.}(2017)\citenamefont {Saleev},
  \citenamefont {Nikolaev}, \citenamefont {Rathmann}, \citenamefont
  {Augustyniak}, \citenamefont {Bagdasarian}, \citenamefont {Bai},
  \citenamefont {Barion}, \citenamefont {Berz}, \citenamefont {Chekmenev},
  \citenamefont {Ciullo}, \citenamefont {Dymov}, \citenamefont {Eversmann},
  \citenamefont {Gaisser}, \citenamefont {Gebel}, \citenamefont {Grigoryev},
  \citenamefont {Grzonka}, \citenamefont {Guidoboni}, \citenamefont
  {Heberling}, \citenamefont {Hejny}, \citenamefont {Hempelmann}, \citenamefont
  {Hetzel}, \citenamefont {Hinder}, \citenamefont {Kacharava}, \citenamefont
  {Kamerdzhiev}, \citenamefont {Keshelashvili}, \citenamefont {Koop},
  \citenamefont {Kulikov}, \citenamefont {Lehrach}, \citenamefont {Lenisa},
  \citenamefont {Lomidze}, \citenamefont {Lorentz}, \citenamefont {Maanen},
  \citenamefont {Macharashvili}, \citenamefont {Magiera}, \citenamefont
  {Mchedlishvili}, \citenamefont {Mey}, \citenamefont {M\"uller}, \citenamefont
  {Nass}, \citenamefont {Pesce}, \citenamefont {Prasuhn}, \citenamefont
  {Pretz}, \citenamefont {Rosenthal}, \citenamefont {Schmidt}, \citenamefont
  {Semertzidis}, \citenamefont {Senichev}, \citenamefont {Shmakova},
  \citenamefont {Silenko}, \citenamefont {Slim}, \citenamefont {Soltner},
  \citenamefont {Stahl}, \citenamefont {Stassen}, \citenamefont {Stephenson},
  \citenamefont {Stockhorst}, \citenamefont {Str\"oher}, \citenamefont
  {Tabidze}, \citenamefont {Tagliente}, \citenamefont {Talman}, \citenamefont
  {Engblom}, \citenamefont {Trinkel}, \citenamefont {Uzikov}, \citenamefont
  {Valdau}, \citenamefont {Valetov}, \citenamefont {Vassiliev}, \citenamefont
  {Weidemann}, \citenamefont {Wro\ifmmode~\acute{n}\else \'{n}\fi{}ska},
  \citenamefont {W\"ustner}, \citenamefont {Zupra\ifmmode~\acute{n}\else
  \'{n}\fi{}ski},\ and\ \citenamefont {Zurek}}]{SpinTuneMapping}%
  \BibitemOpen
  \bibfield  {author} {\bibinfo {author} {\bibfnamefont {A.}~\bibnamefont
  {Saleev}}, \bibinfo {author} {\bibfnamefont {N.~N.}\ \bibnamefont
  {Nikolaev}}, \bibinfo {author} {\bibfnamefont {F.}~\bibnamefont {Rathmann}},
  \bibinfo {author} {\bibfnamefont {W.}~\bibnamefont {Augustyniak}}, \bibinfo
  {author} {\bibfnamefont {Z.}~\bibnamefont {Bagdasarian}}, \bibinfo {author}
  {\bibfnamefont {M.}~\bibnamefont {Bai}}, \bibinfo {author} {\bibfnamefont
  {L.}~\bibnamefont {Barion}}, \bibinfo {author} {\bibfnamefont
  {M.}~\bibnamefont {Berz}}, \bibinfo {author} {\bibfnamefont {S.}~\bibnamefont
  {Chekmenev}}, \bibinfo {author} {\bibfnamefont {G.}~\bibnamefont {Ciullo}},
  \bibinfo {author} {\bibfnamefont {S.}~\bibnamefont {Dymov}}, \bibinfo
  {author} {\bibfnamefont {D.}~\bibnamefont {Eversmann}}, \bibinfo {author}
  {\bibfnamefont {M.}~\bibnamefont {Gaisser}}, \bibinfo {author} {\bibfnamefont
  {R.}~\bibnamefont {Gebel}}, \bibinfo {author} {\bibfnamefont
  {K.}~\bibnamefont {Grigoryev}}, \bibinfo {author} {\bibfnamefont
  {D.}~\bibnamefont {Grzonka}}, \bibinfo {author} {\bibfnamefont
  {G.}~\bibnamefont {Guidoboni}}, \bibinfo {author} {\bibfnamefont
  {D.}~\bibnamefont {Heberling}}, \bibinfo {author} {\bibfnamefont
  {V.}~\bibnamefont {Hejny}}, \bibinfo {author} {\bibfnamefont
  {N.}~\bibnamefont {Hempelmann}}, \bibinfo {author} {\bibfnamefont
  {J.}~\bibnamefont {Hetzel}}, \bibinfo {author} {\bibfnamefont
  {F.}~\bibnamefont {Hinder}}, \bibinfo {author} {\bibfnamefont
  {A.}~\bibnamefont {Kacharava}}, \bibinfo {author} {\bibfnamefont
  {V.}~\bibnamefont {Kamerdzhiev}}, \bibinfo {author} {\bibfnamefont
  {I.}~\bibnamefont {Keshelashvili}}, \bibinfo {author} {\bibfnamefont
  {I.}~\bibnamefont {Koop}}, \bibinfo {author} {\bibfnamefont {A.}~\bibnamefont
  {Kulikov}}, \bibinfo {author} {\bibfnamefont {A.}~\bibnamefont {Lehrach}},
  \bibinfo {author} {\bibfnamefont {P.}~\bibnamefont {Lenisa}}, \bibinfo
  {author} {\bibfnamefont {N.}~\bibnamefont {Lomidze}}, \bibinfo {author}
  {\bibfnamefont {B.}~\bibnamefont {Lorentz}}, \bibinfo {author} {\bibfnamefont
  {P.}~\bibnamefont {Maanen}}, \bibinfo {author} {\bibfnamefont
  {G.}~\bibnamefont {Macharashvili}}, \bibinfo {author} {\bibfnamefont
  {A.}~\bibnamefont {Magiera}}, \bibinfo {author} {\bibfnamefont
  {D.}~\bibnamefont {Mchedlishvili}}, \bibinfo {author} {\bibfnamefont
  {S.}~\bibnamefont {Mey}}, \bibinfo {author} {\bibfnamefont {F.}~\bibnamefont
  {M\"uller}}, \bibinfo {author} {\bibfnamefont {A.}~\bibnamefont {Nass}},
  \bibinfo {author} {\bibfnamefont {A.}~\bibnamefont {Pesce}}, \bibinfo
  {author} {\bibfnamefont {D.}~\bibnamefont {Prasuhn}}, \bibinfo {author}
  {\bibfnamefont {J.}~\bibnamefont {Pretz}}, \bibinfo {author} {\bibfnamefont
  {M.}~\bibnamefont {Rosenthal}}, \bibinfo {author} {\bibfnamefont
  {V.}~\bibnamefont {Schmidt}}, \bibinfo {author} {\bibfnamefont
  {Y.}~\bibnamefont {Semertzidis}}, \bibinfo {author} {\bibfnamefont
  {Y.}~\bibnamefont {Senichev}}, \bibinfo {author} {\bibfnamefont
  {V.}~\bibnamefont {Shmakova}}, \bibinfo {author} {\bibfnamefont
  {A.}~\bibnamefont {Silenko}}, \bibinfo {author} {\bibfnamefont
  {J.}~\bibnamefont {Slim}}, \bibinfo {author} {\bibfnamefont {H.}~\bibnamefont
  {Soltner}}, \bibinfo {author} {\bibfnamefont {A.}~\bibnamefont {Stahl}},
  \bibinfo {author} {\bibfnamefont {R.}~\bibnamefont {Stassen}}, \bibinfo
  {author} {\bibfnamefont {E.}~\bibnamefont {Stephenson}}, \bibinfo {author}
  {\bibfnamefont {H.}~\bibnamefont {Stockhorst}}, \bibinfo {author}
  {\bibfnamefont {H.}~\bibnamefont {Str\"oher}}, \bibinfo {author}
  {\bibfnamefont {M.}~\bibnamefont {Tabidze}}, \bibinfo {author} {\bibfnamefont
  {G.}~\bibnamefont {Tagliente}}, \bibinfo {author} {\bibfnamefont
  {R.}~\bibnamefont {Talman}}, \bibinfo {author} {\bibfnamefont {P.~T.}\
  \bibnamefont {Engblom}}, \bibinfo {author} {\bibfnamefont {F.}~\bibnamefont
  {Trinkel}}, \bibinfo {author} {\bibfnamefont {Y.}~\bibnamefont {Uzikov}},
  \bibinfo {author} {\bibfnamefont {Y.}~\bibnamefont {Valdau}}, \bibinfo
  {author} {\bibfnamefont {E.}~\bibnamefont {Valetov}}, \bibinfo {author}
  {\bibfnamefont {A.}~\bibnamefont {Vassiliev}}, \bibinfo {author}
  {\bibfnamefont {C.}~\bibnamefont {Weidemann}}, \bibinfo {author}
  {\bibfnamefont {A.}~\bibnamefont {Wro\ifmmode~\acute{n}\else \'{n}\fi{}ska}},
  \bibinfo {author} {\bibfnamefont {P.}~\bibnamefont {W\"ustner}}, \bibinfo
  {author} {\bibfnamefont {P.}~\bibnamefont {Zupra\ifmmode~\acute{n}\else
  \'{n}\fi{}ski}},\ and\ \bibinfo {author} {\bibfnamefont {M.}~\bibnamefont
  {Zurek}} (\bibinfo {collaboration} {JEDI collaboration}),\ }\href
  {https://doi.org/10.1103/PhysRevAccelBeams.20.072801} {\bibfield  {journal}
  {\bibinfo  {journal} {Phys. Rev. Accel. Beams}\ }\textbf {\bibinfo {volume}
  {20}},\ \bibinfo {pages} {072801} (\bibinfo {year} {2017})}\BibitemShut
  {NoStop}%
\bibitem [{\citenamefont {Wagner}\ \emph {et~al.}(2021)\citenamefont {Wagner},
  \citenamefont {Nass}, \citenamefont {Pretz}, \citenamefont {Abusaif},
  \citenamefont {Aggarwal}, \citenamefont {Andres}, \citenamefont {Bekman},
  \citenamefont {Canale}, \citenamefont {Ciepal}, \citenamefont {Ciullo},
  \citenamefont {Dahmen}, \citenamefont {Dymov}, \citenamefont {Ehrlich},
  \citenamefont {Gebel}, \citenamefont {Grigoryev}, \citenamefont {Grzonka},
  \citenamefont {Hejny}, \citenamefont {Hetzel}, \citenamefont {Kacharava},
  \citenamefont {Kamerdzhiev}, \citenamefont {Karanth}, \citenamefont
  {Keshelashvili}, \citenamefont {Kononov}, \citenamefont {Kulikov},
  \citenamefont {Laihem}, \citenamefont {Lehrach}, \citenamefont {Lenisa},
  \citenamefont {Lomidze}, \citenamefont {Magiera}, \citenamefont
  {Mchedlishvili}, \citenamefont {Müller}, \citenamefont {Nikolaev},
  \citenamefont {Pesce}, \citenamefont {Poncza}, \citenamefont {Rathmann},
  \citenamefont {Retzlaff}, \citenamefont {Saleev}, \citenamefont {Schmühl},
  \citenamefont {Shergelashvili}, \citenamefont {Shmakova}, \citenamefont
  {Slim}, \citenamefont {Stahl}, \citenamefont {Stephenson}, \citenamefont
  {Ströher}, \citenamefont {Tabidze}, \citenamefont {Tagliente}, \citenamefont
  {Talman}, \citenamefont {Uzikov}, \citenamefont {Valdau},\ and\ \citenamefont
  {Wrońska}}]{BBAJEDI}%
  \BibitemOpen
  \bibfield  {author} {\bibinfo {author} {\bibfnamefont {T.}~\bibnamefont
  {Wagner}}, \bibinfo {author} {\bibfnamefont {A.}~\bibnamefont {Nass}},
  \bibinfo {author} {\bibfnamefont {J.}~\bibnamefont {Pretz}}, \bibinfo
  {author} {\bibfnamefont {F.}~\bibnamefont {Abusaif}}, \bibinfo {author}
  {\bibfnamefont {A.}~\bibnamefont {Aggarwal}}, \bibinfo {author}
  {\bibfnamefont {A.}~\bibnamefont {Andres}}, \bibinfo {author} {\bibfnamefont
  {I.}~\bibnamefont {Bekman}}, \bibinfo {author} {\bibfnamefont
  {N.}~\bibnamefont {Canale}}, \bibinfo {author} {\bibfnamefont
  {I.}~\bibnamefont {Ciepal}}, \bibinfo {author} {\bibfnamefont
  {G.}~\bibnamefont {Ciullo}}, \bibinfo {author} {\bibfnamefont
  {F.}~\bibnamefont {Dahmen}}, \bibinfo {author} {\bibfnamefont
  {S.}~\bibnamefont {Dymov}}, \bibinfo {author} {\bibfnamefont
  {C.}~\bibnamefont {Ehrlich}}, \bibinfo {author} {\bibfnamefont
  {R.}~\bibnamefont {Gebel}}, \bibinfo {author} {\bibfnamefont
  {K.}~\bibnamefont {Grigoryev}}, \bibinfo {author} {\bibfnamefont
  {D.}~\bibnamefont {Grzonka}}, \bibinfo {author} {\bibfnamefont
  {V.}~\bibnamefont {Hejny}}, \bibinfo {author} {\bibfnamefont
  {J.}~\bibnamefont {Hetzel}}, \bibinfo {author} {\bibfnamefont
  {A.}~\bibnamefont {Kacharava}}, \bibinfo {author} {\bibfnamefont
  {V.}~\bibnamefont {Kamerdzhiev}}, \bibinfo {author} {\bibfnamefont
  {S.}~\bibnamefont {Karanth}}, \bibinfo {author} {\bibfnamefont
  {I.}~\bibnamefont {Keshelashvili}}, \bibinfo {author} {\bibfnamefont
  {A.}~\bibnamefont {Kononov}}, \bibinfo {author} {\bibfnamefont
  {A.}~\bibnamefont {Kulikov}}, \bibinfo {author} {\bibfnamefont
  {K.}~\bibnamefont {Laihem}}, \bibinfo {author} {\bibfnamefont
  {A.}~\bibnamefont {Lehrach}}, \bibinfo {author} {\bibfnamefont
  {P.}~\bibnamefont {Lenisa}}, \bibinfo {author} {\bibfnamefont
  {N.}~\bibnamefont {Lomidze}}, \bibinfo {author} {\bibfnamefont
  {A.}~\bibnamefont {Magiera}}, \bibinfo {author} {\bibfnamefont
  {D.}~\bibnamefont {Mchedlishvili}}, \bibinfo {author} {\bibfnamefont
  {F.}~\bibnamefont {Müller}}, \bibinfo {author} {\bibfnamefont
  {N.}~\bibnamefont {Nikolaev}}, \bibinfo {author} {\bibfnamefont
  {A.}~\bibnamefont {Pesce}}, \bibinfo {author} {\bibfnamefont
  {V.}~\bibnamefont {Poncza}}, \bibinfo {author} {\bibfnamefont
  {F.}~\bibnamefont {Rathmann}}, \bibinfo {author} {\bibfnamefont
  {M.}~\bibnamefont {Retzlaff}}, \bibinfo {author} {\bibfnamefont
  {A.}~\bibnamefont {Saleev}}, \bibinfo {author} {\bibfnamefont
  {M.}~\bibnamefont {Schmühl}}, \bibinfo {author} {\bibfnamefont
  {D.}~\bibnamefont {Shergelashvili}}, \bibinfo {author} {\bibfnamefont
  {V.}~\bibnamefont {Shmakova}}, \bibinfo {author} {\bibfnamefont
  {J.}~\bibnamefont {Slim}}, \bibinfo {author} {\bibfnamefont {A.}~\bibnamefont
  {Stahl}}, \bibinfo {author} {\bibfnamefont {E.}~\bibnamefont {Stephenson}},
  \bibinfo {author} {\bibfnamefont {H.}~\bibnamefont {Ströher}}, \bibinfo
  {author} {\bibfnamefont {M.}~\bibnamefont {Tabidze}}, \bibinfo {author}
  {\bibfnamefont {G.}~\bibnamefont {Tagliente}}, \bibinfo {author}
  {\bibfnamefont {R.}~\bibnamefont {Talman}}, \bibinfo {author} {\bibfnamefont
  {Y.}~\bibnamefont {Uzikov}}, \bibinfo {author} {\bibfnamefont
  {Y.}~\bibnamefont {Valdau}},\ and\ \bibinfo {author} {\bibfnamefont
  {A.}~\bibnamefont {Wrońska}},\ }\href
  {https://doi.org/10.1088/1748-0221/16/02/t02001} {\bibfield  {journal}
  {\bibinfo  {journal} {Journal of Instrumentation}\ }\textbf {\bibinfo
  {volume} {16}}\bibinfo  {number} { (02)},\ \bibinfo {pages}
  {T02001–T02001}}\BibitemShut {NoStop}%
\bibitem [{\citenamefont {Slim}\ \emph {et~al.}(2016)\citenamefont {Slim},
  \citenamefont {Gebel}, \citenamefont {Heberling}, \citenamefont {Hinder},
  \citenamefont {H{\"o}lscher}, \citenamefont {Lehrach}, \citenamefont
  {Lorentz}, \citenamefont {Mey}, \citenamefont {Nass}, \citenamefont
  {Rathmann}, \citenamefont {Reifferscheidt}, \citenamefont {Soltner},
  \citenamefont {Straatmann}, \citenamefont {Trinkel},\ and\ \citenamefont
  {Wolters}}]{SlimWFDesign}%
  \BibitemOpen
\bibfield  {number} {  }\bibfield  {author} {\bibinfo {author} {\bibfnamefont
  {J.}~\bibnamefont {Slim}}, \bibinfo {author} {\bibfnamefont {R.}~\bibnamefont
  {Gebel}}, \bibinfo {author} {\bibfnamefont {D.}~\bibnamefont {Heberling}},
  \bibinfo {author} {\bibfnamefont {F.}~\bibnamefont {Hinder}}, \bibinfo
  {author} {\bibfnamefont {D.}~\bibnamefont {H{\"o}lscher}}, \bibinfo {author}
  {\bibfnamefont {A.}~\bibnamefont {Lehrach}}, \bibinfo {author} {\bibfnamefont
  {B.}~\bibnamefont {Lorentz}}, \bibinfo {author} {\bibfnamefont
  {S.}~\bibnamefont {Mey}}, \bibinfo {author} {\bibfnamefont {A.}~\bibnamefont
  {Nass}}, \bibinfo {author} {\bibfnamefont {F.}~\bibnamefont {Rathmann}},
  \bibinfo {author} {\bibfnamefont {L.}~\bibnamefont {Reifferscheidt}},
  \bibinfo {author} {\bibfnamefont {H.}~\bibnamefont {Soltner}}, \bibinfo
  {author} {\bibfnamefont {H.}~\bibnamefont {Straatmann}}, \bibinfo {author}
  {\bibfnamefont {F.}~\bibnamefont {Trinkel}},\ and\ \bibinfo {author}
  {\bibfnamefont {J.}~\bibnamefont {Wolters}},\ }\href
  {https://doi.org/http://dx.doi.org/10.1016/j.nima.2016.05.012} {\bibfield
  {journal} {\bibinfo  {journal} {Nuclear Instruments and Methods in Physics
  Research Section A: Accelerators, Spectrometers, Detectors and Associated
  Equipment}\ }\textbf {\bibinfo {volume} {828}},\ \bibinfo {pages} {116 }
  (\bibinfo {year} {2016})}\BibitemShut {NoStop}%
\bibitem [{\citenamefont {Slim}\ \emph {et~al.}(2020)\citenamefont {Slim},
  \citenamefont {Nass}, \citenamefont {Rathmann}, \citenamefont {Soltner},
  \citenamefont {Tagliente},\ and\ \citenamefont {Heberling}}]{SlimWFCircuit}%
  \BibitemOpen
  \bibfield  {author} {\bibinfo {author} {\bibfnamefont {J.}~\bibnamefont
  {Slim}}, \bibinfo {author} {\bibfnamefont {A.}~\bibnamefont {Nass}}, \bibinfo
  {author} {\bibfnamefont {F.}~\bibnamefont {Rathmann}}, \bibinfo {author}
  {\bibfnamefont {H.}~\bibnamefont {Soltner}}, \bibinfo {author} {\bibfnamefont
  {G.}~\bibnamefont {Tagliente}},\ and\ \bibinfo {author} {\bibfnamefont
  {D.}~\bibnamefont {Heberling}},\ }\href
  {https://doi.org/10.1088/1748-0221/15/03/P03021} {\bibfield  {journal}
  {\bibinfo  {journal} {JINST}\ }\textbf {\bibinfo {volume} {15}}\bibinfo
  {number} { (03)},\ \bibinfo {pages} {P03021}}\BibitemShut {NoStop}%
\bibitem [{\citenamefont {Slim}(2019)}]{SlimComissioning}%
  \BibitemOpen
\bibfield  {number} {  }\bibfield  {author} {\bibinfo {author} {\bibfnamefont
  {J.}~\bibnamefont {Slim}} (\bibinfo {collaboration} {JEDI}),\ }\href
  {https://doi.org/10.1007/s10751-018-1547-6} {\bibfield  {journal} {\bibinfo
  {journal} {Hyperfine Interact.}\ }\textbf {\bibinfo {volume} {240}},\
  \bibinfo {pages} {7} (\bibinfo {year} {2019})}\BibitemShut {NoStop}%
\bibitem [{\citenamefont {Slim}\ \emph {et~al.}(2021)\citenamefont {Slim},
  \citenamefont {Nikolaev}, \citenamefont {Rathmann}, \citenamefont {Wirzba},
  \citenamefont {Nass}, \citenamefont {Hejny}, \citenamefont {Pretz},
  \citenamefont {Soltner}, \citenamefont {Abusaif}, \citenamefont {Aggarwal},
  \citenamefont {Aksentev}, \citenamefont {Andres}, \citenamefont {Barion},
  \citenamefont {Ciullo}, \citenamefont {Dymov}, \citenamefont {Gebel},
  \citenamefont {Gaisser}, \citenamefont {Grigoryev}, \citenamefont {Grzonka},
  \citenamefont {Javakhishvili}, \citenamefont {Kacharava}, \citenamefont
  {Kamerdzhiev}, \citenamefont {Karanth}, \citenamefont {Keshelashvili},
  \citenamefont {Lehrach}, \citenamefont {Lenisa}, \citenamefont {Lomidze},
  \citenamefont {Lorentz}, \citenamefont {Magiera}, \citenamefont
  {Mchedlishvili}, \citenamefont {M\"uller}, \citenamefont {Pesce},
  \citenamefont {Poncza}, \citenamefont {Prasuhn}, \citenamefont {Saleev},
  \citenamefont {Shmakova}, \citenamefont {Str\"oher}, \citenamefont {Tabidze},
  \citenamefont {Tagliente}, \citenamefont {Valdau}, \citenamefont {Wagner},
  \citenamefont {Weidemann}, \citenamefont {Wro\ifmmode~\acute{n}\else
  \'{n}\fi{}ska},\ and\ \citenamefont {\ifmmode~\dot{Z}\else
  \.{Z}\fi{}urek}}]{SlimWFoscillations}%
  \BibitemOpen
  \bibfield  {author} {\bibinfo {author} {\bibfnamefont {J.}~\bibnamefont
  {Slim}}, \bibinfo {author} {\bibfnamefont {N.~N.}\ \bibnamefont {Nikolaev}},
  \bibinfo {author} {\bibfnamefont {F.}~\bibnamefont {Rathmann}}, \bibinfo
  {author} {\bibfnamefont {A.}~\bibnamefont {Wirzba}}, \bibinfo {author}
  {\bibfnamefont {A.}~\bibnamefont {Nass}}, \bibinfo {author} {\bibfnamefont
  {V.}~\bibnamefont {Hejny}}, \bibinfo {author} {\bibfnamefont
  {J.}~\bibnamefont {Pretz}}, \bibinfo {author} {\bibfnamefont
  {H.}~\bibnamefont {Soltner}}, \bibinfo {author} {\bibfnamefont
  {F.}~\bibnamefont {Abusaif}}, \bibinfo {author} {\bibfnamefont
  {A.}~\bibnamefont {Aggarwal}}, \bibinfo {author} {\bibfnamefont
  {A.}~\bibnamefont {Aksentev}}, \bibinfo {author} {\bibfnamefont
  {A.}~\bibnamefont {Andres}}, \bibinfo {author} {\bibfnamefont
  {L.}~\bibnamefont {Barion}}, \bibinfo {author} {\bibfnamefont
  {G.}~\bibnamefont {Ciullo}}, \bibinfo {author} {\bibfnamefont
  {S.}~\bibnamefont {Dymov}}, \bibinfo {author} {\bibfnamefont
  {R.}~\bibnamefont {Gebel}}, \bibinfo {author} {\bibfnamefont
  {M.}~\bibnamefont {Gaisser}}, \bibinfo {author} {\bibfnamefont
  {K.}~\bibnamefont {Grigoryev}}, \bibinfo {author} {\bibfnamefont
  {D.}~\bibnamefont {Grzonka}}, \bibinfo {author} {\bibfnamefont
  {O.}~\bibnamefont {Javakhishvili}}, \bibinfo {author} {\bibfnamefont
  {A.}~\bibnamefont {Kacharava}}, \bibinfo {author} {\bibfnamefont
  {V.}~\bibnamefont {Kamerdzhiev}}, \bibinfo {author} {\bibfnamefont
  {S.}~\bibnamefont {Karanth}}, \bibinfo {author} {\bibfnamefont
  {I.}~\bibnamefont {Keshelashvili}}, \bibinfo {author} {\bibfnamefont
  {A.}~\bibnamefont {Lehrach}}, \bibinfo {author} {\bibfnamefont
  {P.}~\bibnamefont {Lenisa}}, \bibinfo {author} {\bibfnamefont
  {N.}~\bibnamefont {Lomidze}}, \bibinfo {author} {\bibfnamefont
  {B.}~\bibnamefont {Lorentz}}, \bibinfo {author} {\bibfnamefont
  {A.}~\bibnamefont {Magiera}}, \bibinfo {author} {\bibfnamefont
  {D.}~\bibnamefont {Mchedlishvili}}, \bibinfo {author} {\bibfnamefont
  {F.}~\bibnamefont {M\"uller}}, \bibinfo {author} {\bibfnamefont
  {A.}~\bibnamefont {Pesce}}, \bibinfo {author} {\bibfnamefont
  {V.}~\bibnamefont {Poncza}}, \bibinfo {author} {\bibfnamefont
  {D.}~\bibnamefont {Prasuhn}}, \bibinfo {author} {\bibfnamefont
  {A.}~\bibnamefont {Saleev}}, \bibinfo {author} {\bibfnamefont
  {V.}~\bibnamefont {Shmakova}}, \bibinfo {author} {\bibfnamefont
  {H.}~\bibnamefont {Str\"oher}}, \bibinfo {author} {\bibfnamefont
  {M.}~\bibnamefont {Tabidze}}, \bibinfo {author} {\bibfnamefont
  {G.}~\bibnamefont {Tagliente}}, \bibinfo {author} {\bibfnamefont
  {Y.}~\bibnamefont {Valdau}}, \bibinfo {author} {\bibfnamefont
  {T.}~\bibnamefont {Wagner}}, \bibinfo {author} {\bibfnamefont
  {C.}~\bibnamefont {Weidemann}}, \bibinfo {author} {\bibfnamefont
  {A.}~\bibnamefont {Wro\ifmmode~\acute{n}\else \'{n}\fi{}ska}},\ and\ \bibinfo
  {author} {\bibfnamefont {M.}~\bibnamefont {\ifmmode~\dot{Z}\else
  \.{Z}\fi{}urek}} (\bibinfo {collaboration} {JEDI Collaboration}),\ }\href
  {https://doi.org/10.1103/PhysRevAccelBeams.24.124601} {\bibfield  {journal}
  {\bibinfo  {journal} {Phys. Rev. Accel. Beams}\ }\textbf {\bibinfo {volume}
  {24}},\ \bibinfo {pages} {124601} (\bibinfo {year} {2021})}\BibitemShut
  {NoStop}%
\bibitem [{\citenamefont {Slim}\ \emph {et~al.}(2023)\citenamefont {Slim},
  \citenamefont {Nikolaev}, \citenamefont {Rathmann},\ and\ \citenamefont
  {Wirzba}}]{QuantumOscillations}%
  \BibitemOpen
  \bibfield  {author} {\bibinfo {author} {\bibfnamefont {J.}~\bibnamefont
  {Slim}}, \bibinfo {author} {\bibfnamefont {N.~N.}\ \bibnamefont {Nikolaev}},
  \bibinfo {author} {\bibfnamefont {F.}~\bibnamefont {Rathmann}},\ and\
  \bibinfo {author} {\bibfnamefont {A.}~\bibnamefont {Wirzba}},\ }\href
  {https://doi.org/10.1103/PhysRevAccelBeams.26.014201} {\bibfield  {journal}
  {\bibinfo  {journal} {Phys. Rev. Accel. Beams}\ }\textbf {\bibinfo {volume}
  {26}},\ \bibinfo {pages} {014201} (\bibinfo {year} {2023})}\BibitemShut
  {NoStop}%
\bibitem [{\citenamefont {Hejny}\ \emph {et~al.}(2025)\citenamefont {Hejny},
  \citenamefont {Andres}, \citenamefont {Pretz}, \citenamefont {Abusaif},
  \citenamefont {Aggarwal}, \citenamefont {Aksentev}, \citenamefont {Alberdi},
  \citenamefont {Barion}, \citenamefont {Bekman}, \citenamefont {Bey\ss{}},
  \citenamefont {B\"ohme}, \citenamefont {Breitkreutz}, \citenamefont {Canale},
  \citenamefont {Ciullo}, \citenamefont {Dymov}, \citenamefont {Fr\"ohlich},
  \citenamefont {Gebel}, \citenamefont {Gaisser}, \citenamefont {Grigoryev},
  \citenamefont {Grzonka}, \citenamefont {Hetzel}, \citenamefont
  {Javakhishvili}, \citenamefont {Kacharava}, \citenamefont {Kamerdzhiev},
  \citenamefont {Karanth}, \citenamefont {Keshelashvili}, \citenamefont
  {Kononov}, \citenamefont {Laihem}, \citenamefont {Lehrach}, \citenamefont
  {Lenisa}, \citenamefont {Lomidze}, \citenamefont {Lorentz}, \citenamefont
  {Macharashvili}, \citenamefont {Magiera}, \citenamefont {Mchedlishvili},
  \citenamefont {Melnikov}, \citenamefont {M\"uller}, \citenamefont {Nass},
  \citenamefont {Nikolaev}, \citenamefont {Okropiridze}, \citenamefont {Pesce},
  \citenamefont {Piccoli}, \citenamefont {Poncza}, \citenamefont {Prasuhn},
  \citenamefont {Rathmann}, \citenamefont {Saleev}, \citenamefont
  {Shergelashvili}, \citenamefont {Shmakova}, \citenamefont {Shankar},
  \citenamefont {Shurkhno}, \citenamefont {Siddique}, \citenamefont {Silenko},
  \citenamefont {Slim}, \citenamefont {Soltner}, \citenamefont {Stassen},
  \citenamefont {Stephenson}, \citenamefont {Str\"oher}, \citenamefont
  {Tabidze}, \citenamefont {Tagliente}, \citenamefont {Valdau}, \citenamefont
  {Vitz}, \citenamefont {Wagner}, \citenamefont {Wirzba}, \citenamefont
  {Wro\ifmmode~\acute{n}\else \'{n}\fi{}ska}, \citenamefont {W\"ustner},\ and\
  \citenamefont {\ifmmode~\dot{Z}\else \.{Z}\fi{}urek}}]{JEDI_Feedback-2025}%
  \BibitemOpen
  \bibfield  {author} {\bibinfo {author} {\bibfnamefont {V.}~\bibnamefont
  {Hejny}}, \bibinfo {author} {\bibfnamefont {A.}~\bibnamefont {Andres}},
  \bibinfo {author} {\bibfnamefont {J.}~\bibnamefont {Pretz}}, \bibinfo
  {author} {\bibfnamefont {F.}~\bibnamefont {Abusaif}}, \bibinfo {author}
  {\bibfnamefont {A.}~\bibnamefont {Aggarwal}}, \bibinfo {author}
  {\bibfnamefont {A.}~\bibnamefont {Aksentev}}, \bibinfo {author}
  {\bibfnamefont {B.}~\bibnamefont {Alberdi}}, \bibinfo {author} {\bibfnamefont
  {L.}~\bibnamefont {Barion}}, \bibinfo {author} {\bibfnamefont
  {I.}~\bibnamefont {Bekman}}, \bibinfo {author} {\bibfnamefont
  {M.}~\bibnamefont {Bey\ss{}}}, \bibinfo {author} {\bibfnamefont
  {C.}~\bibnamefont {B\"ohme}}, \bibinfo {author} {\bibfnamefont
  {B.}~\bibnamefont {Breitkreutz}}, \bibinfo {author} {\bibfnamefont
  {N.}~\bibnamefont {Canale}}, \bibinfo {author} {\bibfnamefont
  {G.}~\bibnamefont {Ciullo}}, \bibinfo {author} {\bibfnamefont
  {S.}~\bibnamefont {Dymov}}, \bibinfo {author} {\bibfnamefont {N.-O.}\
  \bibnamefont {Fr\"ohlich}}, \bibinfo {author} {\bibfnamefont
  {R.}~\bibnamefont {Gebel}}, \bibinfo {author} {\bibfnamefont
  {M.}~\bibnamefont {Gaisser}}, \bibinfo {author} {\bibfnamefont
  {K.}~\bibnamefont {Grigoryev}}, \bibinfo {author} {\bibfnamefont
  {D.}~\bibnamefont {Grzonka}}, \bibinfo {author} {\bibfnamefont
  {J.}~\bibnamefont {Hetzel}}, \bibinfo {author} {\bibfnamefont
  {O.}~\bibnamefont {Javakhishvili}}, \bibinfo {author} {\bibfnamefont
  {A.}~\bibnamefont {Kacharava}}, \bibinfo {author} {\bibfnamefont
  {V.}~\bibnamefont {Kamerdzhiev}}, \bibinfo {author} {\bibfnamefont
  {S.}~\bibnamefont {Karanth}}, \bibinfo {author} {\bibfnamefont
  {I.}~\bibnamefont {Keshelashvili}}, \bibinfo {author} {\bibfnamefont
  {A.}~\bibnamefont {Kononov}}, \bibinfo {author} {\bibfnamefont
  {K.}~\bibnamefont {Laihem}}, \bibinfo {author} {\bibfnamefont
  {A.}~\bibnamefont {Lehrach}}, \bibinfo {author} {\bibfnamefont
  {P.}~\bibnamefont {Lenisa}}, \bibinfo {author} {\bibfnamefont
  {N.}~\bibnamefont {Lomidze}}, \bibinfo {author} {\bibfnamefont
  {B.}~\bibnamefont {Lorentz}}, \bibinfo {author} {\bibfnamefont
  {G.}~\bibnamefont {Macharashvili}}, \bibinfo {author} {\bibfnamefont
  {A.}~\bibnamefont {Magiera}}, \bibinfo {author} {\bibfnamefont
  {D.}~\bibnamefont {Mchedlishvili}}, \bibinfo {author} {\bibfnamefont
  {A.}~\bibnamefont {Melnikov}}, \bibinfo {author} {\bibfnamefont
  {F.}~\bibnamefont {M\"uller}}, \bibinfo {author} {\bibfnamefont
  {A.}~\bibnamefont {Nass}}, \bibinfo {author} {\bibfnamefont {N.~N.}\
  \bibnamefont {Nikolaev}}, \bibinfo {author} {\bibfnamefont {D.}~\bibnamefont
  {Okropiridze}}, \bibinfo {author} {\bibfnamefont {A.}~\bibnamefont {Pesce}},
  \bibinfo {author} {\bibfnamefont {A.}~\bibnamefont {Piccoli}}, \bibinfo
  {author} {\bibfnamefont {V.}~\bibnamefont {Poncza}}, \bibinfo {author}
  {\bibfnamefont {D.}~\bibnamefont {Prasuhn}}, \bibinfo {author} {\bibfnamefont
  {F.}~\bibnamefont {Rathmann}}, \bibinfo {author} {\bibfnamefont
  {A.}~\bibnamefont {Saleev}}, \bibinfo {author} {\bibfnamefont
  {D.}~\bibnamefont {Shergelashvili}}, \bibinfo {author} {\bibfnamefont
  {V.}~\bibnamefont {Shmakova}}, \bibinfo {author} {\bibfnamefont
  {R.}~\bibnamefont {Shankar}}, \bibinfo {author} {\bibfnamefont
  {N.}~\bibnamefont {Shurkhno}}, \bibinfo {author} {\bibfnamefont
  {S.}~\bibnamefont {Siddique}}, \bibinfo {author} {\bibfnamefont
  {A.}~\bibnamefont {Silenko}}, \bibinfo {author} {\bibfnamefont
  {J.}~\bibnamefont {Slim}}, \bibinfo {author} {\bibfnamefont {H.}~\bibnamefont
  {Soltner}}, \bibinfo {author} {\bibfnamefont {R.}~\bibnamefont {Stassen}},
  \bibinfo {author} {\bibfnamefont {E.~J.}\ \bibnamefont {Stephenson}},
  \bibinfo {author} {\bibfnamefont {H.}~\bibnamefont {Str\"oher}}, \bibinfo
  {author} {\bibfnamefont {M.}~\bibnamefont {Tabidze}}, \bibinfo {author}
  {\bibfnamefont {G.}~\bibnamefont {Tagliente}}, \bibinfo {author}
  {\bibfnamefont {Y.}~\bibnamefont {Valdau}}, \bibinfo {author} {\bibfnamefont
  {M.}~\bibnamefont {Vitz}}, \bibinfo {author} {\bibfnamefont {T.}~\bibnamefont
  {Wagner}}, \bibinfo {author} {\bibfnamefont {A.}~\bibnamefont {Wirzba}},
  \bibinfo {author} {\bibfnamefont {A.}~\bibnamefont
  {Wro\ifmmode~\acute{n}\else \'{n}\fi{}ska}}, \bibinfo {author} {\bibfnamefont
  {P.}~\bibnamefont {W\"ustner}},\ and\ \bibinfo {author} {\bibfnamefont
  {M.}~\bibnamefont {\ifmmode~\dot{Z}\else \.{Z}\fi{}urek}} (\bibinfo
  {collaboration} {JEDI Collaboration}),\ }\href
  {https://doi.org/10.1103/PhysRevAccelBeams.28.062801} {\bibfield  {journal}
  {\bibinfo  {journal} {Phys. Rev. Accel. Beams}\ }\textbf {\bibinfo {volume}
  {28}},\ \bibinfo {pages} {062801} (\bibinfo {year} {2025})}\BibitemShut
  {NoStop}%
\bibitem [{\citenamefont {Andres}\ \emph {et~al.}(2026)\citenamefont {Andres},
  \citenamefont {Hejny}, \citenamefont {Nass}, \citenamefont {Nikolaev},
  \citenamefont {Pretz}, \citenamefont {Rathmann}, \citenamefont {Shmakova},
  \citenamefont {Slim}, \citenamefont {Abusaif}, \citenamefont {Aggarwal},
  \citenamefont {Aksentev}, \citenamefont {Alberdi}, \citenamefont {Barion},
  \citenamefont {Bekman}, \citenamefont {Bey\ss{}}, \citenamefont {B\"ohme},
  \citenamefont {Breitkreutz}, \citenamefont {Canale}, \citenamefont {Ciullo},
  \citenamefont {Dymov}, \citenamefont {Fr\"ohlich}, \citenamefont {Gebel},
  \citenamefont {Gaisser}, \citenamefont {Grigoryev}, \citenamefont {Grzonka},
  \citenamefont {Gu}, \citenamefont {Heberling}, \citenamefont {Hetzel},
  \citenamefont {H\"olscher}, \citenamefont {Javakhishvili}, \citenamefont
  {Kacharava}, \citenamefont {Kamerdzhiev}, \citenamefont {Karanth},
  \citenamefont {Keshelashvili}, \citenamefont {Kononov}, \citenamefont
  {Laihem}, \citenamefont {Lehrach}, \citenamefont {Lenisa}, \citenamefont
  {Lomidze}, \citenamefont {Lorentz}, \citenamefont {Macharashvili},
  \citenamefont {Magiera}, \citenamefont {Margos}, \citenamefont
  {Mchedlishvili}, \citenamefont {Melnikov}, \citenamefont {M\"uller},
  \citenamefont {Okropiridze}, \citenamefont {Pesce}, \citenamefont {Piccoli},
  \citenamefont {Poncza}, \citenamefont {Prasuhn}, \citenamefont {Saleev},
  \citenamefont {Shergelashvili}, \citenamefont {Shankar}, \citenamefont
  {Shurkhno}, \citenamefont {Siddique}, \citenamefont {Silenko}, \citenamefont
  {Soltner}, \citenamefont {Stassen}, \citenamefont {Stephenson}, \citenamefont
  {Str\"oher}, \citenamefont {Tabidze}, \citenamefont {Tagliente},
  \citenamefont {Tempel}, \citenamefont {Valdau}, \citenamefont {Vitz},
  \citenamefont {Wagner}, \citenamefont {Wirzba}, \citenamefont
  {Wro\ifmmode~\acute{n}\else \'{n}\fi{}ska}, \citenamefont {W\"ustner},\ and\
  \citenamefont {\ifmmode~\dot{Z}\else \.{Z}\fi{}urek}}]{EDM-JEDI}%
  \BibitemOpen
  \bibfield  {author} {\bibinfo {author} {\bibfnamefont {A.}~\bibnamefont
  {Andres}}, \bibinfo {author} {\bibfnamefont {V.}~\bibnamefont {Hejny}},
  \bibinfo {author} {\bibfnamefont {A.}~\bibnamefont {Nass}}, \bibinfo {author}
  {\bibfnamefont {N.~N.}\ \bibnamefont {Nikolaev}}, \bibinfo {author}
  {\bibfnamefont {J.}~\bibnamefont {Pretz}}, \bibinfo {author} {\bibfnamefont
  {F.}~\bibnamefont {Rathmann}}, \bibinfo {author} {\bibfnamefont
  {V.}~\bibnamefont {Shmakova}}, \bibinfo {author} {\bibfnamefont
  {J.}~\bibnamefont {Slim}}, \bibinfo {author} {\bibfnamefont {F.}~\bibnamefont
  {Abusaif}}, \bibinfo {author} {\bibfnamefont {A.}~\bibnamefont {Aggarwal}},
  \bibinfo {author} {\bibfnamefont {A.}~\bibnamefont {Aksentev}}, \bibinfo
  {author} {\bibfnamefont {B.}~\bibnamefont {Alberdi}}, \bibinfo {author}
  {\bibfnamefont {L.}~\bibnamefont {Barion}}, \bibinfo {author} {\bibfnamefont
  {I.}~\bibnamefont {Bekman}}, \bibinfo {author} {\bibfnamefont
  {M.}~\bibnamefont {Bey\ss{}}}, \bibinfo {author} {\bibfnamefont
  {C.}~\bibnamefont {B\"ohme}}, \bibinfo {author} {\bibfnamefont
  {B.}~\bibnamefont {Breitkreutz}}, \bibinfo {author} {\bibfnamefont
  {N.}~\bibnamefont {Canale}}, \bibinfo {author} {\bibfnamefont
  {G.}~\bibnamefont {Ciullo}}, \bibinfo {author} {\bibfnamefont
  {S.}~\bibnamefont {Dymov}}, \bibinfo {author} {\bibfnamefont {N.-O.}\
  \bibnamefont {Fr\"ohlich}}, \bibinfo {author} {\bibfnamefont
  {R.}~\bibnamefont {Gebel}}, \bibinfo {author} {\bibfnamefont
  {M.}~\bibnamefont {Gaisser}}, \bibinfo {author} {\bibfnamefont
  {K.}~\bibnamefont {Grigoryev}}, \bibinfo {author} {\bibfnamefont
  {D.}~\bibnamefont {Grzonka}}, \bibinfo {author} {\bibfnamefont
  {D.}~\bibnamefont {Gu}}, \bibinfo {author} {\bibfnamefont {D.}~\bibnamefont
  {Heberling}}, \bibinfo {author} {\bibfnamefont {J.}~\bibnamefont {Hetzel}},
  \bibinfo {author} {\bibfnamefont {D.}~\bibnamefont {H\"olscher}}, \bibinfo
  {author} {\bibfnamefont {O.}~\bibnamefont {Javakhishvili}}, \bibinfo {author}
  {\bibfnamefont {A.}~\bibnamefont {Kacharava}}, \bibinfo {author}
  {\bibfnamefont {V.}~\bibnamefont {Kamerdzhiev}}, \bibinfo {author}
  {\bibfnamefont {S.}~\bibnamefont {Karanth}}, \bibinfo {author} {\bibfnamefont
  {I.}~\bibnamefont {Keshelashvili}}, \bibinfo {author} {\bibfnamefont
  {A.}~\bibnamefont {Kononov}}, \bibinfo {author} {\bibfnamefont
  {K.}~\bibnamefont {Laihem}}, \bibinfo {author} {\bibfnamefont
  {A.}~\bibnamefont {Lehrach}}, \bibinfo {author} {\bibfnamefont
  {P.}~\bibnamefont {Lenisa}}, \bibinfo {author} {\bibfnamefont
  {N.}~\bibnamefont {Lomidze}}, \bibinfo {author} {\bibfnamefont
  {B.}~\bibnamefont {Lorentz}}, \bibinfo {author} {\bibfnamefont
  {G.}~\bibnamefont {Macharashvili}}, \bibinfo {author} {\bibfnamefont
  {A.}~\bibnamefont {Magiera}}, \bibinfo {author} {\bibfnamefont
  {M.}~\bibnamefont {Margos}}, \bibinfo {author} {\bibfnamefont
  {D.}~\bibnamefont {Mchedlishvili}}, \bibinfo {author} {\bibfnamefont
  {A.}~\bibnamefont {Melnikov}}, \bibinfo {author} {\bibfnamefont
  {F.}~\bibnamefont {M\"uller}}, \bibinfo {author} {\bibfnamefont
  {D.}~\bibnamefont {Okropiridze}}, \bibinfo {author} {\bibfnamefont
  {A.}~\bibnamefont {Pesce}}, \bibinfo {author} {\bibfnamefont
  {A.}~\bibnamefont {Piccoli}}, \bibinfo {author} {\bibfnamefont
  {V.}~\bibnamefont {Poncza}}, \bibinfo {author} {\bibfnamefont
  {D.}~\bibnamefont {Prasuhn}}, \bibinfo {author} {\bibfnamefont
  {A.}~\bibnamefont {Saleev}}, \bibinfo {author} {\bibfnamefont
  {D.}~\bibnamefont {Shergelashvili}}, \bibinfo {author} {\bibfnamefont
  {R.}~\bibnamefont {Shankar}}, \bibinfo {author} {\bibfnamefont
  {N.}~\bibnamefont {Shurkhno}}, \bibinfo {author} {\bibfnamefont
  {S.}~\bibnamefont {Siddique}}, \bibinfo {author} {\bibfnamefont
  {A.}~\bibnamefont {Silenko}}, \bibinfo {author} {\bibfnamefont
  {H.}~\bibnamefont {Soltner}}, \bibinfo {author} {\bibfnamefont
  {R.}~\bibnamefont {Stassen}}, \bibinfo {author} {\bibfnamefont {E.~J.}\
  \bibnamefont {Stephenson}}, \bibinfo {author} {\bibfnamefont
  {H.}~\bibnamefont {Str\"oher}}, \bibinfo {author} {\bibfnamefont
  {M.}~\bibnamefont {Tabidze}}, \bibinfo {author} {\bibfnamefont
  {G.}~\bibnamefont {Tagliente}}, \bibinfo {author} {\bibfnamefont
  {V.}~\bibnamefont {Tempel}}, \bibinfo {author} {\bibfnamefont
  {Y.}~\bibnamefont {Valdau}}, \bibinfo {author} {\bibfnamefont
  {M.}~\bibnamefont {Vitz}}, \bibinfo {author} {\bibfnamefont {T.}~\bibnamefont
  {Wagner}}, \bibinfo {author} {\bibfnamefont {A.}~\bibnamefont {Wirzba}},
  \bibinfo {author} {\bibfnamefont {A.}~\bibnamefont
  {Wro\ifmmode~\acute{n}\else \'{n}\fi{}ska}}, \bibinfo {author} {\bibfnamefont
  {P.}~\bibnamefont {W\"ustner}},\ and\ \bibinfo {author} {\bibfnamefont
  {M.}~\bibnamefont {\ifmmode~\dot{Z}\else \.{Z}\fi{}urek}} (\bibinfo
  {collaboration} {JEDI Collaboration}),\ }\href
  {https://doi.org/10.1103/ns3s-ld4k} {\bibfield  {journal} {\bibinfo
  {journal} {Phys. Rev. Lett.}\ }\textbf {\bibinfo {volume} {136}},\ \bibinfo
  {pages} {241801} (\bibinfo {year} {2026})}\BibitemShut {NoStop}%
\bibitem [{\citenamefont {Sitnik}\ \emph {et~al.}(2002)\citenamefont {Sitnik},
  \citenamefont {Volkov}, \citenamefont {Kirillov}, \citenamefont {Piskunov},\
  and\ \citenamefont {Plis}}]{Sitnik}%
  \BibitemOpen
  \bibfield  {author} {\bibinfo {author} {\bibfnamefont {I.~M.}\ \bibnamefont
  {Sitnik}}, \bibinfo {author} {\bibfnamefont {V.~I.}\ \bibnamefont {Volkov}},
  \bibinfo {author} {\bibfnamefont {D.~A.}\ \bibnamefont {Kirillov}}, \bibinfo
  {author} {\bibfnamefont {N.~M.}\ \bibnamefont {Piskunov}},\ and\ \bibinfo
  {author} {\bibfnamefont {Y.~A.}\ \bibnamefont {Plis}},\ }\href@noop {}
  {\bibfield  {journal} {\bibinfo  {journal} {PEPAN Lett.}\ }\textbf {\bibinfo
  {volume} {2}},\ \bibinfo {pages} {22} (\bibinfo {year} {2002})}\BibitemShut
  {NoStop}%
\bibitem [{\citenamefont {Milstein}\ \emph {et~al.}(2020)\citenamefont
  {Milstein}, \citenamefont {Nikolaev},\ and\ \citenamefont
  {Salnikov}}]{milstein2020parity}%
  \BibitemOpen
  \bibfield  {author} {\bibinfo {author} {\bibfnamefont {A.}~\bibnamefont
  {Milstein}}, \bibinfo {author} {\bibfnamefont {N.}~\bibnamefont {Nikolaev}},\
  and\ \bibinfo {author} {\bibfnamefont {S.}~\bibnamefont {Salnikov}},\
  }\href@noop {} {\bibfield  {journal} {\bibinfo  {journal} {JETP Letters}\
  }\textbf {\bibinfo {volume} {112}},\ \bibinfo {pages} {332} (\bibinfo {year}
  {2020})}\BibitemShut {NoStop}%
\bibitem [{\citenamefont {Koop}\ \emph {et~al.}(2021)\citenamefont {Koop},
  \citenamefont {Milstein}, \citenamefont {Nikolaev}, \citenamefont {Popov},
  \citenamefont {Salnikov}, \citenamefont {Shatunov},\ and\ \citenamefont
  {Shatunov}}]{Koop2021tests}%
  \BibitemOpen
  \bibfield  {author} {\bibinfo {author} {\bibfnamefont {I.}~\bibnamefont
  {Koop}}, \bibinfo {author} {\bibfnamefont {A.}~\bibnamefont {Milstein}},
  \bibinfo {author} {\bibfnamefont {N.}~\bibnamefont {Nikolaev}}, \bibinfo
  {author} {\bibfnamefont {A.}~\bibnamefont {Popov}}, \bibinfo {author}
  {\bibfnamefont {S.}~\bibnamefont {Salnikov}}, \bibinfo {author}
  {\bibfnamefont {P.~Y.}\ \bibnamefont {Shatunov}},\ and\ \bibinfo {author}
  {\bibfnamefont {Y.~M.}\ \bibnamefont {Shatunov}},\ }\href
  {https://doi.org/10.1134/S1063779621040365} {\bibfield  {journal} {\bibinfo
  {journal} {Physics of Particles and Nuclei}\ }\textbf {\bibinfo {volume}
  {52}},\ \bibinfo {pages} {549} (\bibinfo {year} {2021})}\BibitemShut
  {NoStop}%
\bibitem [{\citenamefont {Nikolaev}(2022)}]{NikolaevPTR}%
  \BibitemOpen
  \bibfield  {author} {\bibinfo {author} {\bibfnamefont {N.~N.}\ \bibnamefont
  {Nikolaev}},\ }\href {https://doi.org/10.1134/S0021364022600653} {\bibfield
  {journal} {\bibinfo  {journal} {JETP Letters}\ }\textbf {\bibinfo {volume}
  {115}},\ \bibinfo {pages} {639} (\bibinfo {year} {2022})}\BibitemShut
  {NoStop}%
\bibitem [{\citenamefont {Silenko}(2022)}]{Silenko:2021qgc}%
  \BibitemOpen
  \bibfield  {author} {\bibinfo {author} {\bibfnamefont {A.~J.}\ \bibnamefont
  {Silenko}},\ }\href {https://doi.org/10.1140/epjc/s10052-022-10827-7}
  {\bibfield  {journal} {\bibinfo  {journal} {Eur. Phys. J. C}\ }\textbf
  {\bibinfo {volume} {82}},\ \bibinfo {pages} {856} (\bibinfo {year} {2022})},\
  \Eprint {https://arxiv.org/abs/2109.05576} {arXiv:2109.05576 [hep-th]}
  \BibitemShut {NoStop}%
\bibitem [{\citenamefont {Silenko}(2015)}]{SilenkoTensor2015}%
  \BibitemOpen
  \bibfield  {author} {\bibinfo {author} {\bibfnamefont {A.~J.}\ \bibnamefont
  {Silenko}},\ }\href {https://doi.org/10.1088/0954-3899/42/7/075109}
  {\bibfield  {journal} {\bibinfo  {journal} {J. Phys. G}\ }\textbf {\bibinfo
  {volume} {42}},\ \bibinfo {pages} {075109} (\bibinfo {year} {2015})},\
  \Eprint {https://arxiv.org/abs/1503.03005} {arXiv:1503.03005 [hep-ph]}
  \BibitemShut {NoStop}%
\bibitem [{\citenamefont {Kolokolchikov}\ \emph {et~al.}(2023)\citenamefont
  {Kolokolchikov}, \citenamefont {Aksentyev}, \citenamefont {Melnikov},\ and\
  \citenamefont {Senichev}}]{Chromaticity_Kolokolchikov:2023}%
  \BibitemOpen
  \bibfield  {author} {\bibinfo {author} {\bibfnamefont {S.}~\bibnamefont
  {Kolokolchikov}}, \bibinfo {author} {\bibfnamefont {A.}~\bibnamefont
  {Aksentyev}}, \bibinfo {author} {\bibfnamefont {A.}~\bibnamefont
  {Melnikov}},\ and\ \bibinfo {author} {\bibfnamefont {Y.}~\bibnamefont
  {Senichev}},\ }\href {https://doi.org/10.1088/1742-6596/2687/2/022027}
  {\bibfield  {journal} {\bibinfo  {journal} {JACoW}\ }\textbf {\bibinfo
  {volume} {IPAC2023}},\ \bibinfo {pages} {MOPA070} (\bibinfo {year}
  {2023})}\BibitemShut {NoStop}%
\bibitem [{\citenamefont {Weidemann}\ \emph {et~al.}(2015)\citenamefont
  {Weidemann}, \citenamefont {Rathmann}, \citenamefont {Stein}, \citenamefont
  {Lorentz}, \citenamefont {Bagdasarian}, \citenamefont {Barion}, \citenamefont
  {Barsov}, \citenamefont {Bechstedt}, \citenamefont {Bertelli}, \citenamefont
  {Chiladze}, \citenamefont {Ciullo}, \citenamefont {Contalbrigo},
  \citenamefont {Dymov}, \citenamefont {Engels}, \citenamefont {Gaisser},
  \citenamefont {Gebel}, \citenamefont {Goslawski}, \citenamefont {Grigoriev},
  \citenamefont {Guidoboni}, \citenamefont {Kacharava}, \citenamefont
  {Kamerdzhiev}, \citenamefont {Khoukaz}, \citenamefont {Kulikov},
  \citenamefont {Lehrach}, \citenamefont {Lenisa}, \citenamefont {Lomidze},
  \citenamefont {Macharashvili}, \citenamefont {Maier}, \citenamefont {Martin},
  \citenamefont {Mchedlishvili}, \citenamefont {Meyer}, \citenamefont
  {Merzliakov}, \citenamefont {Mielke}, \citenamefont {Mikirtychiants},
  \citenamefont {Mikirtychiants}, \citenamefont {Nass}, \citenamefont
  {Nikolaev}, \citenamefont {Oellers}, \citenamefont {Papenbrock},
  \citenamefont {Pesce}, \citenamefont {Prasuhn}, \citenamefont {Retzlaff},
  \citenamefont {Schleichert}, \citenamefont {Schr\"oer}, \citenamefont
  {Seyfarth}, \citenamefont {Soltner}, \citenamefont {Statera}, \citenamefont
  {Steffens}, \citenamefont {Stockhorst}, \citenamefont {Str\"oher},
  \citenamefont {Tabidze}, \citenamefont {Tagliente}, \citenamefont {Engblom},
  \citenamefont {Trusov}, \citenamefont {Valdau}, \citenamefont {Vasiliev},\
  and\ \citenamefont {W\"ustner}}]{MachineDevelopmentSF2}%
  \BibitemOpen
  \bibfield  {author} {\bibinfo {author} {\bibfnamefont {C.}~\bibnamefont
  {Weidemann}}, \bibinfo {author} {\bibfnamefont {F.}~\bibnamefont {Rathmann}},
  \bibinfo {author} {\bibfnamefont {H.~J.}\ \bibnamefont {Stein}}, \bibinfo
  {author} {\bibfnamefont {B.}~\bibnamefont {Lorentz}}, \bibinfo {author}
  {\bibfnamefont {Z.}~\bibnamefont {Bagdasarian}}, \bibinfo {author}
  {\bibfnamefont {L.}~\bibnamefont {Barion}}, \bibinfo {author} {\bibfnamefont
  {S.}~\bibnamefont {Barsov}}, \bibinfo {author} {\bibfnamefont
  {U.}~\bibnamefont {Bechstedt}}, \bibinfo {author} {\bibfnamefont
  {S.}~\bibnamefont {Bertelli}}, \bibinfo {author} {\bibfnamefont
  {D.}~\bibnamefont {Chiladze}}, \bibinfo {author} {\bibfnamefont
  {G.}~\bibnamefont {Ciullo}}, \bibinfo {author} {\bibfnamefont
  {M.}~\bibnamefont {Contalbrigo}}, \bibinfo {author} {\bibfnamefont
  {S.}~\bibnamefont {Dymov}}, \bibinfo {author} {\bibfnamefont
  {R.}~\bibnamefont {Engels}}, \bibinfo {author} {\bibfnamefont
  {M.}~\bibnamefont {Gaisser}}, \bibinfo {author} {\bibfnamefont
  {R.}~\bibnamefont {Gebel}}, \bibinfo {author} {\bibfnamefont
  {P.}~\bibnamefont {Goslawski}}, \bibinfo {author} {\bibfnamefont
  {K.}~\bibnamefont {Grigoriev}}, \bibinfo {author} {\bibfnamefont
  {G.}~\bibnamefont {Guidoboni}}, \bibinfo {author} {\bibfnamefont
  {A.}~\bibnamefont {Kacharava}}, \bibinfo {author} {\bibfnamefont
  {V.}~\bibnamefont {Kamerdzhiev}}, \bibinfo {author} {\bibfnamefont
  {A.}~\bibnamefont {Khoukaz}}, \bibinfo {author} {\bibfnamefont
  {A.}~\bibnamefont {Kulikov}}, \bibinfo {author} {\bibfnamefont
  {A.}~\bibnamefont {Lehrach}}, \bibinfo {author} {\bibfnamefont
  {P.}~\bibnamefont {Lenisa}}, \bibinfo {author} {\bibfnamefont
  {N.}~\bibnamefont {Lomidze}}, \bibinfo {author} {\bibfnamefont
  {G.}~\bibnamefont {Macharashvili}}, \bibinfo {author} {\bibfnamefont
  {R.}~\bibnamefont {Maier}}, \bibinfo {author} {\bibfnamefont
  {S.}~\bibnamefont {Martin}}, \bibinfo {author} {\bibfnamefont
  {D.}~\bibnamefont {Mchedlishvili}}, \bibinfo {author} {\bibfnamefont {H.~O.}\
  \bibnamefont {Meyer}}, \bibinfo {author} {\bibfnamefont {S.}~\bibnamefont
  {Merzliakov}}, \bibinfo {author} {\bibfnamefont {M.}~\bibnamefont {Mielke}},
  \bibinfo {author} {\bibfnamefont {M.}~\bibnamefont {Mikirtychiants}},
  \bibinfo {author} {\bibfnamefont {S.}~\bibnamefont {Mikirtychiants}},
  \bibinfo {author} {\bibfnamefont {A.}~\bibnamefont {Nass}}, \bibinfo {author}
  {\bibfnamefont {N.~N.}\ \bibnamefont {Nikolaev}}, \bibinfo {author}
  {\bibfnamefont {D.}~\bibnamefont {Oellers}}, \bibinfo {author} {\bibfnamefont
  {M.}~\bibnamefont {Papenbrock}}, \bibinfo {author} {\bibfnamefont
  {A.}~\bibnamefont {Pesce}}, \bibinfo {author} {\bibfnamefont
  {D.}~\bibnamefont {Prasuhn}}, \bibinfo {author} {\bibfnamefont
  {M.}~\bibnamefont {Retzlaff}}, \bibinfo {author} {\bibfnamefont
  {R.}~\bibnamefont {Schleichert}}, \bibinfo {author} {\bibfnamefont
  {D.}~\bibnamefont {Schr\"oer}}, \bibinfo {author} {\bibfnamefont
  {H.}~\bibnamefont {Seyfarth}}, \bibinfo {author} {\bibfnamefont
  {H.}~\bibnamefont {Soltner}}, \bibinfo {author} {\bibfnamefont
  {M.}~\bibnamefont {Statera}}, \bibinfo {author} {\bibfnamefont
  {E.}~\bibnamefont {Steffens}}, \bibinfo {author} {\bibfnamefont
  {H.}~\bibnamefont {Stockhorst}}, \bibinfo {author} {\bibfnamefont
  {H.}~\bibnamefont {Str\"oher}}, \bibinfo {author} {\bibfnamefont
  {M.}~\bibnamefont {Tabidze}}, \bibinfo {author} {\bibfnamefont
  {G.}~\bibnamefont {Tagliente}}, \bibinfo {author} {\bibfnamefont {P.~T.}\
  \bibnamefont {Engblom}}, \bibinfo {author} {\bibfnamefont {S.}~\bibnamefont
  {Trusov}}, \bibinfo {author} {\bibfnamefont {Y.}~\bibnamefont {Valdau}},
  \bibinfo {author} {\bibfnamefont {A.}~\bibnamefont {Vasiliev}},\ and\
  \bibinfo {author} {\bibfnamefont {P.}~\bibnamefont {W\"ustner}},\ }\href
  {https://doi.org/10.1103/PhysRevSTAB.18.020101} {\bibfield  {journal}
  {\bibinfo  {journal} {Phys. Rev. ST Accel. Beams}\ }\textbf {\bibinfo
  {volume} {18}},\ \bibinfo {pages} {020101} (\bibinfo {year}
  {2015})}\BibitemShut {NoStop}%
\bibitem [{\citenamefont {Kozlov}\ \emph {et~al.}(2022)\citenamefont {Kozlov},
  \citenamefont {Kostromin}, \citenamefont {Melnikov}, \citenamefont {Meshkov},
  \citenamefont {Smirnov}, \citenamefont {Tuzikov}, \citenamefont {Filippov},\
  and\ \citenamefont {Shandov}}]{NICA_BEAM_MESHKOV}%
  \BibitemOpen
  \bibfield  {author} {\bibinfo {author} {\bibfnamefont {O.~S.}\ \bibnamefont
  {Kozlov}}, \bibinfo {author} {\bibfnamefont {S.~A.}\ \bibnamefont
  {Kostromin}}, \bibinfo {author} {\bibfnamefont {S.~A.}\ \bibnamefont
  {Melnikov}}, \bibinfo {author} {\bibfnamefont {I.~N.}\ \bibnamefont
  {Meshkov}}, \bibinfo {author} {\bibfnamefont {V.~L.}\ \bibnamefont
  {Smirnov}}, \bibinfo {author} {\bibfnamefont {A.~V.}\ \bibnamefont
  {Tuzikov}}, \bibinfo {author} {\bibfnamefont {A.~V.}\ \bibnamefont
  {Filippov}},\ and\ \bibinfo {author} {\bibfnamefont {M.~M.}\ \bibnamefont
  {Shandov}},\ }\href {https://doi.org/10.1134/S1063779622050057} {\bibfield
  {journal} {\bibinfo  {journal} {Phys. Part. Nucl.}\ }\textbf {\bibinfo
  {volume} {53}},\ \bibinfo {pages} {1021} (\bibinfo {year}
  {2022})}\BibitemShut {NoStop}%
\bibitem [{\citenamefont {Anderle}\ \emph {et~al.}(2021)\citenamefont
  {Anderle}, \citenamefont {Bertone}, \citenamefont {Cao}, \citenamefont
  {Chang}, \citenamefont {Chang}, \citenamefont {Chen}, \citenamefont {Chen},
  \citenamefont {Chen}, \citenamefont {Cui}, \citenamefont {Dai}, \citenamefont
  {Deng}, \citenamefont {Ding}, \citenamefont {Feng}, \citenamefont {Gong},
  \citenamefont {Gui}, \citenamefont {Guo}, \citenamefont {Han}, \citenamefont
  {He}, \citenamefont {Hou}, \citenamefont {Huang}, \citenamefont {Huang},
  \citenamefont {KumeričKi}, \citenamefont {Kaptari}, \citenamefont {Li},
  \citenamefont {Li}, \citenamefont {Li}, \citenamefont {Li}, \citenamefont
  {Liang}, \citenamefont {Liang}, \citenamefont {Liu}, \citenamefont {Liu},
  \citenamefont {Liu}, \citenamefont {Liu}, \citenamefont {Liu}, \citenamefont
  {Liu}, \citenamefont {Liu}, \citenamefont {Luo}, \citenamefont {Lyu},
  \citenamefont {Ma}, \citenamefont {Ma}, \citenamefont {Ma}, \citenamefont
  {Ma}, \citenamefont {Mao}, \citenamefont {Mezrag}, \citenamefont {Moutarde},
  \citenamefont {Ping}, \citenamefont {Qin}, \citenamefont {Ren}, \citenamefont
  {Roberts}, \citenamefont {Rojo}, \citenamefont {Shen}, \citenamefont {Shi},
  \citenamefont {Song}, \citenamefont {Sun}, \citenamefont {Sznajder},
  \citenamefont {Wang}, \citenamefont {Wang}, \citenamefont {Wang},
  \citenamefont {Wang}, \citenamefont {Wang}, \citenamefont {Wang},
  \citenamefont {Wang}, \citenamefont {Wang}, \citenamefont {Wang},
  \citenamefont {Wu}, \citenamefont {Wu}, \citenamefont {Xia}, \citenamefont
  {Xiao}, \citenamefont {Xiao}, \citenamefont {Xie}, \citenamefont {Xie},
  \citenamefont {Xing}, \citenamefont {Xu}, \citenamefont {Xu}, \citenamefont
  {Xu}, \citenamefont {Yan}, \citenamefont {Yan}, \citenamefont {Yan},
  \citenamefont {Yan}, \citenamefont {Yang}, \citenamefont {Yang},
  \citenamefont {Yang}, \citenamefont {Yao}, \citenamefont {Ye}, \citenamefont
  {Yin}, \citenamefont {Yuan}, \citenamefont {Zhan}, \citenamefont {Zhang},
  \citenamefont {Zhang}, \citenamefont {Zhang}, \citenamefont {Zhang},
  \citenamefont {Chang}, \citenamefont {Zhang}, \citenamefont {Zhao},
  \citenamefont {Chao}, \citenamefont {Zhao}, \citenamefont {Zhao},
  \citenamefont {Zhao}, \citenamefont {Zheng}, \citenamefont {Zhou},
  \citenamefont {Zhou}, \citenamefont {Zhou}, \citenamefont {Zou},\ and\
  \citenamefont {Zou}}]{EicC_wp}%
  \BibitemOpen
  \bibfield  {author} {\bibinfo {author} {\bibfnamefont {D.~P.}\ \bibnamefont
  {Anderle}}, \bibinfo {author} {\bibfnamefont {V.}~\bibnamefont {Bertone}},
  \bibinfo {author} {\bibfnamefont {X.}~\bibnamefont {Cao}}, \bibinfo {author}
  {\bibfnamefont {L.}~\bibnamefont {Chang}}, \bibinfo {author} {\bibfnamefont
  {N.}~\bibnamefont {Chang}}, \bibinfo {author} {\bibfnamefont
  {G.}~\bibnamefont {Chen}}, \bibinfo {author} {\bibfnamefont {X.}~\bibnamefont
  {Chen}}, \bibinfo {author} {\bibfnamefont {Z.}~\bibnamefont {Chen}}, \bibinfo
  {author} {\bibfnamefont {Z.}~\bibnamefont {Cui}}, \bibinfo {author}
  {\bibfnamefont {L.}~\bibnamefont {Dai}}, \bibinfo {author} {\bibfnamefont
  {W.}~\bibnamefont {Deng}}, \bibinfo {author} {\bibfnamefont {M.}~\bibnamefont
  {Ding}}, \bibinfo {author} {\bibfnamefont {X.}~\bibnamefont {Feng}}, \bibinfo
  {author} {\bibfnamefont {C.}~\bibnamefont {Gong}}, \bibinfo {author}
  {\bibfnamefont {L.}~\bibnamefont {Gui}}, \bibinfo {author} {\bibfnamefont
  {F.-K.}\ \bibnamefont {Guo}}, \bibinfo {author} {\bibfnamefont
  {C.}~\bibnamefont {Han}}, \bibinfo {author} {\bibfnamefont {J.}~\bibnamefont
  {He}}, \bibinfo {author} {\bibfnamefont {T.-J.}\ \bibnamefont {Hou}},
  \bibinfo {author} {\bibfnamefont {H.}~\bibnamefont {Huang}}, \bibinfo
  {author} {\bibfnamefont {Y.}~\bibnamefont {Huang}}, \bibinfo {author}
  {\bibfnamefont {K.}~\bibnamefont {KumeričKi}}, \bibinfo {author}
  {\bibfnamefont {L.~P.}\ \bibnamefont {Kaptari}}, \bibinfo {author}
  {\bibfnamefont {D.}~\bibnamefont {Li}}, \bibinfo {author} {\bibfnamefont
  {H.}~\bibnamefont {Li}}, \bibinfo {author} {\bibfnamefont {M.}~\bibnamefont
  {Li}}, \bibinfo {author} {\bibfnamefont {X.}~\bibnamefont {Li}}, \bibinfo
  {author} {\bibfnamefont {Y.}~\bibnamefont {Liang}}, \bibinfo {author}
  {\bibfnamefont {Z.}~\bibnamefont {Liang}}, \bibinfo {author} {\bibfnamefont
  {C.}~\bibnamefont {Liu}}, \bibinfo {author} {\bibfnamefont {C.}~\bibnamefont
  {Liu}}, \bibinfo {author} {\bibfnamefont {G.}~\bibnamefont {Liu}}, \bibinfo
  {author} {\bibfnamefont {J.}~\bibnamefont {Liu}}, \bibinfo {author}
  {\bibfnamefont {L.}~\bibnamefont {Liu}}, \bibinfo {author} {\bibfnamefont
  {X.}~\bibnamefont {Liu}}, \bibinfo {author} {\bibfnamefont {T.}~\bibnamefont
  {Liu}}, \bibinfo {author} {\bibfnamefont {X.}~\bibnamefont {Luo}}, \bibinfo
  {author} {\bibfnamefont {Z.}~\bibnamefont {Lyu}}, \bibinfo {author}
  {\bibfnamefont {B.}~\bibnamefont {Ma}}, \bibinfo {author} {\bibfnamefont
  {F.}~\bibnamefont {Ma}}, \bibinfo {author} {\bibfnamefont {J.}~\bibnamefont
  {Ma}}, \bibinfo {author} {\bibfnamefont {Y.}~\bibnamefont {Ma}}, \bibinfo
  {author} {\bibfnamefont {L.}~\bibnamefont {Mao}}, \bibinfo {author}
  {\bibfnamefont {C.}~\bibnamefont {Mezrag}}, \bibinfo {author} {\bibfnamefont
  {H.}~\bibnamefont {Moutarde}}, \bibinfo {author} {\bibfnamefont
  {J.}~\bibnamefont {Ping}}, \bibinfo {author} {\bibfnamefont {S.}~\bibnamefont
  {Qin}}, \bibinfo {author} {\bibfnamefont {H.}~\bibnamefont {Ren}}, \bibinfo
  {author} {\bibfnamefont {C.~D.}\ \bibnamefont {Roberts}}, \bibinfo {author}
  {\bibfnamefont {J.}~\bibnamefont {Rojo}}, \bibinfo {author} {\bibfnamefont
  {G.}~\bibnamefont {Shen}}, \bibinfo {author} {\bibfnamefont {C.}~\bibnamefont
  {Shi}}, \bibinfo {author} {\bibfnamefont {Q.}~\bibnamefont {Song}}, \bibinfo
  {author} {\bibfnamefont {H.}~\bibnamefont {Sun}}, \bibinfo {author}
  {\bibfnamefont {P.}~\bibnamefont {Sznajder}}, \bibinfo {author}
  {\bibfnamefont {E.}~\bibnamefont {Wang}}, \bibinfo {author} {\bibfnamefont
  {F.}~\bibnamefont {Wang}}, \bibinfo {author} {\bibfnamefont {Q.}~\bibnamefont
  {Wang}}, \bibinfo {author} {\bibfnamefont {R.}~\bibnamefont {Wang}}, \bibinfo
  {author} {\bibfnamefont {R.}~\bibnamefont {Wang}}, \bibinfo {author}
  {\bibfnamefont {T.}~\bibnamefont {Wang}}, \bibinfo {author} {\bibfnamefont
  {W.}~\bibnamefont {Wang}}, \bibinfo {author} {\bibfnamefont {X.}~\bibnamefont
  {Wang}}, \bibinfo {author} {\bibfnamefont {X.}~\bibnamefont {Wang}}, \bibinfo
  {author} {\bibfnamefont {J.}~\bibnamefont {Wu}}, \bibinfo {author}
  {\bibfnamefont {X.}~\bibnamefont {Wu}}, \bibinfo {author} {\bibfnamefont
  {L.}~\bibnamefont {Xia}}, \bibinfo {author} {\bibfnamefont {B.}~\bibnamefont
  {Xiao}}, \bibinfo {author} {\bibfnamefont {G.}~\bibnamefont {Xiao}}, \bibinfo
  {author} {\bibfnamefont {J.-J.}\ \bibnamefont {Xie}}, \bibinfo {author}
  {\bibfnamefont {Y.}~\bibnamefont {Xie}}, \bibinfo {author} {\bibfnamefont
  {H.}~\bibnamefont {Xing}}, \bibinfo {author} {\bibfnamefont {H.}~\bibnamefont
  {Xu}}, \bibinfo {author} {\bibfnamefont {N.}~\bibnamefont {Xu}}, \bibinfo
  {author} {\bibfnamefont {S.}~\bibnamefont {Xu}}, \bibinfo {author}
  {\bibfnamefont {M.}~\bibnamefont {Yan}}, \bibinfo {author} {\bibfnamefont
  {W.}~\bibnamefont {Yan}}, \bibinfo {author} {\bibfnamefont {W.}~\bibnamefont
  {Yan}}, \bibinfo {author} {\bibfnamefont {X.}~\bibnamefont {Yan}}, \bibinfo
  {author} {\bibfnamefont {J.}~\bibnamefont {Yang}}, \bibinfo {author}
  {\bibfnamefont {Y.-B.}\ \bibnamefont {Yang}}, \bibinfo {author}
  {\bibfnamefont {Z.}~\bibnamefont {Yang}}, \bibinfo {author} {\bibfnamefont
  {D.}~\bibnamefont {Yao}}, \bibinfo {author} {\bibfnamefont {Z.}~\bibnamefont
  {Ye}}, \bibinfo {author} {\bibfnamefont {P.}~\bibnamefont {Yin}}, \bibinfo
  {author} {\bibfnamefont {C.-P.}\ \bibnamefont {Yuan}}, \bibinfo {author}
  {\bibfnamefont {W.}~\bibnamefont {Zhan}}, \bibinfo {author} {\bibfnamefont
  {J.}~\bibnamefont {Zhang}}, \bibinfo {author} {\bibfnamefont
  {J.}~\bibnamefont {Zhang}}, \bibinfo {author} {\bibfnamefont
  {P.}~\bibnamefont {Zhang}}, \bibinfo {author} {\bibfnamefont
  {Y.}~\bibnamefont {Zhang}}, \bibinfo {author} {\bibfnamefont {C.-H.}\
  \bibnamefont {Chang}}, \bibinfo {author} {\bibfnamefont {Z.}~\bibnamefont
  {Zhang}}, \bibinfo {author} {\bibfnamefont {H.}~\bibnamefont {Zhao}},
  \bibinfo {author} {\bibfnamefont {K.-T.}\ \bibnamefont {Chao}}, \bibinfo
  {author} {\bibfnamefont {Q.}~\bibnamefont {Zhao}}, \bibinfo {author}
  {\bibfnamefont {Y.}~\bibnamefont {Zhao}}, \bibinfo {author} {\bibfnamefont
  {Z.}~\bibnamefont {Zhao}}, \bibinfo {author} {\bibfnamefont {L.}~\bibnamefont
  {Zheng}}, \bibinfo {author} {\bibfnamefont {J.}~\bibnamefont {Zhou}},
  \bibinfo {author} {\bibfnamefont {X.}~\bibnamefont {Zhou}}, \bibinfo {author}
  {\bibfnamefont {X.}~\bibnamefont {Zhou}}, \bibinfo {author} {\bibfnamefont
  {B.}~\bibnamefont {Zou}},\ and\ \bibinfo {author} {\bibfnamefont
  {L.}~\bibnamefont {Zou}},\ }\bibfield  {journal} {\bibinfo  {journal}
  {Frontiers of Physics}\ }\textbf {\bibinfo {volume} {16}},\ \href
  {https://doi.org/10.1007/s11467-021-1062-0} {10.1007/s11467-021-1062-0}
  (\bibinfo {year} {2021})\BibitemShut {NoStop}%
\bibitem [{\citenamefont {Froissart}\ and\ \citenamefont
  {Stora}(1960)}]{FroissartScan}%
  \BibitemOpen
  \bibfield  {author} {\bibinfo {author} {\bibfnamefont {M.}~\bibnamefont
  {Froissart}}\ and\ \bibinfo {author} {\bibfnamefont {R.}~\bibnamefont
  {Stora}},\ }\href
  {https://doi.org/https://doi.org/10.1016/0029-554X(60)90033-1} {\bibfield
  {journal} {\bibinfo  {journal} {Nuclear Instruments and Methods}\ }\textbf
  {\bibinfo {volume} {7}},\ \bibinfo {pages} {297} (\bibinfo {year}
  {1960})}\BibitemShut {NoStop}%
\bibitem [{\citenamefont {Franco}\ and\ \citenamefont
  {Glauber}(1969)}]{FrancoGlauber1969}%
  \BibitemOpen
  \bibfield  {author} {\bibinfo {author} {\bibfnamefont {V.}~\bibnamefont
  {Franco}}\ and\ \bibinfo {author} {\bibfnamefont {R.~J.}\ \bibnamefont
  {Glauber}},\ }\href {https://doi.org/10.1103/PhysRevLett.22.370} {\bibfield
  {journal} {\bibinfo  {journal} {Phys. Rev. Lett.}\ }\textbf {\bibinfo
  {volume} {22}},\ \bibinfo {pages} {370} (\bibinfo {year} {1969})}\BibitemShut
  {NoStop}%
\bibitem [{\citenamefont {de~Vries}\ \emph {et~al.}(2020)\citenamefont
  {de~Vries}, \citenamefont {Epelbaum}, \citenamefont {Girlanda}, \citenamefont
  {Gnech}, \citenamefont {Mereghetti},\ and\ \citenamefont
  {Viviani}}]{JordyTVPVreview}%
  \BibitemOpen
  \bibfield  {author} {\bibinfo {author} {\bibfnamefont {J.}~\bibnamefont
  {de~Vries}}, \bibinfo {author} {\bibfnamefont {E.}~\bibnamefont {Epelbaum}},
  \bibinfo {author} {\bibfnamefont {L.}~\bibnamefont {Girlanda}}, \bibinfo
  {author} {\bibfnamefont {A.}~\bibnamefont {Gnech}}, \bibinfo {author}
  {\bibfnamefont {E.}~\bibnamefont {Mereghetti}},\ and\ \bibinfo {author}
  {\bibfnamefont {M.}~\bibnamefont {Viviani}},\ }\href@noop {} {\bibfield
  {journal} {\bibinfo  {journal} {Frontiers in Physics}\ }\textbf {\bibinfo
  {volume} {8}},\ \bibinfo {pages} {218} (\bibinfo {year} {2020})},\ \Eprint
  {https://arxiv.org/abs/2001.09050} {arXiv:2001.09050} \BibitemShut {NoStop}%
\bibitem [{\citenamefont {Barabanov}(1997)}]{Barabanov:1995fa}%
  \BibitemOpen
  \bibfield  {author} {\bibinfo {author} {\bibfnamefont {A.~L.}\ \bibnamefont
  {Barabanov}},\ }\href {https://doi.org/10.1016/S0375-9474(96)00454-X}
  {\bibfield  {journal} {\bibinfo  {journal} {Nucl. Phys. A}\ }\textbf
  {\bibinfo {volume} {614}},\ \bibinfo {pages} {1} (\bibinfo {year} {1997})},\
  \Eprint {https://arxiv.org/abs/nucl-th/9512020} {arXiv:nucl-th/9512020}
  \BibitemShut {NoStop}%
%%CITATION = NUCL-TH/9512020;%%
\bibitem [{\citenamefont {Bunakov}\ and\ \citenamefont
  {Novikov}(1998)}]{Bunakov:1998xh}%
  \BibitemOpen
  \bibfield  {author} {\bibinfo {author} {\bibfnamefont {V.}~\bibnamefont
  {Bunakov}}\ and\ \bibinfo {author} {\bibfnamefont {Y.}~\bibnamefont
  {Novikov}},\ }\href {https://doi.org/10.1016/S0370-2693(98)00462-6}
  {\bibfield  {journal} {\bibinfo  {journal} {Phys. Lett.}\ }\textbf {\bibinfo
  {volume} {B429}},\ \bibinfo {pages} {7} (\bibinfo {year} {1998})},\ \Eprint
  {https://arxiv.org/abs/nucl-th/9811059} {arXiv:nucl-th/9811059} \BibitemShut
  {NoStop}%
%%CITATION = NUCL-TH/9811059;%%
\bibitem [{\citenamefont {Eversheim}\ \emph {et~al.}(2017)\citenamefont
  {Eversheim}, \citenamefont {Valdau},\ and\ \citenamefont
  {Lorentz}}]{EversheimTV}%
  \BibitemOpen
  \bibfield  {author} {\bibinfo {author} {\bibfnamefont {D.}~\bibnamefont
  {Eversheim}}, \bibinfo {author} {\bibfnamefont {Y.}~\bibnamefont {Valdau}},\
  and\ \bibinfo {author} {\bibfnamefont {B.}~\bibnamefont {Lorentz}},\ }\href
  {https://doi.org/10.22323/1.281.0177} {\bibfield  {journal} {\bibinfo
  {journal} {PoS}\ }\textbf {\bibinfo {volume} {INPC2016}},\ \bibinfo {pages}
  {177} (\bibinfo {year} {2017})}\BibitemShut {NoStop}%
\bibitem [{\citenamefont {Aksentyev}\ \emph {et~al.}(2017)\citenamefont
  {Aksentyev}, \citenamefont {Eversheim}, \citenamefont {Lorentz},\ and\
  \citenamefont {Valdau}}]{Aksentyev:2017dnk}%
  \BibitemOpen
  \bibfield  {author} {\bibinfo {author} {\bibfnamefont {A.}~\bibnamefont
  {Aksentyev}}, \bibinfo {author} {\bibfnamefont {D.}~\bibnamefont
  {Eversheim}}, \bibinfo {author} {\bibfnamefont {B.}~\bibnamefont {Lorentz}},\
  and\ \bibinfo {author} {\bibfnamefont {Y.}~\bibnamefont {Valdau}} (\bibinfo
  {collaboration} {PAX}),\ }\bibfield  {booktitle} {\emph {\bibinfo {booktitle}
  {{Proceedings, 2nd Jagiellonian Symposium of Fundamental and Applied
  Subatomic Physics (J-SYMPOSIUM 2017): Krak\'ow, Poland, June 4-9, 2017}}},\
  }\href {https://doi.org/10.5506/APhysPolB.48.1925} {\bibfield  {journal}
  {\bibinfo  {journal} {Acta Phys. Polon. B}\ }\textbf {\bibinfo {volume}
  {48}},\ \bibinfo {pages} {1925} (\bibinfo {year} {2017})}\BibitemShut
  {NoStop}%
%%CITATION = APPOA,B48,1925;%%
\bibitem [{\citenamefont {Rathmann}\ \emph {et~al.}(1998)\citenamefont
  {Rathmann}, \citenamefont {von Przewoski}, \citenamefont {Dezarn},
  \citenamefont {Doskow}, \citenamefont {Dzemidzic}, \citenamefont {Haeberli},
  \citenamefont {Hardie}, \citenamefont {Lorentz}, \citenamefont {Meyer},
  \citenamefont {Pancella}, \citenamefont {Pollock}, \citenamefont {Rinckel},
  \citenamefont {Sperisen},\ and\ \citenamefont
  {Wise}}]{Rathmann-TargetSpinFlip}%
  \BibitemOpen
  \bibfield  {author} {\bibinfo {author} {\bibfnamefont {F.}~\bibnamefont
  {Rathmann}}, \bibinfo {author} {\bibfnamefont {B.}~\bibnamefont {von
  Przewoski}}, \bibinfo {author} {\bibfnamefont {W.~A.}\ \bibnamefont
  {Dezarn}}, \bibinfo {author} {\bibfnamefont {J.}~\bibnamefont {Doskow}},
  \bibinfo {author} {\bibfnamefont {M.}~\bibnamefont {Dzemidzic}}, \bibinfo
  {author} {\bibfnamefont {W.}~\bibnamefont {Haeberli}}, \bibinfo {author}
  {\bibfnamefont {J.~G.}\ \bibnamefont {Hardie}}, \bibinfo {author}
  {\bibfnamefont {B.}~\bibnamefont {Lorentz}}, \bibinfo {author} {\bibfnamefont
  {H.~O.}\ \bibnamefont {Meyer}}, \bibinfo {author} {\bibfnamefont {P.~V.}\
  \bibnamefont {Pancella}}, \bibinfo {author} {\bibfnamefont {R.~E.}\
  \bibnamefont {Pollock}}, \bibinfo {author} {\bibfnamefont {T.}~\bibnamefont
  {Rinckel}}, \bibinfo {author} {\bibfnamefont {F.}~\bibnamefont {Sperisen}},\
  and\ \bibinfo {author} {\bibfnamefont {T.}~\bibnamefont {Wise}},\ }\href
  {https://doi.org/10.1103/PhysRevC.58.658} {\bibfield  {journal} {\bibinfo
  {journal} {Phys. Rev. C}\ }\textbf {\bibinfo {volume} {58}},\ \bibinfo
  {pages} {658} (\bibinfo {year} {1998})}\BibitemShut {NoStop}%
\bibitem [{\citenamefont {Ciullo}\ \emph {et~al.}(2011)\citenamefont {Ciullo},
  \citenamefont {Barion}, \citenamefont {Barschel}, \citenamefont {Grigoriev},
  \citenamefont {Lenisa}, \citenamefont {Nass}, \citenamefont {Sarkadi},
  \citenamefont {Statera}, \citenamefont {Steffens},\ and\ \citenamefont
  {Tagliente}}]{PAXtarget2}%
  \BibitemOpen
  \bibfield  {author} {\bibinfo {author} {\bibfnamefont {G.}~\bibnamefont
  {Ciullo}}, \bibinfo {author} {\bibfnamefont {L.}~\bibnamefont {Barion}},
  \bibinfo {author} {\bibfnamefont {C.}~\bibnamefont {Barschel}}, \bibinfo
  {author} {\bibfnamefont {K.}~\bibnamefont {Grigoriev}}, \bibinfo {author}
  {\bibfnamefont {P.}~\bibnamefont {Lenisa}}, \bibinfo {author} {\bibfnamefont
  {A.}~\bibnamefont {Nass}}, \bibinfo {author} {\bibfnamefont {J.}~\bibnamefont
  {Sarkadi}}, \bibinfo {author} {\bibfnamefont {M.}~\bibnamefont {Statera}},
  \bibinfo {author} {\bibfnamefont {E.}~\bibnamefont {Steffens}},\ and\
  \bibinfo {author} {\bibfnamefont {G.}~\bibnamefont {Tagliente}},\ }\bibfield
  {booktitle} {\emph {\bibinfo {booktitle} {{Spin physics. Proceedings, 19th
  International Symposium, SPIN 2010, Juelich, Germany, September 27-October 2,
  2010}}},\ }\href {https://doi.org/10.1088/1742-6596/295/1/012150} {\bibfield
  {journal} {\bibinfo  {journal} {J. Phys. Conf. Ser.}\ }\textbf {\bibinfo
  {volume} {295}},\ \bibinfo {pages} {012150} (\bibinfo {year}
  {2011})}\BibitemShut {NoStop}%
%%CITATION = 00462,295,012150;%%
\bibitem [{\citenamefont {Huffman}\ \emph {et~al.}(1997)\citenamefont
  {Huffman}, \citenamefont {Roberson}, \citenamefont {Wilburn}, \citenamefont
  {Gould}, \citenamefont {Haase}, \citenamefont {Keith}, \citenamefont
  {Raichle}, \citenamefont {Seely},\ and\ \citenamefont
  {Walston}}]{Huffman:1996ix}%
  \BibitemOpen
  \bibfield  {author} {\bibinfo {author} {\bibfnamefont {P.~R.}\ \bibnamefont
  {Huffman}}, \bibinfo {author} {\bibfnamefont {N.~R.}\ \bibnamefont
  {Roberson}}, \bibinfo {author} {\bibfnamefont {W.~S.}\ \bibnamefont
  {Wilburn}}, \bibinfo {author} {\bibfnamefont {C.~R.}\ \bibnamefont {Gould}},
  \bibinfo {author} {\bibfnamefont {D.~G.}\ \bibnamefont {Haase}}, \bibinfo
  {author} {\bibfnamefont {C.~D.}\ \bibnamefont {Keith}}, \bibinfo {author}
  {\bibfnamefont {B.~W.}\ \bibnamefont {Raichle}}, \bibinfo {author}
  {\bibfnamefont {M.~L.}\ \bibnamefont {Seely}},\ and\ \bibinfo {author}
  {\bibfnamefont {J.~R.}\ \bibnamefont {Walston}},\ }\href
  {https://doi.org/10.1103/PhysRevC.55.2684} {\bibfield  {journal} {\bibinfo
  {journal} {Phys. Rev. C}\ }\textbf {\bibinfo {volume} {55}},\ \bibinfo
  {pages} {2684} (\bibinfo {year} {1997})},\ \Eprint
  {https://arxiv.org/abs/nucl-ex/9605005} {arXiv:nucl-ex/9605005} \BibitemShut
  {NoStop}%
\bibitem [{\citenamefont {Lv}\ and\ \citenamefont {Gou}(2025)}]{IMP_PIT}%
  \BibitemOpen
  \bibfield  {author} {\bibinfo {author} {\bibfnamefont {X.}~\bibnamefont
  {Lv}}\ and\ \bibinfo {author} {\bibfnamefont {B.}~\bibnamefont {Gou}},\ }in\
  \href@noop {} {\emph {\bibinfo {booktitle} {Proceedings of 26th International
  Symposium on Spin Physics -- PoS(SPIN2025)}}}\ (\bibinfo {address} {Tsingtao,
  China},\ \bibinfo {year} {2025})\BibitemShut {NoStop}%
\bibitem [{\citenamefont {Rekalo}\ \emph {et~al.}(1998)\citenamefont {Rekalo},
  \citenamefont {Piskunov},\ and\ \citenamefont {Sitnik}}]{Rekalo:1997fh}%
  \BibitemOpen
  \bibfield  {author} {\bibinfo {author} {\bibfnamefont {M.~P.}\ \bibnamefont
  {Rekalo}}, \bibinfo {author} {\bibfnamefont {N.~M.}\ \bibnamefont
  {Piskunov}},\ and\ \bibinfo {author} {\bibfnamefont {I.~M.}\ \bibnamefont
  {Sitnik}},\ }\href {https://doi.org/10.1007/s006010050070} {\bibfield
  {journal} {\bibinfo  {journal} {Few Body Syst.}\ }\textbf {\bibinfo {volume}
  {23}},\ \bibinfo {pages} {187} (\bibinfo {year} {1998})}\BibitemShut
  {NoStop}%
\bibitem [{\citenamefont {Okun}(1982)}]{okun1963slaboe}%
  \BibitemOpen
  \bibfield  {author} {\bibinfo {author} {\bibfnamefont {L.~B.}\ \bibnamefont
  {Okun}},\ }\href {https://doi.org/10.1142/9162} {\emph {\bibinfo {title}
  {{Leptons and Quarks}: {Special Edition Commemorating the Discovery of the
  Higgs Boson}}}}\ (\bibinfo  {publisher} {North-Holland},\ \bibinfo {address}
  {Amsterdam, Netherlands},\ \bibinfo {year} {1982})\BibitemShut {NoStop}%
\bibitem [{\citenamefont {Overseth}\ and\ \citenamefont
  {Pakvasa}(1969)}]{OversethFSI}%
  \BibitemOpen
  \bibfield  {author} {\bibinfo {author} {\bibfnamefont {O.~E.}\ \bibnamefont
  {Overseth}}\ and\ \bibinfo {author} {\bibfnamefont {S.}~\bibnamefont
  {Pakvasa}},\ }\href {https://doi.org/10.1103/PhysRev.184.1663} {\bibfield
  {journal} {\bibinfo  {journal} {Phys. Rev.}\ }\textbf {\bibinfo {volume}
  {184}},\ \bibinfo {pages} {1663} (\bibinfo {year} {1969})}\BibitemShut
  {NoStop}%
\bibitem [{\citenamefont {Simonius}(1975)}]{SimoniusRhoNN}%
  \BibitemOpen
  \bibfield  {author} {\bibinfo {author} {\bibfnamefont {M.}~\bibnamefont
  {Simonius}},\ }\href {https://doi.org/10.1016/0370-2693(75)90624-3}
  {\bibfield  {journal} {\bibinfo  {journal} {Phys. Lett. B}\ }\textbf
  {\bibinfo {volume} {58}},\ \bibinfo {pages} {147} (\bibinfo {year}
  {1975})}\BibitemShut {NoStop}%
\bibitem [{\citenamefont {Beyer}(1993)}]{Beyer93}%
  \BibitemOpen
  \bibfield  {author} {\bibinfo {author} {\bibfnamefont {M.}~\bibnamefont
  {Beyer}},\ }\href {https://doi.org/10.1016/0375-9474(93)90137-M} {\bibfield
  {journal} {\bibinfo  {journal} {Nucl. Phys. A}\ }\textbf {\bibinfo {volume}
  {560}},\ \bibinfo {pages} {895} (\bibinfo {year} {1993})},\ \Eprint
  {https://arxiv.org/abs/nucl-th/9302002} {arXiv:nucl-th/9302002} \BibitemShut
  {NoStop}%
\bibitem [{\citenamefont {Platonova}\ and\ \citenamefont
  {Kukulin}(2010)}]{Platonova:2010wjt}%
  \BibitemOpen
  \bibfield  {author} {\bibinfo {author} {\bibfnamefont {M.~N.}\ \bibnamefont
  {Platonova}}\ and\ \bibinfo {author} {\bibfnamefont {V.~I.}\ \bibnamefont
  {Kukulin}},\ }\href {https://doi.org/10.1103/PhysRevC.81.014004} {\bibfield
  {journal} {\bibinfo  {journal} {Phys. Rev. C}\ }\textbf {\bibinfo {volume}
  {81}},\ \bibinfo {pages} {014004} (\bibinfo {year} {2010})},\ \bibinfo {note}
  {[Erratum: Phys.Rev.C 94, 069902 (2016)]},\ \Eprint
  {https://arxiv.org/abs/1612.08694} {arXiv:1612.08694 [nucl-th]} \BibitemShut
  {NoStop}%
\bibitem [{\citenamefont {Sorensen}(1979)}]{Sorensen:1978vk}%
  \BibitemOpen
  \bibfield  {author} {\bibinfo {author} {\bibfnamefont {C.}~\bibnamefont
  {Sorensen}},\ }\href {https://doi.org/10.1103/PhysRevD.19.1444} {\bibfield
  {journal} {\bibinfo  {journal} {Phys. Rev. D}\ }\textbf {\bibinfo {volume}
  {19}},\ \bibinfo {pages} {1444} (\bibinfo {year} {1979})}\BibitemShut
  {NoStop}%
\bibitem [{\citenamefont {Kurylov}\ \emph {et~al.}(2001)\citenamefont
  {Kurylov}, \citenamefont {McLaughlin},\ and\ \citenamefont
  {Ramsey-Musolf}}]{Kurylov:2000}%
  \BibitemOpen
  \bibfield  {author} {\bibinfo {author} {\bibfnamefont {A.}~\bibnamefont
  {Kurylov}}, \bibinfo {author} {\bibfnamefont {G.~C.}\ \bibnamefont
  {McLaughlin}},\ and\ \bibinfo {author} {\bibfnamefont {M.~J.}\ \bibnamefont
  {Ramsey-Musolf}},\ }\href {https://doi.org/10.1103/PhysRevD.63.076007}
  {\bibfield  {journal} {\bibinfo  {journal} {Phys. Rev. D}\ }\textbf {\bibinfo
  {volume} {63}},\ \bibinfo {pages} {076007} (\bibinfo {year} {2001})},\
  \Eprint {https://arxiv.org/abs/hep-ph/0011185} {arXiv:hep-ph/0011185}
  \BibitemShut {NoStop}%
\bibitem [{\citenamefont {Uzikov}(2016)}]{Uzikov:2016bja}%
  \BibitemOpen
  \bibfield  {author} {\bibinfo {author} {\bibfnamefont {Y.~N.}\ \bibnamefont
  {Uzikov}},\ }\href {https://doi.org/10.1051/epjconf/201611304027} {\bibfield
  {journal} {\bibinfo  {journal} {EPJ Web Conf.}\ }\textbf {\bibinfo {volume}
  {113}},\ \bibinfo {pages} {04027} (\bibinfo {year} {2016})}\BibitemShut
  {NoStop}%
\bibitem [{\citenamefont {Song}\ \emph {et~al.}(2011)\citenamefont {Song},
  \citenamefont {Lazauskas},\ and\ \citenamefont {Gudkov}}]{Song:2011jh}%
  \BibitemOpen
  \bibfield  {author} {\bibinfo {author} {\bibfnamefont {Y.-H.}\ \bibnamefont
  {Song}}, \bibinfo {author} {\bibfnamefont {R.}~\bibnamefont {Lazauskas}},\
  and\ \bibinfo {author} {\bibfnamefont {V.}~\bibnamefont {Gudkov}},\ }\href
  {https://doi.org/10.1103/PhysRevC.84.025501} {\bibfield  {journal} {\bibinfo
  {journal} {Phys. Rev. C}\ }\textbf {\bibinfo {volume} {84}},\ \bibinfo
  {pages} {025501} (\bibinfo {year} {2011})},\ \bibinfo {note} {[Erratum:
  Phys.Rev.C 93, 049901 (2016)]},\ \Eprint {https://arxiv.org/abs/1105.1327}
  {arXiv:1105.1327 [nucl-th]} \BibitemShut {NoStop}%
\bibitem [{\citenamefont {Uzikov}\ and\ \citenamefont
  {Platonova}(2023)}]{UzikovHe3D}%
  \BibitemOpen
  \bibfield  {author} {\bibinfo {author} {\bibfnamefont {Y.~N.}\ \bibnamefont
  {Uzikov}}\ and\ \bibinfo {author} {\bibfnamefont {M.~N.}\ \bibnamefont
  {Platonova}},\ }\href {https://doi.org/10.1134/S0021364023603044} {\bibfield
  {journal} {\bibinfo  {journal} {JETP Lett.}\ }\textbf {\bibinfo {volume}
  {118}},\ \bibinfo {pages} {785} (\bibinfo {year} {2023})},\ \Eprint
  {https://arxiv.org/abs/2311.10841} {arXiv:2311.10841 [nucl-th]} \BibitemShut
  {NoStop}%
\bibitem [{\citenamefont {Uzikov}\ \emph {et~al.}(2024)\citenamefont {Uzikov},
  \citenamefont {Platonova}, \citenamefont {Klimochkina},\ and\ \citenamefont
  {Kornev}}]{Uzikov_T-odd-DD}%
  \BibitemOpen
  \bibfield  {author} {\bibinfo {author} {\bibfnamefont {Y.}~\bibnamefont
  {Uzikov}}, \bibinfo {author} {\bibfnamefont {M.}~\bibnamefont {Platonova}},
  \bibinfo {author} {\bibfnamefont {A.}~\bibnamefont {Klimochkina}},\ and\
  \bibinfo {author} {\bibfnamefont {A.}~\bibnamefont {Kornev}},\ }\href
  {https://doi.org/10.1142/S0218301324410039} {\bibfield  {journal} {\bibinfo
  {journal} {Int. J. Mod. Phys. E}\ }\textbf {\bibinfo {volume} {33}},\
  \bibinfo {pages} {2441003} (\bibinfo {year} {2024})}\BibitemShut {NoStop}%
\bibitem [{\citenamefont {Kopeliovich}\ and\ \citenamefont
  {Lapidus}(1974)}]{KopLap1974}%
  \BibitemOpen
  \bibfield  {author} {\bibinfo {author} {\bibfnamefont {B.~Z.}\ \bibnamefont
  {Kopeliovich}}\ and\ \bibinfo {author} {\bibfnamefont {L.~I.}\ \bibnamefont
  {Lapidus}},\ }\href@noop {} {\bibfield  {journal} {\bibinfo  {journal} {Sov.
  J. Nucl. Phys}\ }\textbf {\bibinfo {volume} {19}},\ \bibinfo {pages} {114}
  (\bibinfo {year} {1974})}\BibitemShut {NoStop}%
\bibitem [{\citenamefont {Buttimore}\ \emph {et~al.}(1999)\citenamefont
  {Buttimore}, \citenamefont {Kopeliovich}, \citenamefont {Leader},
  \citenamefont {Soffer},\ and\ \citenamefont {Trueman}}]{CNI-Theory}%
  \BibitemOpen
  \bibfield  {author} {\bibinfo {author} {\bibfnamefont {N.~H.}\ \bibnamefont
  {Buttimore}}, \bibinfo {author} {\bibfnamefont {B.~Z.}\ \bibnamefont
  {Kopeliovich}}, \bibinfo {author} {\bibfnamefont {E.}~\bibnamefont {Leader}},
  \bibinfo {author} {\bibfnamefont {J.}~\bibnamefont {Soffer}},\ and\ \bibinfo
  {author} {\bibfnamefont {T.~L.}\ \bibnamefont {Trueman}},\ }\href
  {https://doi.org/10.1103/PhysRevD.59.114010} {\bibfield  {journal} {\bibinfo
  {journal} {Phys. Rev. D}\ }\textbf {\bibinfo {volume} {59}},\ \bibinfo
  {pages} {114010} (\bibinfo {year} {1999})},\ \Eprint
  {https://arxiv.org/abs/hep-ph/9901339} {arXiv:hep-ph/9901339} \BibitemShut
  {NoStop}%
\bibitem [{\citenamefont {Nakagawa}\ \emph {et~al.}(2008)\citenamefont
  {Nakagawa}, \citenamefont {Alekseev}, \citenamefont {Bazilevsky},
  \citenamefont {Bravar}, \citenamefont {Bunce}, \citenamefont {Dhawan},
  \citenamefont {Eyser}, \citenamefont {Gill}, \citenamefont {Haeberli},
  \citenamefont {Huang}, \citenamefont {Makdisi}, \citenamefont {Nass},
  \citenamefont {Okada}, \citenamefont {Stephenson}, \citenamefont {Svinda},
  \citenamefont {Wise}, \citenamefont {Wood},\ and\ \citenamefont
  {Zelenski}}]{CNI-at-RHIC}%
  \BibitemOpen
  \bibfield  {author} {\bibinfo {author} {\bibfnamefont {I.}~\bibnamefont
  {Nakagawa}}, \bibinfo {author} {\bibfnamefont {I.}~\bibnamefont {Alekseev}},
  \bibinfo {author} {\bibfnamefont {A.}~\bibnamefont {Bazilevsky}}, \bibinfo
  {author} {\bibfnamefont {A.}~\bibnamefont {Bravar}}, \bibinfo {author}
  {\bibfnamefont {G.}~\bibnamefont {Bunce}}, \bibinfo {author} {\bibfnamefont
  {S.}~\bibnamefont {Dhawan}}, \bibinfo {author} {\bibfnamefont {K.~O.}\
  \bibnamefont {Eyser}}, \bibinfo {author} {\bibfnamefont {R.}~\bibnamefont
  {Gill}}, \bibinfo {author} {\bibfnamefont {W.}~\bibnamefont {Haeberli}},
  \bibinfo {author} {\bibfnamefont {H.}~\bibnamefont {Huang}}, \bibinfo
  {author} {\bibfnamefont {Y.}~\bibnamefont {Makdisi}}, \bibinfo {author}
  {\bibfnamefont {A.}~\bibnamefont {Nass}}, \bibinfo {author} {\bibfnamefont
  {H.}~\bibnamefont {Okada}}, \bibinfo {author} {\bibfnamefont
  {E.}~\bibnamefont {Stephenson}}, \bibinfo {author} {\bibfnamefont {D.~N.}\
  \bibnamefont {Svinda}}, \bibinfo {author} {\bibfnamefont {T.}~\bibnamefont
  {Wise}}, \bibinfo {author} {\bibfnamefont {J.}~\bibnamefont {Wood}},\ and\
  \bibinfo {author} {\bibfnamefont {A.}~\bibnamefont {Zelenski}},\ }\href
  {https://doi.org/10.1140/epjst/e2008-00801-1} {\bibfield  {journal} {\bibinfo
   {journal} {Eur. Phys. J. ST}\ }\textbf {\bibinfo {volume} {162}},\ \bibinfo
  {pages} {259} (\bibinfo {year} {2008})}\BibitemShut {NoStop}%
\bibitem [{\citenamefont {Varma}\ and\ \citenamefont
  {Franco}(1977)}]{Franco1977dC}%
  \BibitemOpen
  \bibfield  {author} {\bibinfo {author} {\bibfnamefont {G.~K.}\ \bibnamefont
  {Varma}}\ and\ \bibinfo {author} {\bibfnamefont {V.}~\bibnamefont {Franco}},\
  }\href {https://doi.org/10.1103/PhysRevC.15.813} {\bibfield  {journal}
  {\bibinfo  {journal} {Phys. Rev. C}\ }\textbf {\bibinfo {volume} {15}},\
  \bibinfo {pages} {813} (\bibinfo {year} {1977})}\BibitemShut {NoStop}%
\bibitem [{\citenamefont {Seyfarth}\ \emph {et~al.}(2010)\citenamefont
  {Seyfarth}, \citenamefont {Engels}, \citenamefont {Rathmann}, \citenamefont
  {Str\"oher}, \citenamefont {Baryshevsky}, \citenamefont {Rouba},
  \citenamefont {D\"uweke}, \citenamefont {Emmerich}, \citenamefont {Imig},
  \citenamefont {Grigoryev}, \citenamefont {Mikirtychiants},\ and\
  \citenamefont {Vasilyev}}]{TensorCarbon}%
  \BibitemOpen
  \bibfield  {author} {\bibinfo {author} {\bibfnamefont {H.}~\bibnamefont
  {Seyfarth}}, \bibinfo {author} {\bibfnamefont {R.}~\bibnamefont {Engels}},
  \bibinfo {author} {\bibfnamefont {F.}~\bibnamefont {Rathmann}}, \bibinfo
  {author} {\bibfnamefont {H.}~\bibnamefont {Str\"oher}}, \bibinfo {author}
  {\bibfnamefont {V.}~\bibnamefont {Baryshevsky}}, \bibinfo {author}
  {\bibfnamefont {A.}~\bibnamefont {Rouba}}, \bibinfo {author} {\bibfnamefont
  {C.}~\bibnamefont {D\"uweke}}, \bibinfo {author} {\bibfnamefont
  {R.}~\bibnamefont {Emmerich}}, \bibinfo {author} {\bibfnamefont
  {A.}~\bibnamefont {Imig}}, \bibinfo {author} {\bibfnamefont {K.}~\bibnamefont
  {Grigoryev}}, \bibinfo {author} {\bibfnamefont {M.}~\bibnamefont
  {Mikirtychiants}},\ and\ \bibinfo {author} {\bibfnamefont {A.}~\bibnamefont
  {Vasilyev}},\ }\href {https://doi.org/10.1103/PhysRevLett.104.222501}
  {\bibfield  {journal} {\bibinfo  {journal} {Phys. Rev. Lett.}\ }\textbf
  {\bibinfo {volume} {104}},\ \bibinfo {pages} {222501} (\bibinfo {year}
  {2010})}\BibitemShut {NoStop}%
\bibitem [{\citenamefont {Ramsey}\ \emph {et~al.}(1953)\citenamefont {Ramsey},
  \citenamefont {Malenka},\ and\ \citenamefont
  {Kruse}}]{ramsey1953polarizability}%
  \BibitemOpen
  \bibfield  {author} {\bibinfo {author} {\bibfnamefont {N.}~\bibnamefont
  {Ramsey}}, \bibinfo {author} {\bibfnamefont {B.}~\bibnamefont {Malenka}},\
  and\ \bibinfo {author} {\bibfnamefont {U.}~\bibnamefont {Kruse}},\
  }\href@noop {} {\bibfield  {journal} {\bibinfo  {journal} {Physical Review}\
  }\textbf {\bibinfo {volume} {91}},\ \bibinfo {pages} {1162} (\bibinfo {year}
  {1953})}\BibitemShut {NoStop}%
\bibitem [{\citenamefont {Petrunkin}(1981)}]{Petrunkin:1981me}%
  \BibitemOpen
  \bibfield  {author} {\bibinfo {author} {\bibfnamefont {V.~A.}\ \bibnamefont
  {Petrunkin}},\ }\href@noop {} {\bibfield  {journal} {\bibinfo  {journal}
  {Fiz. Elem. Chast. Atom. Yadra}\ }\textbf {\bibinfo {volume} {12}},\ \bibinfo
  {pages} {692} (\bibinfo {year} {1981})}\BibitemShut {NoStop}%
\bibitem [{\citenamefont {Silenko}()}]{Silenko2007Dpol}%
  \BibitemOpen
  \bibfield  {author} {\bibinfo {author} {\bibfnamefont {A.~J.}\ \bibnamefont
  {Silenko}},\ }\href {https://doi.org/10.1103/PhysRevC.77.021001} {\bibfield
  {journal} {\bibinfo  {journal} {Phys. Rev. C}\ }\textbf {\bibinfo {volume}
  {77}},\ \bibinfo {pages} {021001}},\ \Eprint
  {https://arxiv.org/abs/0711.2390} {arXiv:0711.2390 [nucl-th]} \BibitemShut
  {NoStop}%
\bibitem [{\citenamefont {Silenko}(2013)}]{SilenkoPointlike}%
  \BibitemOpen
  \bibfield  {author} {\bibinfo {author} {\bibfnamefont {A.~J.}\ \bibnamefont
  {Silenko}},\ }\href {https://doi.org/10.1103/PhysRevD.87.073015} {\bibfield
  {journal} {\bibinfo  {journal} {Phys. Rev. D}\ }\textbf {\bibinfo {volume}
  {87}},\ \bibinfo {pages} {073015} (\bibinfo {year} {2013})},\ \Eprint
  {https://arxiv.org/abs/1303.6574} {arXiv:1303.6574 [hep-ph]} \BibitemShut
  {NoStop}%
\bibitem [{\citenamefont {Chen}\ \emph {et~al.}(1998)\citenamefont {Chen},
  \citenamefont {Griesshammer}, \citenamefont {Savage},\ and\ \citenamefont
  {Springer}}]{CGS}%
  \BibitemOpen
  \bibfield  {author} {\bibinfo {author} {\bibfnamefont {J.-W.}\ \bibnamefont
  {Chen}}, \bibinfo {author} {\bibfnamefont {H.~W.}\ \bibnamefont
  {Griesshammer}}, \bibinfo {author} {\bibfnamefont {M.~J.}\ \bibnamefont
  {Savage}},\ and\ \bibinfo {author} {\bibfnamefont {R.~P.}\ \bibnamefont
  {Springer}},\ }\href {https://doi.org/10.1016/S0375-9474(98)80012-2}
  {\bibfield  {journal} {\bibinfo  {journal} {Nucl. Phys. A}\ }\textbf
  {\bibinfo {volume} {644}},\ \bibinfo {pages} {221} (\bibinfo {year}
  {1998})},\ \Eprint {https://arxiv.org/abs/nucl-th/9806080}
  {arXiv:nucl-th/9806080} \BibitemShut {NoStop}%
\bibitem [{\citenamefont {Ji}\ and\ \citenamefont {Li}(2004)}]{JL}%
  \BibitemOpen
  \bibfield  {author} {\bibinfo {author} {\bibfnamefont {X.-d.}\ \bibnamefont
  {Ji}}\ and\ \bibinfo {author} {\bibfnamefont {Y.-c.}\ \bibnamefont {Li}},\
  }\href {https://doi.org/10.1016/j.physletb.2004.04.020} {\bibfield  {journal}
  {\bibinfo  {journal} {Phys. Lett. B}\ }\textbf {\bibinfo {volume} {591}},\
  \bibinfo {pages} {76} (\bibinfo {year} {2004})},\ \Eprint
  {https://arxiv.org/abs/nucl-th/0311035} {arXiv:nucl-th/0311035} \BibitemShut
  {NoStop}%
\bibitem [{\citenamefont {Friar}\ and\ \citenamefont {Payne}(2005)}]{FP}%
  \BibitemOpen
  \bibfield  {author} {\bibinfo {author} {\bibfnamefont {J.~L.}\ \bibnamefont
  {Friar}}\ and\ \bibinfo {author} {\bibfnamefont {G.~L.}\ \bibnamefont
  {Payne}},\ }\href {https://doi.org/10.1103/PhysRevC.72.014004} {\bibfield
  {journal} {\bibinfo  {journal} {Phys. Rev. C}\ }\textbf {\bibinfo {volume}
  {72}},\ \bibinfo {pages} {014004} (\bibinfo {year} {2005})},\ \Eprint
  {https://arxiv.org/abs/nucl-th/0503045} {arXiv:nucl-th/0503045} \BibitemShut
  {NoStop}%
\bibitem [{\citenamefont {Silenko}(2007)}]{Silenko:2006polarzability}%
  \BibitemOpen
  \bibfield  {author} {\bibinfo {author} {\bibfnamefont {A.~J.}\ \bibnamefont
  {Silenko}},\ }\href {https://doi.org/10.1103/PhysRevC.75.014003} {\bibfield
  {journal} {\bibinfo  {journal} {Phys. Rev. C}\ }\textbf {\bibinfo {volume}
  {75}},\ \bibinfo {pages} {014003} (\bibinfo {year} {2007})},\ \Eprint
  {https://arxiv.org/abs/nucl-th/0605016} {arXiv:nucl-th/0605016} \BibitemShut
  {NoStop}%
\bibitem [{\citenamefont {Anastassopoulos}\ \emph {et~al.}(2016)\citenamefont
  {Anastassopoulos}, \citenamefont {Andrianov}, \citenamefont {Baartman},
  \citenamefont {Baessler}, \citenamefont {Bai}, \citenamefont {Benante},
  \citenamefont {Berz}, \citenamefont {Blaskiewicz}, \citenamefont {Bowcock},
  \citenamefont {Brown}, \citenamefont {Casey}, \citenamefont {Conte},
  \citenamefont {Crnkovic}, \citenamefont {D’Imperio}, \citenamefont
  {Fanourakis}, \citenamefont {Fedotov}, \citenamefont {Fierlinger},
  \citenamefont {Fischer}, \citenamefont {Gaisser}, \citenamefont {Giomataris},
  \citenamefont {Grosse-Perdekamp}, \citenamefont {Guidoboni}, \citenamefont
  {Hacıömeroğlu}, \citenamefont {Hoffstaetter}, \citenamefont {Huang},
  \citenamefont {Incagli}, \citenamefont {Ivanov}, \citenamefont {Kawall},
  \citenamefont {Kim}, \citenamefont {King}, \citenamefont {Koop},
  \citenamefont {Lazarus}, \citenamefont {Lebedev}, \citenamefont {Lee},
  \citenamefont {Lee}, \citenamefont {Lee}, \citenamefont {Lehrach},
  \citenamefont {Lenisa}, \citenamefont {Sandri}, \citenamefont {Luccio},
  \citenamefont {Lyapin}, \citenamefont {MacKay}, \citenamefont {Maier},
  \citenamefont {Makino}, \citenamefont {Malitsky}, \citenamefont {Marciano},
  \citenamefont {Meng}, \citenamefont {Meot}, \citenamefont {Metodiev},
  \citenamefont {Miceli}, \citenamefont {Moricciani}, \citenamefont {Morse},
  \citenamefont {Nagaitsev}, \citenamefont {Nayak}, \citenamefont {Orlov},
  \citenamefont {Ozben}, \citenamefont {Park}, \citenamefont {Pesce},
  \citenamefont {Petrakou}, \citenamefont {Pile}, \citenamefont {Podobedov},
  \citenamefont {Polychronakos}, \citenamefont {Pretz}, \citenamefont
  {Ptitsyn}, \citenamefont {Ramberg}, \citenamefont {Raparia}, \citenamefont
  {Rathmann}, \citenamefont {Rescia}, \citenamefont {Roser}, \citenamefont
  {Sayed}, \citenamefont {Semertzidis}, \citenamefont {Senichev}, \citenamefont
  {Sidorin}, \citenamefont {Silenko}, \citenamefont {Simos}, \citenamefont
  {Stahl}, \citenamefont {Stephenson}, \citenamefont {Str\"oher}, \citenamefont
  {Syphers}, \citenamefont {Talman}, \citenamefont {Talman}, \citenamefont
  {Tishchenko}, \citenamefont {Touramanis}, \citenamefont {Tsoupas},
  \citenamefont {Venanzoni}, \citenamefont {Vetter}, \citenamefont {Vlassis},
  \citenamefont {Won}, \citenamefont {Zavattini}, \citenamefont {Zelenski},\
  and\ \citenamefont {Zioutas}}]{srEDM}%
  \BibitemOpen
  \bibfield  {author} {\bibinfo {author} {\bibfnamefont {V.}~\bibnamefont
  {Anastassopoulos}}, \bibinfo {author} {\bibfnamefont {S.}~\bibnamefont
  {Andrianov}}, \bibinfo {author} {\bibfnamefont {R.}~\bibnamefont {Baartman}},
  \bibinfo {author} {\bibfnamefont {S.}~\bibnamefont {Baessler}}, \bibinfo
  {author} {\bibfnamefont {M.}~\bibnamefont {Bai}}, \bibinfo {author}
  {\bibfnamefont {J.}~\bibnamefont {Benante}}, \bibinfo {author} {\bibfnamefont
  {M.}~\bibnamefont {Berz}}, \bibinfo {author} {\bibfnamefont {M.}~\bibnamefont
  {Blaskiewicz}}, \bibinfo {author} {\bibfnamefont {T.}~\bibnamefont
  {Bowcock}}, \bibinfo {author} {\bibfnamefont {K.}~\bibnamefont {Brown}},
  \bibinfo {author} {\bibfnamefont {B.}~\bibnamefont {Casey}}, \bibinfo
  {author} {\bibfnamefont {M.}~\bibnamefont {Conte}}, \bibinfo {author}
  {\bibfnamefont {J.~D.}\ \bibnamefont {Crnkovic}}, \bibinfo {author}
  {\bibfnamefont {N.}~\bibnamefont {D’Imperio}}, \bibinfo {author}
  {\bibfnamefont {G.}~\bibnamefont {Fanourakis}}, \bibinfo {author}
  {\bibfnamefont {A.}~\bibnamefont {Fedotov}}, \bibinfo {author} {\bibfnamefont
  {P.}~\bibnamefont {Fierlinger}}, \bibinfo {author} {\bibfnamefont
  {W.}~\bibnamefont {Fischer}}, \bibinfo {author} {\bibfnamefont {M.~O.}\
  \bibnamefont {Gaisser}}, \bibinfo {author} {\bibfnamefont {Y.}~\bibnamefont
  {Giomataris}}, \bibinfo {author} {\bibfnamefont {M.}~\bibnamefont
  {Grosse-Perdekamp}}, \bibinfo {author} {\bibfnamefont {G.}~\bibnamefont
  {Guidoboni}}, \bibinfo {author} {\bibfnamefont {S.}~\bibnamefont
  {Hacıömeroğlu}}, \bibinfo {author} {\bibfnamefont {G.}~\bibnamefont
  {Hoffstaetter}}, \bibinfo {author} {\bibfnamefont {H.}~\bibnamefont {Huang}},
  \bibinfo {author} {\bibfnamefont {M.}~\bibnamefont {Incagli}}, \bibinfo
  {author} {\bibfnamefont {A.}~\bibnamefont {Ivanov}}, \bibinfo {author}
  {\bibfnamefont {D.}~\bibnamefont {Kawall}}, \bibinfo {author} {\bibfnamefont
  {Y.~I.}\ \bibnamefont {Kim}}, \bibinfo {author} {\bibfnamefont
  {B.}~\bibnamefont {King}}, \bibinfo {author} {\bibfnamefont {I.~A.}\
  \bibnamefont {Koop}}, \bibinfo {author} {\bibfnamefont {D.~M.}\ \bibnamefont
  {Lazarus}}, \bibinfo {author} {\bibfnamefont {V.}~\bibnamefont {Lebedev}},
  \bibinfo {author} {\bibfnamefont {M.~J.}\ \bibnamefont {Lee}}, \bibinfo
  {author} {\bibfnamefont {S.}~\bibnamefont {Lee}}, \bibinfo {author}
  {\bibfnamefont {Y.~H.}\ \bibnamefont {Lee}}, \bibinfo {author} {\bibfnamefont
  {A.}~\bibnamefont {Lehrach}}, \bibinfo {author} {\bibfnamefont
  {P.}~\bibnamefont {Lenisa}}, \bibinfo {author} {\bibfnamefont {P.~L.}\
  \bibnamefont {Sandri}}, \bibinfo {author} {\bibfnamefont {A.~U.}\
  \bibnamefont {Luccio}}, \bibinfo {author} {\bibfnamefont {A.}~\bibnamefont
  {Lyapin}}, \bibinfo {author} {\bibfnamefont {W.}~\bibnamefont {MacKay}},
  \bibinfo {author} {\bibfnamefont {R.}~\bibnamefont {Maier}}, \bibinfo
  {author} {\bibfnamefont {K.}~\bibnamefont {Makino}}, \bibinfo {author}
  {\bibfnamefont {N.}~\bibnamefont {Malitsky}}, \bibinfo {author}
  {\bibfnamefont {W.~J.}\ \bibnamefont {Marciano}}, \bibinfo {author}
  {\bibfnamefont {W.}~\bibnamefont {Meng}}, \bibinfo {author} {\bibfnamefont
  {F.}~\bibnamefont {Meot}}, \bibinfo {author} {\bibfnamefont {E.~M.}\
  \bibnamefont {Metodiev}}, \bibinfo {author} {\bibfnamefont {L.}~\bibnamefont
  {Miceli}}, \bibinfo {author} {\bibfnamefont {D.}~\bibnamefont {Moricciani}},
  \bibinfo {author} {\bibfnamefont {W.~M.}\ \bibnamefont {Morse}}, \bibinfo
  {author} {\bibfnamefont {S.}~\bibnamefont {Nagaitsev}}, \bibinfo {author}
  {\bibfnamefont {S.~K.}\ \bibnamefont {Nayak}}, \bibinfo {author}
  {\bibfnamefont {Y.~F.}\ \bibnamefont {Orlov}}, \bibinfo {author}
  {\bibfnamefont {C.~S.}\ \bibnamefont {Ozben}}, \bibinfo {author}
  {\bibfnamefont {S.~T.}\ \bibnamefont {Park}}, \bibinfo {author}
  {\bibfnamefont {A.}~\bibnamefont {Pesce}}, \bibinfo {author} {\bibfnamefont
  {E.}~\bibnamefont {Petrakou}}, \bibinfo {author} {\bibfnamefont
  {P.}~\bibnamefont {Pile}}, \bibinfo {author} {\bibfnamefont {B.}~\bibnamefont
  {Podobedov}}, \bibinfo {author} {\bibfnamefont {V.}~\bibnamefont
  {Polychronakos}}, \bibinfo {author} {\bibfnamefont {J.}~\bibnamefont
  {Pretz}}, \bibinfo {author} {\bibfnamefont {V.}~\bibnamefont {Ptitsyn}},
  \bibinfo {author} {\bibfnamefont {E.}~\bibnamefont {Ramberg}}, \bibinfo
  {author} {\bibfnamefont {D.}~\bibnamefont {Raparia}}, \bibinfo {author}
  {\bibfnamefont {F.}~\bibnamefont {Rathmann}}, \bibinfo {author}
  {\bibfnamefont {S.}~\bibnamefont {Rescia}}, \bibinfo {author} {\bibfnamefont
  {T.}~\bibnamefont {Roser}}, \bibinfo {author} {\bibfnamefont {H.~K.}\
  \bibnamefont {Sayed}}, \bibinfo {author} {\bibfnamefont {Y.~K.}\ \bibnamefont
  {Semertzidis}}, \bibinfo {author} {\bibfnamefont {Y.}~\bibnamefont
  {Senichev}}, \bibinfo {author} {\bibfnamefont {A.}~\bibnamefont {Sidorin}},
  \bibinfo {author} {\bibfnamefont {A.}~\bibnamefont {Silenko}}, \bibinfo
  {author} {\bibfnamefont {N.}~\bibnamefont {Simos}}, \bibinfo {author}
  {\bibfnamefont {A.}~\bibnamefont {Stahl}}, \bibinfo {author} {\bibfnamefont
  {E.~J.}\ \bibnamefont {Stephenson}}, \bibinfo {author} {\bibfnamefont
  {H.}~\bibnamefont {Str\"oher}}, \bibinfo {author} {\bibfnamefont {M.~J.}\
  \bibnamefont {Syphers}}, \bibinfo {author} {\bibfnamefont {J.}~\bibnamefont
  {Talman}}, \bibinfo {author} {\bibfnamefont {R.~M.}\ \bibnamefont {Talman}},
  \bibinfo {author} {\bibfnamefont {V.}~\bibnamefont {Tishchenko}}, \bibinfo
  {author} {\bibfnamefont {C.}~\bibnamefont {Touramanis}}, \bibinfo {author}
  {\bibfnamefont {N.}~\bibnamefont {Tsoupas}}, \bibinfo {author} {\bibfnamefont
  {G.}~\bibnamefont {Venanzoni}}, \bibinfo {author} {\bibfnamefont
  {K.}~\bibnamefont {Vetter}}, \bibinfo {author} {\bibfnamefont
  {S.}~\bibnamefont {Vlassis}}, \bibinfo {author} {\bibfnamefont
  {E.}~\bibnamefont {Won}}, \bibinfo {author} {\bibfnamefont {G.}~\bibnamefont
  {Zavattini}}, \bibinfo {author} {\bibfnamefont {A.}~\bibnamefont
  {Zelenski}},\ and\ \bibinfo {author} {\bibfnamefont {K.}~\bibnamefont
  {Zioutas}},\ }\href {https://doi.org/https://doi.org/10.1063/1.4967465}
  {\bibfield  {journal} {\bibinfo  {journal} {Review of Scientific
  Instruments}\ }\textbf {\bibinfo {volume} {87}},\ \bibinfo {pages} {115116}
  (\bibinfo {year} {2016})}\BibitemShut {NoStop}%
\bibitem [{\citenamefont {Baryshevsky}(1992)}]{BaryshevskyBirefringence}%
  \BibitemOpen
  \bibfield  {author} {\bibinfo {author} {\bibfnamefont {V.}~\bibnamefont
  {Baryshevsky}},\ }\href
  {https://doi.org/https://doi.org/10.1016/0375-9601(92)90672-9} {\bibfield
  {journal} {\bibinfo  {journal} {Physics Letters A}\ }\textbf {\bibinfo
  {volume} {171}},\ \bibinfo {pages} {431 } (\bibinfo {year}
  {1992})}\BibitemShut {NoStop}%
\bibitem [{\citenamefont {Baryshevsky}\ and\ \citenamefont
  {Gurinovich}(2005)}]{Baryshevsky2005}%
  \BibitemOpen
  \bibfield  {author} {\bibinfo {author} {\bibfnamefont {V.~G.}\ \bibnamefont
  {Baryshevsky}}\ and\ \bibinfo {author} {\bibfnamefont {A.~A.}\ \bibnamefont
  {Gurinovich}},\ }\href {https://arxiv.org/abs/hep-ph/0506135} {\bibinfo
  {title} {Spin rotation and birefringence effect for a particle in a high
  energy storage ring and measurement of the real part of the coherent elastic
  zero-angle scattering amplitude, electric and magnetic polarizabilities}}
  (\bibinfo {year} {2005}),\ \Eprint {https://arxiv.org/abs/hep-ph/0506135}
  {arXiv:hep-ph/0506135 [hep-ph]} \BibitemShut {NoStop}%
\bibitem [{\citenamefont {Silenko}(2009)}]{Silenko:2009zq}%
  \BibitemOpen
  \bibfield  {author} {\bibinfo {author} {\bibfnamefont {A.~J.}\ \bibnamefont
  {Silenko}},\ }\href {https://doi.org/10.1103/PhysRevC.80.044315} {\bibfield
  {journal} {\bibinfo  {journal} {Phys. Rev. C}\ }\textbf {\bibinfo {volume}
  {80}},\ \bibinfo {pages} {044315} (\bibinfo {year} {2009})},\ \Eprint
  {https://arxiv.org/abs/0906.4274} {arXiv:0906.4274 [nucl-th]} \BibitemShut
  {NoStop}%
\bibitem [{\citenamefont {Steffens}\ and\ \citenamefont
  {Haeberli}(2003)}]{Erhard_Steffens_2003}%
  \BibitemOpen
  \bibfield  {author} {\bibinfo {author} {\bibfnamefont {E.}~\bibnamefont
  {Steffens}}\ and\ \bibinfo {author} {\bibfnamefont {W.}~\bibnamefont
  {Haeberli}},\ }\href {https://doi.org/10.1088/0034-4885/66/11/R02} {\bibfield
   {journal} {\bibinfo  {journal} {Reports on Progress in Physics}\ }\textbf
  {\bibinfo {volume} {66}},\ \bibinfo {pages} {R02} (\bibinfo {year}
  {2003})}\BibitemShut {NoStop}%
\end{thebibliography}%

\end{document}